\documentclass[cernpreprint, coverpage=false, atlasdraft=false, UKenglish, texmf, orcidlogo]{atlasdoc}
\usepackage{atlaspackage}
\usepackage[backref=false]{atlasbiblatex}

\usepackage{atlasphysics}

\usepackage{fancyvrb}
\usepackage{shortvrb}
\usepackage{cprotect}
\usepackage{textcomp}

\usepackage[export]{adjustbox}
\usepackage{multirow}
\usepackage[notextcomp]{stix}
\usepackage{tabularray}

\usepackage{lscape}

\graphicspath{{logos/}{figures/}}

\usepackage{ANA-SOFT-2022-02-PAPER-defs}

\AtlasTitle{Software and computing for Run 3 of the ATLAS experiment at the LHC}

\AtlasAbstract{%
The ATLAS experiment has developed extensive software and distributed computing systems for Run 3
of the LHC. These systems are described in detail, including software infrastructure
and workflows, distributed data and workload management, database infrastructure,
and validation. The use of these systems to prepare the data for physics analysis
and assess its quality are described, along with the software tools used for data
analysis itself. An outlook for the development of these projects towards Run 4 is
also provided.
}

\author{The ATLAS Collaboration}

\AtlasRefCode{SOFT-2022-02}

\PreprintIdNumber{CERN-EP-2024-100}

\AtlasDate{\today}

\arXivId{2404.06335}

\AtlasJournal{EPJC}
\AtlasJournalRef{\EPJC 85 (2025) 234}
\AtlasDOI{10.1140/epjc/s10052-024-13701-w}

\hypersetup{pdftitle={ATLAS document},pdfauthor={The ATLAS Collaboration}}

\begin{document}

\maketitle

\setcounter{tocdepth}{3}
\tableofcontents

\section{Introduction}
\label{sec:intro}

ATLAS~\cite{PERF-2007-01,GENR-2019-02}, one of the general-purpose experiments at the Large Hadron Collider (LHC), has recently begun its third collision data taking campaign (\RunThr) with proton--proton and heavy-ion collisions. The data collected since its operational start in 2010 are used for a broad physics programme, from precision measurements of the Standard Model, including the discovery of the Higgs boson~\cite{HIGG-2012-27,CMS-HIG-12-028}, to searches for various manifestations of new physics. This paper describes the Software and Computing infrastructure of the ATLAS experiment, in its current state in the midst of \RunThr, along with some of the major upgrades that were done in preparation for \RunThr.

An overview of the ATLAS detector, the operation of the experiment, and the software and computing resources necessary to support it, is given in Section~\ref{sec:resources}.
The experiment relies on resources provided by the Worldwide LHC Computing Grid (WLCG)~\cite{CERN-LHCC-2005-024,CERN-LHCC-2014-014}, and has developed an extensive set of tools to capture additional resources.
The scale of these resources is laid out in Section~\ref{sec:scope}, along with a brief overview of the ATLAS operational data-taking processes and timescales.
A description of the ATLAS experiment's detector is given in Section~\ref{sec:detector}.
From the moment the data leave the detector, they undergo a series of processing steps, calibrations, and re-calibrations.
An overview of the software chain necessary to support the processing of both real and simulated particle collision data is presented in Section~\ref{sec:softwareflow}.
Following this processing the data are transferred around the world, their quality is examined carefully, and they are processed into many derived data formats in preparation for physics analysis. This process, including the chain of custody, is described in Section~\ref{sec:dataflow}.

A detailed discussion of the ATLAS software that supports data production, processing, and simulation is undertaken in Section~\ref{sec:generalCore}.
The ATLAS software has evolved considerably since collision data were first recorded in 2009, growing to match the complexity of the data analysis programme. The software infrastructure, described in Section~\ref{sec:core}, has undergone major changes during this period, most recently to support multithreading. The software also includes a revised configuration layer, detailed in Section~\ref{sec:core:configuration}, and infrastructure for the handling of detector conditions information, which is described in Section~\ref{sec:core:conditions}.
The Event Data Model (EDM), described further in Section~\ref{sec:edm}, has also evolved to better support downstream data analysis.
The modelling of the data requires a detailed detector description, which is described in Section~\ref{sec:dd}.
Many modern data analyses and portions of the upstream data processing software rely on machine learning tools; the integration of and support for these in ATLAS is described in Section~\ref{sec:ml}.

An extensive Monte Carlo (MC) simulation and data processing chain was developed to support the experiment, and is described in Section ~\ref{sec:transforms}.
This chain includes `event generation', i.e.\ the modelling of the initial proton--proton, proton--ion, or ion--ion collision (described in Section~\ref{sec:evgen}), detector simulation (described in Section~\ref{sec:sim}), and `digitisation' (modelling of the detector electronics, described in Section~\ref{sec:digi}). Both the real detector data and MC simulation are passed through a common reconstruction process (hereafter referred to as \emph{the reconstruction}, described in Section~\ref{sec:reco}) and first stage of processing for analysis known as `derivation making' (described in Section~\ref{sec:derivations}).
The forward systems are treated with special workflows through many of these steps, which are summarised in Section~\ref{sec:forward}.
Section~\ref{sec:upgrade} describes the support of all of this software for future configurations of the detector, which must be maintained to provide accurate projections of physics analyses and prepare for future data-taking runs.

The data and MC simulation are extensively vetted and validated before being declared good for analysis,
and before the start of any new production campaign, as discussed in Section~\ref{sec:processing}.
The assessment of data quality is described in Section~\ref{sec:DQ}.
The preparation for both new campaigns of MC simulation production at the beginning of a data-taking period,
and the large-scale reprocessing of data at the end of a data-taking period,
are complex processes that involve many different steps and require significant time.
The software validation process for these campaigns is explained in Section~\ref{sec:validation},
and the computing performance of the various steps of the \RunThr software chain is detailed in Section~\ref{sec:perf}.
The steps required to begin a new MC simulation or data reprocessing campaign are laid out in Section~\ref{sec:campaigns}.

Section~\ref{sec:geninfrastructure} presents the significant technical infrastructure required for the experiment to efficiently develop and deploy software and operate at scale. The infrastructure for development, building, and deployment of the complex software stack is described in Section~\ref{sec:infrastructure}. The database infrastructure, handling calibrations and evolving conditions information, is described in Section~\ref{sec:databases}. There are several systems for handling of metadata at various stages of processing; these are described in Section~\ref{sec:metadata}.

The ATLAS distributed computing systems are described in Section~\ref{sec:distcomp}.
The management of the MC simulation and real detector data, as well as all subsequent data forms and formats, is described in Section~\ref{sec:ddm}.
Section~\ref{sec:prodsys} describes the workflow management system that was developed to orchestrate the running of ATLAS software around the world, including the use of resources beyond those provided by the WLCG.
These are all subject to extensive monitoring to ensure their efficient use; the monitoring and analytics applied are detailed in Section~\ref{sec:analytics}.

Provisions for downstream data analysis are discussed in Section~\ref{sec:downstream}.
After being processed into derivations, physics data are manipulated using a large suite of data analysis tools. An overview of these tools is given in Section~\ref{sec:analysis}.
Data visualisation techniques are used with both simulated and real data to validate the performance of the detector simulation, check for unusual detector conditions or reconstruction features, study interesting events, and educate the public about the physics programme of ATLAS.
Images that show physics events and processes within the ATLAS detector,
called \textit{event displays}, are prepared for these purposes.
The use of event displays and the software developed to produce them are introduced in Section~\ref{sec:eventdisplay}.
Instruction and support for ATLAS members making use of all of these tools is provided through documentation and regular tutorials, which are discussed in Section~\ref{sec:training}.

While the approaches developed to date are sufficient for the challenges of \RunThr,
plans for significant upgrades to prepare for the High-Luminosity LHC (HL-LHC), scheduled to begin operations in 2029, are already underway.
Significant revisions to the software and computing itself are also in preparation. Some of these are described in Section~\ref{sec:outlook}, along with the outlook for the future of software and computing in the experiment.


%
\section{Overview and resources} %
\label{sec:resources}

This section provides an introduction to the ATLAS experiment, including operational data-taking and resource requirements, and an overview of the detector itself.
An executive summary of the software chain needed to process data collected by the detector (or generated by MC simulations)
to a point where they are in a format appropriate for downstream physics analysis is also presented, along with a high-level description of how data are collected from LHC particle collisions.

\subsection{Resource scope}
\label{sec:scope}
The ATLAS experiment comprises almost 6000 members, about 3000 of whom are authors of ATLAS scientific papers.
This section provides a brief overview of the data-taking environment for ATLAS.
The scope of the computing and software resources required to support its operation, and the needs of the collaboration at large, is introduced.

About 140 full-time equivalents (FTE) of effort, divided among about 450 people, is spent on `central'
software development and distributed computing support,\footnote{In addition to central software development and computing support, further efforts are needed to maintain and develop detector-sub-system-specific software, or software to support dedicated physics analysis areas.}
and a further 150 FTEs of effort is spent on maintaining the around 100 WLCG computing centres constituting the distributed computing infrastructure supporting ATLAS.
A small fraction of that effort, around 10--20 FTEs, includes real-time \emph{shift} work for monitoring resources, tests, and software updates.

\subsubsection{ATLAS operational overview}
\label{sec:atlasOper}
In 2022, the LHC began a third prolonged period of data-taking, called \RunThr, following on from \RunOne (2009--2012) and \RunTwo (2015--2018).
The terms Run 1/2/3 refer to the multi-year operational data collection campaigns of the ATLAS experiment.
Each \emph{Run} is followed by a \emph{Long Shutdown} (LS) during which significant upgrades and repairs to the detector can be made; the most recent of these was Long Shutdown~2 (LS2, 2019--2021).
Each Run is divided into years, which are further subdivided into data-taking \emph{periods}, wherein detector and collider conditions are generally stable.
The upgrades of the detector are referred to in \emph{phases}, where Phase-0 indicates the upgrade during LS1, Phase-I the upgrade during LS2~\cite{ATLAS-TDR-20,ATLAS-TDR-22,ATLAS-TDR-23,ATLAS-TDR-24}, and Phase-II the upgrade during the upcoming LS3~\cite{ATLAS-TDR-25,ATLAS-TDR-26,ATLAS-TDR-27,ATLAS-TDR-28,ATLAS-TDR-29,ATLAS-TDR-30,ATLAS-TDR-31}.
This last upgrade will be significant, in preparation for the high-luminosity LHC (HL-LHC) era wherein the instantaneous luminosity of the accelerator will be increased roughly 3-fold.

A period during which beams are circulating in the LHC ring uninterrupted is called a \emph{fill},
and ATLAS will typically record collision data continuously during an LHC fill in an ATLAS \emph{run}.\footnote{This overlap of terminology is unfortunate but ubiquitous at the LHC.
Throughout this paper, Run 1/2/3 refers to the multi-year running time, and `run' refers to a single, several-hour data taking period when beams are continuously circulating in the LHC.}
A run is divided into individual \emph{luminosity blocks}, or lumi blocks, which are roughly one minute long and represent the smallest unit of stable conditions for data analysis.

During proton--proton collision data taking, protons collide in ATLAS about two billion times per second.
The protons in the beams are grouped into bunches, and the collisions occur during \emph{bunch crossings}, which happen about 30 million times per second.
Each bunch crossing results in a physics \emph{event}, and each crossing is labelled with a unique \emph{bunch crossing identifier} (BCID).
A single proton--proton collision from a given bunch crossing is chosen as the collision of interest.
Other proton--proton interactions from within the same bunch crossing, and from neighbouring bunch crossings, can present a \emph{background} to the chosen collision.
These other proton--proton interactions are collectively referred to as \emph{pile-up}, and the treatment of pile-up in the processing of data and MC is discussed in Section~\ref{sec:transforms}.
The average number of pile-up collisions in a BCID is denoted by $\langle\mu\rangle$.
Of these 30 million events per second, about 3500 complete events are written out in full by ATLAS, selected by a multi-stage trigger system~\cite{ATLAS-TRIG-2022-03}.
Small parts of many more events are recorded as well for calibration purposes and specialised data analyses.

During \RunTwo, ATLAS recorded about 18~billion complete events.
Generally, 2--3 times more simulated events are produced than real detector events are recorded.
This volume of MC simulation is required to keep statistical uncertainties in predictions small, and to provide capacity for simulation of new physics signals.

\subsubsection{Computing resources}
\label{sec:computing_resources}

ATLAS distributed computing resources comprise about one million cores of computing, 350~PB of disk, and almost 450~PB of tape storage. The computing is distributed over a variety of sites:

\begin{enumerate}
\item The CERN \emph{Tier-0} site, where the initial data processing (known as \emph{prompt} data processing) is done, and where software releases are built and tested every night.
\item WLCG sites, which include 11 ATLAS Tier-1 sites and around 70 ATLAS Tier-2 sites. The distinction between the two is described in Ref.~\cite{WLCGMoU}; generally speaking, Tier-1 sites are larger and include archival tape storage.
\item A variety of high-performance computing centres, some of which are provided via the standard WLCG pledge mechanism (i.e. they are accounted for as a part of the Tier-1 and Tier-2 resources), and some of which come independent of the WLCG (see Section~\ref{sec:prodsys:hpc}).
\item The high-level trigger processing farm, located near the detector to minimize data transfer latency, which can be used for offline data processing when the experiment is not taking data (see Section~\ref{sec:prodsys:simP1}).
\item BOINC~\cite{DBLP:journals/corr/abs-1903-01699,boinc1,boinc2} resources from volunteer computing provided by ATLAS@Home (see Section~\ref{sec:prodsys:boinc}).
\item Commercial cloud resources, integrated into the experiment like standard WLCG sites where possible (see Section~\ref{subsec:ddm:rad}).
\end{enumerate}

These distributed computing resources are used to process all of the real and simulated data produced by the ATLAS experiment. The resources pledged by the WLCG sites are tracked in the Computing Resource Information Catalogue (CRIC)~\cite{cric}. Generally speaking, all of the disk and tape used for long-term storage is provided as pledge, and about half of the CPU used by the experiment is provided as a part of pledge. Because a significant portion of the CPU is unpledged, there are large variations ($+300k/-100k$ cores from the average) in the amount of CPU available at any given time.

\subsubsection{Software resources}
\label{sec:sw_resources}

An extensive software suite~\cite{ATL-SOFT-PUB-2021-001} is used in the reconstruction and analysis of real
and simulated data, in detector operations, and in the trigger and data acquisition systems of the experiment.\footnote{Ref.~\cite{ATL-SOFT-PUB-2021-001} and the references therein also describe some of the firmware that is used in the ATLAS data acquisition system; that software is not described further here.}
The terms \textit{online} and \textit{offline} are often used to distinguish between software and services that are run during real-time data-taking for the trigger and data acquisition systems (\textit{online software}), and those that are used for subsequent data processing (\textit{offline software}).

The overall offline software framework, Athena~\cite{athena}, is described in Section~\ref{sec:core}.
The collaboration's software is all open source, published in a publicly accessible GitLab repository~\cite{athenaGitlab}.
It is available under the \textsc{Apache} 2.0 License~\cite{apacheLicence}, with Copyright held by CERN on behalf of the collaboration.
Every ATLAS Collaboration member can contribute to the software through \emph{merge requests} to the active branches of the software repository.
These merge requests are reviewed by a team of review shifters and experts,
to ensure that only high--quality code that conforms with ATLAS conventions and passes a comprehensive spectrum of Continuous Integration (CI) tests enters the repository.
More detail on ATLAS software development and release processes can be found in Section~\ref{sec:infrastructure}.


\subsection{ATLAS detector}
\label{sec:detector}

\newcommand{\AtlasCoordFootnote}{%
ATLAS uses a right-handed coordinate system with its origin at the nominal interaction point (IP)
in the centre of the detector and the \(z\)-axis along the beam pipe.
The \(x\)-axis points from the IP to the centre of the LHC ring,
and the \(y\)-axis points upwards.
Cylindrical coordinates \((r,\phi)\) are used in the transverse plane,
\(\phi\) being the azimuthal angle around the \(z\)-axis.
The pseudorapidity is defined in terms of the polar angle \(\theta\) as \(\eta = -\ln \tan(\theta/2)\) and is equal to the rapidity
$ y = \frac{1}{2} \ln \left( \frac{E + p_z c}{E - p_z c} \right) $ in the relativistic limit.
Angular distance is measured in units of \(\Delta R \equiv \sqrt{(\Delta\eta)^{2} + (\Delta\phi)^{2}}\) for geometric quantities,
and in units of \(\Delta R_y \equiv \sqrt{(\Delta y)^{2} + (\Delta\phi)^{2}}\), where $y$ is the rapidity, for physical quantities like momenta.}

The ATLAS detector~\cite{PERF-2007-01,GENR-2019-02} at the LHC covers nearly the entire solid angle around the collision point.\footnote{\AtlasCoordFootnote}
It is located at Point~1 of the LHC; `Point~1' therefore appears in the name of several software projects.
The ATLAS detector consists of an inner tracking detector surrounded by a thin superconducting solenoid, electromagnetic and hadron calorimeters,
and a muon spectrometer incorporating three large superconducting air-core toroidal magnets.

The inner-detector system (ID) is immersed in a \SI{2}{\tesla} axial magnetic field
and provides charged-particle tracking within \(\abseta = 2.5\).
This system provides essential information for the reconstruction of \emph{physics objects} such as electrons and photons~\cite{EGAM-2021-01}, muons~\cite{MUON-2022-01},
$\tau$-leptons~\cite{ATLAS-CONF-2017-029}, and jets~\cite{JETM-2018-05},
as well as for identification of jets containing b-hadrons~\cite{FTAG-2018-01}, and for event-level quantities that use charged-particle tracks as input.
The high-granularity silicon pixel detector covers the interaction region and typically provides four measurements (often called `hits') per track,
the first hit normally being in the insertable B-layer (IBL) installed before \RunTwo~\cite{ATLAS-TDR-19,PIX-2018-001}.
It is followed by the silicon microstrip tracker (SCT), which usually provides eight measurements per track.
These silicon detectors are complemented by the transition radiation tracker (TRT),
which enables radially extended track reconstruction up to \(\abseta = 2.0\).
The TRT also provides electron identification information
based on the fraction of hits (typically 30 in total) above a higher energy-deposit threshold corresponding to transition radiation, which is absorbed by the gas mixture filling the TRT straws.
During \RunOne, several leaks in TRT active-gas exhaust pipes developed.
With the number of leaks expected to increase with higher luminosity operation, continuing operation with the baseline xenon-based gas mixture in the TRT became unaffordable.
Leaking modules were updated to operate with an argon-based gas mixture for \RunTwo,
and for \RunThr\ the new argon-based gas mixture is used for the entire barrel and parts of the endcaps.
While particle identification performance is largely preserved in the endcap regions,
it is significantly reduced in the barrel region due to poor absorption of transition radiation photons by the argon gas.

The calorimeter system covers the pseudorapidity range \(\abseta < 4.9\).
Within the region \(\abseta< 3.2\), electromagnetic calorimetry is provided by barrel and
endcap high-granularity lead/liquid-argon (LAr) calorimeters,
with an additional thin LAr presampler covering \(\abseta < 1.8\)
to correct for energy loss in material upstream of the calorimeters.
Hadronic calorimetry is provided by the steel/scintillator-tile calorimeter,
segmented into three barrel structures within \(\abseta = 1.7\), and two copper/LAr hadron endcap calorimeters.
The solid angle coverage is completed with forward copper/LAr and tungsten/LAr calorimeter modules
optimised for electromagnetic and hadronic energy measurements respectively.
The calorimeters play an important role in the reconstruction of physics objects such as photons,
electrons, $\tau$-leptons and jets, as well as event-level quantities such as missing transverse momentum (with magnitude denoted by \met)~\cite{JETM-2020-03}.

The muon spectrometer (MS) comprises separate trigger and
high-precision tracking chambers measuring the deflection of muons in a magnetic field generated by the superconducting air-core toroidal magnets.
The field integral of the toroids ranges between \num{2.0} and \SI{6.0}{\tesla\metre}
across most of the detector.
Three layers of precision chambers, each consisting of layers of monitored drift tubes (MDTs), cover the region \(\abseta < 2.7\),
except in the innermost layer of the endcap region, where the background is highest and layers of small-strip thin-gap chambers and Micromegas chambers both provide precision tracking in the region \(1.3 < \abseta < 2.7\).
The muon trigger system covers \(\abseta < 2.4\) with resistive-plate chambers (RPCs) in the barrel, and thin-gap chambers (TGCs) in the endcap regions, and small-strip thin-gap chambers and Micromegas chambers in the innermost layer of the endcap.
During \RunTwo the innermost endcap stations, covering \(1.3 < \abseta < 1.8\), were cathode-strip chambers (CSCs).
These were replaced by new small wheels (NSW)~\cite{ATLAS-TDR-20} during LS2,
to improve tracking efficiency and resolution in the high rate environment of \RunThr.
The ATLAS software supports the simulation and reconstruction of data from both \RunTwo and \RunThr, and therefore both the detector technologies are available.

Four forward detector systems are installed around the interaction point.
The luminosity is measured mainly by the LUCID--2 detector that records Cherenkov light produced in the quartz windows of photomultipliers located close to the beampipe.
The Zero-Degree Calorimeter (ZDC), located about 140~m from the interaction point, measures neutral particles.
It is used primarily during heavy ion data taking, both in the trigger and offline event selection.
The ATLAS Forward Proton (AFP) detector, located 210~m from the interaction point, is used primarily to study diffractive physics at low instantaneous luminosity.
The Absolute Luminosity For ATLAS (ALFA) forward proton spectrometer comprises Roman pot detectors placed about 240~m from the interaction point on both the sides of the detector.
It is used both for luminosity information and to measure the total proton--proton scattering cross section.

Events are selected by the first-level trigger system implemented in custom hardware,
followed by selections made by algorithms implemented in software in the high-level trigger (HLT)~\cite{TRIG-2016-01,ATLAS-TRIG-2022-03}.
The hardware trigger accepts events from the \SI{40}{\MHz} bunch crossings at a rate below \SI{100}{\kHz},
which the high-level trigger further reduces to record complete events to disk at an average rate of about \SI{3}{\kHz}.

The \RunThr detector configuration benefits from several upgrades compared with that of \RunTwo to maintain high detector performance at the higher pile-up levels of \RunThr. The improvements from the NSW provide higher redundancy and a large reduction in fake muon triggers. The trigger system also benefits from new LAr digital electronics with significantly increased granularity. Other updates and further details are provided in Ref.~\cite{GENR-2019-02}.


\subsection{Software workflow}
\label{sec:softwareflow}

\begin{figure}[h]
\centering
\includegraphics[width=0.98\textwidth]{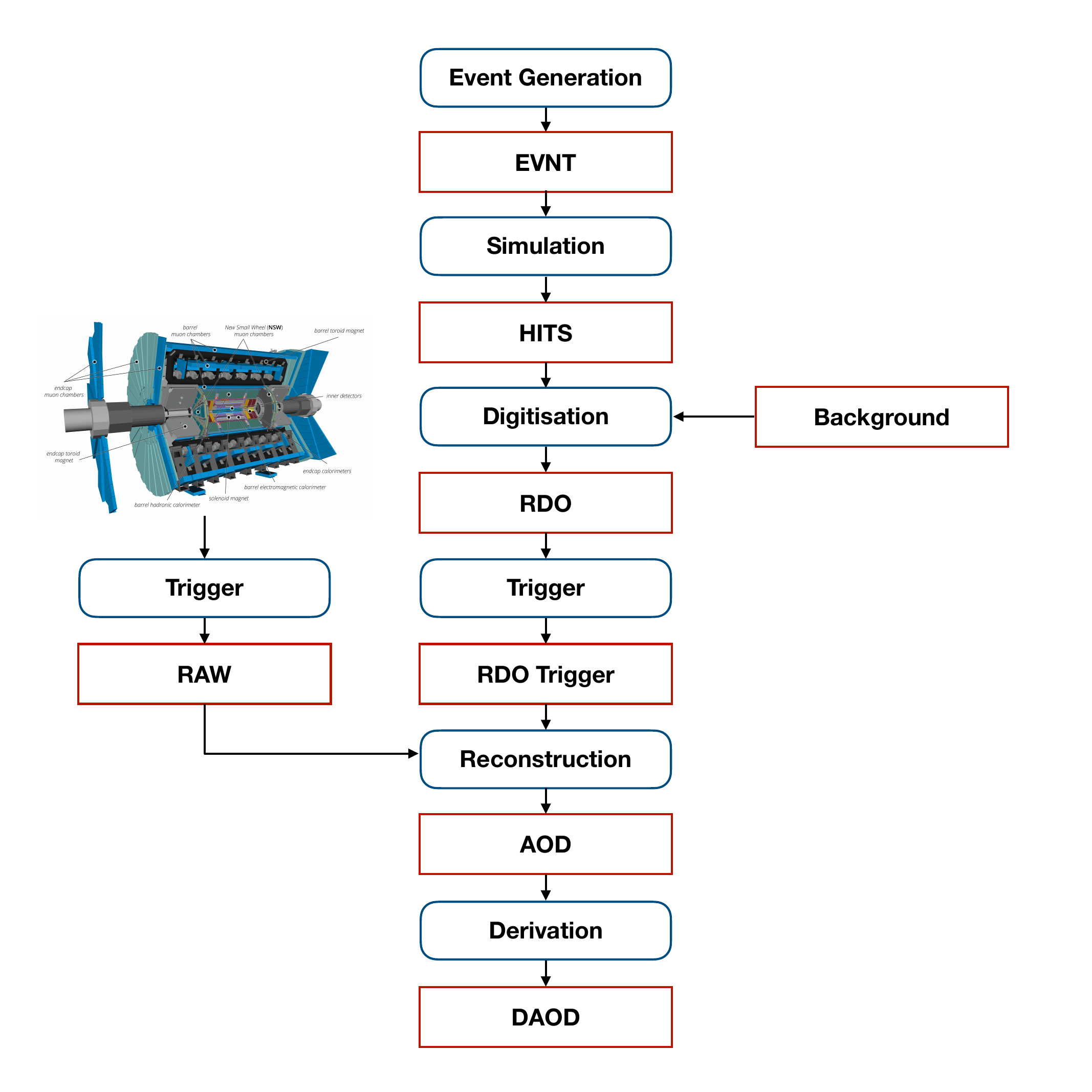}
\caption{The standard software workflow of the ATLAS experiment. Processing steps are represented by blue ovals, with output formats represented as red boxes. The various steps and data formats are described in the text. The background entering digitisation may be additional simulated HITS files, pre-digitised RDO files, or specially processed RAW detector data.}
\label{fig:swflow:flowchart}
\end{figure}

The software workflows for the MC simulation production and the main stages of the data processing are depicted in Figure~\ref{fig:swflow:flowchart}. The workflow for data involves other steps that are discussed in Section~\ref{sec:dataflow}. A detailed description of the data and MC simulation production chain is then given in Section~\ref{sec:transforms}.

The first stage of MC simulation production is called \emph{event generation}. Here, an event generator configuration is run to create HepMC output~\cite{DOBBS200141}. This is written to an output EVNT file, containing the particles output by the event generator together with metadata expressing the configuration of the generator job. The process of event generation is described in Section~\ref{sec:evgen}.

Following event generation, the resulting particles are passed through a detector \emph{simulation}~\cite{Agostinelli:2002hh,Geant42,Geant43,SOFT-2010-01,SIMU-2018-04}. The output from this processing step is stored as a HITS file.\footnote{In this paper, \emph{HIT} will refer to an energy deposition during detector simulation, which is written to a HITS file, while \emph{hit} will refer to a position measurement along a charged particle trajectory or other physical detector measurement.}
The HITS output contains records of energy deposits from each particle, with associated timing information, for each sub-system;
\emph{truth} information about the simulated behaviour of particles during the event, appended to the generator particle record;
and additional metadata concerning the configuration of the simulation job. The process of detector simulation is described in Section~\ref{sec:sim}.

The simulated energy deposits are then run through a \emph{digitisation} step, in which the detector electronics are modelled,
resulting in RDO (Raw Data Object) output, which is conceptually similar to the raw data collected from the detector.
During this stage, pile-up is modelled by the addition of background events, using one of several technical options:
the overlay of many simulated events, a single pre-digitised collection of simulated events, or a specially recorded data event.
The RDO output files also pass along the truth records from the simulation and include further records of the correspondence between true particles and detector signals,
as well as additional metadata. This digitisation process is detailed in Section~\ref{sec:digi}.

The RDO file is then passed through a \emph{trigger simulation} stage, where a \emph{menu} of triggers specific to the MC simulation process is simulated and decisions and key trigger object collections are added to the output.
Both the hardware and high-level triggers are reproduced.
This step may be performed using an older software release to appropriately simulate a trigger that was run online some years ago and is no longer supported in the newest releases.
The output of this step is written to an \emph{RDO Trigger} file, which is an RDO with trigger information and metadata added to the file.
The trigger is described in significantly more detail in Ref.~\cite{ATLAS-TRIG-2022-03}.

The data coming from the detector, called RAW, are written in a custom bytestream format~\cite{ATLAS-TDR-23}. These data are most often processed directly.
They can also be filtered to create derived RAW (DRAW) datasets for processing with special settings (e.g.\ lower momentum thresholds or modified reconstruction configurations for special analyses).

From this point on, the workflows for data and MC simulation are the same, with the next step being \emph{reconstruction}. The detector signals are converted into physics objects (electron candidates, muon candidates, and so on). The output is written to an AOD (Analysis Object Data) file; in special cases, a larger ESD (Event Summary Data) file or other custom format might be written. The reconstruction process is described in Section~\ref{sec:reco}.

The next stage after this main reconstruction step is the \emph{derivation} production. This is primarily a data-reduction operation, but it also includes the reconstruction of some secondary physics objects for which only the inputs were reconstructed and stored in the AOD. For example, jets are built during derivation-making from particle flow objects stored in the AOD; similarly, heavy flavour tagging is performed based on these jets and tracks stored in the AOD (see Section~\ref{sec:reco} for more about these reconstruction steps and objects). The step also achieves data reduction by the calculation of variables that consolidate or summarise several inputs. Many different derivations might be written out; these all share a common file format and are called DAODs, but they differ in event selection and amount of information written for each physics object. Derivations can also be made directly from the EVNT files if generator output is being analysed. A single derivation format might serve one or several analyses, or might be used for calibration purposes. Derivations are discussed in Section~\ref{sec:derivations}.

Depending on the number of events in the output files and to provide large, uniform files better suited to large storage systems, a \emph{merge} step is run after some processing steps.
This merging is typically optimised within the production system to provide the desired number of events per output file.

Generally speaking, these steps are all performed centrally using WLCG resources, called \emph{the Grid}, and any subsequent steps of an analysis are under the control of the individual analyser or data analysis team.
Teams might choose to create further-reduced data formats on the Grid, or to transfer their derivations to a local processing centre for reduction and further analysis.
These later steps are described in Section~\ref{sec:downstream}.


\subsection{Data flow}
\label{sec:dataflow}

The flow of data from the ATLAS detector for prompt reconstruction,
and the procedures employed to ensure these are data of sufficient quality for physics analysis,
are described in the context of \RunTwo in Ref.~\cite{DAPR-2018-01}. The ATLAS data flow in \RunThr is similar.
Most data selected by the trigger are promptly reconstructed at the ATLAS Tier-0 computing facility,
before being distributed to the Grid for production of derivations (as described in Section~\ref{sec:derivations}) and further analysis.
Data reprocessing, which involves repeating the data reconstruction process completely (for reasons discussed in Section~\ref{sec:campaigns-data}),
normally takes place directly on the Grid rather than at the Tier-0 site to take advantage of the larger pool of available resources.

The data are organised into \emph{streams}, each stream being a set of defined trigger selections and prescales (fractions of events to accept for a given trigger) that contain all events recorded to disk after satisfying the selections.

\begin{itemize}
\item \emph{Physics} streams contain data that are potentially interesting for physics analysis;
\item \emph{Calibration} streams exist for particular calibration purposes;
\item the \emph{Express} stream is a special stream that contains a representative subset of the physics streams;
\item \emph{Delayed} streams are a special class of physics streams that contain data that may be processed at a later time (e.g.\ at the end of the year), rather than promptly;
\item the \emph{Debug} stream includes a small fraction (typically of order $10^-7$) of events that are flagged for further investigation, for example because the trigger processing failed or timed out; and
\item \emph{TLA} (Trigger-Level Analysis) streams include events for which only a small subset of the event data (e.g.\ only the jets identified by the trigger) is preserved. This technique allows events to be written at a high rate without consuming significant resources.
\end{itemize}

Before processing, all these data are stored in the RAW format.
To provide a backup, each run's RAW data are stored at the Tier-0 site and at one Tier-1 site.

\begin{figure}[h]
\centering
\includegraphics[width=0.98\textwidth]{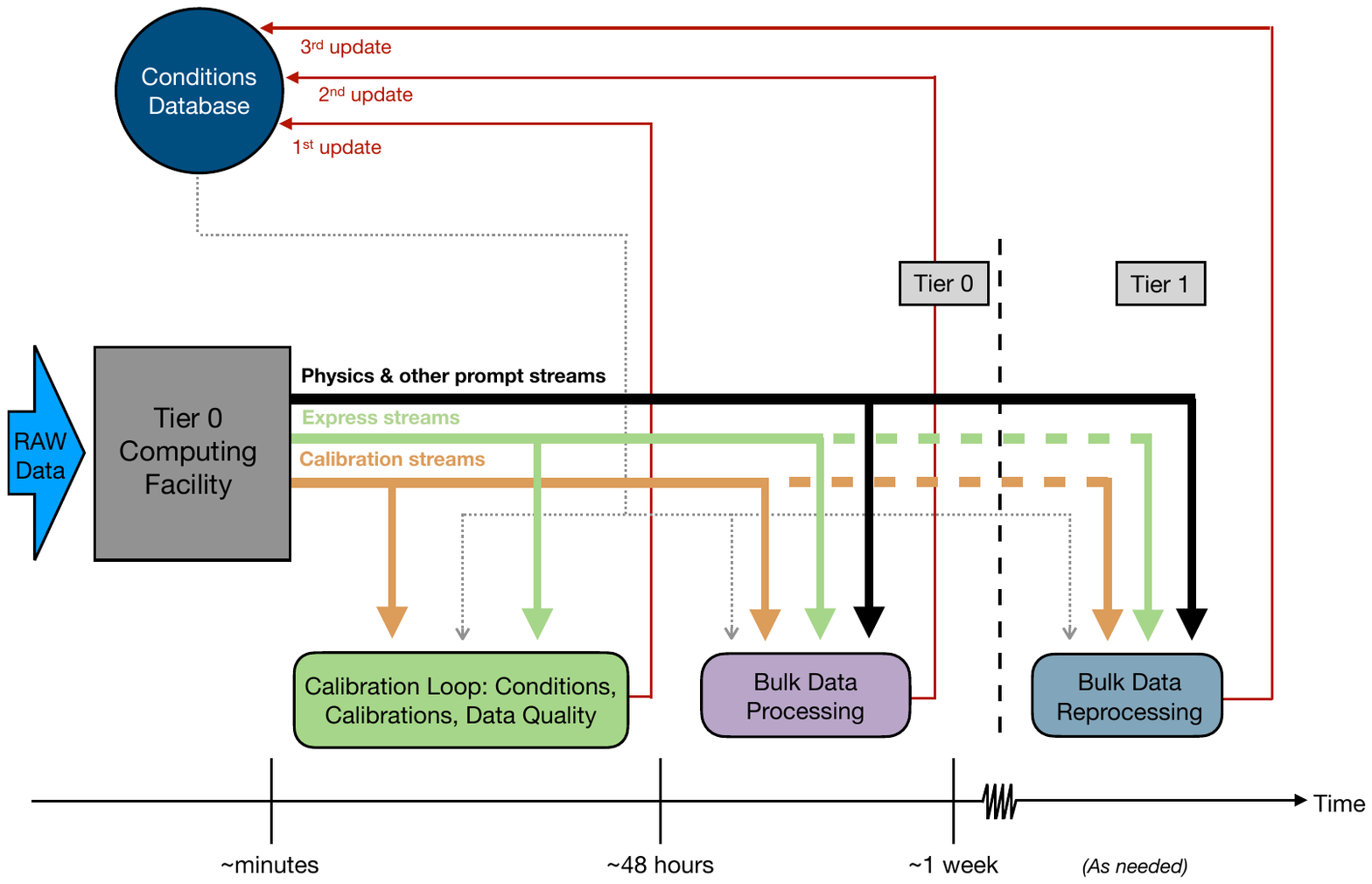}
\caption{Schematic diagram illustrating the nominal flow of ATLAS data from initial reconstruction of a subset of streams during the calibration loop and the bulk processing of promptly reconstructed streams at the Tier-0 site, to future bulk reprocessing of data at Tier-1 Grid sites.}
\label{fig:dataflow}
\end{figure}

The nominal flow of data from through the reconstruction chain is illustrated in Figure~\ref{fig:dataflow}.
Before the relatively large physics streams are processed, a \emph{calibration loop} (which typically lasts for 48 hours) begins.
During this time the express stream and several calibration streams are processed at the Tier-0 site.
The data from this rapid processing are used to update calibration constants and detector conditions in the conditions database (see Section~\ref{subsec:databases:conditions}) for items such as the alignment of the ATLAS ID system, the measured LHC beam spot position at ATLAS, and channel calibration, electronic noise, data corruption or disabled modules within detector sub-system components.
These data are additionally used for operational data quality monitoring, as discussed in Section~\ref{sec:DQ}.
The processes applied in the calibration loop are detailed in Ref.~\cite{DAPR-2018-01}.
Once the calibration loop is completed, this improved conditions information is then applied in the prompt processing of the physics streams.
Further conditions updates can be made following the bulk data processing, making use of the entire promptly processed data.
Generally, high precision data analyses operate exclusively on reprocessed data, while some preliminary or lower-precision analyses might use data immediately after the first processing.

%
%
%
%


%
\section{Core software components}
\label{sec:generalCore}

The ATLAS software framework supports data production and processing, MC generation and simulation, and downstream analysis of the ATLAS detector data.
The codebase is divided into approximately 2000 packages, within which groups place code with a common functionality or aim.
These packages are gathered into several partially-overlapping subsets called \emph{Projects}, which can be compiled together.
The broadest selection form the basis for Athena, the general-purpose offline software project.
Other projects contain more limited sets of packages and support specific use-cases such as Simulation, as discussed further in Section~\ref{subsec:genInfra:cmake}.
ATLAS software has external software dependencies on roughly 200 high--energy physics, data science, and general Linux software packages.
In addition, the offline software depends on approximately 200 packages from the ATLAS online detector software system (TDAQ common; see Section~\ref{subsec:genInfra:cmake}).

The Athena code, consisting of over 50,000 unique files, is mostly C++, configured using \textsc{Python} and built using \textsc{CMake}~\cite{cmake} (see Section~\ref{subsec:genInfra:cmake}).
The full breakdown of the code base by language from a snapshot in time of the repository (which is continuously evolving) is shown in Table~\ref{table:cloc}~\cite{cloc}.
Most of the custom configuration code listed in the table is for data quality monitoring.
The XML files hold a mixture of configuration information (e.g.\ for the trigger system), data (e.g.\ for the muon geometry), and configuration for dictionaries (class descriptions for I/O) in \textsc{ROOT}~\cite{ROOT,Antcheva:2009zz}.

\begin{table}[!htb]
\centering
\caption{Number of files, comment lines, and code lines in the Athena software repository, divided by programming language. This represents a specific snapshot of the repository, which is continuously evolving.}
\label{table:cloc}
\begin{tabular}{lrrrr}
\toprule
Language & Files & Comment & Code \\
\midrule
C++                                  & 17,273     & 457,373   &   2,608,231 \\
\textsc{Python}                      &  9,478     & 211,655   &   1,009,088 \\
C/C++ Header                         & 20,475     & 469,490   &     843,679 \\
Custom Configuration                 &   307      &       0   &     368,828 \\
XML                                  &   954      &  12,800   &     204,169 \\
Shell                                & 1,243      &  12,283   &      48,782 \\
\textsc{CMake} / make                & 2,070      &  11,021   &      35,751 \\
Fortran                              &   166      &   7,674   &      24,024 \\
Web (HTML, CSS, PHP)                 &    44      &     289   &       7,085 \\
CUDA                                 &    28      &     648   &       5,445 \\
Other                                &   171      &   3,235   &      24,027 \\
\midrule
Total                                & 52,191     & 1,186,288 &   5,178,472 \\
\bottomrule
\end{tabular}
\end{table}

The core infrastructure of the software is presented in Section~\ref{sec:core}, and the configuration infrastructure is described in Section~\ref{sec:core:configuration}.
During execution, the software often needs access to conditions data (e.g.\ alignments and calibrations), which is discussed in Section~\ref{sec:core:conditions}.
The data processed by this software must be represented in a way that ensures common access interfaces, internal consistency and long-term maintainability.
The ATLAS \emph{event data model} was developed for this purpose, and is presented in Section~\ref{sec:edm}.
To ensure an accurate description of the ATLAS detector is available,
a detector description system was developed, as described in Section~\ref{sec:dd}.
Machine learning tools are increasingly being used at various stages of the data processing chain.
A summary of those most commonly in use within the ATLAS Collaboration is given in Section~\ref{sec:ml}.

\subsection{Core software}
\label{sec:core}

\begin{figure}[tb]
\centering
\includegraphics[width=12cm,height=5.8cm,clip]{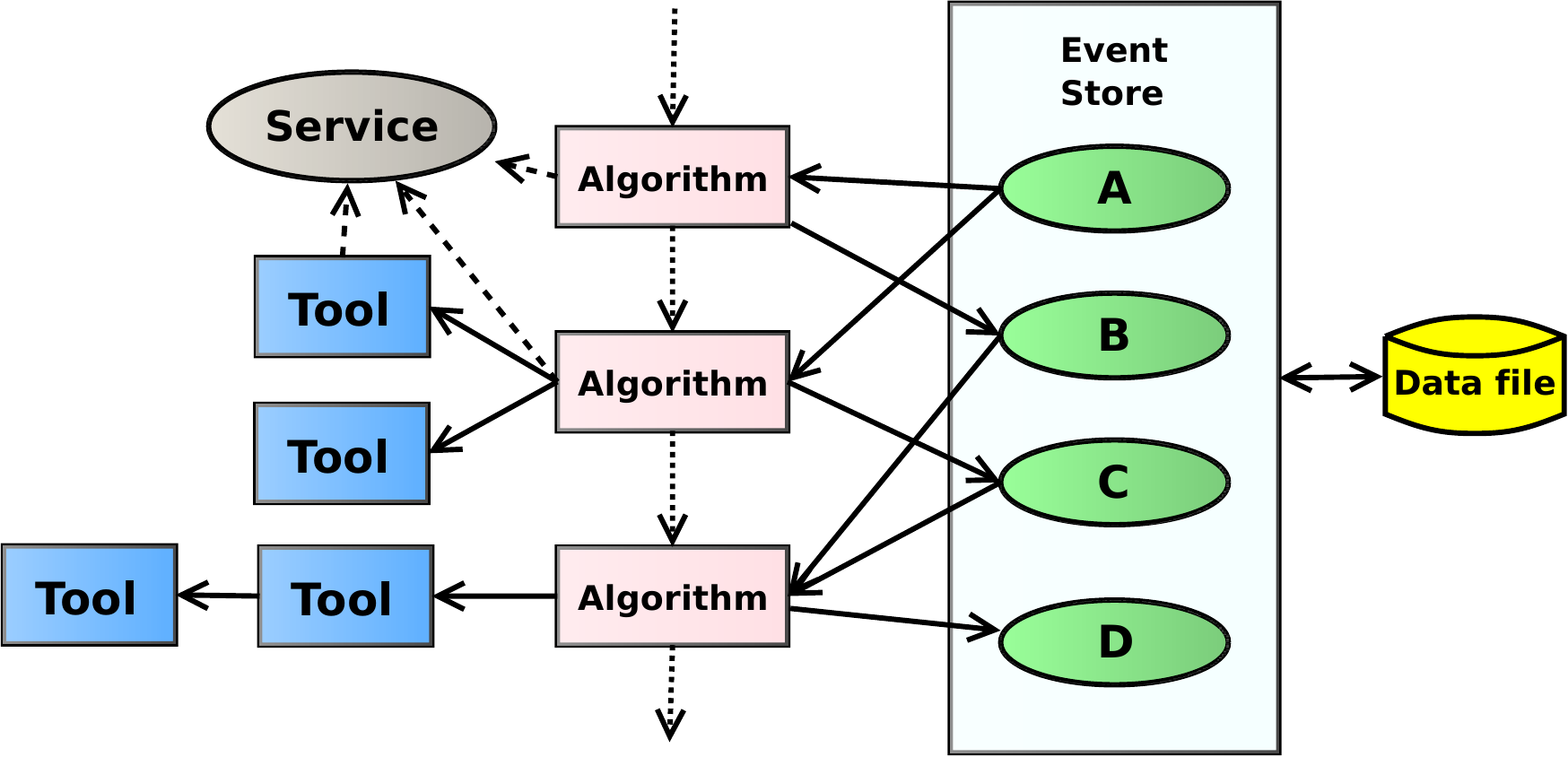}
\caption{General structure of an Athena application.
The dotted line indicates the application control flow.
The solid lines on the right indicate data flow; on the left they indicate ownership.
Dashed lines indicate a non-owning reference between components.}
\label{fig:framework}       %
\end{figure}

Athena is based on the Gaudi
project~\cite{GAUDI}. Gaudi itself is developed jointly with LHCb and other
experiments.
An Athena application consists of dynamically-loadable \emph{components},
which implement the concepts of Algorithms, Services, and Tools;
see \cref{fig:framework}.
Algorithms process data
which reside in a shared \emph{event store}; they read objects (identified by type
and a string key) from the store and write new objects back to the store.
Ideally, an Algorithm itself does not contain any event data
and does not communicate with any other Algorithm except via the event store
(exceptions are mostly special-purpose algorithms that have not been
fully migrated to multithreading).
Services are objects used by multiple other components; examples include the
event store itself,
error logging, and random number generation.  Tools serve as helpers
for other components.  They may be uniquely owned by Algorithms, Services, or other Tools.
All three component types may declare \emph{Properties} to be initialized
during job configuration in a uniform manner (see also Section~\ref{sec:core:configuration}).

The framework may run in one of four modes:

\begin{itemize}
\item A serial mode (Athena), in which one event is processed at a time
and the order in which Algorithms run is fixed during job configuration;
\item A multi-process mode (AthenaMP), in which several worker jobs are forked after
initialization is completed, and memory is shared among the workers as long
as it is not modified, significantly reducing the total memory footprint
of the job while achieving throughput very similar to multiple serial jobs~\cite{ATL-SOFT-PROC-2015-020};
\item A multithreaded mode (AthenaMT), in which multiple
events may be processed concurrently and Algorithms may be executed in parallel
in separate threads. In this latter mode, there is a separate instance
of the event store or \emph{slot} for each event that may be processed
concurrently.  Parallelism may be both \emph{intra-event}, in which Algorithms
processing data from the same event run in parallel, and \emph{inter-event},
in which Algorithms process data from different events in parallel. This significantly
reduces memory requirements beyond what is possible with AthenaMP, while providing
comparable throughput for computationally-intensive workloads~\cite{ATL-SOFT-PUB-2021-002}; or
\item A hybrid multi-process/multithreaded mode, which is used in the HLT in
particular to maximize throughput for the memory capacity of the HLT farm
configuration~\cite{ATLAS-TRIG-2022-03}. The choice of the number of threads
per worker process is configurable.
\end{itemize}

For context, over a recent six-month period, about 18\% of the CPU on the Grid
was consumed by serial (single-core) jobs, of which 40\% was user jobs and 50\%
was event generation. About 12\% of CPU was consumed by multi-process jobs, and
the remaining 70\% was consumed by multi-threaded jobs. On the Grid during the
same period, about 60\% of job slots were 8-core, 13\% were 64-core, 18\% were
single-core, and the remainder were of various other breadths; the
number of cores per slot is largely determined by the site administrators.

Each Algorithm declares the set of data objects it reads from and writes to the
event store, allowing the construction of
a dependency graph for Algorithms during the initialization of the application.
In addition to this `data flow' information,
`control flow' rules may be used to explicitly state the sequence in which
some collection of Algorithms must run.  This can be used to implement
filtering, allowing event processing to stop early if some condition
is not satisfied.
The `Avalanche' scheduler~\cite{HiveScheduler1,HiveScheduler2} looks
for Algorithms that have all their inputs available and control flow rules satisfied
and queues them for execution, using the task facility of the
Intel Threading Building Blocks (TBB)~\cite{tbb} library.
The scheduler is designed to have a low latency and to scale well to
many threads.

The data dependencies of Algorithms and Tools are declared via
their access mechanism to the data~\cite{handledocs}.
To read or write an object of type \texttt{T} from the event store, an Algorithm
declares a Property of one of the special types \texttt{SG::ReadHandleKey<T>} or \texttt{SG::WriteHandleKey<T>}.
The value of the Property is the name of the data object in the event store.
During execution of an Algorithm, these key objects may be combined with
an \texttt{EventContext} object (see below) to form smart pointers that are used
to access the data objects.  These declarations then form the input
that the scheduler uses to build the data flow graph.  Tools may declare
such dependencies as well as Algorithms; the Tool dependencies are then propagated up
to the owning Algorithm.

To ease the adoption of multithreading, a given Algorithm instance by default cannot be
scheduled simultaneously in more than one thread.  When an Algorithm is
migrated to run multithreaded, it is preferably
declared \emph{reentrant}.  In that case, the Algorithm instance
may execute concurrently in multiple threads; accordingly, most Algorithm execute methods
are declared \texttt{const} to reduce the likelihood of multithreading-related issues.
In a typical reconstruction job, only a few percent of the
roughly 1000 scheduled algorithms are not reentrant. Some of the Algorithms
that cannot be made reentrant are declared as \emph{clonable}.
In this case, multiple instances of the Algorithm can
be made, and distinct instances may be scheduled in multiple threads.  Most
Algorithms that are declared clonable are related to detector simulation.

A specific event among the ones currently being processed is identified
via an instance of the \texttt{EventContext} class.  This contains the identifying
numbers of the event (a `run' and `event' number),
the number of the event slot used by this event,
and a direct reference to the event store implementing that slot.
When the scheduler starts execution of an Algorithm, it passes to it
the \texttt{EventContext} for the event to be processed.  The \texttt{EventContext} is
then used to access the event store.  While it is preferred to pass
the \texttt{EventContext} explicitly to functions that need it, the current
\texttt{EventContext} is also stored in a thread-local global variable to ease the integration
of older code.

Once an object is recorded in the event store, it should not be modified;
besides the possibility of data races, this can lead to circular dependencies
in the data flow graph that the scheduler cannot resolve.
However, it is common to need to remake an object already existing in an
input file from other data in the file (e.g.\ if a bug is discovered in a
reconstructed physics object that can be re-computed using other information
in the input file).  To allow for this, any objects in the
input file with names that match any declared \texttt{WriteHandleKeys} will be
ignored, rather than read.  One may also need to make a revision to an
object existing in an input file, for example to correct a problem from
an earlier version of the software.  To support this, objects being read
may be renamed.  For example, an object called `\texttt{Electrons}' may be renamed
to `\texttt{Electrons\_in}' on input.  An Algorithm can then read `\texttt{Electrons\_in}',
make a copy while applying a correction and save the copy as `\texttt{Electrons}'.
Later algorithms that use `\texttt{Electrons}' will then retrieve the correct
version without having to be modified.

Besides event data, reconstruction Algorithms may depend on \emph{conditions}
data~\cite{chep18conditions}, like calibrations, alignment, or maps of problematic
detector readout channels. These data often require some special handling, as
described in Section~\ref{sec:core:conditions}.

Persistency of data objects within the ATLAS offline software~\cite{atlas_io} uses \textsc{ROOT} I/O and was built
on top of the LCG \textsc{POOL}~\cite{USLHC:2012ewx} framework. \textsc{POOL} provided high performance
and highly scalable object serialization to self-describing, schema-evolvable,
random-access files. However, the intention of serving multiple experiments
and use-cases with different software stacks caused the project to grow and be
less efficient than desired. After the other experiments abandoned POOL, the
software was streamlined and incorporated into the ATLAS repository.
This software layer enables ATLAS I/O to support persistent navigation, by
providing tokens that contain the storage address of data objects. Given a
token for a particular data object, the infrastructure is able to directly
access that object, read it and restore its state in the transient store.
This mechanism works independent of file boundaries and even storage technologies,
in principle allowing ATLAS jobs to navigate to upstream data or data augmentations
on demand.\footnote{In practice, the complexities of navigation to another file
that might be at a different site have limited the application of this ability.} The
software framework itself does not require overall event data organization in
the transient store, but persistifies a lists of data objects managed by a
\texttt{DataHeader}. This software also controls the placement of data objects in \textsc{ROOT}
\texttt{TTrees} and \texttt{TBranches} and setting properties such as compression, managed by
the ATLAS I/O framework.

In multithreaded operation, some special considerations apply to the
I/O components~\cite{chep18io}.
Unlike other Athena Algorithms and Services, which independently process data for a single event,
the I/O and storage infrastructure handles compressed buffers containing data for many events (10--1000)
at a time.
This limits concurrency when reading and writing event data as locking is required
to assure a particular I/O buffer is accessed by a single thread only,
even as many events may be processed simultaneously.
There is one Service for reading event data, one for reading conditions
data (see Section~\ref{sec:core:conditions}), and one for writing event data.
When writing multiple streams/files, separate output services are used to gain concurrency.
In addition, the column-wise storage of ATLAS data in \textsc{ROOT}~\cite{ROOT} (see Section~\ref{sec:edm})
allows separate branches to be processed by separate input services concurrently.
All these services are each individually serialized but may run concurrently with each other.

As a result, I/O is not a bottleneck for most applications; typically I/O
is below 5\% of the total job time including compression and decompression.
Read-ahead and caching of event data is provided by the \textsc{ROOT} \texttt{TTreeCache}.
To prevent
I/O thrashing when multiple events are being processed simultaneously,
the maximum virtual \texttt{TTree} size is increased to hold trailing I/O buffers.
Additional parallelism on writing is gained by using the implicit
multithreading mode of \textsc{ROOT}.

Gaudi supports \emph{incidents}, a form of structured callback;
a component can at any time raise an incident of some type.
These incidents are handled by the Gaudi IncidentSvc that in turn
makes callbacks to components that have registered their interest
in that particular type of incident.
Incident types include starting and ending events, files, and runs.
Incidents are problematic for multithreading because
they could in principle be asynchronous relative to event processing,
and do not respect event boundaries.
However, as used in Athena, almost all incidents are raised by the event loop
due to discrete state changes.  Therefore, rather than having the
IncidentSvc make callbacks directly, they are instead made from special
Algorithms that run at the beginning and end of event processing.
Incidents are now only sent to Services, not Algorithms;
these Services may retain data separately for each active
event context.
Algorithms  may observe the effects of incidents by calling the Services
that handle them.

Random number generation can be problematic in a multithreaded environment.
Requiring locking or access to thread-local storage for every random number call
can add a considerable performance overhead, but the generator state must
somehow be protected against concurrent access.  Further, to have
reproducible results, the sequence
of random numbers must be independent of the order in which events are
processed or in which Algorithms are scheduled.
In Athena, different streams of random numbers, with separate seeds and generators,
are distinguished
by the name of the Algorithm that uses them.  For each type,
Athena maintains an array of generator states, one for each event slot.
Because no Algorithm can be executing on the same event slot in more
than one thread, the generator state retrieved from this array can be
used without further synchronization.
For improved reproducibility, the generator states are reseeded for each event,
based on the event and run numbers.

In addition to the offline reconstruction, the online software running in the ATLAS HLT
uses the same multithreaded framework~\cite{chep18trigger}.  However, a
key additional requirement for the HLT is the ability to limit reconstruction to within
a set of geometric \emph{regions of interest} (RoI) in the detector.  This is implemented
via an \emph{EventView}. It provides the same interface as the event store,
but provides only a subset of the detector data, corresponding to the RoI.
In this way, Algorithms that access event data can be transparently
restricted to the subset of event data provided by an EventView.
EventViews are created by specialized Algorithms that fill them with
region-specific data and request the scheduler to execute a sequence of Algorithms
in the context of each EventView.  At the completion of processing,
the results from all EventViews are merged and saved to the primary
event data store.
In addition, the raw detector data are managed by a special thread-safe
container such that data for different regions of the detector can be
simultaneously unpacked in different threads as is required
by the trigger Algorithms.

Although Algorithms execute independently of each other
based on their data dependencies and thus can usually
proceed without explicit synchronization, some special data structures used by the
framework and event data model can be shared between threads and thus
require some form of synchronization.  This can be done using locks,
but this may result in bottlenecks when many threads are used;
locking may also involve significant overheads even in the
absence of contention.  In some cases, there are \emph{lockless} methods
for doing synchronization without explicit locking, but these can be quite
complicated and involve their own significant overheads.
However, many of the structures of interest in the framework are
\emph{read-mostly}; that is, reads are frequent and are important for performance,
while modifications are infrequent.  In this case, one can
allow multiple lockless readers along with a single writer, which can be
serialized with a lock.  This is usually much simpler than the general
lockless case while still providing good performance for read-mostly workloads.
ATLAS uses several data structures developed using these
principles~\cite{chep19concurrent} to improve the scalability of core
components of the framework.  These include a variable-length bitmap,
a hash table, and a specialized container that maps from intervals
of validity to conditions data objects (see Section~\ref{sec:core:conditions}),
as well as helpers to manage lazy initialization of mutable class members
without requiring locking.

\subsubsection{Updates for multithreading}
\label{sec:athenaMT}

The adoption of a fully multithreaded framework, deployed for \RunThr, required
numerous changes to the code base. For example:

\begin{itemize}
\item Event and conditions data have to be accessed via handles; non-thread-safe
data caching and back-channels for communication had to be removed.
\item Conditions Algorithms must be used rather than caching derived conditions
data in Algorithms or Tools (see Section~\ref{sec:core:conditions}).
\item Algorithms that modified data objects existing in the event
store had to be redesigned.
\item  Thread-unfriendly constructs, such as non-\texttt{const} static
data and \texttt{const}-correctness violations, had to be avoided.
\item Services must be explicitly thread-safe.
\item Reentrant algorithms must avoid the use of mutable instance
data.
\item Uses of Gaudi incidents needed to be adapted to the multithreaded
scheme.
\item Normal (`private') Gaudi Tools are owned by another Algorithm, Tool,
or Service.  However, Gaudi also supports `public' tools that act as
singletons.  As these mostly overlap with the functionality of Services,
almost all public Tools in Athena were changed to either private Tools
or Services.
\end{itemize}

To assist in finding thread-unfriendly code, ATLAS uses a static code
checker~\cite{chep18athenamt}, implemented as a GCC~\cite{gcc} plugin.
As GCC is the primary compiler used by ATLAS, the plugin can be enabled
for both the central software builds as well as for individual developers.
The plugin can check for  problems such as the use of non-\texttt{const} static
data, \texttt{const}-correctness violations (including the use of mutable members),
casting away \texttt{const}, returning non-\texttt{const} pointer members from a
\texttt{const} member function, and calling non-\texttt{const} methods via a pointer
member from a \texttt{const} member function. Diagnostic messages about such
violations may be suppressed on a case-by-case basis
by the use of macros that expand to custom C++ attributes.
Code authors can tag specific
packages or source files to be checked; in addition,
a configuration file may also
be used to request checking of all files in a particular source subtree.
The checker also enforces several other ATLAS coding rules,
such as naming conventions (see also Section~\ref{sec:genInfra:quality}).

As failures in multithreaded programs can be rare and are often not consistently reproducible,
it is essential to have good diagnostics in case of application crashes~\cite{Krasznahorkay:2871750}.
On a crash, the currently-executing Algorithm(s) for each slot are printed,
and \textsc{ROOT} is used to generate a stack backtrace for each thread.
This procedure can, however, fail if the program state is corrupt.
Therefore, the handler for fatal signals first executes a `fast' stack trace
in the thread in which the error was detected.  This starts by dumping the
contents of the machine registers, which is often invaluable in understanding
the details of the crash.  This is followed by a stack backtrace that is
carefully written to avoid any dynamic memory allocation.  Addresses in this
backtrace are written both in absolute form and as an offset in the
containing shared library, allowing for easy location of the code
in a disassembly of the shared library.  For builds with gcc on
x86\_64 Linux systems, the system stack unwinder is also modified
to allow it in many cases to proceed past corrupt stack frames
(as could be caused, for example, by calling a virtual function on a deleted
C++ object).  A preallocated alternate stack is also declared for the
signal handler.  These measures ensure that a usable backtrace can be produced
most of the time, even in the event of heap or stack corruption.
Techniques used for diagnosing some of the more difficult threading-related
failures included modifying the memory allocator
(\textsc{tcmalloc}~\cite{tcmalloc}) to include extra checking and a custom
\textsc{Valgrind}~\cite{valgrind} checker
to watch for particular memory access patterns associated with a crash (e.g.~identifying
all cases where a specific value was written to memory, irrespective of the address~\cite{Krasznahorkay:2871750}).

\subsection{Athena configuration}
\label{sec:core:configuration}

A typical reconstruction job consists of hundreds of Algorithms, Services, and Tools,
all of which must be properly and consistently configured in order for the job to run.
As there can be complex dependencies between the various components, this is done
using \textsc{Python}~3.

For each component to be configured, a corresponding \textsc{Python} object is created (a
`\texttt{Configurable}') containing the values of the Properties for that component. The \textsc{Python}
\texttt{Configurable} classes are automatically generated during the software build based on
the components and their Properties declared in the C++ code. Sets of components
representing a consistent configuration are collected in an object called a
\texttt{Component Accumulator}~\cite{chep18configuration}.

\textsc{Python} functions set Properties of components and arrange their dependencies as needed.
These functions take \emph{Configuration Flags} as input and return \texttt{Component Accumulator}
objects. The Configuration Flags describe global properties, like whether the input is
simulated data or detector data, or if cosmics or collisions are to be processed.
Generally, flags are used to ensure the configuration of individual components is
done consistently for a particular situation. This approach allows the same \textsc{Python} program
(`RecoSteering') to process inputs from proton--proton or heavy-ion collisions,
simulation, cosmic rays, and so forth.

The flags have auto-configuration capabilities: they can set themselves automatically
based on information found in the input file. For this purpose, the (first) input file is
opened and inspected at the configuration stage. For data reconstruction jobs, the
run number and time stamp of the first event are used to query the conditions database to
determine the appropriate data-taking conditions and configure the reconstruction job
accordingly.

For each piece of a job (e.g.\ calorimeter clustering in reconstruction), there is a \textsc{Python} function
taking configuration flags and producing a \texttt{Component Accumulator}.  With minimal additions,
these small units are standalone-runnable as long as their inputs (in the case of
calorimeter clustering, the calorimeter cells) are in the input file. This design allows
unit-testing of individual components or relatively small sets of components.

The \texttt{Component Accumulator} objects created by the configuration functions can be merged
together to assemble more complex jobs. In this merging process, duplicate components
are reconciled into one instance, with an error reported if they are configured
inconsistently. The process is recursive, so that each component can call functions to add
any of its dependencies, add the component being configured, merge them all together, and
return the complete \texttt{Component Accumulator} object. This process allows the configuration of the
full job (e.g.\ reconstruction or simulation) to be built in a modular way. The \textsc{Python} program
producing the configuration of the full reconstruction calls functions configuring subsets
(for example jet-finding), passing the configuration flags as a function parameter and
merges the resulting \texttt{Component Accumulator} objects.

Once the final configuration is built, it can be used directly to create the C++ components
for the job and run the application. Alternatively, it can be saved to a \textsc{Python} pickle file
for later use.

This \texttt{Component Accumulator}-based approach was developed during LS2 and put into
production for the 2023 data-taking year. The configuration system used up to 2022 was
also \textsc{Python}-based but was not as modular, leading to difficulties in maintenance,
extension, simplification, and debugging. The old system
relied on fragments of \textsc{Python} code, called \emph{job options}, which could be stitched together
to drive the configuration of an Athena job. This type of configuration persists in event
generation (see Section~\ref{sec:evgen}), where hundreds of thousands of individual generator configurations are
defined in short \textsc{Python} snippets that are executed as a part of the configuration of a job. Many of these snippets are programmatically generated. Each snippet corresponds to a separate physics process; these are described further in Section~\ref{sec:evgen}. The migration of the myriad configurations of event generators to use the \texttt{Component Accumulator} is ongoing.

For standard workflows like detector simulation or reconstruction, rather than writing a
\texttt{Component Accumulator}-based \textsc{Python} script for each job, \emph{Job Transforms} are used to
provide a convenient command-line configuration. The input type and output type must be
specified on the command line, and almost all other settings are optional. The job transform
then determines what software steps are to be run based on a graph from input to output
file type and configures the job by passing command-line parameters to the appropriate
\texttt{Component Accumulator} functions. In the production system (see Section~\ref{sec:prodsys}),
all jobs are run via these job transforms, so that only the command-line settings need to
be stored to fully reproduce the configuration of the job. These production configurations are
stored in the AMI metadata system (see Section~\ref{sec:metadata}). The job transform also
includes some convenience features, including running a monitoring program that records
memory and I/O usage of the job, running input and output file validation, automatic log file
parsing (e.g.\ to search for indications of errors), and producing a job summary report that
can be easily parsed in the production system.

\subsection{Conditions data handling}
\label{sec:core:conditions}

Conditions data are valid over some range of events or time, called an \emph{Interval of Validity} (IoV).
Conditions data are presented to Athena algorithms in the form of \emph{Conditions Objects}. In Athena,
two types of Conditions Objects are distinguished: \emph{Raw} and \emph{Derived}. Raw objects are
constructed using the data retrieved from the ATLAS COOL~\cite{Verducci:2008zzb,USLHC:2012ewx} Conditions
Database by a specialized Athena service called \texttt{IOVDbSvc}. Each of these objects corresponds to one COOL
\emph{Folder} in the Conditions Database. Often it is necessary to apply some transformations to
conditions data. The objects that are created by such transformations are called Derived
Conditions Objects. Derived objects are constructed by taking one or several other conditions objects ---
either raw or derived --- as input. Athena supports conditions objects of arbitrary type, although in most
cases conditions objects are represented as a collection of key--value pairs where the keys are strings and
the values are simple C++ types or vectors of them.

Since the multithreaded framework may be concurrently processing multiple events, it must be able to
manage having potentially several versions of a conditions object active at any one time. To satisfy this
requirement, conditions data are kept in a separate transient \emph{conditions store} analogous to the
event store; however, objects recorded in this store are containers (i.e.\ \emph{Conditions Containers}) that can
hold multiple versions of a conditions object of the same type. Elements in each Conditions Container are
indexed by their IoV. IoVs within a single Conditions Container are non-overlapping, and
hence one can uniquely identify an object within the container by providing an input time or (luminosity block,
run number) pair.

Condition Objects are inserted into Condition Containers by \emph{Conditions Algorithms}, which are a specialised
set of Athena Algorithms that operate on conditions data. All COOL folders needed for a
given Athena job are registered at the configuration stage with a special Conditions Algorithm called
\texttt{CondInputLoader}. This algorithm is executed at the start of processing an event. It loops over all COOL
folders registered to it and for each of them checks that the corresponding Condition Container has a
conditions object that is valid for the given event. If this is not the case for some containers, then new
raw conditions objects are retrieved from the \texttt{IOVDbSvc} (which means either fetching new data from the
Conditions Database or retrieving it from the \texttt{IOVDbFolder} cache) and inserted into the container. After
that, the framework schedules the execution of those Conditions Algorithms that are responsible for
creating the derived conditions objects that depend on the newly constructed raw objects, using the same data flow mechanism as in regular Algorithms.

Algorithms, both regular and conditions, and the Tools they use access objects in the conditions store
in a similar manner to that for event data.  A Property of the \texttt{SG::ReadCondHandleKey<T>} declares a
dependency on the conditions object, and may be used to initialize a smart pointer that can use the
event identification information in the \texttt{EventContext} to look up the proper version of the Conditions
Object in the store.  Similarly, \texttt{SG::WriteCondHandleKey<T>} is used to record new objects in the
conditions store that are created by a Conditions Algorithm. This provides the framework with the
dependency information needed to ensure that Conditions Algorithms execute before any downstream Algorithms
that require the data that they produce.

Since Conditions Algorithms can only insert new elements into Conditions Containers, the size of the
conditions store grows as the job progresses, and this may result in a significant increase in the job's
memory footprint over time. However, not all container elements are required at any one time, only the
ones that are valid for the events that are currently being processed. To optimise the memory
usage by the conditions store, a garbage collection mechanism removes objects from conditions containers
when they are no longer needed~\cite{chep18conditions}. This mechanism takes into consideration the fact
that events are not necessarily guaranteed to be processed in the same order that they were taken, and
tries to avoid the need for multiple instantiation of the same conditions object during the job.


\subsection{Event data model}
\label{sec:edm}

The structure of ATLAS data is defined at various stages along the processing chain,
which is discussed in detail in Section~\ref{sec:transforms}.
The scale of the ATLAS experiment, the collaboration, and the data it collects and produces,
is such that common data objects and interfaces are crucial to ensure maintainability and internal consistency.
The ATLAS EDM is a collection of interfaces,
classes and types that combine to provide a representation of an ATLAS event.
It provides a commonality across detector sub-systems, allowing common tools to be factored out and shared.
It also permits the use of common software between online and offline data processing environments.

The object stores used by Athena,  such as the event store and the conditions
store, are implemented by instances of the Service \texttt{StoreGateSvc},
which provides a type-safe mapping of string-based identifiers to arbitrary C++ objects.
For event data, containers of objects are usually represented by the
type \texttt{DataVector<T>}~\cite{chep15edm}, which is much like \texttt{std::vector<T*>},
but with several additions:

\begin{description}
\item[Optional ownership] A \texttt{DataVector} may own the elements to which it
points, which are then deleted when they are erased from the vector.
An optional argument to the \texttt{DataVector} constructor is used to control
whether or not a given \texttt{DataVector} owns its elements.
A \texttt{DataVector} that does not
own its elements is sometimes called a \emph{view vector}, as it may be used
to create a `view' of elements from other \texttt{DataVector} instances.

\item[Container covariance] In these declarations,

\begin{Verbatim}[numbers=left, frame=single, commandchars=@\#\^]
class FourVector {};
class Particle : public FourVector {};
DATAVECTOR_BASE (Particle, FourVector);
\end{Verbatim}
the use of the \texttt{DATAVECTOR\_BASE} macro causes the class  \texttt{DataVector<Particle>}
to derive from \texttt{DataVector<FourVector>}.
This makes it possible to write, for example, a generic algorithm operating
on a container of \texttt{FourVector} objects.  The event store is also aware
of this, so that objects of type \texttt{DataVector<Particle>} may be retrieved
as type \texttt{DataVector<FourVector>}.

\item[Auxiliary data] Named data of arbitrary type can be attached to elements
of a \texttt{DataVector}.  These data are stored as contiguous blocks
of memory (resulting in improved locality of reference) and are accessed
via an abstract interface of a separate \emph{auxiliary store} object.
\end{description}

\begin{figure}[tb]
\centering
\includegraphics[width=14cm,height=6cm]{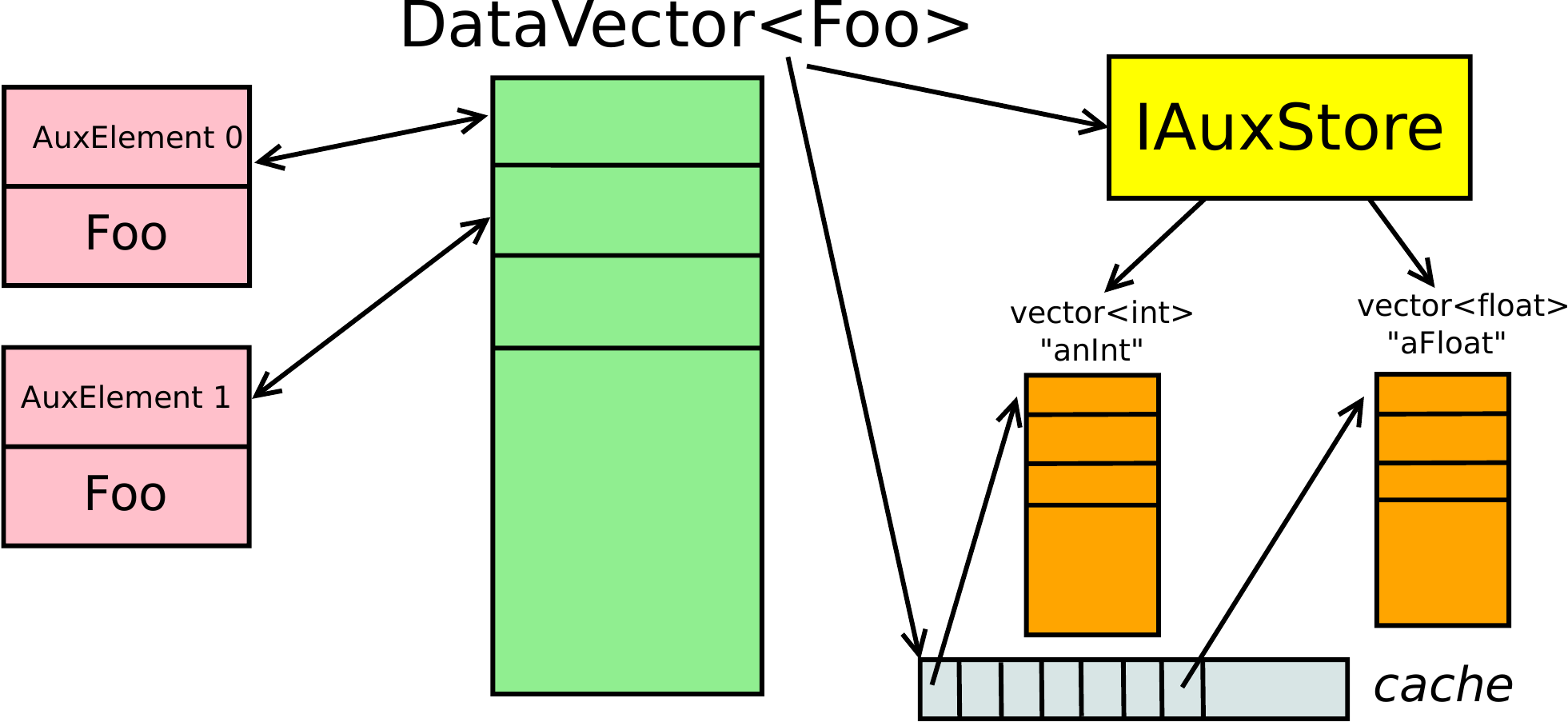}
\cprotect\caption{A \texttt{DataVector<Foo>} contains pointers to individual \texttt{Foo} objects,
which derive from \texttt{AuxElement}.  Each \texttt{Foo} object contains a pointer
back to the \texttt{DataVector} and its index within the vector.
The auxiliary data are stored in a separate object that derives
from \texttt{IAuxStore}; the \texttt{DataVector} has a pointer to this
auxiliary store object.  The auxiliary store contains contiguous
blocks of memory for each available variable, here `\texttt{anInt}'
and `\texttt{aFloat}'.  These are usually managed using \texttt{std::vector}
but this is not required by the design.
For efficiency, the \texttt{DataVector} maintains a cache array of
pointers to the start of the block for each variable; this array
is indexed by the integer identifiers for the variables.}
\label{fig:auxstore}
\end{figure}

The design of the auxiliary data is summarised
in Figure~\ref{fig:auxstore}.  A \texttt{DataVector} object contains pointers
to its elements, each of which contains a pointer back to the
container and its index within the container.  A \texttt{DataVector} also
has a pointer to an object implementing the abstract
interface \texttt{IAuxStore}.  This auxiliary store object manages the variables,
which are internally identified by small integers and which must be stored
in a contiguous block of memory (such as a \texttt{std::vector}).
When an auxiliary variable is accessed for a given element of the container,
the \texttt{DataVector} retrieves the vector for that variable from the store,
saving it in a cache indexed by the variable's identifier.
The element's index is then used to find the proper entry in the variable
vector.

For objects that are intended to form the input to analysis, almost
all the object data are stored as auxiliary data rather than as class members.
These are sometimes referred to as \textit{analysis data objects}, or \emph{xAOD} objects.
Other object types, which usually represent intermediate or more detailed results of the
reconstruction, generally do not use auxiliary data.

\begin{figure}[tb]
\centering
\includegraphics[width=14cm,height=6cm]{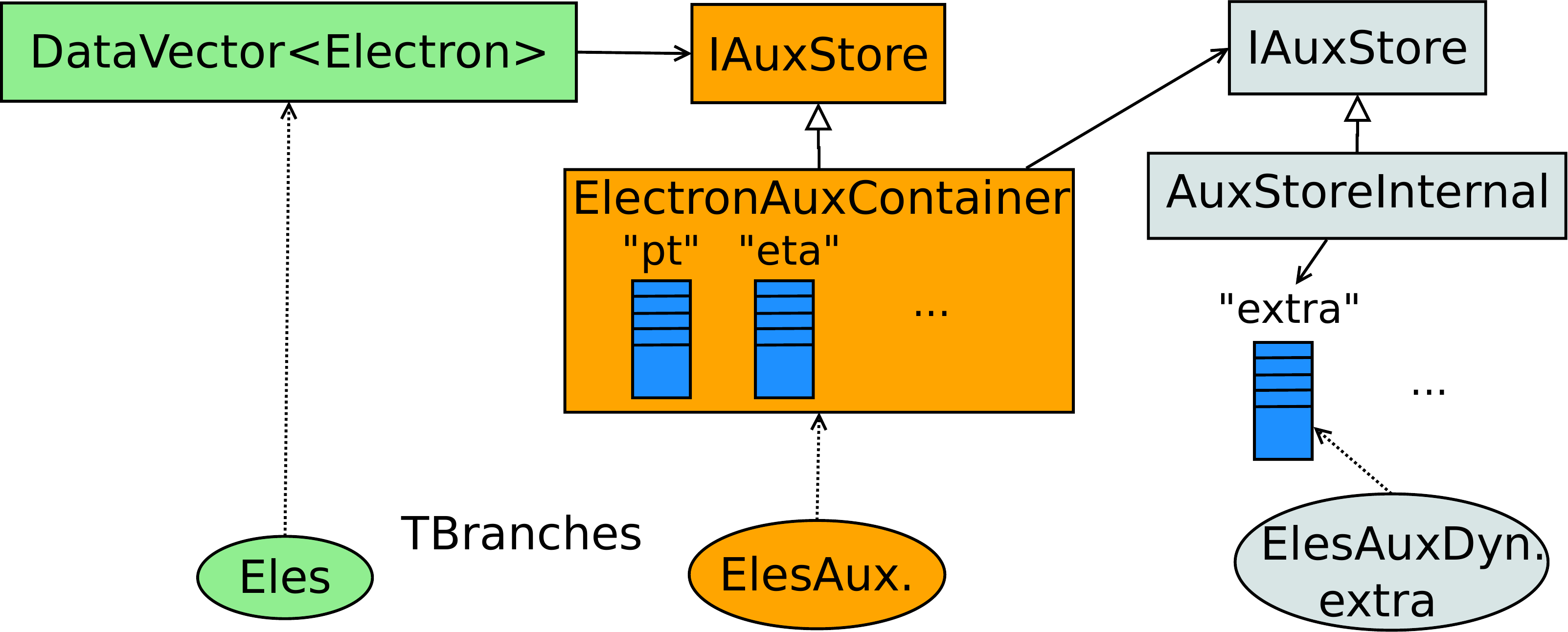}%
\cprotect\caption{Illustration of static and dynamic stores used for xAOD objects.
In this figure, an object representing a container of electrons of type
\texttt{DataVector<Electron>} is stored in the event data store with
name `\texttt{Eles}', which is also the name used for the \textsc{ROOT} \texttt{TBranch}
when these objects are saved to \textsc{ROOT}.  This container references
a static auxiliary store object of type \texttt{ElectronAuxContainer}
(deriving from the abstract interface \texttt{IAuxStore}).  This object
has members \texttt{pt}, \texttt{eta}, and so forth that are usually \texttt{std::vector}
instances holding the actual data.  This static auxiliary store
is also saved in the event data store with name `\texttt{ElesAux.}', which
is again the name of the corresponding \textsc{ROOT} \texttt{TBranch} into which
the \texttt{ElectronAuxContainer} object is saved.  The static auxiliary store
can also own a reference to a dynamic auxiliary store, here of type
\texttt{AuxStoreInternal}, also deriving from \texttt{IAuxStore}.  The dynamic store
holds vectors representing additional variables that are attached
to the container.  In this example, there is an additional variable (a decoration)
named `\texttt{extra}'.  When written to \textsc{ROOT}, this appears as
an additional \texttt{TBranch} named `\texttt{ElesAuxDyn.extra}'. Solid lines
with open arrows indicate inheritance relationships. Solid lines show references
between objects. Dotted lines indicate the mapping between names and objects in
the event store.}
\label{fig:statvsdyn}
\end{figure}

\begin{sloppypar}
For a given xAOD object type, for example \texttt{xAOD::Electron}, the type
\texttt{DataVector<xAOD::Electron>} is aliased to \texttt{xAOD::ElectronContainer}.
An additional class \texttt{xAOD::ElectronAuxContainer} then implements the
\texttt{IAuxStore} interface and contains the `static' data always associated with
an \texttt{xAOD::Electron}.  Its members are \texttt{std::vector} instances, one for
each variable.  An \texttt{AuxContainer} class can forward requests for variables
that it does not contain to another \texttt{IAuxStore} instance.  In this case,
this is usually an object of type \texttt{IAuxStoreInternal}, which manages
a dynamic map of variables.  Thus, an xAOD object has a certain set of
variables that it should always contain, but additional variables,
called \emph{decorations}, may be added dynamically.  This design is shown
in Figure~\ref{fig:statvsdyn}.
Like the objects themselves, the scheduler knows about decorations
read and written by Algorithms and can take that into account
in scheduling decisions.  The auxiliary store object is saved
separately in the event store: for an object named \texttt{Electrons}, the
auxiliary store object is named \texttt{ElectronsAux}.
The four-momenta are stored in the objects, while information about identification
and classification is included in the auxiliary store.
This design allows flexibility in the data that is provided to analyses,
letting users select exactly those variables they require, thereby reducing disk
space and confusion at the cost of some modest CPU overhead. The software workflow
steps preceding the analysis stage are much more standard and programmatic, and this
flexibility is not normally needed.
\end{sloppypar}

A key feature of this design is that the data in the auxiliary store
is accessed via an abstract interface.  This feature allows
using different implementations of the auxiliary store without having
to make changes to the \texttt{DataVector} class.  For example:

\begin{itemize}
\item Rather than having separate static and dynamic stores, an xAOD object
can transparently use only a dynamic store, making all variables dynamic.
This is useful when xAOD data are used as the input for analysis.
\item Data objects produced during data-taking by the HLT can use
an auxiliary store implementation that is specialised
for storage in RAW data files.
\item When objects are being read, their dynamic variables are managed
by an auxiliary store implementation that defers the actual reading
of the data until it is used for the first time.
\item This mechanism is also used to implement \emph{shallow copies}, which allow
data to be shared between multiple objects.  A shallow copy of a \texttt{DataVector}
will produce a new \texttt{DataVector} that has an auxiliary store of type
\texttt{ShallowAuxContainer}, which maintains a reference to the original store.
Variables that are written or modified are managed by the
\texttt{ShallowAuxContainer} using a copy-on-write mechanism, while attempts to read variables
not in the \texttt{ShallowAuxContainer} are forwarded to the original store.
This is useful for cases where one wants to make a copy of an object
and change a few variables, allowing storage to be shared for the
unchanged variables.
\end{itemize}

To represent references between objects in the event store, special link
classes are used. \texttt{DataLink<Obj>} represents a link to an object of type
\texttt{Obj} in the event store (e.g.\ a charged particle track might link to
its component hits).  The object is identified by an integer hash of the
object name and type; this integer is what is written when a \texttt{DataLink}
is saved.  The mapping between hashes and (name, type) pairs is saved
in the file metadata.  Similarly, \texttt{ElementLink<~DataVector<T>~>} represents
a reference to a particular element in a \texttt{DataVector}.  It consists
of a hash as for \texttt{DataLink} identifying the container, along with the index
of the desired element in the vector.

When a container is being written to persistent storage, it is possible
to select specific elements to be written.  This process is called
\emph{thinning}\footnote{While thinning can in principle be applied to any
container, in practice it has to date only been needed for and applied to
\texttt{DataVector} containers.} (see also Section~\ref{sec:derivationintro}).
For each container that is to be thinned, there should
be an Algorithm that writes to the event store a special object containing
a bit map of the entries in the container that should be written.  The name
of this object encodes both the name of the container to be written and
the output file to which it is to be written.  When a container is written,
the I/O code will test for the presence of this thinning object and implement
thinning if it exists.  The transient version of the container in the event
store is not itself modified by thinning. It is also possible to write only
a select subset of decorations, which is a process known as \emph{slimming}.

The I/O process works differently for xAOD and non-xAOD objects.
For a non-xAOD transient object of type \texttt{Obj},
there is a corresponding \emph{persistent} class
of type \texttt{Obj\_p1}. In the case that the contents of the class change beyond
what can be handled by \textsc{ROOT}'s automatic schema evolution, additional versions
\texttt{Obj\_p2}, \texttt{Obj\_p3}, and so on may be defined. Classes that rely on
polymorphism use a more complicated scheme.
When an object is to be written,
a specialised converter class copies data from the \texttt{Obj} instance to the
\texttt{Obj\_p}N instance (where N is the latest version).  During this copy,
thinning requests are applied.  If \texttt{Obj} is a container that is being
thinned, the requested elements are omitted, and if \texttt{Obj} contains
\texttt{ElementLink}s to a container being thinned, then the indices of those
links are adjusted to preserve the references.
The resulting persistent object is then written as a branch in a \textsc{ROOT}
\texttt{TTree} representing the event data.
When an object is read, the version of the persistent class present
in the input file is used to select the proper converter class to copy
data back from the persistent object to the transient object.
This approach allows any version of Athena to read many old data formats, while it
writes only the latest.

\begin{sloppypar}
For xAOD objects, the \texttt{DataVector<xAOD::Obj>} itself is saved as if it were
a \texttt{std::vector<xAOD::Obj>} with a custom \textsc{ROOT} collection proxy.
Since for most xAOD objects the elements themselves do not
contain any data, this effectively just records the length of the vector.
As discussed above, object data are stored in the separate auxiliary
store objects.  These are versioned: for an object of type
\texttt{DataVector<xAOD::Obj>} (aliased to \texttt{xAOD::ObjContainer}) there are
auxiliary store classes \texttt{xAOD::ObjAuxContainer\_v1}, \texttt{xAOD::ObjAuxContainer\_v2},
and so on, with the most recent one aliased as \texttt{xAOD::ObjAuxContainer}.
This object is saved separately to the \textsc{ROOT} event \texttt{TTree} as a single object.
Dynamic variables are then saved to separate \textsc{ROOT} \texttt{TBranch} objects,
one branch per object.  Although there is no separate persistent form
for xAOD objects, copies are still made of the objects before giving
them to \textsc{ROOT} to be written.  This design allows the implementation of thinning,
as well as transformations such as making all variables dynamic.
On input, if an older version of the auxiliary store object is found
in the input file, a converter class is used to convert the data to an
instance of the current version of the store.
\end{sloppypar}


\subsection{Detector description}
\label{sec:dd}

The ATLAS detector is described in Section~\ref{sec:detector}.
The software description of the detector underlies much of ATLAS software and the workflows that depend on it.
In simulation, an accurate description is required to model the interactions between particles and detector components, both active (i.e.\ instrumented for readout) and inactive.
In both the simulation and reconstruction, the detector description is used to translate detector hits into various relevant coordinate systems.
Since this translation depends upon the time-dependent alignment of detector elements, such as silicon sensors, muon chambers and the like,
the software description of the detector needs to follow conditions data, in addition to static data describing the default position of such elements.
The implementation of the detector description for Runs~1--3 is presented in Section~\ref{sec:dd13},
and new tools developed to support the detector description in \RunFour are presented in Section~\ref{sec:dd4}.

\subsubsection{Detector description in Runs 1, 2, and 3}
\label{sec:dd13}
For about two decades the software description of the ATLAS detector has relied on the GeoModel class library~\cite{Boudreau:865601}.
In brief, this library uses a scene-graph approach, building a hierarchical tree representing the geometry that permits a compact
in-memory description of the detector geometry. Users construct a directed acyclic graph consisting of:

\begin{itemize}
\item Physical Volumes (volumes with a size and shape),
\item Transformations (translations and rotations to place the physical volumes),
\item Name tags (which assign character strings to physical volumes), and
\item Identifier tags (which associate integers to physical volumes),
\end{itemize}

and special graph nodes that are built to allow repeated, systematic placement of multiple volumes following a set of rules:

\begin{itemize}
\item Serial Transformers (to programmatically translate and rotate a series of volumes),
\item Serial Denominators (to programmatically name volumes), and
\item Serial Identifiers (to programmatically identify volumes, e.g.\ by index number).
\end{itemize}

The GeoModel class library relies on polymorphism for flexibility and extensibility: other geometrical objects or properties thereof can be added according to need.

Serial Transformers allow the embedding of almost-arbitrary recipes for generating transformations to be encoded within the tree. Serial Denominators and Serial Identifiers specify policies for naming and identifying physical volumes. \texttt{FullPhysicalVolumes} are specializations of Physical Volumes used for active detector elements; they cache the position of the detector element in the world coordinate system, after computing it from the sequence of cascading local transformations. Alignable Transformations are specializations of Transformations that can be adjusted or tweaked according to evolving alignment conditions.

During its first two decades of use, very few modifications to the GeoModel were required. The most intricate change was the adaptation of the Alignable Transforms to the multithreading environment of today's offline software, which implies that detector elements simultaneously cache \emph{multiple} global-to-local transformations,
to enable the concurrent processing of multiple events that may
have different alignments.

Primary numbers for the description of the ATLAS detector are stored as tabular data in a relational database (\textsc{Oracle}\textregistered). This database consists of more than 1000 database tables to describe all geometric layouts used over the lifetime of the experiment, including both tables of geometry information populated by experts using a variety of tools and scripts and tables used for versioning of the geometries. A small fraction of these tables are typically updated when a new layout is constructed. Specific ranges in the data tables are assigned alphanumeric tags and the tags are finally collected into one overall tag for the whole detector configuration. These tags may be developed, locked and placed into production, and eventually obsoleted by indicating the last Athena release that supports them.\footnote{This obsoletion mechanism allows the removal of obsolete code and a clear sign of what detector layouts and geometries must be supported by all systems.} The contents of the database can be easily consulted through a dedicated web-based browser, and is replicated in a lightweight SQLite~\cite{SQLite} file with a size of around 75~MB.

The same database holds tables corresponding to materials for all of the detector elements. A precise elemental composition of
some elements is quite important for the simulation of radiation in the detector. For example, including layers of boronated
polyethylene is critical to understanding low-energy neutron flux. The understanding of the detector is sufficiently precise
that even element densities are important. For example, the evaporative cooling system in the inner detector produces visibly
different numbers of hadronic interactions in different detector regions, owing to the transition from denser liquid-coolant
to less-dense gas. In some cases, the material of the detector changes over the course of data taking; this is the case for the
gas in some modules of the TRT. Which modules contain which gas is therefore encoded in the conditions database.

The detector geometry system must not only provide a best-knowledge geometry for each year of data taking, but must also support
several alternative configurations and geometries that are important to the collaboration. For example, to support cosmic-ray
data taking and MC simulation, layouts of the detector that include the concrete cavern, steel gangways and other infrastructure
around the detector, and even the bedrock above the detector are supported. Several of the forward detector systems can be
described using GeoModel, but they are not normally included when building a layout of the detector for simulation, for example.
During commissioning of the detector, some detector
elements might be significantly displaced from their nominal positions (e.g.\ the calorimeter endcap might be displaced to allow
access to the inner detector). These layouts have also been simulated to provide early understanding of cosmic-ray data
for detector groups. Elements of the ATLAS detector have also been placed in many different test beams, some of which are supported
by the standard detector description system. These have proven useful to the \Geant{} Collaboration~\cite{Agostinelli:2002hh,Geant42,Geant43} towards the understanding of
the tuning of physics models, and have therefore been exported in XML format for their use outside of ATLAS.

This detector description provides the reference geometry for the detector. However, for several applications it must be translated
into a different format.
For the detector simulation (see Section~\ref{sec:sim}), the geometry is loaded into \Geant{} during the job initialization. This
also means that the alignment of the detector during a simulation job does not change. For charged particle tracking (see Section~\ref{sec:trackReco})
and fast track simulation (see Section~\ref{sec:sim:outlook}), a simplified \emph{tracking geometry} is created that maps the
geometry into simplified concentric cylindrical shells for much faster navigation and transportation.

Because of its importance to the work of the experiment, constant efforts are underway to improve the geometry description of the
detector. Some of these, like efforts around radiation simulation (see Ref.~\cite{GENR-2019-02}), require better understanding of materials and details like the
heavy metals inside on-detector electronics. In other cases, this includes going back to design drawings, discussing with engineers,
or examining pictures taken during installation to build CAD representations of the detector that can then be compared with the
existing geometry, or to simply identify missing material in the detector description and add it directly. This is both difficult
and important in some regions of the detector like cable trays, where reality often deviates from design.

One final, critical element of the detector description is the magnetic field map. A link is provided in the conditions database
to select the external field map file to be loaded during an Athena job. The map itself is built from a combination of probe
measurements and simulation of the currents in the solenoid and toroidal magnets, including the distortion effects from material
in the cavern. These maps are generated for several standard magnet configurations (e.g.\ nominal field, solenoid off, and toroidal
magnets off). The field strength is scaled based on the currents recorded during operation. The field value is based on
interpolation between values at points provided within the map; a thread-local cache holds the values at the eight map points
closest to the position being queried, such that a re-evaluation of the field value nearby (as occurs frequently during detector
simulation) is fast.

\subsubsection{Detector description in \RunFour}
\label{sec:dd4}

The GeoModel class library will continue to be used in \RunFour. However, new tools were developed since about 2019 to greatly simplify the workflow of development and maintenance of the ATLAS detector model on the one hand, and its incorporation into offline workflows on the other. The `GeoModel Toolkit'~\cite{GeoModelURL} constitutes an integrated development environment for detector modelling. It consists of:

\begin{itemize}
\item The same GeoModel class library as used today in ATLAS;
\item Mechanisms to save and restore detector geometries to data files;
\item A mechanism for dynamically loading plugins that build geometry;
\item \textsc{gmex}, the `geometry explorer' for fast visual debugging of geometry;
\item \textsc{gmcat}, a tool for assembling geometry files from plugins or file input;
\item \textsc{gmstatistics}, for performance benchmarking;
\item \textsc{gmclash}, \Geant{}-based detection of geometry overlaps (clashes) which cause unpredictable behaviour during simulation;
\item \textsc{gmgeantino}, a command line tool for generation of Geantino\footnote{A Geantino is a fictitious particle implemented in \Geant{} that has transportation processes but no physics properties. It is often used to integrate the material traversed by a particle traveling in a detector.} maps;
\item Tools for gdml-to-geomodel conversion (and vice versa); and
\item \textsc{fullsimlight}, a lightweight \Geant{}-based simulator and \textsc{fsl}, its graphical user interface.
\end{itemize}

The introduction of these tools is the result of lessons learned in Runs 1, 2 and 3, as well as the availability of newer technologies. The main implication for the detector description workflow is that the geometry description can be implemented in a lightweight, portable, and modular environment, taking primary numbers from XML files in a git-managed database, following which an SQLite file is created and fed to ATLAS simulation, reconstruction, and other offline tasks. A transition to these new tools, and the new workflow, is now underway.

Already during LHC \RunThr, the GeoModel toolkit is built and compiled outside of ATLAS and linked as an external package in Athena-based workflows (see Section~\ref{subsec:genInfra:cmake}). It does not depend on ATLAS and is available for use in other experiments as well.


\subsection{Machine learning and software infrastructure}
\label{sec:ml}

Machine learning (ML) is used in a wide range of applications in the ATLAS Collaboration, including physics analysis, simulation and physics object identification in the reconstruction and trigger system.
Models trained for classification and regression, as well as generative models, are core components of the software and analysis chain.
This section describes the infrastructure and tools for ML used within the collaboration.

Classification models are the most commonly employed.
For physics analyses, these models are used to categorise events into signal-like or background-like samples,
efficiently distinguishing events of interest.
Often multivariate event classifiers based on boosted decision trees (BDTs) are used~\cite{TOPQ-2014-03,TOPQ-2012-20,TOPQ-2015-16,TOPQ-2016-12,EXOT-2017-34,TOPQ-2017-03,EXOT-2016-08,EXOT-2017-25,EXOT-2017-28,TOPQ-2018-05,HIGG-2018-51,TOPQ-2020-10,HDBS-2019-22,EXOT-2019-23}.
BDTs have also been used for regression~\cite{PERF-2013-05}.
Other analyses have explored the use of neural networks (NNs) for classification, for example in top-quark physics analyses~\cite{TOPQ-2014-13,TOPQ-2012-21,TOPQ-2015-15,TOPQ-2016-14,TOPQ-2015-05,TOPQ-2016-08,TOPQ-2018-01,TOPQ-2016-06} to separate signal and background processes.
Classification models are also used extensively to accurately identify physics objects, with applications in the identification of hadronically-decaying $\tau$-leptons~\cite{PERF-2013-06}, boosted jets~\cite{JETM-2018-03} and heavy-flavour jets~\cite{FTAG-2018-01}.
More recently, recurrent neural networks (RNNs)~\cite{RNN} are exploited for their ability to harness sequential characteristics to take into account correlations between track impact parameters, thereby improving physics performance for jet flavour tagging~\cite{FTAG-2019-07} and $\tau$-lepton reconstruction~\cite{ATL-PHYS-PUB-2022-044} (see Section~\ref{sec:reco}).

Artificial neural networks have also been used for lower-level reconstruction tasks.
For example, to identify merged clusters in the pixel detector,
which are created by multiple charged particles traversing the silicon,
a clustering algorithm is implemented to identify and split the clusters and provide improved position estimates for the individual cluster pieces~\cite{PERF-2012-05}.

Generative models have predominantly been explored for use in simulation tasks.
Originally developed for image generation, generative adversarial networks (GANs)~\cite{goodfellow2014generative} have shown particular promise for calorimeter simulation~\cite{de_Oliveira_2017,PhysRevLett.120.042003},
where energy depositions from particles interacting with a calorimeter can be represented as images.
GANs have recently been integrated into the ATLAS fast simulation chain (see Section~\ref{sim:fastsim} and Ref.~\cite{SIMU-2018-04}), where
they are used to approximate the calorimeter response to pions.

In the past, most simple models used in the collaboration were trained with and supported by the TMVA \textsc{ROOT} library~\cite{TMVA}.
With the rise of more modern and complex ML methods, in particular with the advent of deep learning, the use of other libraries to train models has become more prevalent.
The most popular libraries for training models are \textsc{TensorFlow}~\cite{tensorflow}, using the \textsc{Keras}~\cite{chollet2015keras} frontend, and \textsc{PyTorch}~\cite{pytorch}. However, \textsc{XGBoost}~\cite{xgboost}, \textsc{lightGBM}~\cite{lgbm} and other frameworks are still commonly used for the training of BDTs.
Many neural networks used in physics analysis rely on the \textsc{NeuroBayes} package~\cite{FEINDT2006190}.
Support for inference from many modern ML frameworks is provided in the ATLAS software environment.
To avoid maintaining and supporting multiple competing ML toolkits for inference, the ATLAS Collaboration has moved to using general inference frameworks.

The first general purpose inference framework integrated into the ATLAS software environment was \textsc{LWTNN}~\cite{lwtnn}, which uses the \textsc{boost} library~\cite{boost} as a JSON parser and \textsc{eigen}~\cite{eigen} for tensor operations.
Users can convert their trained NN models into JSON format, defining the weights and layer operations.
This is then handled by a lightweight class for running inference on single events.
For the inference of BDTs, a custom wrapper was written that converts models trained with TMVA, \textsc{LightGBM} or \textsc{XGBoost} into a common \textsc{ROOT} |TTree| format for inference at runtime.
However, as BDTs are mostly used in physics analyses, other libraries are compiled in standalone ntuple processing frameworks to provide inference.
Due to the rise of more complex models, which use layers and operations that are not supported by \textsc{LWTNN}, the \textsc{ONNX Runtime} (\textsc{ORT}) library~\cite{onnx} was chosen as a general inference library.
Most modern ML libraries can save models in an \textsc{ORT} format, or are supported with conversion tools. These include, but are not limited to, \textsc{PyTorch}, \textsc{TensorFlow}, \textsc{LightGBM} and \textsc{XGBoost}.
Using only a single inference library to support many models trained in different frameworks significantly reduces the maintenance load.
\textsc{ORT} is observed to provide faster inference than previous implementations and support for batched inference.
Although deep NNs are reliant on GPUs for training, inference is performed almost exclusively on CPUs.
In addition to supporting inference of modern ML models, training infrastructure is available through WLCG with support for job submissions using \PanDA~\cite{panda}, including access to many GPUs (see Section~\ref{sec:infrastructure}).
However, most models used by the collaboration are trained on private resources due to the modest resource requirements and satisfactory availability of institutional GPU resources.


%
\section{Data and Monte Carlo production and processing}
\label{sec:transforms}

There are several steps in the standard ATLAS software workflow, as described in Section~\ref{sec:softwareflow}.
This section goes into detail about each step of the workflow, including improvements that were made for \RunThr.
Here the focus is on those steps of the workflow that are run centrally, while steps that are normally run by individual
users are described in Section~\ref{sec:downstream}.
The forward detector systems are dealt with somewhat separately, often outside of the normal workflows, because they
are often used for specific, special purposes. Their treatment is described in Section~\ref{sec:forward}.
The support within the current \RunThr software for future hardware upgrades to the detector is described in Section~\ref{sec:upgrade}.

\subsection{Event generation}
\label{sec:evgen}

Simulated detector events play a critical role in numerous areas of the ATLAS experiment,
from understanding the detector response to serving as the signal and background estimates
for the large range of measurements and searches performed by ATLAS.
MC event generators are at the core of these simulated datasets
and are the first step needed to produce simulated datasets.
All MC sample sets used in data analysis are centrally produced and managed in ATLAS.
The first step of MC sample production is the generation of the event record from a theoretical framework implemented in an event generator program.
Events produced at this level are written into files called EVNT files.
While many of the MC samples are passed through the full detector simulation, digitisation, and reconstruction steps (described in the following sections),
some samples with alternative configurations, such as samples with systematic variations of one or more theoretical or phenomenological parameters, are often not.

The ATLAS Collaboration makes use of a wide range of MC event generator programs:
\POWHEG~\cite{Alioli:2010xd}, \MGNLO~\cite{Alwall:2014hca} (\MADGRAPH), \SHERPA~\cite{Hoeche:2012yf}, \PYTHIA~\cite{Sjostrand:2014zea}, and \HERWIG~\cite{Bellm:2017jjp} are used for a wide variety of processes.
Some generators, including \SHERPA, \PYTHIA, and \HERWIG, can produce complete events alone, including a calculation of the relevant process (e.g.\ $W$-boson production), parton showering, hadronisation, and particle decays.
Several generators, particularly those that implement higher-order calculations, only produce final states with a few partons (often referred to as hard-scatter or matrix-element events) and require other programs to produce realistic events with stable particles.
The physics details of such further processing can be quite involved (relating to resummation or avoidance of double-counting),
but essentially they must all be interfaced to another program, such as \PYTHIA or \HERWIG, for the parton shower, hadronisation, and particle decays.
A few generators run as secondary Athena algorithms after the main generator code, to refine specific aspects of the event generation.
The main such generators, often referred to as \emph{afterburners}, are \PHOTOSpp~\cite{Golonka:2005pn} for refinement of QED final state radiation (FSR),
\textsc{Tauola++}~\cite{Tauola} for precise $\tau$-lepton decay modelling,
and \EVTGEN~\cite{Lange:2001uf} for refinement of heavy-flavour hadron decays, in particular for $B$-physics and flavour tagging purposes.

The generators providing the most events and samples in ATLAS are \SHERPA, \POWHEG, and \MADGRAPH, with showering performed using
\PYTHIA or \HERWIG. However, a wide variety of bespoke, limited-use, or customised generators are used for special samples, including:

\begin{itemize}
\item A beam-halo generator that reads files provided by the LHC machine group detailing the expected result of the showering of LHC beams
through upstream collimators, useful for background estimations.
\item A cavern-background generator, used for re-simulation of low-momentum neutrons and photons that permeate the cavern during operation.
\item A cosmic-ray generator based on Refs.~\cite{CosmicFlux1,CosmicFlux2}, used for the simulation of cosmic rays, including their propagation through
the bedrock above the detector.
\item \textsc{EPOS}~\cite{EPOS}, a generator of perturbative QCD events that is most often used to model minimum bias interactions.
\item \textsc{Hijing}~\cite{Hijing1,Hijing2}, a generator specifically used for the modelling of heavy-ion collisions.
\item \textsc{Hto4l}~\cite{Hto4l1,Hto4l2}, a generator for events with the Standard Model Higgs boson decaying into four leptons.
\item \textsc{Hydjet}~\cite{Hydjet}, an alternative generator for heavy-ion collisions.
\item A \textsc{Python}-based particle gun, useful for single-particle event generation and various tests of detector performance.
\item \textsc{Profecy4f}~\cite{Prophecy4f}, a generator for events with a Standard Model Higgs boson decaying into four fermions.
\item \textsc{Protos}~\cite{Protos}, a leading-order event generator for some new physics models involving the top quark.
\item \textsc{Pythia8B}, a modified version of \PYTHIA that re-decays heavy-flavour hadrons to produce samples enriched in specific hadrons and decays.
\item \textsc{QGSJet}~\cite{QGSJet}, a generator used for alternative modelling of minimum bias interactions.
\item \textsc{STARlight}~\cite{STARlight}, an event generator for ultra-peripheral heavy-ion collisions.
\item \textsc{SuperChic}~\cite{SuperChic}, an event generator for exclusive and photon-induced processes.
\end{itemize}

The reading of events in Les Houches Event Format (LHEF)~\cite{LHEF} is also supported as a means to aid
the usage of additional generators like \textsc{gg2VV}~\cite{gg2VV}, \textsc{DYNNLO}~\cite{DYNNLO}, and \textsc{CHARYBDIS}~\cite{CHARYBDIS}. Many new physics
models are generated using Universal FeynRules Output (UFO) models~\cite{UFO}, in particular through
integration with \MADGRAPH.
This plethora of event generators and their capabilities is shown schematically in Table~\ref{table:evgen:support}, along
with the output formats for each step.

\begin{table}[!htb]
\small
\centering
\caption{Event generators supported in Athena and their general capabilities. At top, the output formats used at each step. Les Houches Event Format (LHEF) output is used for the transmission of matrix element events to generators that implement parton showers; this output format is not supported by all event generators.}
\label{table:evgen:support}
\resizebox{\textwidth}{!}{
\begin{tblr}{Q[r,l,0.28\textwidth]Q[c,c,0.18\textwidth]Q[c,c,0.18\textwidth]Q[c,c,0.18\textwidth]Q[c,c,0.18\textwidth]}
\SetCell[r=2]{c} \textbf{Output Format} & Les Houches &               & HepMC /      & HepMC / \\
& Events      &               & EVNT         & EVNT    \\
& \SetCell[c=4]{l} $\hspace{15mm}\enclosediamond\hspace{4mm}\xrightarrow{\makebox[2cm]{}}\hspace{5mm}\enclosediamond\hspace{4mm}\xrightarrow{\makebox[20mm]{}}\hspace{5mm}\enclosediamond\hspace{4mm}\xrightarrow{\makebox[20mm]{}}\hspace{5mm}\enclosediamond$ \\
\hline[1pt]
\textbf{Generator} & \textbf{Matrix Element} & \textbf{Parton Shower} & \textbf{Stable Particles} & \textbf{Afterburner} \\
\hline[0.7pt]
\textsc{Beam Halo Generator} & & & \checkmark & \\
\hline
\textsc{Cavern Background Generator} & & & \checkmark & \\
\hline
\textsc{Cosmic Ray Generator} & & & \checkmark & \\
\hline
\textsc{EPOS} & Only Minimum Bias & \checkmark & \checkmark & \\
\hline
\EVTGEN &  &  &  & \checkmark\\
\hline
\Herwig & $2\rightarrow 2$ LO and NLO & \checkmark & \checkmark & \\
\hline
\textsc{Hijing} & Only Minimum Bias & \checkmark & \checkmark & \\
\hline
\textsc{Hto4l} & \checkmark &  &  & \\
\hline
\textsc{Hydjet} & Only Minimum Bias & \checkmark & \checkmark & \\
\hline
\MGNLO & \checkmark & &  & \\
\hline
\textsc{ParticleGun} &  &  & \checkmark & \\
\hline
\PHOTOS &  &  &  & \checkmark \\
\hline
\POWHEGBOX & \checkmark &  &  & \\
\hline
\PROPHECY & \checkmark &  &  & \\
\hline
\PROTOS & \checkmark &  &  & \\
\hline
\Pythia & Only $2\to 2$ & \checkmark & \checkmark & \\
\hline
\Pythia[8B] &  &  & \checkmark & \\
\hline
\textsc{QGSJet} & Only $2\to 2$ jets & \checkmark & \checkmark & \\
\hline
\Sherpa & \checkmark & \checkmark & \checkmark & \\
\hline
\textsc{STARlight} & & \checkmark & \checkmark & \\
\hline
\textsc{SuperChic} & \checkmark &  &  & \\
\hline
\textsc{Tauola} &  &  &  & \checkmark \\
\hline
Other Generators (via LHEF) &  \checkmark &  &  & \\
\hline
Other Generators (via HepMC) &  &  & \checkmark &  \\
\hline[1pt]
\end{tblr}}
\end{table}

Although the generation for the newest (\RunThr) campaigns is ongoing, for illustrative purposes the approximate numbers of events and
datasets (unique configurations) generated with various event generators thus far in \RunTwo is shown in Table~\ref{table:evgen_statistics}.
Additionally, the number of simulated events with the standard \RunTwo configuration is included in the table.
In many cases, the number of events per generator includes multiple generator versions
that were produced throughout the entire \RunTwo period as improved theoretical predictions became available.
Among the largest datasets are the single vector boson samples produced using the \SHERPA, \MADGRAPH and \POWHEG event generators.
These samples are used throughout ATLAS, all the way from detector calibrations to measurements and searches,
and as such, large samples are needed across a wide phase-space.
As can be seen from Table~\ref{table:evgen_statistics}, most generated events are not passed through the detector simulation.
These events are primarily used as inputs for low filter efficiency final states, including final states with extra heavy flavour jets or high missing transverse momentum,
which are selected from the inclusively generated phase-space.
The generator with the most datasets is \MADGRAPH,
because it is typically used for searches for new physics scanning a model parameter space.
While this results in many datasets,
the contribution to the total number of events is small compared with the high-rate Standard Model processes described above.

\begin{table}[!htb]
\centering
\caption{Number of datasets (with unique configurations) and events (in billions) generated with various generators thus far during the MC simulation campaign of \RunTwo.}
\label{table:evgen_statistics}
\begin{tabular}{lrrr}
\toprule
Event Generator & Datasets & Generated Events ($\times 10^9$) & Simulated Events ($\times 10^9$) \\
\midrule
\SHERPA     &   3887 & 89.7 & 27.6    \\
\POWHEG     &   6747 & 55.7 & 15.9    \\
\MADGRAPH   & 251023 & 52.2 & 12.5   \\
\PYTHIA     &   6240 & 13.8  & 7.5 \\
\Pythia[8B] &    422 &  5.1 & 2.0  \\
\HERWIG     &    813 &  4.3 & 2.4  \\
Others     &   9851 &  3.5 & 0.5 \\
\midrule
Total        & 280935 & 224.4 & 68.4  \\
\bottomrule
\end{tabular}
\end{table}

Many event generators can provide estimates of systematic uncertainties arising from
variations of the renormalisation and factorisation scales, and variations in the Parton Distribution Function (PDF) sets, via modified event weights.
Moreover, some generators such as \SHERPA are now capable of providing cross-sections calculated at next-to-leading order (NLO)
in the electroweak coupling constant, allowing analyses to assess the impact of NLO electroweak corrections
in their phase-space by modifying an event weight~\cite{Biedermann:2017yoi}.
While these extra \emph{on-the-fly weights} incur extra CPU time during event generation,
it is small --- typically around 10\% additional CPU for more than 100 event weights~\cite{EventWeightCPU} ---
particularly when compared with generating each variation standalone.
Additionally, the generation of systematic variations as event weights within the nominal sample allows these variations
to be available in the simulated and reconstructed datasets with no additional CPU cost.
They also significantly reduce statistical fluctuations when evaluating systematic uncertainties,
because the kinematics of the nominal and varied events are identical.
For standard samples, around 200--300 systematic weights are stored per event.
For systematic variations that can theoretically be estimated by using on-the-fly weights, the weights are saved; for other variations dedicated samples need to be prepared.

Most of the MC event generators are provided by dedicated teams outside ATLAS.
This software is built using the \emph{Layered-Stack} that is used in GENSER (GENerator SERvice) installations (see also Section~\ref{subsec:genInfra:cmake}).
One of the features of the layered stack is that there cannot be multiple generator versions in one LCG release.
Whenever a new version of a generator appears that will be used in ATLAS (an almost weekly occurance), a new layer is requested, in which the generator version is updated but all other externals remain fixed.
This significantly reduces the overhead and build times for providing new generator versions to the collaboration.
GENSER provides a new layer for all the platforms supported by a given LCG configuration.
After installation and validation by GENSER the layer is included into a nightly build and validated by ATLAS.
For a few generators, like \POWHEG, builds are maintained by ATLAS directly.

Regular validation of event generation is performed as a part of the ATLAS Release Testing (see Section~\ref{subsec:genInfra:releaseTesting}).
In these tests, small samples of events are generated for select physics processes.
The output of each test is compared automatically to the reference output from the previous test.
In the case of a generator version change, a more thorough physics validation based on \textsc{Rivet}~\cite{Bierlich:2019rhm} analyses is performed~\cite{ATL-PHYS-PUB-2024-013}.
The procedure of layer creation and validation usually takes a week. However, for very urgent installations, it can be shortened to one or two days.
The steering of the MC event generators is passed from the Athena framework to the event generators.
The generators are instantiated and configured in job options (see Section~\ref{sec:core:configuration}).
The job options contain relevant information about the physics process and define the generator type and its configurations.
Major generator packages contain single job fragments to configure common aspects of the job to ensure consistency across different sample sets.
The event generation is invoked by Athena as part of the standard algorithm event loop, and the generated events that are created are converted into HepMC format.
In \RunThr, the HepMC event record used is HepMC3~\cite{Buckley_2021}.
An attractive feature that motivated the migration from HepMC2 to HepMC3 is to enable event numbers with up to 64-bit precision.
In HepMC2, event numbers were limited in size to 32 bits, requiring careful workaround solutions for the largest of samples the collaboration needed to produce.

After generation, standardised filter algorithms can be inserted into the Athena algorithm sequence. These are used to select only events that meet certain criteria, e.g.\ charged leptons, photons or jets with certain transverse momentum (\pt) and $\eta$ properties, certain flavours of $W$-boson, $Z$-boson or top-quark decays or other event shape variables. A combination of several filters is also possible.
Significant validation of the generated events is run during the Athena job, checking for un-decayed partons, stable particles not known to \GEANT, highly-displaced particle decays (e.g.\ \KzeroL decays), energy imbalance, and other features that might indicate incorrect physics output or create problems in downstream software.

Beyond the nominal workflow described above, events can also be filtered in a two-step approach if an inclusive sample must be split up, for example by the flavour of jets or number of leptons.
In this workflow, first an inclusive dataset is generated and saved to disk.
Then, a dedicated filtering job is run that takes as input the inclusive dataset, and an output dataset with events that satisfy the event filter are saved to disk.
Finally, the output dataset is then passed through detector simulation and subsequent steps.

%


\subsection{Detector simulation}
\label{sec:sim}

The ATLAS simulation begins from events in EVNT files and simulates the propagation through the material of the ATLAS detector of all generated particles that escape the beam pipe, including any decays and interactions with detector material that may occur.  All energy deposits in sensitive volumes of the ATLAS sub-detectors, along with additional truth information about particles created or destroyed during simulation, are written to a file format called HITS.

There are two main simulation approaches employed by ATLAS: `Full Simulation', discussed in Sections~\ref{sim:fullsim}, and `Fast Simulation' discussed in Section~\ref{sim:fastsim}. About 50\% of all MC simulations in \RunThr are fast simulations. Section~\ref{sim:fullsim} includes some discussions of the variation optimisations used to speed up `Full Simulation'. Section~\ref{sim:fastsim} includes a brief discussion of further improvements to the ATLAS fast simulation that were developed for \RunThr. Some of the basic concepts applied in the ATLAS simulation are described in Ref.~\cite{SOFT-2010-01}; this section focuses on improvements, and particularly those improvements deployed for \RunThr.

Simulated data are divided into \emph{MC Campaigns}, numbered based on the year in which the production began, and with each supporting conditions that match specific data-taking periods.
The MC16 campaign supported \RunTwo analyses in the older Release 21 version of the ATLAS software, MC20 supports analysis of reprocessed \RunTwo data in Release 22,
MC21 was the initial \RunThr-like MC simulation campaign for testing and early validation, and MC23 supports detailed analysis of \RunThr data.
Letters are used to distinguish the modelling of different data-taking years: MC20a corresponds to 2015 and 2016 data, MC20d corresponds to 2017 data, and MC20e corresponds to 2018 data, for example.

\subsubsection{Full simulation}
\label{sim:fullsim}

ATLAS uses the \Geant{} toolkit~\cite{Agostinelli:2002hh,Geant42,Geant43} to simulate in detail the interactions of particles with the ATLAS detector. In this section, the \Geant{} setups and optimisation used for \RunThr are described.

\paragraph{\Geant{} version}
\sloppy
The MC16 and MC20 campaigns (see Section~\ref{sec:sw_resources}) of \RunTwo used an ATLAS-patched version of \Geant{}, namely \texttt{10.1.patch03}. For MC21 and MC23, which are the current MC campaigns in \RunThr, the more recent \Geant{} \texttt{10.6.patch03} is used. This version includes an improved hadronic physics model and several updates to the electromagnetic (EM) model. The gamma general process (described under `G4GammaGeneralProcess' below) feature is enabled in this version. A custom stepper (described in Ref.~\cite{SOFT-2010-01}) with a configuration optimised for ATLAS is also enabled, but this was used already since \RunTwo. A retuning of Birk's law and a recalculation of the sampling fractions~\footnote{The sampling fraction is the ratio of the total energy deposit in a cell to the energy deposit in the active material. It depends on the calorimeter specifics and therefore varies between layers.} were completed for this updated \Geant{} version.

\subparagraph{Birks' law tuning}

Birks' Law~\cite{Birks} describes the relation between the energy deposited by particles and the signal of the calorimeters. A retuning of the parameters if this formalism became necessary after the updates of the physics models in \Geant{} 10.6.
The initial parameters of Birks' Law used in the LAr and tile calorimeters were taken from \textsc{Geant3}~\cite{Geant3} at the point when ATLAS switched to using \Geant{}. These values were obtained from experimental data assuming no delta-ray emissions. While correct for \textsc{Geant3}, this is inconsistent with \Geant{} simulation, where delta-ray emissions do occur above reasonable production thresholds. More precisely, it leads to the quenching effect being underestimated, which implies an artificially higher energy response in the simulation.

The parameters were therefore re-tuned such that the ratio of the EM and hadronic response in data and MC simulation matched. This tuning was undertaken for the Tile calorimeter by comparing MC simulation to test beam data using the previously published analysis~\cite{TileTB}. For the EM calorimeters, the value of $E/p$ for charged pions measured in low pile-up collision data collected in 2017 was used~\cite{JETM-2020-03}. The hadronic endcap calorimeter response was found to be unaffected by tuning Birks' Coefficient. No tuning was done for the forward calorimeter due to the lack of available test beam data. The values for Birks' Coefficient used in \RunTwo and \RunThr simulation are shown in Table~\ref{sim:tab:Birks}.

\begin{table}[!htb]
\centering
\caption{Values of $kB$ used in the Birks' law relation $dS/dr = (A \times d\textrm{E}/dr)/(1 + kB \times d\textrm{E}/dr)$ for the sensitive volumes of ATLAS calorimeters in \RunTwo and \RunThr simulation.}
\label{sim:tab:Birks}
\begin{tabular}{lrr}
\hline
Calorimeter &         \RunTwo     & \RunThr \\
& $[ \MeV/(\textrm{g}\times\textrm{cm}^2) ]$ & $[ \MeV/(\textrm{g}\times\textrm{cm}^2) ]$ \\
\noalign{\smallskip}\hline\noalign{\smallskip}
LAr EM barrel and endcap & 0.0486 & 0.05832 \\
Tile            & 0.0130  & 0.02002 \\
\hline
\end{tabular}
\end{table}

\paragraph{Simulation improvements}
\label{sim:fullsimImprovements}

Several improvements were made to the configuration of the simulation for \RunThr that improve the physics performance (i.e.\ the agreement with data) of the MC simulation.
While many small improvements were implemented, the two most significant global improvements are described here.

\subparagraph{Beam spot modelling}

The longitudinal beam spot size varies over the course of a run as the LHC attempts to \emph{level} the instantaneous luminosity during the run as the beams are depleted. In \RunThr, the beam spot size within ATLAS shrinks from 43\,mm to 34\,mm during the first part of each run because of this luminosity levelling. This effect was included in the \RunThr MC simulation. Rather than modelling a continuous distribution, four discrete beam spot sizes (in the $z$ direction) are used: 43, 40, 37 and 34~mm. These values are chosen to be representative of the continuous variation in beam spot sizes during the 2022--2023 \RunThr proton--proton collision data-taking period. The first three values represent the luminosity levelling regime. Levelling is stopped for the last part of the run and the beam spot is approximately constant as the instantaneous luminosity falls. The fourth value represents the beam spot size during this final period. In signal process simulation jobs, distinct lumi block ranges are used to mark off groups of events with different beam spot widths and thereby generate the correct fraction of events with each beam spot size. During the merging of HITS files in the production system, events are sorted according to luminosity block so that events with the same beam spot size are grouped in the merged HITS file.

\subparagraph{Quasi-stable particle simulation}

In the MC16 campaign, and before, all particles decayed by the generator were ignored by \Geant{}, even those that propagated outside the beam-pipe. If such particles propagate past sensitive detector layers before decaying, then there is the potential for missed energy deposits and hence tracking differences compared with data. This has implications for flavour tagging and $\tau$-lepton reconstruction in particular at high \pT.
For the MC20, MC21 and MC23 campaigns this was changed to pass such \emph{quasi-stable} particles to \Geant{} along with their predefined decay chains. Missing particle definitions were added to \Geant{} and the ionisation process was added to all charged particles. This means that all charged quasi-stable particles can undergo energy loss, deposit energy, and be affected by the magnetic field.\footnote{\Geant{} lacks hadronic interaction models for many of these particles, but once they are available, they would naturally be added.} This improves the agreement with data and allows better tuning of flavour tagging and $\tau$-lepton reconstruction algorithms~\cite{FTAG-2020-002}.
Quasi-stable particles are a separate category from long-lived exotic particles, which are not decayed by the generator, but which require extensions to the \Geant{} physics list to be propagated and decayed within \Geant{}.

\paragraph{Full simulation optimisations}
\label{sim:subsec:g4optis}

The performance optimisation of the \Geant{}-based simulation has been a continuous task since before the start of data taking~\cite{Rimoldi:1151298}. The \Geant{} Optimisation Task Force was established in September 2020 and is responsible for optimising the performance of the ATLAS \Geant{} simulation software with the mandate of achieving $> 30\%$ CPU performance speedup for \RunThr relative to the \RunTwo simulation. Several optimisations were implemented, validated and put in production and are described in what follows.

\subparagraph{EM range cuts}

\begin{figure}[tbp]
\centering
\includegraphics[width=0.6\textwidth]{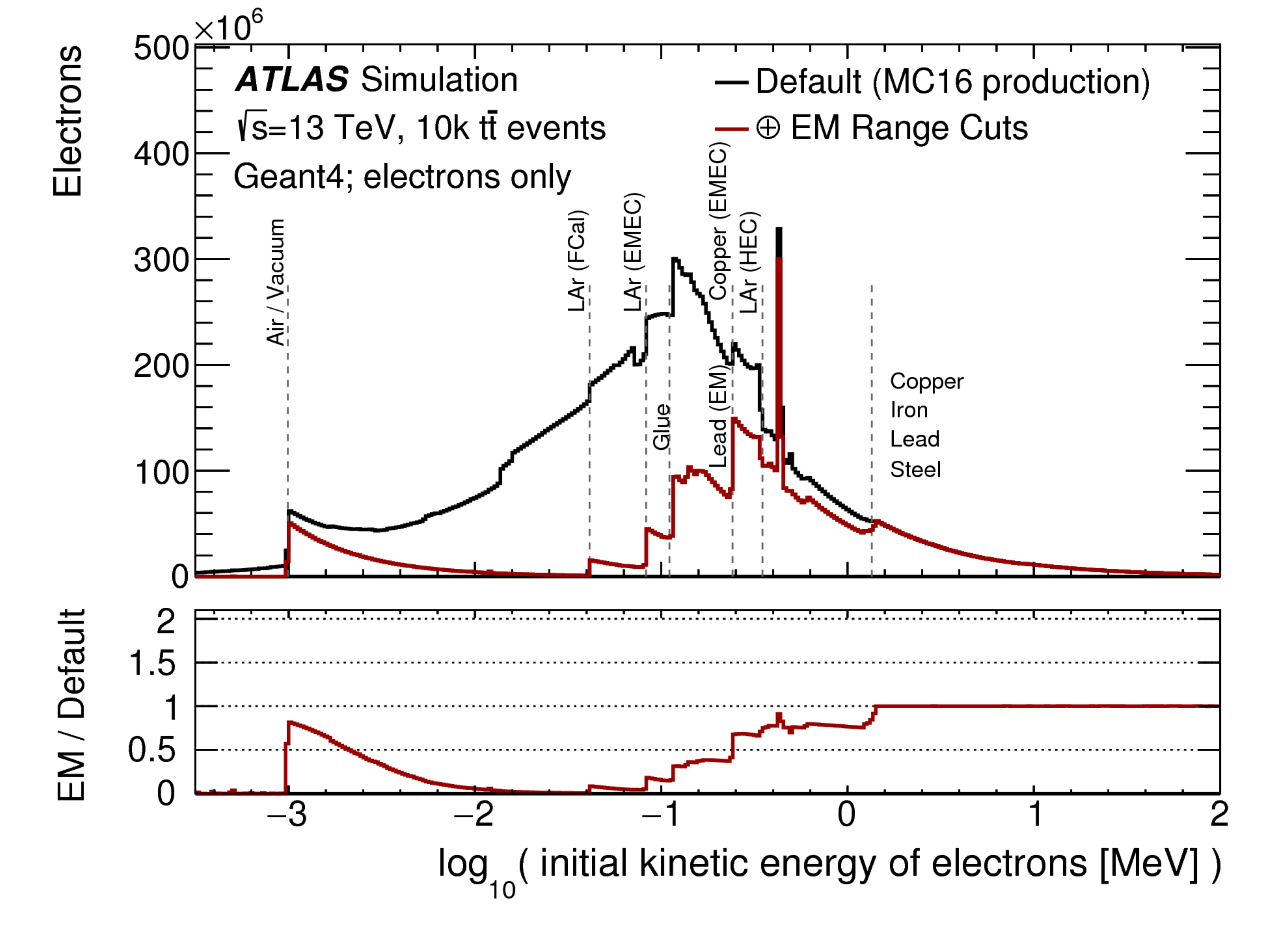}
\caption{Distribution of the initial kinetic energy of electrons in the ATLAS \Geant{} simulation. The black curve shows the distribution for the \RunTwo setup (MC16 production) and the red curve shows the distribution after the addition of range cuts for electromagnetic \Geant{} processes (`conv', `phot', and `compt'). Vertical grey dashed lines indicate range cut values for several key materials and the right-most dashed line indicates an area with multiple range cuts in close proximity for various metals. Figure from Ref.~\cite{SIM-2019-001}.}
\label{fig:sim:rangecuts}
\end{figure}

Several physics processes have very high cross sections at low energies (e.g.\ bremsstrahlung, ionisation, and electron--positron pair production by muons) and it is therefore necessary to implement a production cut so that all particles below the cut are not generated~\cite{Muskinja:2696432}. \Geant{} offers a solution with the \emph{range cuts}, where it is possible to specify a minimum propagation distance (range) for secondary particles. This distance is converted to an energy threshold in each material internally by \Geant{}.  Below this energy threshold secondary particles are not created and their energy is immediately deposited at the end of the production step. Further, it is possible to specify range cuts for each material--volume pair separately for photons, electrons, positrons, and protons. In the \RunTwo simulation, the range cuts were off by default for Compton scattering, conversion and the photoelectric effect, and turning them on with the value of 0.1\,mm already used for electron processes results in a significant decrease in the number of secondaries, as shown in Figure~\ref{fig:sim:rangecuts}~\cite{SIM-2019-001}. This decrease in the number of secondaries led to an 8\% decrease in simulation time.

\subparagraph{Neutron and photon Russian roulette}

\begin{figure}[tbp]
\centering
\subfloat[]{
\label{fig:sim:NRR_a}
\includegraphics[width=0.48\textwidth,valign=c]{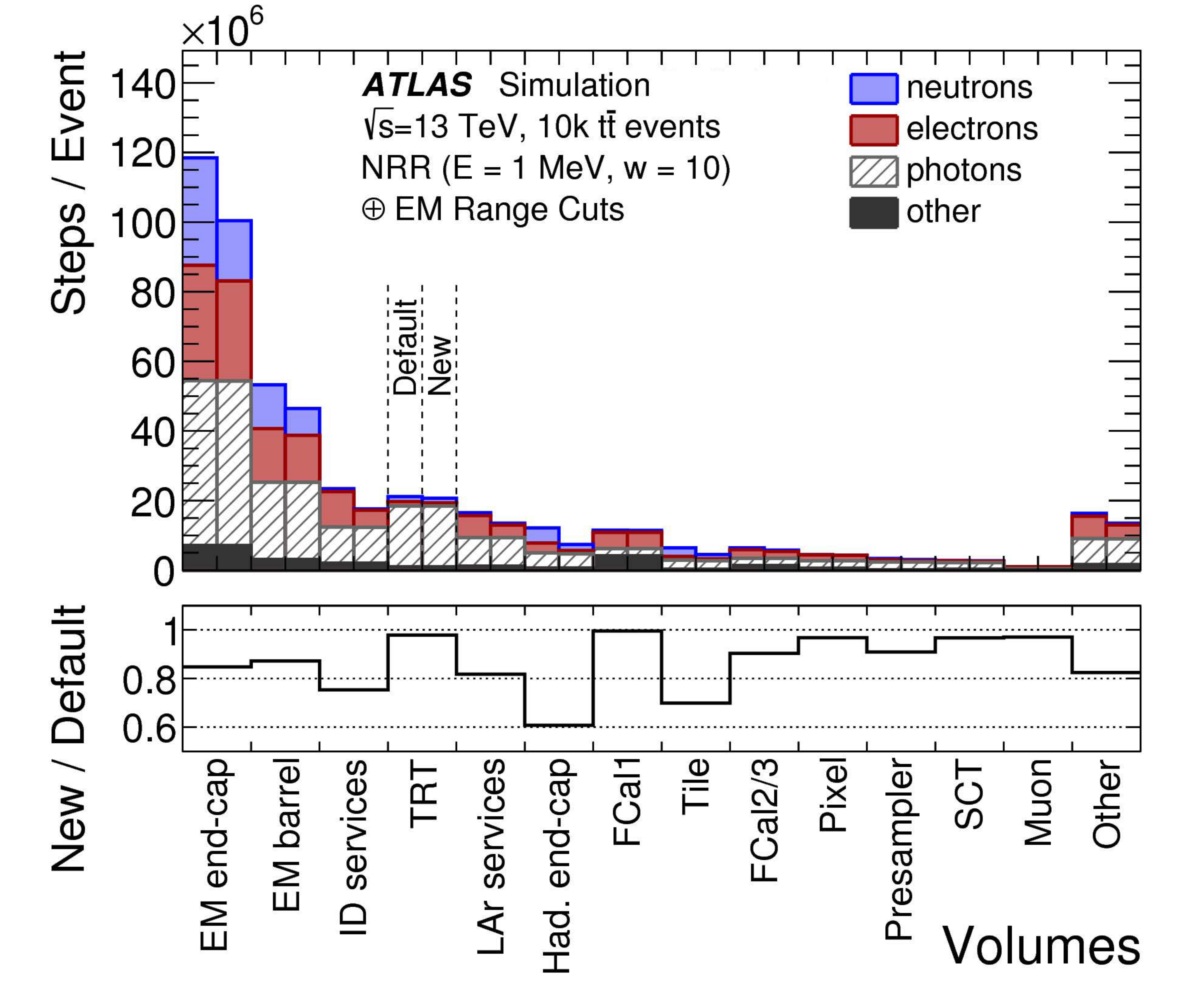}
}
\subfloat[]{
\label{fig:sim:NRR_b}
\includegraphics[width=0.48\textwidth,valign=c]{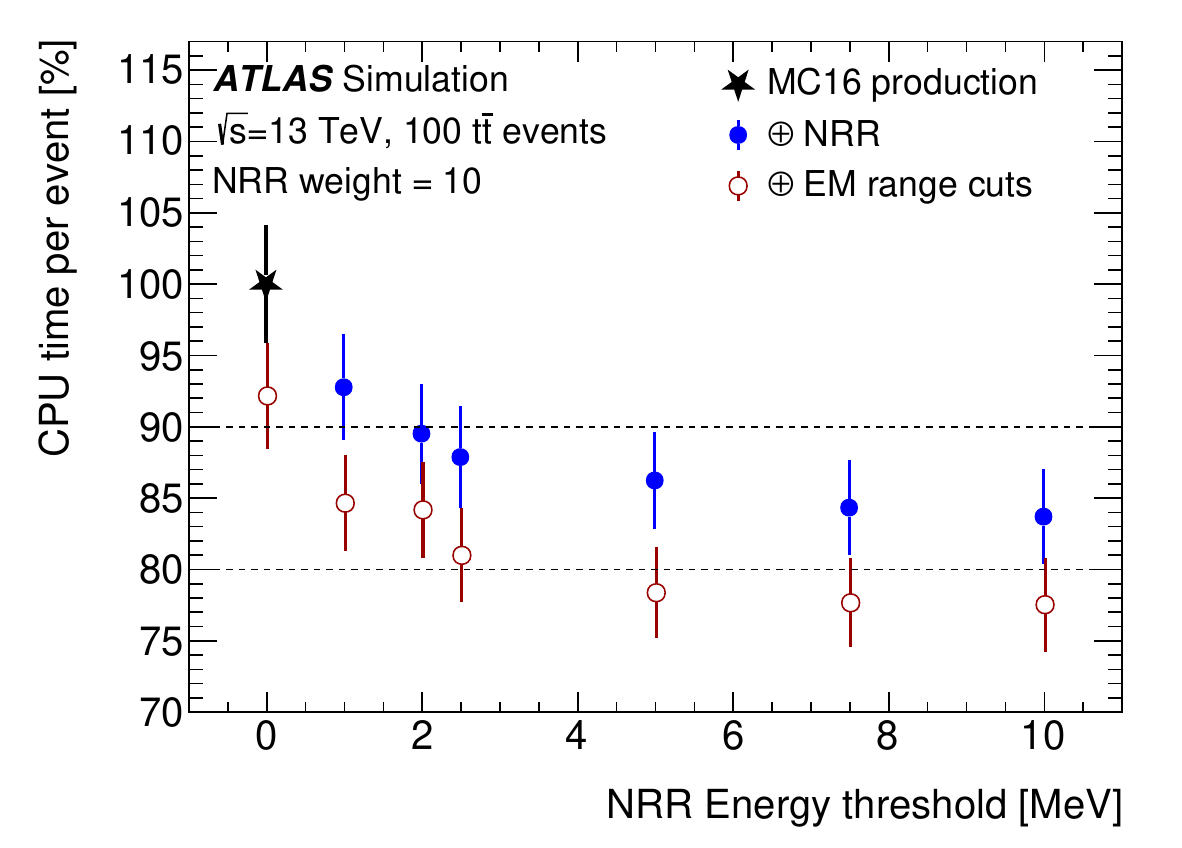}
}
\caption{\label{fig:sim:NRR}(a) Average number of \Geant{} steps per event as a function of the subsystem in the \RunTwo setup (MC16 production). (b) CPU time per event for various thresholds of the NRR algorithm relative to the average \RunTwo MC16 value (black dot) with or without the EM range cuts. Error bars indicate the root-mean-square of the CPU time for the simulated events. Figures from Ref.~\cite{SIM-2019-001}.}
\end{figure}

Neutrons and photons take the most CPU time in the simulation of the electromagnetic calorimeters, which are usually the most resource intensive systems to simulate. This is illustrated in Figure~\ref{fig:sim:NRR}, which shows the number of steps in various volumes for several particle species~\cite{SIM-2019-001}. The idea of the Neutron and Photon Russian Roulette (NRR/PRR) method is to randomly discard particles below an energy threshold and increase the energy deposits of remaining particles accordingly. This strongly reduces the number of secondary particles generated in the showers that are simulated by \Geant{}. The final tuning used in production in \RunThr is a $2~\MeV$ threshold with a weight of 10 for neutrons and a $0.5~\MeV$ threshold with a weight of 10 for photons.
These settings result in a speed-up of about $10\%$ for $t\bar{t}$ MC events.

\subparagraph{G4GammaGeneralProcess}

The \texttt{G4GammaGeneralProcess}, introduced in \Geant{} version 10.6, is a super-process that hides all the physics processes involving photons (e.g.\ Rayleigh scattering, the photoelectric effect, the Compton effect, conversion to electron--positron pairs, and photo-nuclear scattering), providing a single access point to the |G4SteppingManager| that sees only one physics process. As a consequence, only one mean free path must be calculated for a photon, and therefore the number of instructions is reduced at the price of introducing extra physics tables shared between threads. Once validated and included in the production configuration, a speed up of 3\% was observed for $t\bar{t}$ MC events.

\subparagraph{Woodcock tracking}

Woodcock Tracking~\cite{WOODCOCK} is a tracking optimisation technique suited to be applied to highly segmented detectors where the geometry boundaries rather than the physical interactions limit the simulation steps. It is applied on top of the \texttt{G4GammaGeneralProcess}, and the idea is to track particles in a simplified geometry made of the densest material (i.e.\ without boundaries).

Woodcock Tracking introduces an additional, fictitious or $\delta$-interaction, which does not alter the initial state, with a macroscopic cross section $\sigma_{\delta}$ that can be expressed as

\begin{equation}
\sigma_{\delta} (E, \textrm{material}) = \textrm{const}. - \sigma_{\gamma} (E, \textrm{material}),
\end{equation}

\noindent where $\sigma_{\gamma}=\sum_{p} \sigma_{\gamma}^{p} (E,\textrm{material})$ is the total macroscopic cross section, summing up the cross sections for all possible interactions $p$ of a photon in material $\sigma_{\gamma}^{p}$, which itself is already simplified to one process within the \texttt{G4GammaGeneralProcess}.
Using this fictitious interaction, the macroscopic cross section $\sigma(E)$ is now constant and can be written as

\begin{equation}
\sigma(E) = \sigma_{\gamma} (E, \textrm{material}) + \sigma_{\delta} (E, \textrm{material}) = \textrm{const}.,
\end{equation}

\noindent which is constant also and especially if the material changes along a step. Using $\sigma(E)$ to sample the step length $s(E)$ until the next (real $\gamma$ or a $\delta$) interaction eliminates the need to stop at volume or material boundaries.
Therefore, volume or material boundaries can be ignored and the probability of a real interaction $P_{\gamma}$ can be calculated as:

\begin{equation}
P_{\gamma} (E, \textrm{material}) = \sigma_{\gamma} (E, \textrm{material})/\Sigma(E).
\end{equation}

By applying this technique, the number of cross section evaluations as well the number of steps caused by the crossing of a geometric boundary is drastically reduced, without compromising the physics results. Woodcock Tracking was implemented and tested to be used in the EM endcap calorimeter and benchmarks have shown a speedup of 17.5\% for $t\bar{t}$ MC events.

\subparagraph{VecGeom}

\textsc{VecGeom}~\cite{Wenzel:2754098} is a geometry modelling library with hit-detection features designed to support CPU optimisations such as data-level parallelism. It provides optimised and fast geometry primitives and navigation algorithms that perform well especially for complex geometric shapes.
The detector geometry used by \Geant{} is built from classes inheriting from \texttt{G4Solid}, each representing a different shape (or set of shapes). It is possible to replace specific \texttt{G4Solid} classes with their \textsc{VecGeom} equivalents. The optimal set for the ATLAS geometry was found to be cones, tubes, and polycones. The use of \textsc{VecGeom} for these shapes gives a speed-up that varies depending on the CPU model from 2\%--7\% for $t\bar{t}$ MC events.

\subparagraph{\Geant{} static linking}
\label{sec:sim:staticlinking}

\emph{Static linking}~\cite{Marcon:2813807} is a purely technical optimisation that targets the way \Geant{} is linked and used within Athena. By default, Athena builds many small shared libraries, one per package of code, which are dynamically loaded at runtime (see Section~\ref{subsec:genInfra:cmake}). Different build types were compared and tested for performance: dynamic (the default multi-library configuration used in Athena during \RunTwo), single dynamic library, and static linking.\footnote{More information about shared libraries is available in Ref.~\cite{Drep11}.}  Tests with the single dynamic library resulted in a slowdown in the execution time, which can be ascribed to the trampoline/lookup table mechanism of dynamic linking. Each call to a function in a dynamic library takes advantage of a trampoline that reads the memory address of the called method from a lookup table and passes it to the calling function. This results in an increased number of calls and jumps, which slows down the simulation execution. The static linking has proven to be the best performing build type. To enable it in Athena, all the packages that link to \Geant{} were reorganised into one large library that then can be linked statically with \Geant{}. During execution, although the remainder of Athena continues to use dynamic libraries, function calls internal to this static library benefit from using the known function code locations within a statically linked library and avoid the overhead from lookups. Benchmarks have shown a gain of about 5\%--7\% for $t\bar{t}$ MC events.

\subparagraph{EMEC geometry optimisation}

The EM endcap calorimeter (EMEC) is built in a complicated `Spanish Fan' geometry, which could not be efficiently described with the geometric primitives available in early versions of \Geant. The geometry was therefore described with custom solids implementing the geometry algebraically.
In \RunThr, the code describing the EMEC detector was optimised and two new variants were introduced, besides the nominal one called the Wheel variant:

\begin{itemize}
\item Cones: this version reduces the use of |G4Polycone|s by the introduction of an improved shape (|G4ShiftedCone|). In this configuration, the outer wheel is divided into two conical sections.
\item Slices: this variant reduces the time needed for geometry navigation calls by dividing the inner (outer) wheel into 14 (21) thick slices along the $z$-axis.
\end{itemize}

The Slices variant was found to be the best performing, bringing a speedup of about 5\%--6\% for $t\bar{t}$ MC events.

\subparagraph{Magnetic field tailored switch-off}

The solenoid field in the inner detector returns through the iron yoke supporting the tile calorimeter, resulting in a very small residual magnetic field within the bulk of the calorimeter volume.
It is therefore possible to switch off the magnetic field in the LAr calorimeter for all particles except muons without significantly affecting the shower shapes.
In addition to the small field value, the effect of the magnetic field on the electron trajectories is dwarfed by the effect of multiple scattering.
However, the calculation of the showering electron and positron trajectories in the magnetic field still requires significant CPU time. This filter on the detector region and particle type was implemented and integrated to be used in production for \RunThr. Tests with 200 \ttbar-production events have shown a speedup of about 3\% for $t\bar{t}$ MC events.
The same concept could be further exploited for other parts of the detector.

\subparagraph{The frozen showers method}
\label{sec:sim:frozenshowers}

The frozen showers method~\cite{Bar09} speeds up the simulation of showers in the calorimeters by replacing low energy electrons, photons and neutrons with pre-simulated showers. This mostly affects secondaries that are produced in huge numbers in high-energetic showers. The showers are simulated with \Geant{} until the energy of the particles produced in the shower falls below the energy threshold, at which point HITs are generated from the shower library. This approach was already used for the forward calorimeter in the \RunTwo MC simulation production. The energy thresholds are 1~\GeV for electrons, 10~\MeV for photons, and 100~\MeV kinetic energy for neutrons; below these values, showers from libraries are used. The pre-simulated showers are stored in \textsc{ROOT} libraries, binned in $\mid\!\eta\!\mid$ and distance from the closest rod\footnote{The forward calorimeter can be thought of as a large, solid tube with many holes in it. These holes are filled with small rods, and the gaps between the rods and the tube are filled with liquid argon. It is the distance to these rods that is of relevance for the frozen showers.} centre; different libraries are also used for the first and second hadronic compartments of the forward calorimeter. The library is derived from a \ttbar-production event sample simulated with the \Geant{}-based full simulation.

In the first \RunThr MC campaign (MC21) the frozen showers library was derived by scaling the energy of the showers in the \RunTwo library to reproduce the energy scale observed in a \RunThr{} \Geant{}-based full simulation sample. The energy thresholds and the bins were kept the same as in \RunTwo. A new \RunThr library was developed in 2022, with revised bins, providing a more accurate modelling of the \Geant{}-based full simulation without the need for energy scale tuning. The improved library is used in MC23. In all campaigns, the use of frozen showers reduces the CPU required to simulate high-energy (several hundred \GeV) electrons and photons in the forward calorimeter by a factor of three, and reduces the overall simulation time for \ttbar-production events by about 25\%.

\subparagraph{Summary of full simulation optimisations}

\begin{figure}[tbp]
\centering
\includegraphics[width=0.6\textwidth]{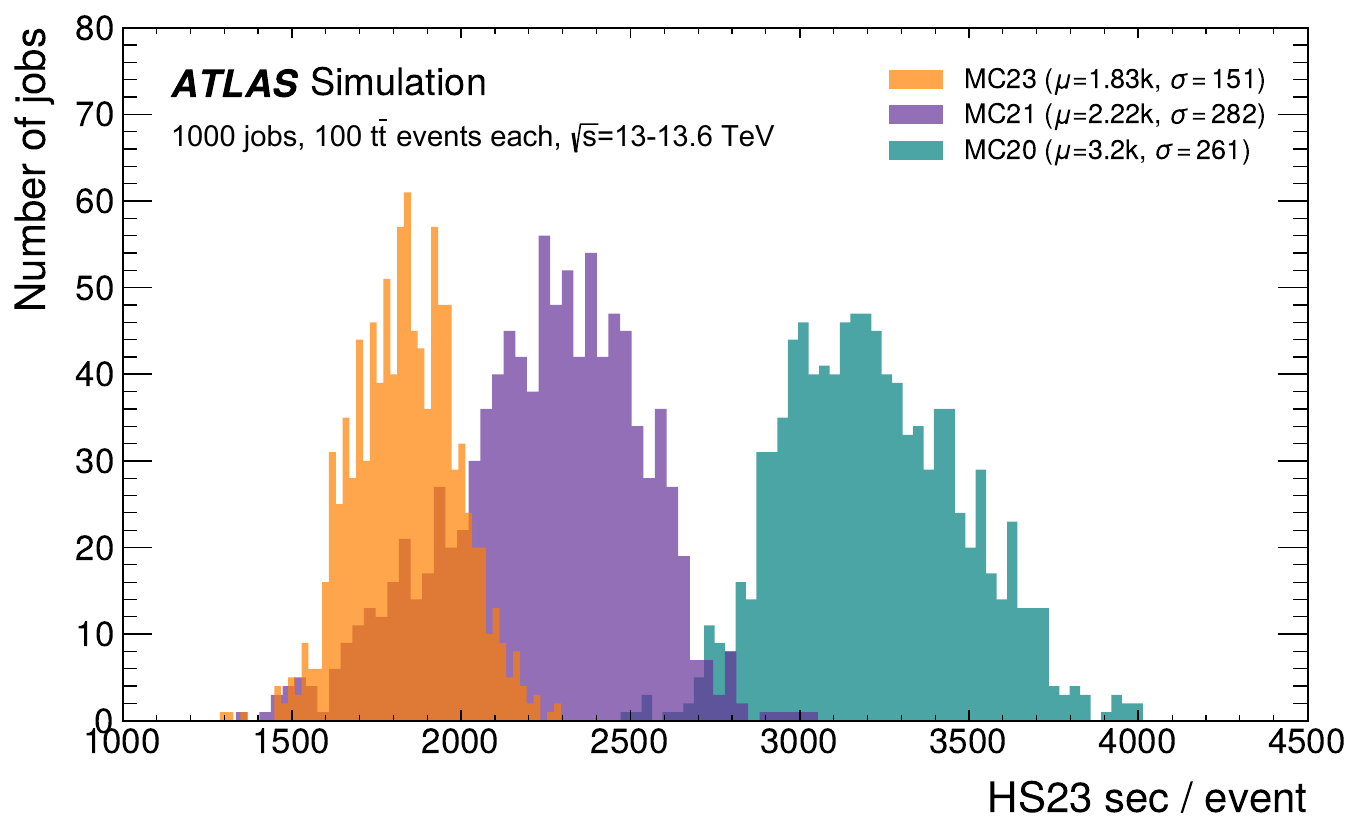}
\caption{Distributions of ATLAS \Geant{} detector simulation HS23 sec per event using the nominal and the optimised versions of the software by the \Geant{} Optimisation Task Force for the MC20, MC21 and MC23 campaigns, corresponding to a centre-of-mass energy of 13--13.6~\TeV. The benchmarks comprise 1000 jobs simulating 100 $t\bar{t}$ events each at the Brookhaven National Lab Tier-1 cluster. The mean value ($\mu$) and standard deviation ($\sigma$) of the distributions are also indicated.}
\label{fig:sim:G4CPUopt}
\end{figure}

The MC21 simulation campaign included the optimisations described above in this Section. The MC23 simulation campaign added the Woodcock Tracking optimisation. The CPU required to simulated one \ttbar-production event was reduced by 48\% (36\%) relative to \RunTwo in the MC23 (MC21) campaign, as shown in Figure~\ref{fig:sim:G4CPUopt}. This reduction allows the simulation of 92\% (55\%) more events with the same CPU resources for MC23 (MC21).

Each of these changes to the simulation is put through a rigorous validation of physics performance (see Section~\ref{sec:validation}). Only those changes that were found to not alter the physics performance, and therefore to not affect the agreement between data and MC simulation, are put into production.

\subsubsection{Fast simulation \AF{}}
\label{sim:fastsim}

\paragraph{Overview}
Even after the optimisations discussed in the previous section, the full simulation of the detector requires considerable CPU resources. For this reason, ATLAS has developed tools to replace the calorimeter shower simulation, which is the most CPU intensive step, with faster simulation methods. In \AF{}~\cite{SIMU-2018-04}, the simulation of hadrons, photons and electrons in the calorimeters is handled by a combination of two fast simulations tools; \texttt{FastCaloSimV2}, which uses a parametric approach, and \texttt{FastCaloGAN}, which uses generative adversarial networks (GANs). \texttt{FastCaloGAN} is among the first tools based on generative models used for production in a large HEP experiment. \texttt{FastCaloGAN} was developed later than \texttt{FastCaloSim}, which struggled to reproduce lateral correlations in hadronic showers, and overcame that challenge using ML methods. Each tool is applied in the kinematic region in which it provides the best performance~\cite{SIMU-2018-04}, after detailed validation and comparisons of physics observables. These tools replace the slow propagation and interactions of incident particles with the direct generation of energy deposits in the calorimeters. For the average ATLAS MC simulation event, \AF{} requires only 20\% of the CPU of the full simulation, where most time is spent on the simulation of the inner detector with \Geant{}.

\AF{} (as deployed in \RunTwo) was initially tuned to reproduce the output of the full simulation for \RunTwo as closely as possible. The update of the \Geant{} version used in \RunThr simulation implies that the shape of EM and hadronic showers is sufficiently different between \RunTwo and \RunThr full simulation samples that the parameterisation used by \texttt{FastCaloSimV2} and the training of the neural networks used by \texttt{FastCaloGAN} needed to be updated accordingly. For the \RunThr version of \AF{}, several qualitative improvements have also been developed, in particular for \texttt{FastCaloGAN}, resulting in an upgraded version of the tool now called \texttt{FastCaloGANV2}. These improvements are listed below.

The training of the GANs and the parameterisation used for \texttt{FastCaloSimV2} are based on single particles that are simulated with \Geant{}. This training is performed separately for various particle types and in fine bins of $\eta$, because the detector geometry and material changes strongly with $\eta$. The parameterisation is separately performed for 17 different energy values; the GANs in \texttt{FastCaloGANV2} are trained for two energy ranges. The inputs for the lateral shower shape modelling are \emph{HITs} (point-like localized energy deposits) for \texttt{FastCaloSimV2} and \emph{voxels} (small, regularly-sized volumes in which energy deposits are integrated) for the GANs. The granularity of voxels was optimised and is finer than that of the calorimeter cells, which improves the modelling.

\begin{figure}[tbp]
\centering
\includegraphics[width=0.95\textwidth]{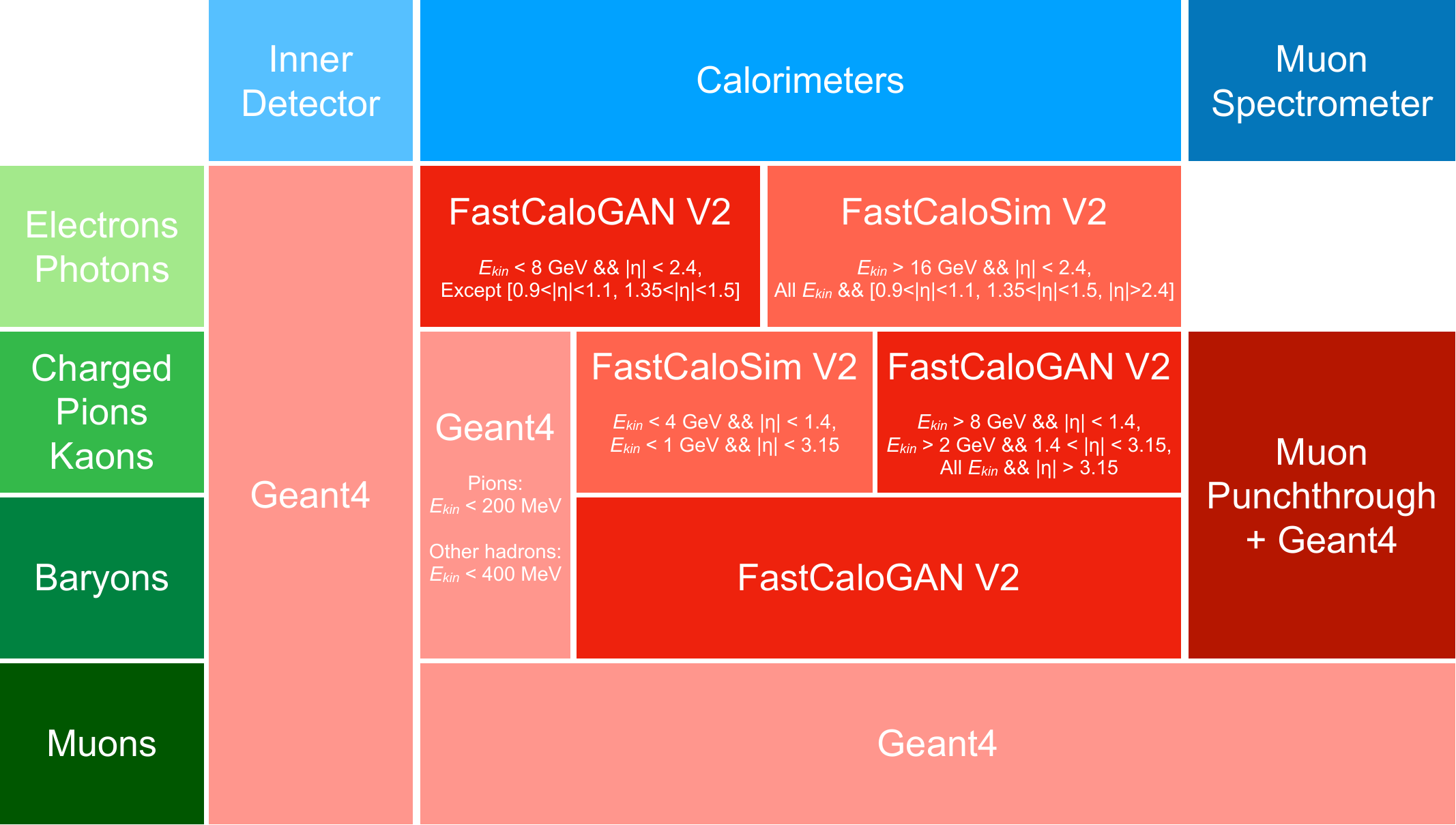}
\caption{The configuration of tools that together form \AF{}. The tools (\Geant{}, \texttt{FastCaloSimV2} and \texttt{FastCaloGANV2}) are combined in a way that the physics performance of \AF{} is closest to \Geant{}, which depends on the particle type, $\eta$ slice and kinetic energy ($E_\textrm{kin}$) range. The inner detector and muon spectrometer are simulated using only \Geant{}.}
\label{fig:sim:af3config}
\end{figure}

For each particle type and energy, the tool that reproduces the \Geant{} simulation output with the best accuracy is used. The combination of the two tools was reoptimised in \RunThr, and is illustrated in Figure~\ref{fig:sim:af3config}. \Geant{} is still used for the simulation of all particles in the inner detector, for muons and very low energy hadrons in the calorimeters, and in the muon spectrometer. \texttt{FastCaloGANV2} is used for simulating baryons (except at very low energies), low-energy photons and electrons, and higher-energy pions. \texttt{FastCaloSimV2} is used for pions with lower-energy and higher-energy electrons and photons.

High-energy hadrons may interact late --- or even not at all --- in the calorimeter. The resulting spray of hadrons into the muon spectrometer is known as \emph{punch-through}. The particles produced in these showers are now modelled with a new tool based on deep neural networks (DNN). This tool can predict the probability of a punch through occurring better than was possible in the \RunTwo version of \AF{}.

\begin{figure}[tbp]
\centering
\includegraphics[width=\textwidth]{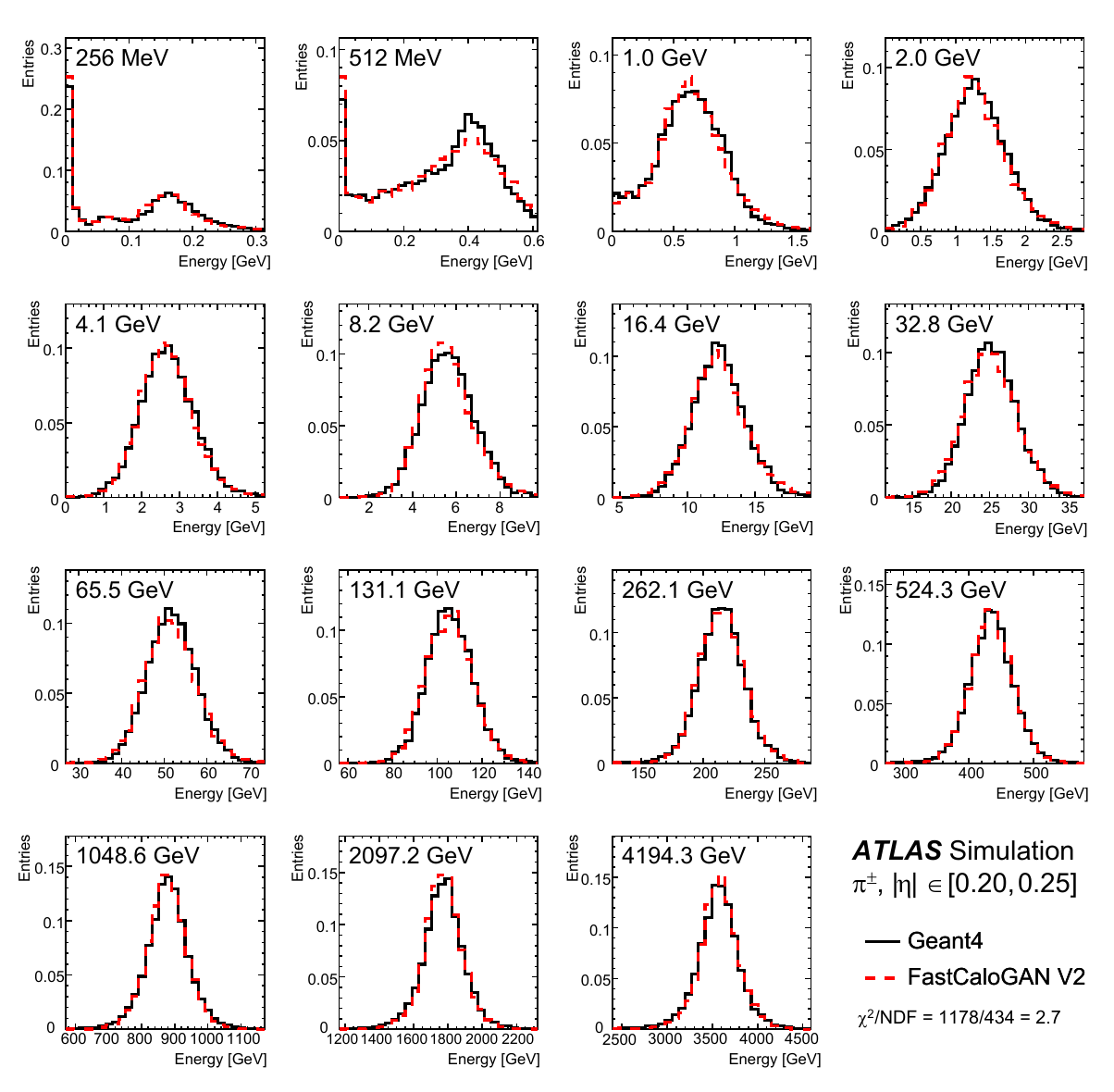}
\caption{\label{fig:sim:af3performance:pions}The performance of \texttt{FastCaloGANV2} for single pions in an $\lvert\eta\rvert$ range of 0.2--0.25. The energy distributions from \Geant{} are well reproduced for all energy values. The single pion truth energy value is indicated in each figure.}
\end{figure}

\begin{figure}[tbp]
\centering
\subfloat[]{
\includegraphics[width=0.42\textwidth,valign=c]{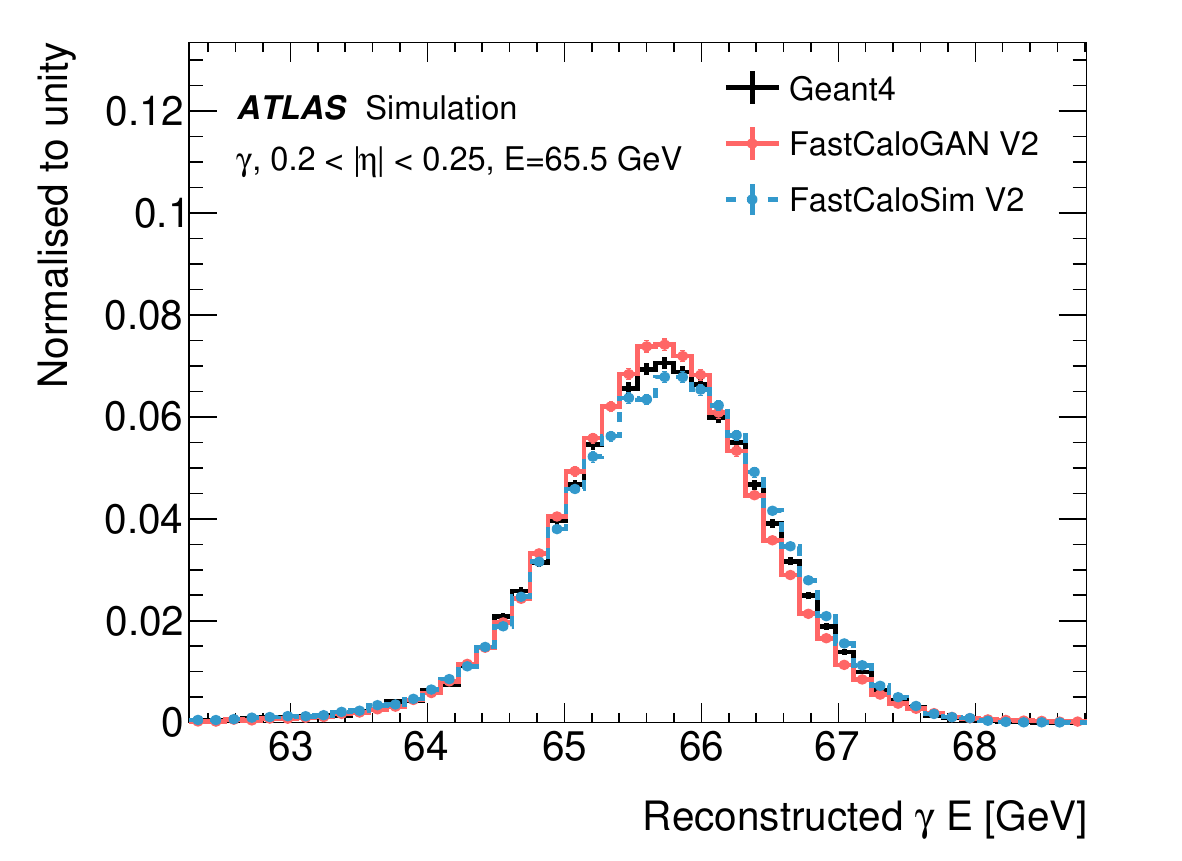}
}
\subfloat[]{
\includegraphics[width=0.42\textwidth,valign=c]{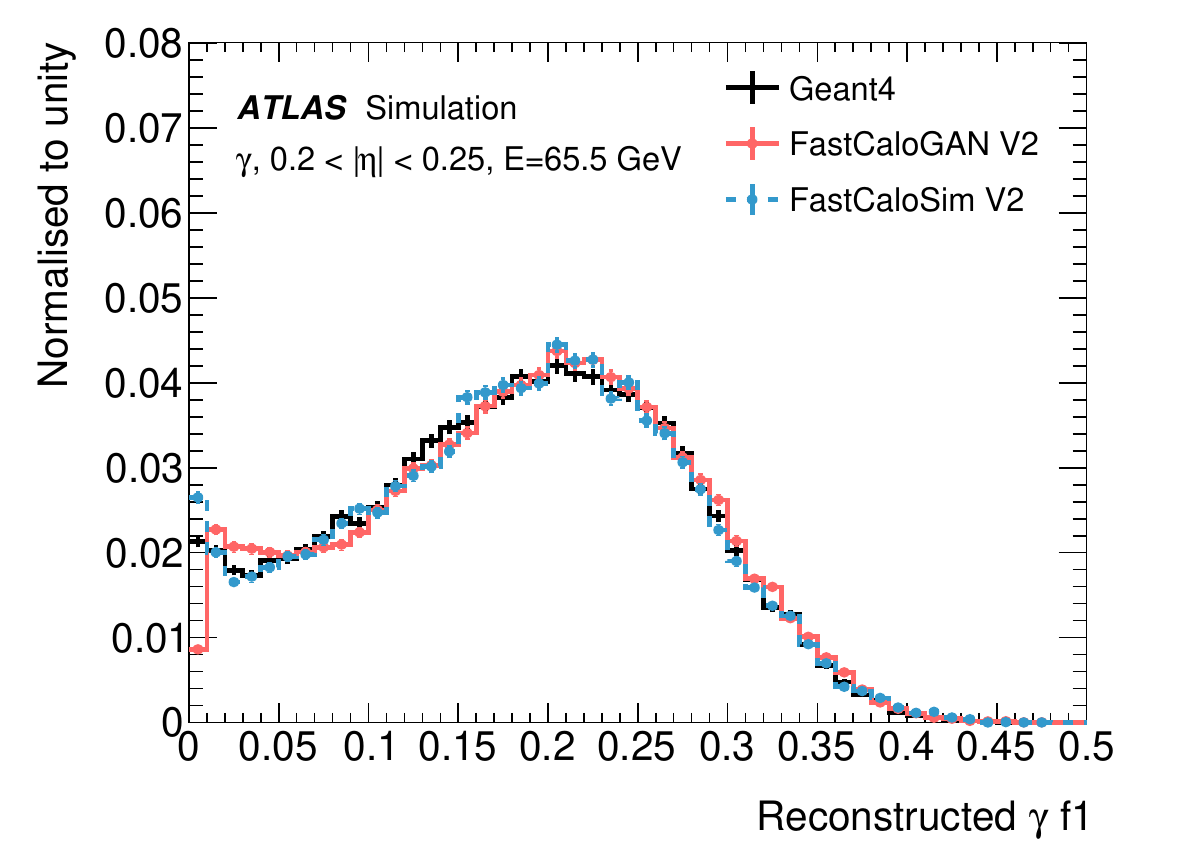}
} \\
\subfloat[]{
\includegraphics[width=0.42\textwidth,valign=c]{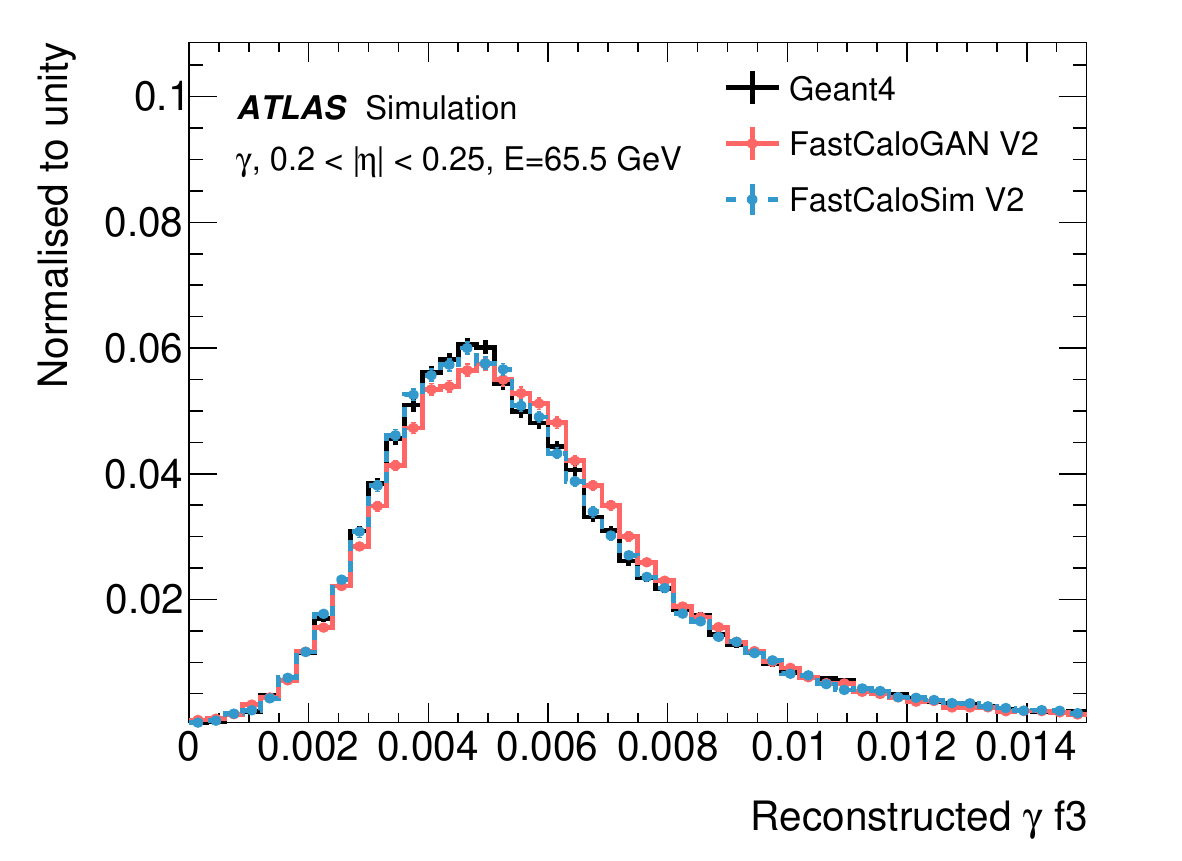}
}
\subfloat[]{
\includegraphics[width=0.42\textwidth,valign=c]{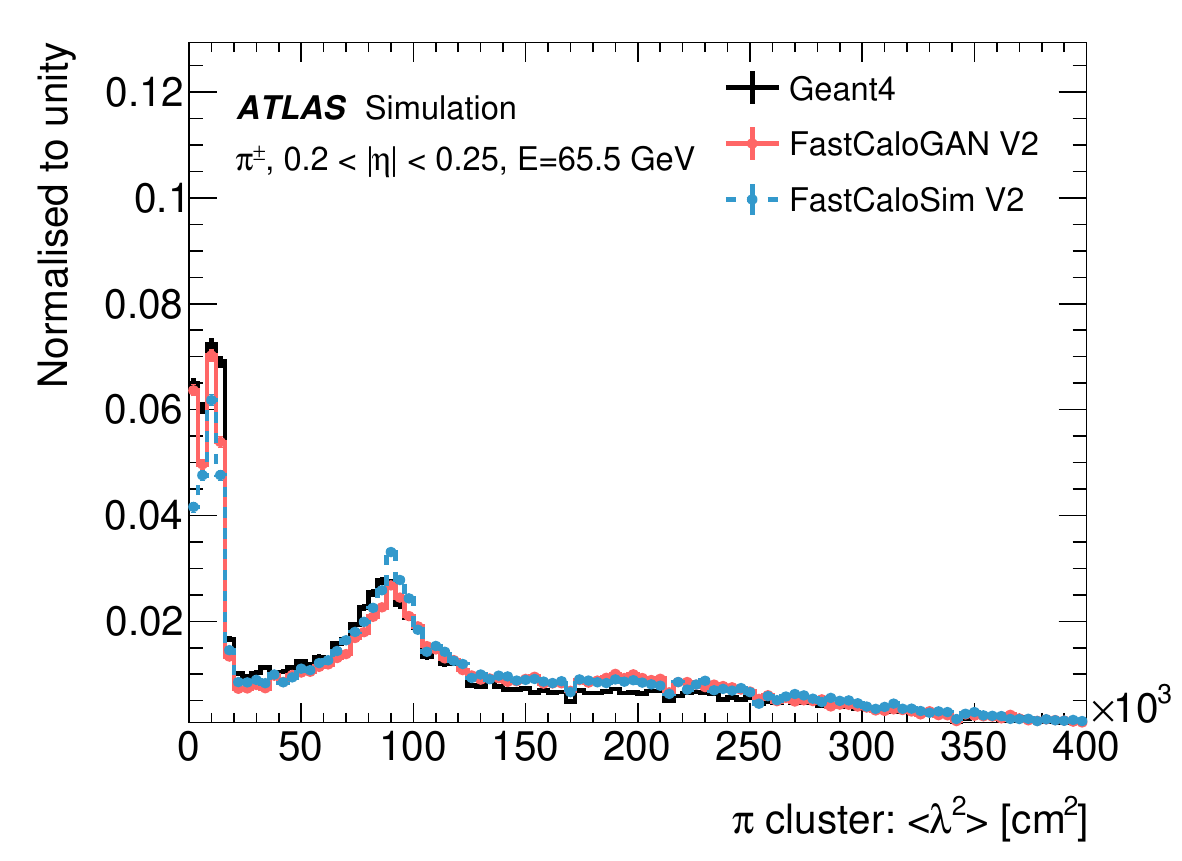}
} \\
\subfloat[]{
\label{fig:pisecondmomentr}
\includegraphics[width=0.42\textwidth,valign=c]{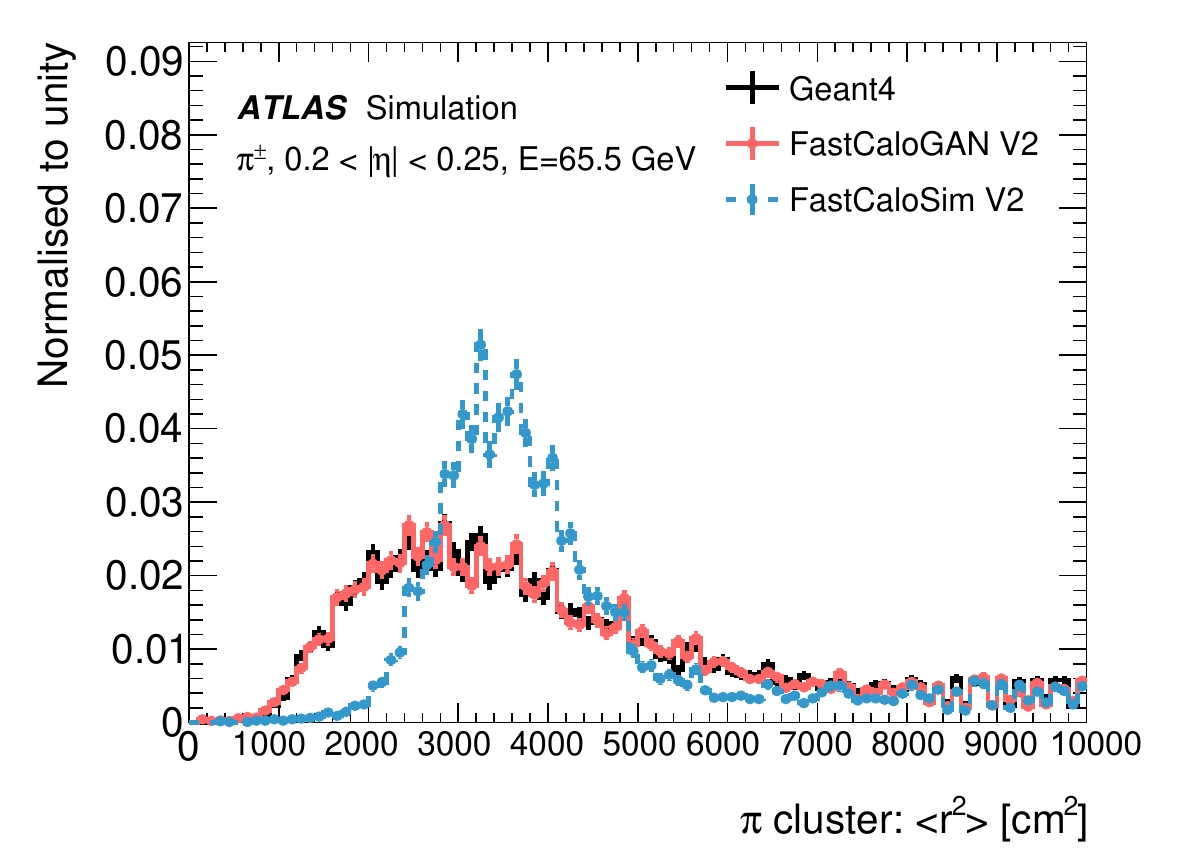}
}
\subfloat[]{
\includegraphics[width=0.42\textwidth,valign=c]{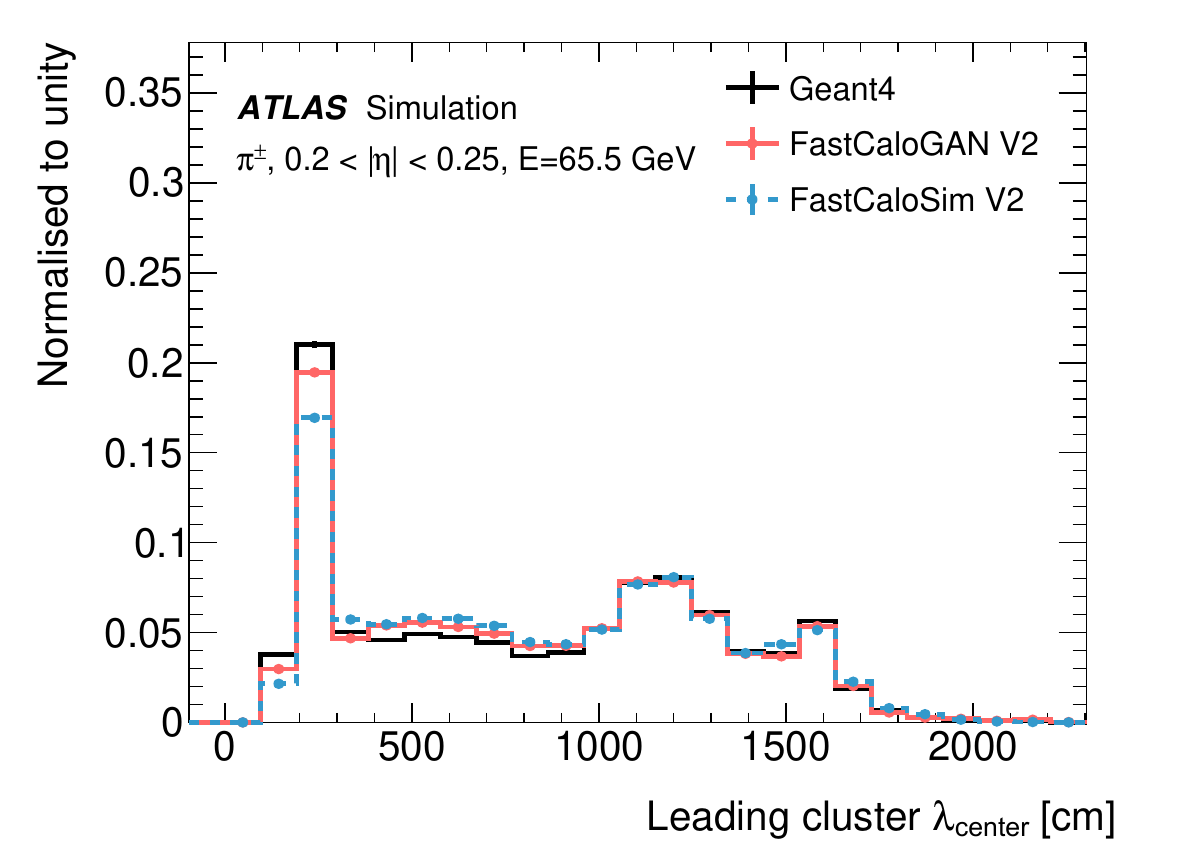}
}
\caption{\label{fig:sim:af3performance:ph_pion}
(a)--(c): The energy of reconstructed central photons with a true energy of 65.5~\GeV. (a) displays the total energy, (b) the energy fraction deposited in the first calorimeter layer (relative to the total energy), and (c) the energy fraction in the third calorimeter layer. (d)-(f): Shower moments for reconstructed central pions with a true energy of 65.5~\GeV. (d) displays the second moment in $\lambda$, (e) the second moment in $r$, and (f) the $\lambda_{center}$. \Geant{} (black) is compared with \texttt{FastCaloGANV2} (red) and \texttt{FastCaloSimV2} (blue).}
\end{figure}

For most distributions of properties of objects used in physics analyses, \AF{} and \Geant{} agree within a few percent. The agreement is much better than that achieved with the previous generation of fast simulation, \AFII{}~\cite{SOFT-2010-01}, with key improvements in the modelling of the forward calorimeters and a better fluctuation model that enables in particular the simulation of substructure within jets~\cite{SIMU-2018-04}. Comparisons of energy distributions obtained with the \RunThr version of \AF{} to those of \Geant{} are presented in Figure~\ref{fig:sim:af3performance:pions} for single pions. Further comparisons for reconstructed single photons and pions are shown in Figure~\ref{fig:sim:af3performance:ph_pion}. The comparisons for single pions display shower moments as defined in Ref.~\cite{showermoments}, where $\lambda$ is the energy-weighted distance of a cell from the shower centre along the shower axis, and $r$ is the energy-weighted distance of a cell from the shower centre perpendicular to the shower axis. Figure~\ref{fig:pisecondmomentr} in particular shows an example of a distribution with significantly better modelling in \texttt{FastCaloGANV2} compared with \texttt{FastCaloSimV2}; these distributions were among those that motivated the choice of the scheme shown in Figure~\ref{fig:sim:af3config}.

\paragraph{\texttt{FastCaloGANV2} improvements for \RunThr}
\label{sec:sim:af3:GANrun3improvements}

Several improvements were made to \texttt{FastCaloGAN} that are incorporated into the upgraded version called \texttt{FastCaloGANV2}, which is used in the \RunThr version of \AF{}:

\begin{itemize}
\item The number of volumes in which \Geant{} energy deposits are grouped (voxels) was optimised, reducing extrapolation in voxel-to-cell energy assignment.

\item The bias in the energy of HITS generated with a small simulation step, described in Ref.~\cite{SIMU-2018-04}, was corrected, resolving a discrepancy in the reconstructed electron and photon total energy.

\item A similar correction is applied to the hadrons but the energy at which the shower is rescaled is derived from the cell energies rather than the \Geant{} HITs.

\item The $\phi$-modulation correction~\cite{SIMU-2018-04} related to the accordion structure of the EM calorimeter is now corrected before training.

\item The architecture of the networks and the hyperparameters were optimised for each particle, energy range, and detector region.

\item Separate trainings are done for low- and high-energy electron and photon showers, and the minimum energy was lowered from 256~\MeV to 64~\MeV, matching the value used in \texttt{FastCaloSimV2}.

\item The training strategy was updated to use the full set of initial particle energy ranges from the first iteration, rather than starting from a single range and slowly adding more energies, as was done previously.

\item The HIT-to-cell assignment was improved by correcting for the average lateral energy distribution within each voxel. More energy is deposited closer to the centre of the shower as occurs in \Geant{}.

\item A minimum energy of 10~\MeV is used for HITs in electron and photon showers as in \texttt{FastCaloSimV2}. No changes were introduced for pions, i.e. the total energy assigned to a voxel is split between the HITs without any minimum energy criteria.

\item The GANs also benefit from a new correction for the longitudinal position of the HITs within a layer. This is implemented using a DNN extrapolation method trained on the energy deposited in each layer.
\end{itemize}

In addition to these improvements, 100 new GANs were trained using protons to parameterise their energy response in the calorimeter; these GANs are then used to simulate baryons. This method replaces the previous approach in which only the total energy was corrected and thus enables consideration of the different shape of the showers between pions and baryons. This new feature significantly improves the simulation of low energy ($<10$~\GeV) baryons. Kaons are still simulated based on pions, since they have similar shower shapes.

\paragraph{Outlook}
\label{sec:sim:outlook}

Several further improvements to the \AF{} tools are being developed. These include: %

\begin{itemize}
\item For fast calorimeter simulation, the use of variational autoencoders and other methods such as flow or diffusion models are investigated and the performance compared with the GANs.
\item A new voxelisation is being investigated with the aim to improve the description of showers in the highly granular regions of the EM calorimeter where the relatively course voxelization is expected to play a more significant role.
\item A low-energy parameterisation is planned to address the simulation of pions with kinetic energy below 200~\MeV, currently done with \Geant{}; this will further speed up the simulation.
\item The memory footprint of \texttt{FastCaloSimV2} will be optimised, and the memory of the GANs will be reduced using ONNX instead of LWTNN for the inference of the models.
\item In the \AF{} infrastructure, the fast simulation will be integrated more tightly with \Geant{} using the \texttt{G4VFastSimulationModel} class. The reduction of custom code is expected to make the fast simulation code less brittle and easier to develop and maintain.
\end{itemize}

In addition, an even faster simulation will be delivered by the FastChain~\cite{Basalaev_2017} project, which includes two research activities that are still planned for deployment in \RunThr. \Fatras{} (Fast Tracker Simulation)~\cite{ATL-SOFT-PUB-2008-001,ATL-SOFT-PUB-2014-001} aims to replace the propagation of particles in the inner detector using \Geant{} with a simplified treatment of particle interactions with the simpler detector description model used in track reconstruction (see Section~\ref{sec:dd}); preliminary results show that the CPU time needed to simulate \ttbar-production events can be reduced by a factor of 30 over that of \AF{}. Another speedup as a part of FastChain is the treatment of pile-up simulation, using a technique called Track Overlay, described in Section~\ref{digi:sec:trackoverlay}.


\subsection{Digitisation}
\label{sec:digi}

After simulation using the \Geant{} toolkit or \AF{}, all energy deposits in ATLAS sub-detectors sensitive volumes are stored in a file format called \texttt{HITS}. The digitisation code reads \texttt{HITS}\footnote{Time information is also included in \texttt{HITS} files. For all sub detectors, the time of flight at the speed of light from the interaction point is subtracted from the \Geant{} HITS' time during digitisation.} and emulates the detector response, producing an output (\emph{digits}) that conceptually mirrors the real data detector response (typically voltages or times on pre-amplifier outputs), with the addition of some truth and metadata information that is then altogether stored in RDO output files. In addition, the inputs for the hardware trigger are produced during the digitisation step. This operation is strongly specific to the different sub-system technologies used in ATLAS. Section~\ref{digi:sec:subsystem} describes the sub-system specific code, with an emphasis on the new features introduced in \RunThr.\\
In digitisation, the treatment of the proton--proton collision of interest, often referred to as the signal or hard-scatter collision, is typically separate from the treatment of the additional proton--proton background collisions (pile-up). In Section~\ref{digi:sec:PU} the treatment of pile-up during the LHC fill in \RunThr collisions and various techniques to merge it with the hard-scatter information are also described.

\subsubsection{Sub-system digitisation code}
\label{digi:sec:subsystem}

\paragraph{Pixel digitisation}
\label{digi:sec:pixel}

The pixel digitisation takes the energy deposits from charged particles in the silicon wafers (as simulated by \Geant{}) and converts them into time-over-threshold (ToT) values on pixel sensor electrodes. The first step of the pixel digitisation process is to divide the ionisation energy deposit from each \Geant{} HIT into a maximum of twenty ionisation energy deposits using the Bichsel model~\cite{RevModPhys.60.663} to correct for possible straggling in thin silicon.

As the integrated luminosity delivered by LHC since the installation of the pixel detector increases, the effects of radiation damage in the silicon bulk on the detector response become increasingly important. The pixel detector digitisation for \RunThr\ includes for the first time in production the effects of radiation damage in the silicon sensors~\cite{IDET-2017-10}. This represents a valuable tool to understand and predict radiation damage effects and their relation to the performance of physics object reconstruction.

The radiation damage digitiser computes the signals induced by the charge carriers produced by ionising particles in \Geant{}~\cite{Agostinelli:2002hh} by using precise electric field, Lorentz angle and weighting potential maps, taking into account carrier-trapping and diffusion effects. The electric field distribution for a given applied bias voltage after irradiation at a given fluence are taken from detailed TCAD (Technology Computer Aided Design) simulations. The predicted electric field is used to calculate the expected time spent by charge carriers to reach the collecting electrode, via the carrier-mobility relation. The induced charge on the pixels is calculated from the initial and trapped positions using the weighting potential map and including charge sharing effects. The total induced charge is converted into a time-over-threshold value used for clustering in the reconstruction.

\begin{figure}[!tbp]
\begin{center}
\includegraphics[width=0.65\textwidth]{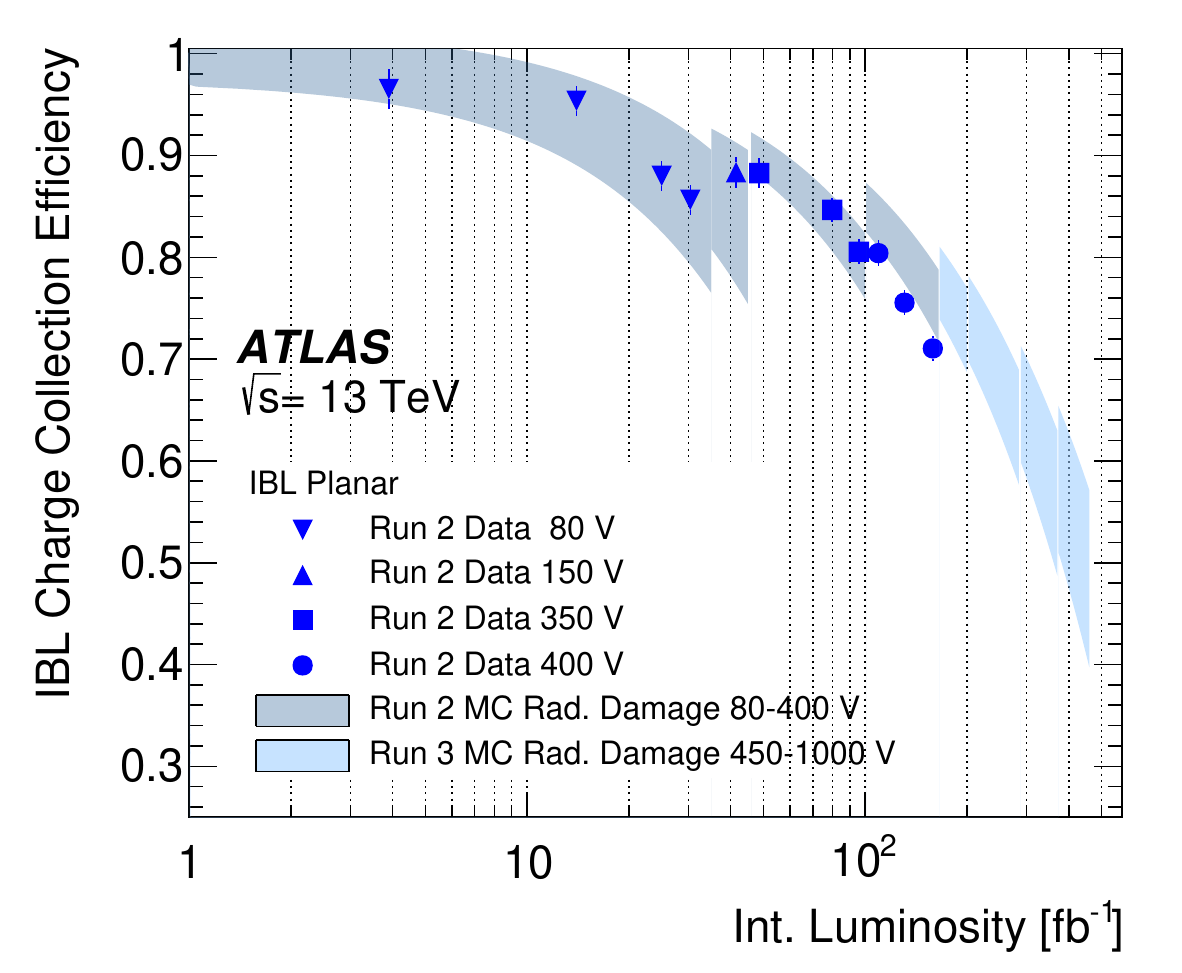}
\caption{\label{fig:IBL_CCE} Charge collection efficiency as a function of integrated luminosity for IBL planar sensors for \RunTwo\ data and the ATLAS radiation damage simulation for \RunTwo\ and \RunThr. The points represent the data and the bands the simulation predictions. The parametric uncertainty in the simulation defining the width of the bands includes variations in the radiation damage model parameters and the uncertainty in the luminosity-to-fluence conversion. Horizontal error bars on the data points due to the luminosity uncertainty are smaller than the size of the markers. Figure from Ref.~\cite{ATL-PHYS-PUB-2022-033}.}
\end{center}
\end{figure}

The IBL charge collection efficiency variation as a function of the integrated luminosity measured in data and that predicted by the radiation damage simulation from the start of \RunTwo\ is presented in Figure~\ref{fig:IBL_CCE}~\cite{ATL-PHYS-PUB-2022-033}. For \RunThr, the track reconstruction algorithms were re-tuned to at least partially mitigate the performance degradation caused by the radiation damage~\cite{ATL-PHYS-PUB-2022-033}.

The charge calibration of the reconstruction involves re-calculating the charge from the ToT values. While ample data are available to calibrate this procedure at moderate ToT values, for very large and very small ToT values an extrapolation or fit function must be used. For \RunThr, the pixel charge calibration procedure for the IBL was modified to avoid the fitting step of the charge and ToT values that may cause biases at low and high charge values. The new IBL calibration procedure implements look-up tables that use the average of the injected charge values on the pixels of a front end chip corresponding to each ToT calibration point. These look-up-tables are stored in the calibration database and applied to detector data. In simulation, the ToT is extracted by linear interpolation of the input charges and rounded to an integer value.

\paragraph{SCT digitisation}

The SCT digitisation process begins with the conversion of the energy deposited on the silicon wafer by each particle into a charge on the readout electrodes. The energy deposited in each \Geant{} tracking step is divided uniformly into 5~$\mu m$ sub-steps and converted into an electron--hole pair for each 3.63~\eV of deposited energy. The charges are drifted towards the electrodes taking into account the diffusion and the Lorentz angle, which are calculated assuming a uniform electric field in the sensor.

The final stage of digitisation is the simulation of the response of the readout electronics. The amplified signal of a charge arriving at a readout strip is calculated for three different readout time intervals, corresponding to three consecutive bunch crossings. The cross-talk between neighbouring strips is also taken into account and estimated from the shape of the main output pulse of the strips. Electronic readout noise is generated independently in each time interval from a Gaussian distribution based on the data measurements and added to the strips. The nominal readout threshold, above which the signal is recorded on a single strip, is set to 1~fC according to the data acquisition configuration.
To reproduce the noise occupancy observed in the data, random strips are added to the readout list from among those with no deposited charge.

\paragraph{TRT digitisation}

The TRT digitisation software~\cite{Kittelmann:2224292} converts HITs produced by the \Geant{} simulation into digits that correspond to the detector readout signal for each straw. Each digit consists of a timed bit-pattern spanning 2.5 bunch crossings. This bit-pattern contains 20 low-level threshold bits (3.125 ns/bit) that are set if the signal exceeds a low threshold in the corresponding time interval. One high-threshold bit is set if the high threshold is exceeded at any time during the 25 to 50 ns time interval corresponding to the bunch crossing of interest.

The signals in each straw are simulated in detail. For charged particle tracks, energy clusters are created and placed randomly along the path of the track in the active gas volume.
The number of clusters is calculated by sampling a Poisson distribution with the most probable value set to the mean free path of the particle. For photons from transition radiation and bremsstrahlung, a single energy cluster is created. The number of drift electrons in a cluster is determined from the energy of the cluster. Some electrons can be recaptured stochastically as they drift towards the anode, where the number of surviving electrons is determined by sampling a Binomial distribution with a survival probability of 0.4. The time taken for each surviving electron to arrive at the anode wire is determined from the electron drift length and includes a detailed mapping of the magnetic field that causes the electron trajectories to bend.
Each electron that arrives at the anode wire cascades and induces a signal whose amplitude is simulated as a random sample of an exponential distribution. The propagation time for the signal to arrive at the front-end electronics and the attenuation in the wire are determined taking into account that half of the signal travels directly to the front-end while the other half travels in the opposite direction and reflects.

The signal from each drift electron is superimposed and convolved with signal shaping functions (separately for low and high threshold and for xenon and argon gas mixtures). Electronic noise is added and the signals are then discriminated against low and high thresholds in appropriate time slices and encoded into the output signal that is saved in the RDOs.

In earlier LHC operations where detector occupancy was relatively low, digits were written to storage in (digit, straw identifier) pairs only for hit straws. To reduce the size of the TRT RDO at higher LHC luminosity where most TRT straws are hit, this scheme was changed to sequentially write out digits for all straws omitting the straw identifier and use a detector map to assign the straw identifier in the reconstruction step. To further reduce RDO size, and CPU time,
the simulation of noise in straws with no hits is removed. To account for inaccuracies in high threshold probability due to the overlay procedure, corrections are applied during overlay by randomly adding high-threshold bits with a probability that increases with occupancy.

\paragraph{LAr digitisation}

The energy deposited in the liquid-argon gaps of the LAr calorimeters~\cite{LARG-2009-01} induces an electrical current that is proportional to the deposited energy. The signal is then amplified and shaped in the front-end readout electronics~\cite{Abreu2010} using three different gains to cover the large dynamic range of the signals of interest. The output is then sampled at the LHC clock frequency and converted to ADC (Analog to Digital Converter) counts. The digitisation~\cite{ATL-LARG-PUB-2007-011,SIMU-2020-01} emulates the detector readout, converting the deposited energy for each bunch crossing to ADC counts, taking into account the ADC gain, the LAr sampling fraction and the energy calibration of each cell. The ADC conversion also considers the time of each event relative to the hard scatter collision time (\emph{cell-timing}).
During MC overlay (see Section~\ref{digi:sec:overlay}), the ADC counts in the presampled RDOs are converted back in raw energy, which is then added to the energy from the hard scatter, for each time sample and cell. The combined energy is then converted back into ADC counts and the electronics noise is added. Optimal filtering coefficients (OFCs)~\cite{Cleland:2002rya} are then used in the reconstruction to compute the energy per cell, as described in Ref.~\cite{LARG-2009-01}.

To save disk space, the identifiers of the channels are not stored; instead, a vector with a length equal to the number of readout channels is created, and for each channel two bits are used to store whether that channel has a signal and, if so, what readout gain was used.\footnote{The detector readout uses three different gain settings to more precisely read out a wide range of energies. The appropriate gain is chosen per cell, based on the energy in that cell.} Another vector stores the ADC values of the channels with a signal. The number of samples is assumed to be the same for all readout channels and so it is only stored once.

As part of the activities of the Phase-I (\RunThr) trigger upgrade, it was suggested in Ref.~\cite{Enari_2019} that the hardware trigger system for the calorimeter (L1Calo) could profit from introducing \emph{Super Cells}~\cite{Schwienhorst}. Super Cells have higher granularity than trigger towers~\cite{ATLAS-TDR-22} (which are $\delta\eta \times\delta\phi = 0.1\times 0.1$ in the detector central region) used during Runs~1 and 2, with dimensions further optimised to describe electron and photon showers. In the digitisation step of the electronics simulation, these are emulated by grouping together the HITs (energy--time pairs) that are related to all cells that form each of the Super Cells. Mapping tools allow such grouping based on the identifiers of the cells. The information is used to simulate pulses in all bunch crossings still prone to produce some signal in the event of interest (i.e.\ the present event at $t=0$), given the length of the liquid argon ionisation pulse in the detector.
The normalised expected pulse shape as recorded in the conditions database is multiplied by the estimated amplitude of the pulse, which is directly related to the energy content of the Super Cell and shifted by the HIT time, producing estimated samples. Samples from multiple HITs composing the same Super Cell are simply added together, forming the pile-up plus hard-scattering digits. Electronics noise, estimates of which are also saved in the conditions database, is added at this step. This way, digits with regular 25~ns samples of the signal are formed and recorded as part of the RDO content.

\paragraph{Tile digitisation}

For the tile calorimeter (TileCal), HITs contain energy deposits in scintillator tiles. Since all normal TileCal cells are read out by two photomultiplier tubes (PMTs), every energy deposit is split into two parts according to the optical model and stored in two independent |TileHit| objects that correspond to the two PMTs.
All energy deposits are accumulated in 0.5~ns time bins, and every |TileHit| contains a vector of time values that corresponds to non-empty time bins and a vector of total energy deposits in every time bin.

The first step in TileCal digitisation is the simulation of photo-statistics effects~\cite{ATL-TILECAL-PUB-2009-002}. According to measurements from beam tests, at nominal PMT gain 70 photo-electrons per 1~\GeV of deposited energy are created in a PMT. Energy stored in a |TileHit| is smeared according to a Poisson distribution with an average value equal to the number of photo-electrons created in a given PMT in a given event.

After that, every energy deposit is converted independently to a pulse using pulse shapes measured at the beam tests in 0.5~ns time steps~\cite{Tile_Readiness}.
Every channel is readout by two 10-bit ADCs, called low gain (LG) and high gain (HG), with a gain ratio of $1:64$ between them. Two pulses are created with amplitudes proportional to the value of the energy deposit, converted from \MeV to ADC counts, taking into account the ADC gain and TileCal sampling fraction.
The position of the pulse maximum is shifted according to the time value from the deposit. All individual pulses are summed up to construct final LG and HG pulses in every PMT, and these pulses are `digitised' at seven fixed times: 0, $\pm 25$, $\pm 50$, and $\pm 75$~ns, and two vectors of seven samples are created. After all the signals are summed up, a constant pedestal value of about 40--60 ADC counts (depending on the cell) is added to the resulting pulse.

To simulate electronics noise, Gaussian-distributed noise of about 1.5 (0.7) ADC counts is added to the HG (LG) samples. After the two final pulses are constructed, the LG ADC is dropped if the maximal amplitude in the HG ADC remains below 1023 ADC counts (which roughly corresponds to 10--12~\GeV). If the HG ADC saturates, the LG ADC is selected. So, at the end of the digitisation step a single |TileDigit| object with seven samples is created for every readout channel.
This |TileDigit| object represents the data that are sent from the on-detector electronics (located in the support drawers for the TileCal) to the off-detector electronics.

The next step is to simulate the behaviour of off-detector electronics, which, similarly to the case of the LAr calorimeter, reconstructs cell energies using OFCs. The results of the OFC application are stored in |TileRawChannel| objects and the |TileRawChannelContainer| with all 12,228 readout channels is written to the output RDO file~\cite{Fullana:816152}.

\paragraph{Muon system digitisation}

The digitisation of the muon spectrometer sub-systems installed for \RunOne (MDT, RPC, TGC and CSC\footnote{As explained in Section~\ref{sec:detector}, the CSC chambers were removed at the end of \RunTwo.}) was documented in Ref.~\cite{ATL-SOFT-PUB-2007-001}. The RPC digitisation code has not changed substantially since then, while some noticeable improvements were made to the TGC digitisation code: a new timing calculation is used in the bunch crossing identification, considering the position dependence of the signal propagation to the front end; the bunch identification is done for a 4-bunch readout from the previous to the next-to-next bunch crossing; the channel cross-talk calculation was improved; and numbers previously read from text files were migrated to the conditions database.

The MDT digitisation code processes \Geant{} |MDTSimHits| to make MDT digits that are a simulation of MDT raw data. The |MDTSimHit| data has an identifier that identifies the specific MDT tube hit, the global time, and the impact parameter of the track relative to the wire in the MDT tube.  The digitised HITs consist of an offline identifier indicating the specific MDT tube hit, the drift time emulating the output of the Time--To--Digital Converter (TDC) and pulse height data emulating that calculated with an ADC. The MDT digitisation code includes a simulation of ionisation clusters in the MDT tube that are propagated to the MDT wire using a time-to-space function. An amplifier response function simulates the signal generated at the wire from drift electrons. The first drift electron which creates a signal over threshold determines the drift time. The TDC signal is calculated by combining the drift time, time-of-flight from the interaction point to the MDT, propagation time of the signal along the MDT wire, and the simulated ATLAS beam clock time. The ADC value is calculated using a simulation of the Wilkinson ADC~\cite{Posch:529410} used in the MDT electronics. The TDCs have a programmable dead time that is set to the maximum dead time of the tube~\cite{MUON-2002-003}; this is accounted for in the digitisation when the HIT with the earliest time sets the beginning of the dead time window, during which additional signals are discarded.

The New Small Wheels (NSW) are composed of Micromegas (MM) and small-strip Thin Gas Chambers (sTGC). Similar to the other subsystems, the digitisation consists of two parts, the first one modelling the response of the detector to the passage of any ionising particle and the second one simulating the response of the readout electronics.

The MM chambers contain micro-mesh gaseous detectors, with a gas gap of a few mm where charges are created and drifted, and an amplification region of 120--130~$\mu\textrm{m}$ between the metallic mesh and the readout electrode. The signal is read from the strips of the readout electrodes, with a 425--450~$\mu\textrm{m}$ pitch. For each strip, if the collected charge exceeds a set threshold, the charge and the time are recorded. Since a detailed simulation of the charge creation, transport, and avalanche for billions of events would be extremely expensive in terms of CPU usage, the MM digitisation instead relies on distributions obtained by simulating the passage of a smaller number of muons through a Micromegas detector using \textsc{Garfield++}~\cite{Garfield}, software specialised for the simulation of gaseous detectors. These distributions, e.g.\ for the diffusion or the number of ionizations per distance, are sampled in the digitisation, leading to good modelling of the detector response. Sampling the distributions also allows tuning of the simulated detector response towards the response of the real detector based on the parameters of the distributions.

The second part of the Micromegas digitisation is performing the simulation of the electronics response, in particular the response of the charge amplification and digitisation process carried out in a custom integrated circuit called the VMM~\cite{vmm}. The actual transfer function of the VMM is used to get the response of the shaping amplifier. Afterward, the peak height and time measurement are performed in the same way as implemented in the VMM. For the modelling of the charge threshold, a linear dependence between the noise level and the strip length is implemented using noise data gathered during the detector commissioning to determine the noise of the shortest and longest strips. Other features of the VMM are also implemented in the digitisation, including the ability to read the signal of a strip below the threshold if its neighbour has a signal above threshold (the \emph{neighbour logic}), and the \emph{address in real time} signal, providing the information of the first VMM channel (out of 64 channels per VMM chip) above the threshold, which is used for the trigger.

The sTGC consists of a multi-wire proportional chamber with three independent and complimentary readout technologies: wires, strips, and pads. When a particle crosses the sTGC gaseous gap, an avalanche forms along the ionising track towards the nearest wire. If the induced charge on the strips and pads exceeds set thresholds, the charge and time are recorded. A detailed timing spectrum and avalanche process on the sTGC anode wires operated at 2.9~kV is simulated using the \textsc{Garfield++}~\cite{Garfield} and \textsc{Magboltz}~\cite{Magboltz} packages. The induced signals on the strips and pads follow the response of the resistive layer as described in Ref.~\cite{DIXIT2006281}. The detailed simulation of the sTGC is then parameterised for fast HIT digitisation in Athena. The energy lost by particles traversing the sTGC gaseous gap is provided by \Geant{} in ATLAS. This energy is converted into an effective ionization charge in the gap. The time of arrival of the first electron cluster onto the wire is parameterised as a function of the distance of closest approach of the particle. The effects of ionization, noise, electronics threshold and avalanche gain fluctuations are modelled with a Polya distribution. The pad and strips closest to the wire with the avalanche fire. The charge on the strips is modelled by a Gaussian distribution of tunable width as a function of the polar angle of the incoming particle. The pad, wire and strip timing spectrum, as well as the strip hit multiplicity, were tuned on data collected during test-beam campaigns~\cite{Smakhtin:2009zz,Abusleme:2015yja}. The sTGC digitisation parameterised model offers an accurate timing performance, an adequate simulation of the charge sharing among adjacent readout strips, and a good representation of the overall spatial resolution. The final time and charge of the hits associated with a global BCID are obtained using a VMM calibration curve that converts a time in nanoseconds and a charge in picocoulombs into a \emph{Time Detector Output} and a \emph{Peak Detector Output} object, respectively.

\subsubsection{Treatment of pile-up and beam size effects}
\label{digi:sec:PU}

\paragraph{MC overlay}
\label{digi:sec:overlay}

In MC23, as in MC20, the technique of MC Overlay~\cite{SIMU-2020-01} is used to add the effect of additional minimum bias collisions to simulated events. In this approach, digitisation is carried out separately for the hard-scatter part. Digitised hard-scatter events are then combined with pre-digitised pile-up events. The libraries of these presampled RDO events are campaign-specific and each amount to about 500 million events per year, corresponding to a size of 1~PB. This method has reduced CPU, memory and I/O requirements relative to the pile-up digitisation technique used in previous campaigns (see, for example, Ref.~\cite{SOFT-2010-01}). The technique is similar to the premixing technique used in CMS~\cite{CMSpremixing}.

Pile-up digitisation is still used to create the presampled pile-up RDO datasets, but this is only done once per MC simulation sub-campaign (e.g.\ MC20a). This step takes as input two equally large datasets of minimum bias events in HITS format: a high-\pt sample containing at least one jet with \pt larger than 35~\GeV, photon with \pt larger than 8~\GeV, or b-hadron that has \pt larger than 5~\GeV that decays to a lepton with \pt larger than 4~\GeV, and a low-\pt sample with the remaining events. The two datasets are sampled according to their relative cross sections, but avoiding the duplication of the high-\pt minimum bias events that might cause visible features during analysis~\cite{SIMU-2020-01}. The presampled RDO events can then be re-used for multiple hard-scatter samples without causing any issues in physics analysis. Each presampled RDO event has a unique hash that can be used to tag whether a presampled RDO event is selected multiple times within an analysis. This can be used to adjust the presampled RDO dataset size and hence the level of re-use for future MC simulation campaigns. The lumi block-ordered hard-scatter HITS and presampled pile-up RDO files are both merged into files with 10,000 events each. Having the same number of events in the two file types has the advantage that when the MC Overlay job reads the Nth event from each file and combines them, the hard scatter and pile-up events have the same beam spot size. The MC Overlay code contains a sanity check that will abort the job if this is not the case.

In addition to the overlay algorithms described for the \RunTwo\ detector sub-systems in Ref.~\cite{SIMU-2020-01}, for \RunThr new overlay algorithms were prepared for the sTGC and MM sub-systems. One further new overlay algorithm was implemented for LAr Super Cells used as input to the Phase-I Trigger.

For the presampled pile-up, the signals in the sTGC (MM strips) from the hard-scatter and pile-up events are combined as follows. In a given channel, if a signal is only present in either the hard-scatter event or the pile-up event, that signal is copied to the output RDO. If a signal is in both, the output RDO depends on the timing of the signals and the sTGC VMM pulse shaping time (MM VMM integration time). If the hard-scatter and pile-up signals are separated in time by more than the set shaping time (integration time), the earliest signal is copied to the output RDO. If the signals are separated by less than the shaping time (integration time, about 200~ns), the two signals are combined by adding the charges and taking the earliest signal time.

As with standard LAr Overlay, LAr Super Cell Overlay is done at the digit level. Hard-scatter HITS are digitised in the standard way, except that no noise is added (as this has already been applied to the background LAr Super Cell digits). |LArDigit|s contain several time-samples of the signal in each channel.
The algorithm loops over the Super Cells. Any hard-scatter contribution is added to the pile-up background contribution, if available, and otherwise it is added to the standard pedestal for that channel. If any of the samples are below zero or above the maximum ADC value, then they are constrained to lie in this range. The results are written into the output |LArDigit| container. The creation of LAr Super Cells from the combined |LArDigits| then proceeds in the standard way.

Beam size effects are also taken into account in the minimum bias simulation. Separate minimum bias background datasets are simulated with each beam spot size (43, 40, 37 and 34~mm, see also Section~\ref{sim:fullsimImprovements}). During the pile-up presampling step~\cite{SIMU-2020-01} minbias files with all four beam spot sizes are used as input. Separate Athena jobs are run for each beam spot size with the appropriate $\langle\mu\rangle$ distribution and number of events per job for that part of the run. The RDO files created are then merged into lumiblock-ordered presampled RDO files.

\paragraph{Track overlay}
\label{digi:sec:trackoverlay}

Track overlay is an evolution of the concept of MC overlay. In MC overlay, simulated and digitised pile-up data (or digitised real detector pile-up data) is overlayed onto the simulated hard-scatter event. In track overlay, the inner detector tracks of simulated pile-up data are first reconstructed, and then these events are overlayed onto the hard-scatter event. Repeated ID reconstruction for the same events is thus avoided, and CPU can be saved (up to 50\% based on preliminary performance tests). Track overlay is not yet applied in the recent MC23 campaigns, but its deployment is still planned for \RunThr.

Tracks in the core of (often) high-\pt jets are affected by the presence of pile-up hits, meaning the individual reconstruction of pile-up and hard-scatter tracks leads to differences relative to the merged reconstruction. Track overlay is therefor not suitable for all events. To remedy that, a neural network is used to decide for individual hard-scatter events whether it is appropriate to use track overlay or MC overlay. This is called \emph{hybrid overlay}. This neural network is trained on the tracks using truth features and variables relating to the track density, as well as pile-up conditions, with the aim to classify tracks into those that can be used for track overlay or not. The per-event ratio of these track populations is then used as a score, to decide whether the event is sent to track or MC overlay.

\begin{figure}[tbp]
\centering
\subfloat[]{
\label{fig:sim:fig:sim:overlayvali1_a}
\includegraphics[width=0.49\textwidth,valign=c]{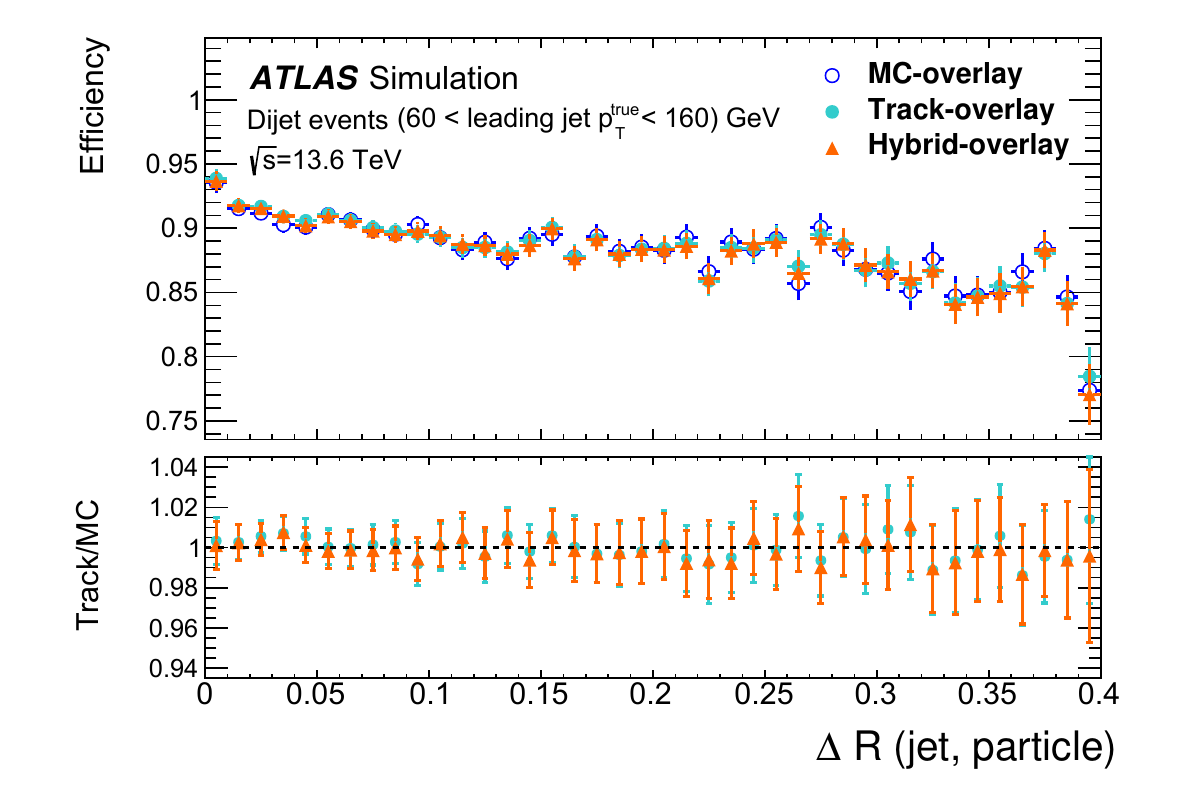}
}
\subfloat[]{
\label{fig:sim:fig:sim:overlayvali1_b}
\includegraphics[width=0.49\textwidth,valign=c]{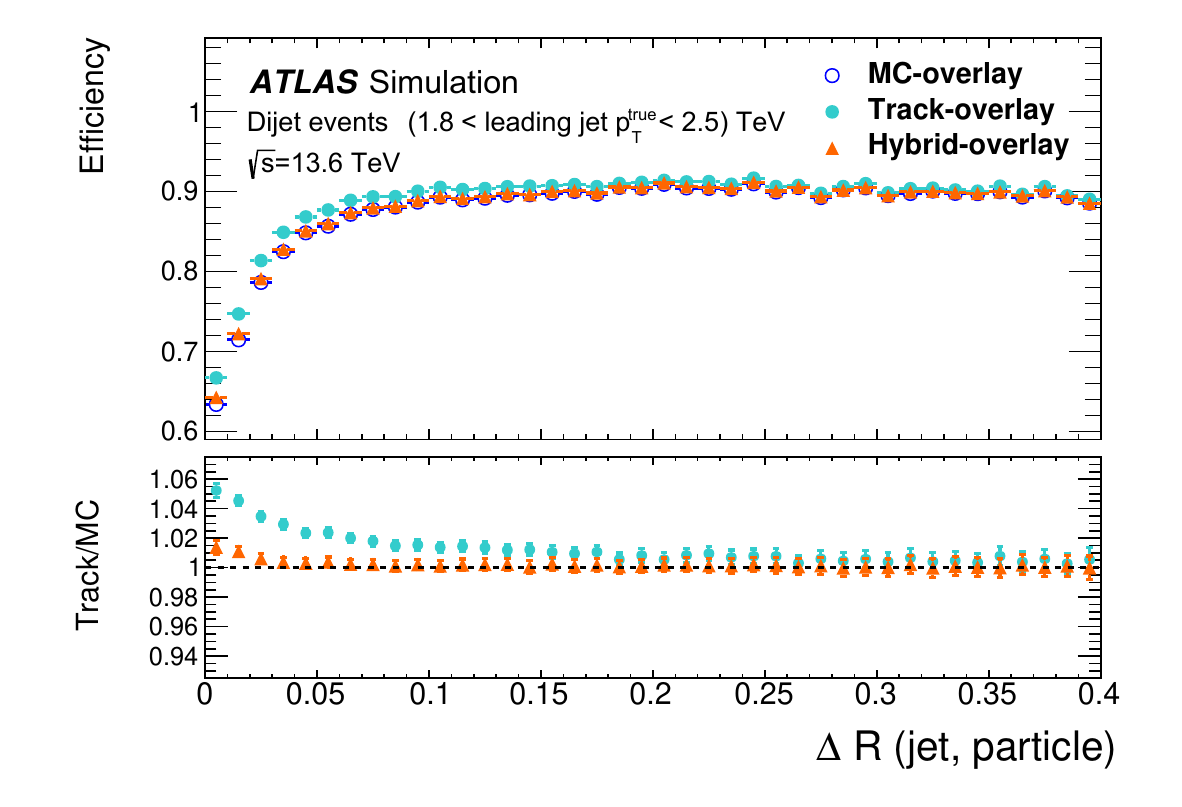}
}
\caption{\label{fig:sim:overlayvali1}The hard-scatter event track reconstruction efficiency for tracks in jets as a function of the distance between the jet axis and truth particles. Dijet events at 13.6~\TeV containing a leading jet with (a) $60~\GeV < p_{\textrm{T}}^\textrm{true} < 160~\GeV$ and (b) $1.8~\TeV < p_{\textrm{T}}^\textrm{true} < 2.5~\TeV$ are reconstructed with MC overlay (empty circles), pure track overlay (filled circles) or the hybrid (filled triangles). For the hybrid case, the fraction of events that is sent to track overlay is 93.5\% (35.3\%) for the low-\pt (high-\pt) sample.}
\end{figure}

\begin{figure}[tbp]
\centering
\includegraphics[width=0.6\textwidth]{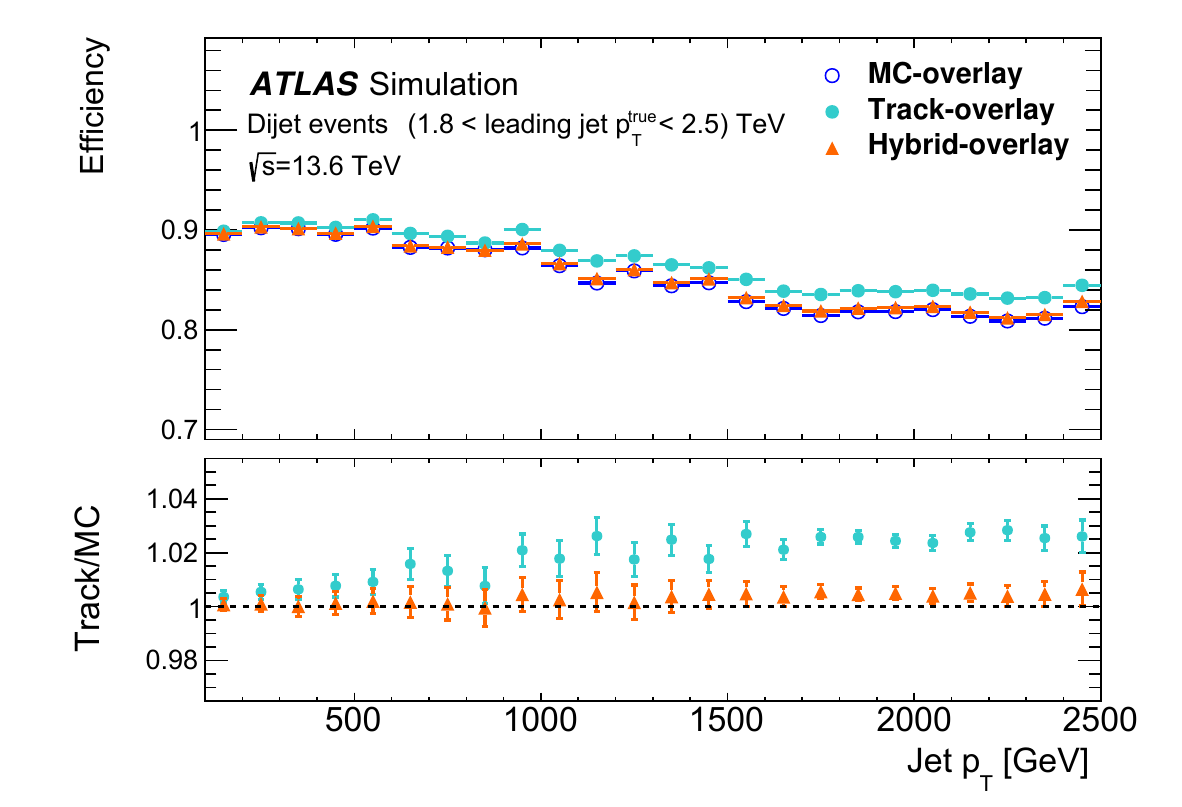}
\caption{\label{fig:sim:overlayvali2}The hard-scatter event track reconstruction efficiency for tracks in jets as a function of the jet transverse momentum. Dijet events at 13.6~\TeV containing a leading jet with $1.8~\TeV < p_{\textrm{T}}^\textrm{true} < 2.5~\TeV$ are reconstructed with MC overlay (empty circles), pure track overlay (filled circles) or the hybrid (filled triangles). For the hybrid case, the fraction of events that is sent to track overlay is 35.3\%.}
\end{figure}

In Figure~\ref{fig:sim:overlayvali1}, the efficiency of track reconstruction is shown as a function of the distance between the reconstructed jet axis and truth particles, for a sample with low or high-\pt jets. The reconstructed jets are particle flow jets with a radius parameter of 0.4 (see also Section~\ref{sec:reco:jets}). In Figure~\ref{fig:sim:overlayvali2}, the track reconstruction efficiency for high-\pt jets is presented as a function of the jet \pt. Track overlay reproduces the efficiency of the MC overlay well for jets with moderate transverse momentum, while close to the centre of very high-\pt jets the efficiency is overestimated. The hybrid case results in a very good agreement with pure MC overlay even at high-\pt. In the studies presented, the ML score cut is 0.742, corresponding to a fraction of events used for track overlay of 93.5\% for low-\pt dijets and 35.3\% for high-\pt dijets.

\paragraph{Data overlay}
\label{digi:sec:dataoverlay}

Particularly for heavy ion data analysis, data overlay is used rather than MC overlay.
In heavy ion collisions, rather than additional collisions, the `background' is primarily
the enormous underlying event from the collision itself. This background is extremely
difficult to correctly model, owing to the complex particle correlations that arise from
the dynamics of the collisions. For that reason, it is extracted directly from data events
and overlaid onto `signal' events.

The data are selected using special triggers based on the total energy in the event. Several
different triggers are combined to produce a complete spectrum of total energy (which is strongly
correlated with centrality in the heavy ion collisions). The events are then mixed together into
batches that represent the entire heavy ion data-taking period, and the MC simulation is
reweighted to reproduce the spectrum of total energy observed in the data. To
ensure accuracy in the overlay step, the data must be read out without the normal suppression of
calorimeter cells that are below threshold.

Simple proton--proton collision events are then generated for a variety of interesting processes.
For each generated proton--proton signal event, a background
heavy ion collision is selected. The generated event is then placed exactly on top of the
reconstructed vertex of the heavy ion data event, so that the particles from the proton--proton
collision appear to also emanate from the same ion--ion collision as the background.
The detector simulation then proceeds with the same detector geometry and conditions as are
used for the reconstruction of the real detector data. At the digitisation stage, the signals
from the simulated proton--proton collision and the data heavy ion event are combined, in much
the same way as is done for MC overlay. The reconstruction then proceeds as normal. The result
is an event that has a simulated signal, for example a $Z$ boson decaying into two leptons,
embedded in a heavy ion collision.

Because the data conditions are used, the heavy ion data overlay procedure perfectly emulates
certain detector features, including noise, disabled detector channels, and non-collision
backgrounds. However, because the simulation must use conditions from the real
detector, misalignments can induce overlaps in the volumes simulated with \Geant{} --- there
is no requirement in the alignment that volumes not overlap. While for precision proton--proton
collision measurements these overlaps are sufficiently concerning that they have prevented the
up-take of data overlay, for heavy ion events where the events are much busier and most
analysis is done using ratios of quantities to cancel out systematic effects, the risk is much
reduced.


\subsection{Reconstruction}
\label{sec:reco}

Following the digitisation step, simulated events are passed through a trigger simulation~\cite{ATLAS-TRIG-2022-03}.
The next step for both the simulated and real particle collision data is reconstruction.
Both the simulated and real data pass through the same set of algorithms, except algorithms specifically treating truth information (e.g.\ event generator records).
The reconstruction process provides analysis object data (AOD) outputs, representing a summary of the reconstructed event.
These outputs are produced in an xAOD format (see Section~\ref{sec:edm}), which is readable by both Athena and \textsc{ROOT},
easing manipulation for subsequent user analysis.
In practice, xAODs are further processed to create \emph{derived}-xAODs (DAODs), or \emph{derivations}, before being used as input to physics analysis, as described in Section~\ref{sec:derivations}.

The ATLAS reconstruction software was updated for the migration to a multithreaded framework\footnote{The first piece of ATLAS software to work in a multithreaded environment was the calorimeter energy clustering, which was used as demonstrator to showcase the feasibility of the transition.} for \RunThr.

This required numerous changes, as outlined in Section~\ref{sec:athenaMT}.

The ATLAS experiment has two main categories of detector sub-systems:
\begin{itemize}
\item Tracking detectors (the inner detector and muon spectrometer), which measure charged particle momentum; and
\item Calorimeters (LAr and Tile calorimeters), which measure energy deposited by traversing particles.
\end{itemize}

Information from these systems are used to reconstruct physics objects such as charged particle tracks~\cite{PERF-2015-08}, primary vertices~\cite{PERF-2015-01}, calorimeter energy clusters~\cite{PERF-2014-07}, electrons, photons, muons, $\tau$-leptons, jets, and event-level quantities such as \met.
These objects and their associated quantities (for example the objects' energy) are often referred to as object \emph{collections}.
A single object collection will contain a set of consistently defined objects of a given type.
For example, track objects are reconstructed using inputs from the ID with various thresholds and selection criteria applied.
While a standard track collection is sufficient for many physics analyses, searches for long-lived particles may require a track collection containing large-radius tracks (see Section~\ref{sec:trackReco}).

The reconstruction of physics objects and event-level quantities during proton--proton physics data-taking is described in Sections~\ref{sec:clusterReco}--\ref{sec:tauReco}.
The LHC also delivers heavy-ion collisions to ATLAS during dedicated data-taking periods.
The reconstruction process during these periods differs from `standard' proton--proton collision reconstruction, and is described in Section~\ref{sec:hiReco}.
Special configurations of the muon and inner detector track reconstruction can also be used during the collection of cosmic ray data, particularly when no beams are circulating in the collider.

\subsubsection{Calorimeter energy clusters}
\label{sec:clusterReco}

The calorimeter energy clustering runs first in the reconstruction of an event, as it can be used to `seed' track reconstruction, in particular for back-tracking to recover photon conversions (see Section~\ref{sec:trackReco}).
Calorimeter reconstruction begins from individual cells, and is described in Ref.~\cite{LARG-2009-01}.
The digitisation process for simulated calorimeter data is described in Section~\ref{sec:digi}.
For real data, the energy deposited in each calorimeter cell is calculated by the online system through the application of OFCs to the time-sampled ionization pulse.
The bytestream data written by the ATLAS data-acquisition system contains energies for each of the 191,720 calorimeter channels.
For channels of the LAr calorimeter above a pre-set threshold,
the cell-timing, a quality factor~\cite{LARG-2013-01} and the raw ADC samples are also written out.
The energy for these channels is then recalculated offline, allowing for update of calibration constants during offline processing.

Calibration constants required to derive the energy from the ADC samples include the \emph{pedestal},\footnote{The baseline signal level of each LAr channel and the noise characteristics of the associated readout electronics is called the \emph{pedestal} and is measured in dedicated \emph{pedestal runs}.}
OFCs and an ADC-to-\MeV conversion factor.\footnote{The ADC-to-\MeV conversion factor includes the amplification of the readout electronics, which is regularly re-measured in pulser runs for each channel.}
In total, about 68~MB of calibration constants are used to calculate the energy deposit in \MeV for all 182,468 LAr cells,
each of which has four associated ADC values, one from each of the four samples readout from sampling the cell's ionisation pulse.

The next step of the reconstruction chain involves building a |CaloCell| container.
This is a container of all cells in both the LAr and Tile calorimeters, ordered by a geometric identifier such that the same physical cell is always in the same container position.
During the cell-building process, several higher-level corrections are applied.
The most important of these is the correction of fluctuations (trips) of the high-voltage (HV).
The actual voltage of each of the 4,837 HV lines is measured by the Detector Control System (DCS)~\cite{DCS} and stored in a database.
The offline software calculates energy-correction factors based on these voltages.
The correction factors change every time the voltage changes, which can happen multiple times in a single luminosity block.\footnote{The high frequency of HV changes is partly due to imperfect smoothing by the Detector Control System.}
In serial reconstruction (as used in \RunTwo), IoV callbacks trigger the recalculation of correction factors at event boundaries.
In multithreaded reconstruction (used in \RunThr), multiple sets of sets of HV-correction factors have to be kept in memory simultaneously,
because events belonging to different HV-periods are processed concurrently (see Section~\ref{sec:core:conditions}).

Each LAr cell is supplied by at least two HV lines, so there is redundancy in the case of a trip of one of the HV lines.
If one line trips, the energy deposit is estimated by using an appropriate correction factor.
The expected noise is then re-scaled accordingly, to account for the higher correction factor.
When updating to run multithreaded reconstruction, the re-scaling of noise proved to be challenging owing to the dependence on both the conditions data and high voltage information.
Input conditions data are indexed by run and luminosity block number, while the voltage is recorded by the DCS and indexed by time-stamp.

Once built, |CaloCells| are gathered into clusters that contain the shower induced by a traversing particle.
For \RunThr\ the topological clustering algorithm~\cite{PERF-2014-07} is most commonly used to complete this task.
Input cells are classified according to their signal-over-noise ratio,
which depends on the high-voltage-corrected noise for each cell.

\begin{figure}
\begin{center}
\includegraphics[width=0.55\textwidth]{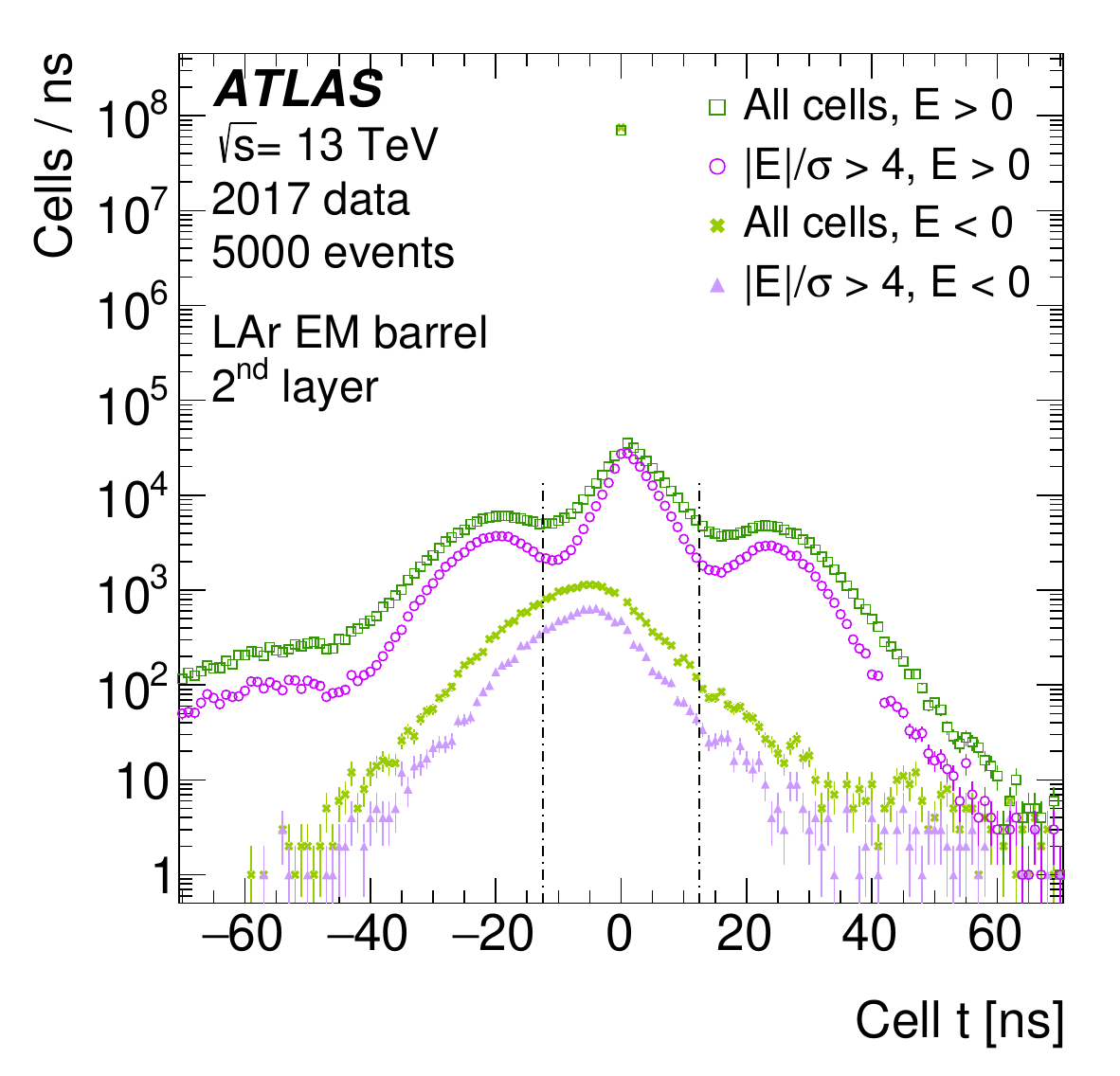}
\caption{The reconstructed signal time distribution for calorimeter cells in the second layer of the LAr EM barrel calorimeter. Both the inclusive cell time spectrum (green empty squares and solid crosses) and the one for seed candidates with signal-over-noise ratio greater than four (empty circles and solid triangles) are shown, separately for cells with positive and negative reconstructed energy. The vertical lines represent the cell time rejection limits of $\pm12.5$~ns. Figure from Ref.~\cite{JETM-2023-01}.}
\label{fig:clustertimecut}
\end{center}
\end{figure}

A reconstruction time requirement is applied to cells to reduce the effect of out-of-time pile-up~\cite{JETM-2023-01}.
The effect of this requirement is shown in Figure~\ref{fig:clustertimecut}~\cite{JETM-2023-01}.
In the time distribution of positive energy cells, secondary contributions are visible at $\pm25$~ns, one bunch crossing from the collision of interest.
These are out-of-time pile-up contributions that are suppressed thanks to the calorimeter energy cluster time requirement.

The topological clusters, or \emph{TopoClusters}, output by this algorithm serve as seeds for electron and photon reconstruction, and are provided as inputs for jet-building.

\subsubsection{Inner detector tracks}
\label{sec:trackReco}

Precise measurements of charged-particle trajectories are vital to a successful physics programme.
Charged particles passing through the ID deposit energy through processes such as ionisation and radiative loss,
and their trajectories can be reconstructed to form tracks.
Energy deposits in neighbouring detector channels are clustered to form three-dimensional space points referred to as \emph{hits}.
Neural networks are used to refine the estimates of where the particles pass through active material, and to split individual clusters if they were produced by several coincident particles.
During reconstruction, tracks are constructed as a set of hits identified as being consistent with coming from the same single charged particle, and then fit to determine the corresponding particle trajectory.

\begin{figure}
\begin{center}
\includegraphics[width=0.45\textwidth]{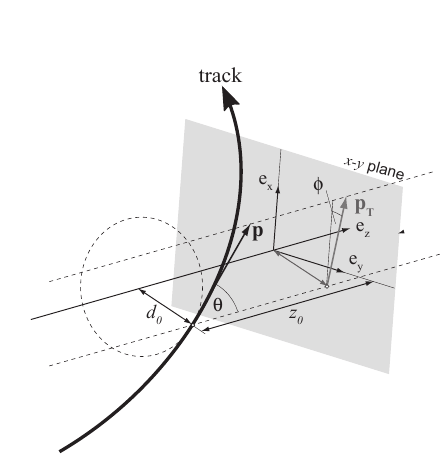}
\caption{Illustration of the five global track parameters, \dzero, $z_{0}$, $\phi$, $\theta$, $q/p$~\cite{ATL-SOFT-PUB-2007-003}.
These are defined relative to a reference point, the perigee, or point of closest approach to the beam line.}
\label{fig:trackparameters}
\end{center}
\end{figure}

Particle trajectories are described by five parameters (\dzero, $z_{0}$, $\phi$, $\theta$, $q/p$) defined relative to a reference point (the perigee).
These parameters are depicted in Figure~\ref{fig:trackparameters}~\cite{ATL-SOFT-PUB-2007-003}.
Here \dzero\ and $z_{0}$ are the transverse and longitudinal impact parameters, $\phi$ and $\theta$ are the azimuthal and polar angle, and $q/p$ is the charge divided by the momentum.
The default reference is the beam line centred at the beam spot~\cite{PERF-2015-01}.

Improvements were made to the ATLAS track reconstruction software ahead of \RunThr to prepare for proton collisions with an average pile-up of about $50$~\cite{IDTR-2022-04}.
Such a busy environment demands highly performant software capable of efficiently exploiting available computing resources to provide prompt reconstruction of particle collision events.
The main changes to the ATLAS track reconstruction software in preparation for \RunThr are documented in Ref.~\cite{IDTR-2022-04}.

The track reconstruction strategy followed by the ATLAS experiment pivots on a single \emph{inside-out} track reconstruction sequence requiring eight hits to form tracks with $\pt > 500$~\MeV.
Additional inside-out and outside-in or \emph{back-tracking} sequences targeting specific signatures such as long-lived particles (LLP) and photon conversions complement the main sequence.

An inside-out sequence begins with a \emph{seeding} process that forms triplets from the individual silicon detectors. Then, a combinatorial Kalman Filter~\cite{kalman} extends the seeds along search roads through the detector elements creating multiple track candidates to maximise reconstruction efficiency.
Limits on hit usage and sharing, which adapt to local conditions using a set of neural networks~\cite{PERF-2015-08}, are enforced through the ambiguity-solving stage to maintain a low rate of false positive track assignments (a low `fake rate'). Next, surviving track candidates are passed to the global $\chi^2$ method for a high-precision track parameter estimate. Finally, tracks are extended when possible, with TRT measurements improving momentum resolution and particle identification.

\begin{figure}
\begin{center}
\includegraphics[width=0.55\textwidth]{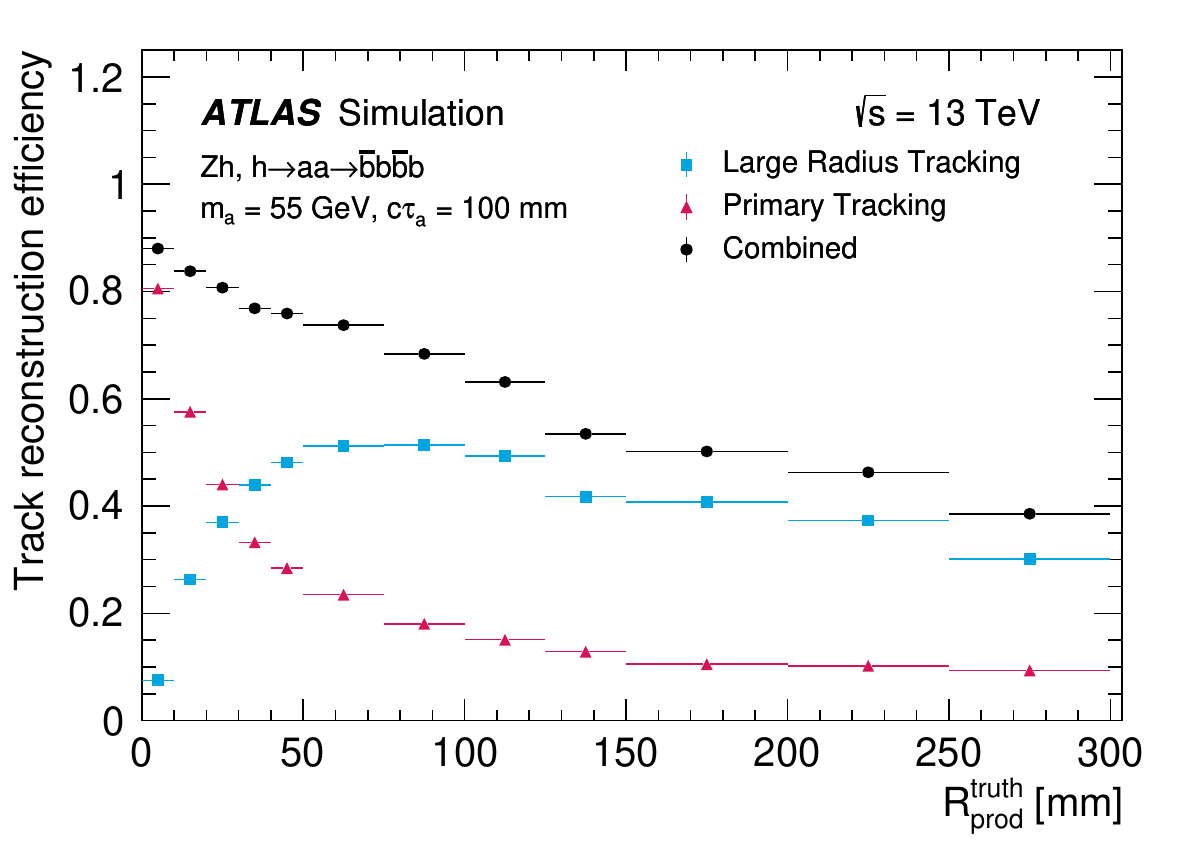}
\caption{The primary, LRT, and combined track reconstruction efficiencies for displaced charged particles produced by the decay of a LLP in a Higgs portal signal model. Efficiencies are shown as a function of true decay position (which is the charged particle production position) in $R_\text{prod}^\text{truth}=\sqrt{x^2+y^2}$. Truth particles are required to have $\pt>1.2$~\GeV and $\abseta<2.5$. Figure from Ref.~\cite{IDTR-2021-03}.}
\label{fig:lrtefficiency}
\end{center}
\end{figure}

Subsequent inside-out and back-tracking sequences are tuned to particular signatures and do not consider hits already associated with reconstructed tracks.
For example, large-radius tracking (LRT) considerably improves the efficiency for reconstructing the decays of LLPs displaced by more than 5~mm transversely from the interaction point by increasing the allowed range for the track candidate impact parameters~\cite{ATL-PHYS-PUB-2017-014}. Stricter selection criteria combat the increased combinatorics to ensure a fast processing time per event, while a back-tracking sequence maintains the secondary track and photon conversion efficiency. The back-tracking is seeded from TRT and SCT segments built in a narrow geometric cone (a \emph{region of interest}) around a calorimeter energy cluster.
During \RunTwo, LRT was only applied to a subset of events, but the performance improvements in both the standard and large-radius track reconstruction in preparation for \RunThr have resulted in the ability to run LRT on all events, thanks to computational speed-ups of a factor of 40--60, depending on $\langle\mu\rangle$~\cite{IDTR-2021-03}.
This development provides a potential boost in accessible phase-space for LLP searches, and increases the efficiency of LLP reconstruction and analysis workflows.
The improvement in efficiency is shown for the case of a Higgs portal signal model in which long-lived pseudo-scalar bosons ($a$) from the Higgs boson decay into charged particles is shown in Figure~\ref{fig:lrtefficiency}~\cite{IDTR-2021-03}.

Specific sequence configurations target particles with $\pt<500$~\MeV\ separately in low and high pile-up events.
Heavy ion collisions warrant a specialised configuration as well (see Section~\ref{sec:hiReco}).

The main inside-out track reconstruction iteration reconstructs primary tracks with an isolated track efficiency of approximately 70 to 85\% depending on the track \pt and $\eta$~\cite{IDTR-2021-01}.
Following high-quality selection criteria, the rate of reconstructed tracks remains extremely linear up to high $\langle\mu\rangle$, suggesting the rate of spuriously reconstructed tracks remains at a sub-percent level on average for typical operating conditions.
Within a jet with \pt of 1~\TeV, where very high measurement densities make track reconstruction challenging, reconstruction efficiencies are maintained at approximately 90\% of that for isolated tracks.
Photon-conversion reconstruction efficiency is around 70\% over a broad range of $\ET$, with some degradation as a function of $\langle\mu\rangle$~\cite{EGAM-2018-01}.
Low pile-up reconstruction is discussed in Ref.~\cite{STDM-2015-17}. The track-based alignment of the detector sensors is described in Ref.~\cite{IDTR-2019-05}.

\subsubsection{Primary vertices}

Primary vertices (PV) represent the reconstructed position of an individual particle collision.
For a given bunch crossing the PV with the largest sum of track $\pt^2$ is labelled the \emph{hard-scatter vertex}.
This identifies the location of what is considered to be the primary event and the interaction of interest for the given bunch crossing.
The location of the hard-scatter vertex also serves as a reference point for down-stream reconstruction algorithms.

PVs start as candidate vertex locations estimated by using a Gaussian-distributed resolution model for the track impact parameters.
During the track fit, candidate vertices compete for tracks to reduce the chance of nearby proton--proton interactions being reconstructed as a single merged vertex.
For \RunThr, the primary offline vertex reconstruction in ATLAS is done using the open-source, experiment-independent ACTS toolkit~\cite{ai2021common, acts}.
The higher instantaneous luminosity in \RunThr\ means that primary vertex reconstruction is more challenging than ever before.
To prepare for this, two new tools were developed to preserve primary vertex reconstruction efficiency~\cite{ATL-PHYS-PUB-2019-015}.
A Gaussian track density seed finder (GS) and adaptive multi-vertex finder (AMVF) replace the iterative vertex finder (IVF)~\cite{PERF-2015-01} used during \RunTwo.
For the beam spot determination IVF is still used, and once the beam spot is known AMVF is used.

PV reconstruction performance depends on the event topology and is described in Ref.~\cite{IDTR-2021-01}.
In Standard Model \ttbar events, the reconstruction efficiency for the vertex including the \ttbar\ production is close to 100\% in \RunThr conditions.
For the softer pile-up vertices, the reconstruction acceptance is around 70\%, and the reconstruction efficiency is around 80\% at $\langle\mu\rangle=60$.
The identification efficiency in \ttbar events (i.e.\ efficiency for correctly identifying the \ttbar production collision as the hard scatter) is
around 95\% at $\langle\mu\rangle=60$~\cite{ATL-PHYS-PUB-2019-015}.
The vertex position resolution along the beam line is estimated to be 18~$\mu m$ in \ttbar events and is somewhat degraded for lower multiplicity processes.

\subsubsection{Electrons and photons}

The process by which electrons and photons are reconstructed at ATLAS is described in Ref.~\cite{EGAM-2018-01}.
Electron and photon reconstruction begins from calorimeter TopoClusters.
In order for a TopoCluster to be considered, at least 50\% of the total TopoCluster energy must be from cells in the EM calorimeter.
The track parameter estimates for electron candidates are improved by a Gaussian Sum Filter (GSF)~\cite{PERF-2017-01},
which accounts for significant electron energy loss due to bremsstrahlung interactions with the detector material.
The inner detector tracks with the GSF applied are also used to build photon-conversion vertices.
Both the conversion vertices and the refitted tracks are then matched to the selected TopoClusters.

Following this track--cluster matching process, electron and photon \emph{superclusters} are built.
The electron and photon superclustering algorithms~\cite{ATL-PHYS-PUB-2017-022,EGAM-2018-01} replace previously used sliding-window algorithms~\cite{ATL-LARG-PUB-2008-002} (which resulted in fixed-size clusters) in favour of the creation of dynamic superclusters of variable size.
This improves the ability to account for radiation loss in the ID due to bremsstrahlung by enabling the recovery of low-energy photons, which are then paired with their associated electron or converted photon via the supercluster.
First, seed clusters for electron and photon supercluster building are identified, and following this satellite cluster candidates are identified from nearby clusters.
Positional corrections are applied to the resulting superclusters, and ID tracks are matched to electron superclusters, while conversion vertices are matched to photon superclusters.
Track-matched superclusters form electrons, superclusters matched to conversion vertices form converted photons, and superclusters with no conversion vertex or track match form unconverted photons.

In addition to the improvements brought by switching to use dynamic clustering for electron and photon reconstruction,
optimisations and simplifications to reconstruction software in preparation for \RunThr\ resulted in significant performance gains.
For example, the use of a dedicated track extrapolation vastly reduced the CPU consumption of one of the most demanding algorithms in the electron reconstruction chain.
The algorithm time was reduced by more than an order of magnitude, and the overall electron and photon reconstruction time was reduced by more than 25\%.
The GSF was also improved and sped up by about a factor of two.
Optimisation and tuning of conversion identification was necessary for \RunThr\ to address the new gas configuration of the TRT, as discussed in Section~\ref{sec:detector}.
To improve support for analyses focusing on long-lived particles, an electron collection built using large-radius tracks was introduced for \RunThr.

\subsubsection{Muons}

During LS2, the muon small wheel was replaced by the NSW.
The NSW deploys two detector technologies, Micromegas and sTGCs, as noted in Section~\ref{sec:detector}.
Both the technologies provide excellent trigger and tracking performance in high-rate environments~\cite{ATLAS-TDR-20}.

Muon reconstruction in \RunThr\ starts from inside the muon spectrometer, as described in Ref.~\cite{PERF-2015-10}.
MS tracks are created, extrapolated to the hard-scatter vertex,
and refitted with a loose interaction point constraint taking into account the energy loss in the calorimeter.

The first step of muon reconstruction is segment finding.
A single physical muon chamber might comprise several layers of one or more detector technologies, physically attached to one another.
Rather than directly reconstructing muons from the hits from all of these detectors, first the hits within each chamber are gathered into segments.
In standard muon reconstruction, these segments serve as a first cleaning of electronic noise and cavern background.
In standard proton--proton collision reconstruction, they are required to roughly correspond to a feasible muon trajectory.

Alignment effects are implemented using an event data model object known as an \emph{Alignment Effect on Track} (AEOT)~\cite{Leight:2018ogy}.
This object holds the list of hits it affects and the standard deviations on the constraints (angular or sagitta-based) to apply to those hits.
A refitting algorithm (the |MuonRefitTool|) constructs the appropriate AEOTs for the hits used in the track, based on misalignment information stored in the conditions database and processed by a specialized tool (the |AlignmentErrorTool|), and adds them to the vector of measurements that defines the track.
After a muon is identified, a track refit is then done, with the AEOTs included in the derivative matrix that enters the computation of the fit $\chi^{2}$,
so that their effect is directly incorporated into the minimization process.

Several different algorithms are used to identify and reconstruct muons, with separate muon collections being available in xAOD files.
\emph{Combined} (CB) muon reconstruction relies on the successful combination of an MS track with an ID track~\cite{MUON-2022-01}.
Inner detector tracks are sought in a cone around the extrapolated MS track. This strategy follows an outside-in approach where the information from the MS is used to seed the muon reconstruction, looking for confirmation in the ID detector.
The \emph{inside-out} algorithm~\cite{MUON-2018-03} starts from tracks reconstructed in the inner detector and extrapolates them to the MS.
In this strategy, information from the ID is used to seed the muon reconstruction,
with confirmation later being sought in the MS detector.
The ID tracks with $\pt > 2$~\GeV\ are extrapolated outwards, with MS hits found along the extrapolated trajectory being associated with the given track.
For \RunThr, the inside-out algorithm was significantly sped-up by removing ID tracks from consideration if they had already successfully created a CB muon.
Lower-\pt\ muons that do not have enough energy to traverse the entire MS may be reconstructed as a \emph{Segment-Tagged} (ST) muon, which does not require a fully reconstructed muon track, but only muon segments associated with an ID track.
Finally, \emph{Calo-Tagged} (CT) muons are identified selecting ID tracks matched with a calorimeter energy deposition compatible with a minimally ionizing particle.
A new addition to the software with the start of \RunThr\ is the |CaloMuonScore| algorithm.
This extrapolates particle tracks from the ID to the MS,
forms three-dimensional representations of energy deposits in the calorimeter,
and runs a convolutional neural network on these energy deposit patterns to assign a likelihood score corresponding to whether or not the particle is a muon.

\begin{figure}
\begin{center}
\centering
\subfloat[]{
\includegraphics[width=0.32\textwidth,valign=c]{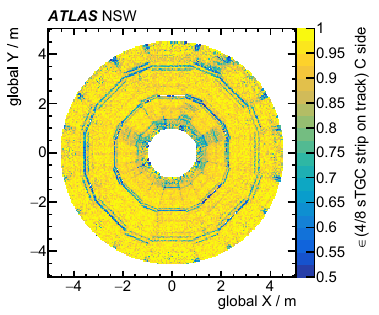}
}
\subfloat[]{
\includegraphics[width=0.32\textwidth,valign=c]{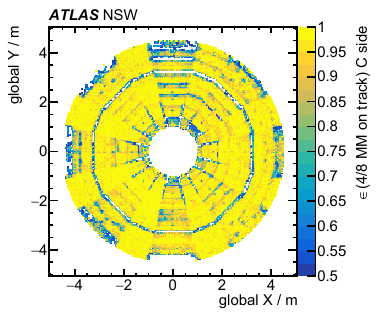}
}
\subfloat[]{
\includegraphics[width=0.32\textwidth,valign=c]{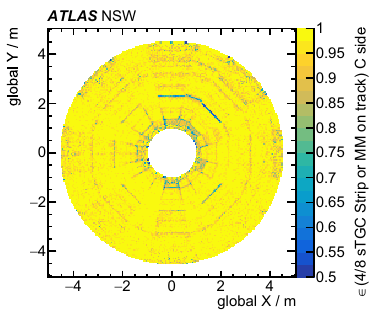}
}

\caption{\label{fig:nswhits}Efficiency for having at least four out of eight layers of either (a) sTGC strips, (b) Micromegas strips and (c) sTGC or Micromegas strips associated with a combined (with the inner detector) or standalone (muon spectrometer only) track with $\pT>15$~\GeV passing through the NSW on the C side during proton--proton collision data taking on May 14th 2023. Regions of low efficiency correspond to detector or readout issues during data taking. Figures from Ref.~\cite{MDET-2023-05}.}
\end{center}
\end{figure}

The NSW is still being commissioned. It is fully integrated into the offline reconstruction, and standard data analysis using \RunThr data now includes muons with NSW hits.
The efficiency for having a NSW segment (at least four of eight layers with either technology) associated with a combined (with the inner detector) or standalone (muon spectrometer only) track with $\pT>15$~\GeV is shown for a 2023 data taking run in Figure~\ref{fig:nswhits}~\cite{MDET-2023-05}.

\subsubsection{Particle flow and jets}
\label{sec:reco:jets}

The reconstruction of jets, collimated groups of hadrons emerging from the proton--proton collision, makes use of iterative clustering algorithms, taking various detector signals as inputs.
The most commonly used algorithm is the \antikt~\cite{Cacciari:2008gp} algorithm.
The jet-finding inputs, or constituents, may be TopoClusters, particle flow objects~\cite{PERF-2015-09}, particles from an event generator (for simulated samples) or any other object representable as a momentum four-vector.
The clustering algorithms themselves use the \textsc{fastjet}~\cite{Fastjet} implementation,
meaning the constituents must be translated from ATLAS data types into \textsc{fastjet} |PseudoJet| objects for interfacing.

The particle flow procedure was updated ahead of \RunThr.
To exploit the fact that particle flow objects are built from the same calorimeter TopoClusters and inner detector tracks as other objects,
links --- known as global particle flow (GPF) links --- between the particle flow objects and electron, photon, muon and $\tau$-lepton objects were introduced.
By default, particle flow uses all tracks, satisfying some basic quality criteria and \pt\ requirement, and TopoClusters.
The tracks in this context are assumed to be pions, and the energy in TopoClusters corresponding to extrapolated charged particle tracks is subtracted from the particle flow objects on that basis.
The introduction of GPF links allows users to revisit this decision,
since the list of TopoClusters matched to each track (along with their subtracted energies) is available from the charged particle flow objects.

The first step of jet reconstruction is to prepare the jet constituents, which may entail procedures such as calibration or filtering, e.g.\ to suppress the impact of pile-up.
This is executed by a |JetConstituentModSequence|, which runs a configurable series of Tools that modify the collections of constituents.
These Tools each implement the |IJetConstituentModifier| interface.

Next, a collection of |PseudoJet| objects is constructed from the constituents, such that they can be input into \textsc{fastjet} for clustering.
This is done by a |PseudoJetAlgorithm|, which reads in the constituent collection, creates a corresponding collection of |PseudoJet|s, and records them to the event store.
Any constituent type that implements the |IParticle| interface (i.e.\ has a momentum four-vector) can be used.

The interfacing to \textsc{fastjet} and translation of the resulting jets into the xAOD |Jet| type is handled by a |JetClusterer|.
This is a Tool that reads in the collection of |PseudoJet|s, runs the actual clustering algorithm, and builds a |JetContainer| from the output.
The |JetClusterer| does not necessarily record these jets to the event store, as further modifications to the collection may be desired first.
Such modifications (e.g.\ calibration, filtering, and calculation of additional variables beyond the four-momentum) are carried out by tools that implement the |IJetModifier| interface.
Usually, these modifications must be done before the jet collection is recorded to the event store to preserve const-correctness for thread safety.
An |IJetModifier| that only adds new variables to jets as decorations (as opposed to modifying existing quantities) implements the |IJetDecorator| interface,
which is derived from |IJetModifier| but guarantees that the jets are treated as const.
This permits safely running |IJetDecorator|s on the jet collection after it is recorded to the event store,
which is useful for downstream operations such as flavour tagging or analysis-specific jet variable calculations.

Jet grooming is a technique that involves selectively modifying the constituents of a jet, and can include procedures such as trimming~\cite{trimming} and pruning~\cite{Ellis_2010}.
Jet grooming requires a more specialized treatment compared with other jet modifications, as it generally requires that the clustering procedure is repeated, with further interfacing with \textsc{fastjet}.
A dedicated |JetGroomer| class is provided to help this.
This is a Tool that takes as input the ungroomed jet collection and its input |PseudoJet| collection, and outputs a |PseudoJet| collection representing the groomed jets (which are then translated into a new |JetContainer|). Because many different grooming techniques exist, a different class derived from |JetGroomer| is provided for each one that implements the specifics of that method.

To simplify the jet reconstruction procedure for the user and streamline configuration, an Algorithm (|JetRecAlg|) is provided that wraps some of the previous steps.
It has three steps: first, creating a jet collection using an |IJetProvider|; second, running an arbitrary sequence of |IJetModifier|s on it; and third, recording the final collection to the event store.
|IJetProvider| is a unified interface for all Tools that create a jet collection. This includes |JetClusterer| and |JetGroomer| as well as |JetCopier|, a Tool that provides a jet collection by copying an existing one.

The same jet reconstruction code runs as part of the reconstruction and derivation-making (see Section~\ref{sec:derivations}).
In practice, the jets that are used for data analysis are built during the derivation step.
Jet-building can also be done even further down the processing chain as a part of an analysis, assuming the input constituents are available.
This might be done for analyses that do some jet reclustering, for example. The Algorithms and Tools used for jet reconstruction are sufficiently flexible to allow this.

\subsubsection{Missing transverse momentum}
\label{sec:reco:met}

Missing transverse momentum (\met) is a valuable observable that is used widely in particle physics analyses.
This observable serves to approximate the transverse momentum carried by particles that remain undetected as they traverse the detector.
It is essentially the negative vector sum of the transverse momenta of all objects in an event,
but the calculation is sensitive to the definitions and calibrations of these objects, and the treatment of overlaps between them.
This makes the reconstruction of \met~\cite{PERF-2016-07} particularly complex.
Object definitions can vary considerably, depending on how individual analyses optimise and prioritise their object selections.
For example, a set of tracks matched to calorimeter TopoClusters may be considered as an electron in one analysis,
but as part of a jet in another analysis, meaning different calibrations would be applied in the two cases.
To accommodate the wide spectrum of analysis needs, missing transverse momentum is only computed during the data analysis, according to its own event description.
At the reconstruction step, therefore, the aim is to ensure that the objects required to permit this calculation during the analysis are provided with the necessary information.
To make this possible, a compact representation of all possible overlaps between objects in the event is required to avoid double-counting in the \met\ calculation.
This is implemented in the |MissingETAssociationMap|~\cite{METSoftware}, which provides all information needed to compute the \met\ using any arbitrary object selection.
Jets are used as the basis objects for this representation: the map consists of a |MissingETAssociation| object for each jet in the event,
plus one `miscellaneous' association to capture detector signals not associated with any jet.
Each |MissingETAssociation| object contains information about which other objects share which detector signals with that jet, and with each other.
The reconstruction of \met\ is therefore divided into two steps: the construction of the association map,
and the usage of this association map to compute the final \met\ based on the object selection definitions of the given analysis.

Constructing the association map amounts to calculating the overlaps between all objects that could potentially go into the final event description.
This is done for each type of reconstructed object sequentially using |METAssociator| Tools.
Each type of object (jets, electrons, muons, photons, $\tau$-leptons, and `soft' detector signals --- signals that too low-energy to be associated with a reconstructed object) has a class derived from |METAssociator| that extracts the corresponding detector signals and computes the appropriate overlaps.
These overlaps are recorded in the association map, and the map is recorded in the event store.

The second step of computing the final missing transverse momentum is carried out using the |METMaker| Tool.
This takes as input the association map and reconstructed objects with analysis-specific selections applied.
First, a `term' of the overall vector sum is constructed for each type of reconstructed object by sequentially providing the corresponding reconstructed object collections.
The order in which these are provided defines a priority ordering for overlap removal, and only objects satisfying this overlap removal are included in the \met\ sum.
Keeping track of the overlaps during this process is done using a |MissingETAssociationHelper| object, which implements a transient thread-local cache recording this information.
Updates to ensure that this is a thread-safe process were made ahead of \RunThr.
Lastly, the |METMaker| adds each individual overlap-removed term to compute the final result.

This second step is applied by each analysis, downstream of the standard reconstruction, in order to account for analysis-specific selections.
Analyses are thereby able to determine whether some object should be considered an electron, photon, or jet, for example, or whether a muon is sufficiently well reconstructed to be included, and to use exactly those objects with their corresponding calibrations in the \met\ definition.
The analysis tools described in Section~\ref{sec:uncertaintyTools} help ensure that uncertainties on these objects are correctly propagated into the \met, and the \met\ tools provide an additional calibration and uncertainty on the soft detector signals described above.

\subsubsection{Flavour tagging}
\label{sec:reco:ftag}

Flavour tagging~\cite{FTAG-2019-07} is the process of distinguishing jets that contain $b$- and $c$-hadrons from those that do not.
This procedure is applied in two stages.
In the first stage, low-level algorithms are used to reconstruct the features of jets resulting from heavy-flavour decays.
These features are then used as inputs to multivariate classifiers, which constitute the set of higher-level algorithms used in the second stage of the flavour-tagging process.
During the flavour-tagging process, multiple jet collections can be tagged simultaneously either during reconstruction or derivation-making (see Section~\ref{sec:derivations}).
The primary purpose of flavour tagging, when applied during reconstruction, is to enable the monitoring of flavour-tagging outputs for data quality monitoring purposes~\cite{DAPR-2018-01} (see Section~\ref{sec:DQ}).
Flavour tagging is applied to jets built from both particle flow objects and calorimeter TopoClusters during reconstruction,
but the results of this process are not saved to the output xAOD file.
The flavour tagging process is then applied to particle flow and variable-radius jets during derivation-making, where outputs are saved in the output DAOD file.

Flavour tagging relies on simulated samples to enable the training of the tagging algorithms, or \emph{flavour taggers}.
In simulated samples, jet flavour \emph{labels} are assigned according to the presence of a hadron from the event generator within $\Delta R_y(\mathrm{hadron}, \mathrm{jet}) = 0.3$ of the jet four-momentum.
If a $b$-hadron is found, the jet is labelled a $b$-jet.
Without a $b$-hadron, if a $c$-hadron is found, the jet is called a $c$-jet.
If no $b$- or $c$-hadrons are found, but a $\tau$-lepton is found in the jet, it is labelled a $\tau$-jet.
Otherwise the jet is labelled a \emph{light}-jet.
This information is used for training the flavour taggers to efficiently identify $b$-jets and sufficiently reject background jets.

The first step of the flavour tagging process is to associate tracks to the jets in an exclusive way.
This is done by setting a maximum angular separation, $\Delta R_y$, between the jet and track four-momenta.
Given that $b$-hadrons with a higher \pt will decay to form more collimated jets than those with lower \pt,
the maximum allowed angular separation between the tracks and the jets decreases as a function of jet \pt.
The precise relation used is $\Delta R_y < 0.239 + \exp{\left(-1.22-0.0164/\GeV \times \pt\right)}$,
with $\Delta R_y \sim 0.45$ for jets with $\pt > 20$~\GeV and $\Delta R_y \sim 0.25$ for jets with $\pt > 200$~\GeV.
If a track is within the allowed distance from more than one jet, it is assigned to the jet with a smaller $\Delta R_y(\mathrm{track}, \mathrm{jet})$.

After track--jet association, flavour tagging is done on each jet independently.
It is possible to use multiple taggers without the need to create separate containers for the results of each.
Associated tracks are used to reconstruct secondary vertices, using two algorithms: \texttt{JetFitter}~\cite{ATL-PHYS-PUB-2018-025} and \texttt{SV1}~\cite{ATL-PHYS-PUB-2017-011}.
The parameters describing the secondary vertices from heavy flavour decays, together with the hit content and impact parameters of the associated tracks relative to the primary vertex,
are used by higher-level algorithms to compute the final flavour-tagging discriminant.
These higher-level algorithms include binned-likelihood estimators \texttt{IPxD},
and several neural networks that are trained using a \textsc{Python} software stack that runs outside of Athena~\cite{Umami}
and is based primarily on \textsc{NumPy}~\cite{harris2020array}, \textsc{HDF5}~\cite{HDF5}, \textsc{Keras}~\cite{chollet2015keras} / \textsc{TensorFlow}~\cite{tensorflow}, \textsc{PyTorch}~\cite{pytorch}, \textsc{PyTorch Lightning}~\cite{pytorchlightning}, and \textsc{Deep Graph Library}~\cite{wang2020deep}.
HDF5 datasets to quantify systematic uncertainties and for training are written directly from DAODs, using Athena (or one of the lighter projects described in Section~\ref{subsec:genInfra:cmake}), and are used as input to the \textsc{Python}-based training software.
Trained networks are then loaded using the \textsc{ONNX Runtime} library or \textsc{LWTNN} for inference within Athena.

\subsubsection{$\tau$-leptons}
\label{sec:tauReco}

The $\tau$-lepton reconstruction algorithms focus on hadronic $\tau$-lepton decays and the associated visible decay products, which form a \tauhadvis candidate.
Several improvements were made to $\tau$-lepton identification, classification and background rejection algorithms for \RunThr,
largely driven by the development of neural-network based approaches.
These approaches are discussed in Ref.~\cite{ATL-PHYS-PUB-2022-044},
alongside a detailed description of hadronic $\tau$-lepton reconstruction, identification and calibration for \RunThr.

Hadronic $\tau$-lepton decays display characteristic displacement, multiplicity and kinematic properties, with a branching ratio of approximately 65\%.
To reconstruct \tauhadvis objects corresponding to the visible decay products of a hadronically decaying $\tau$-lepton, \tauhadvis candidates are seeded by jets, reconstructed from calorimeter TopoClusters, using the \antikt algorithm with a radius parameter $R = 0.4$.
A $\tau$-lepton will travel away from the interaction point before decaying.
To identify the production vertex for each \tauhadvis candidate,
a dedicated $\tau$--vertex association algorithm is used~\cite{PERF-2013-06}.
The identified vertex serves as the basis of the coordinate system in which $\tau$-lepton identification variables are calculated.
The track parameters of tracks associated with the \tauhadvis candidate are recalculated relative to the $\tau$-lepton production vertex.

The tracks associated with a \tauhadvis candidate are processed by a track classifier,
and then categorised as either $\tau$ tracks (i.e.\ corresponding to charged $\tau$-lepton decay products),
conversion tracks (from electrons and positrons from photon conversions),
isolation tracks (likely originating from quark- or gluon-initiated jets) or fake tracks (mis-reconstructed tracks or pile-up tracks).
While a BDT-based method for $\tau$-lepton track classification was used during \RunTwo,
for \RunThr a novel method that uses a recurrent neural network (RNN) was developed~\cite{ATL-PHYS-PUB-2022-044}.

A separate $\tau$-lepton identification algorithm was developed during \RunTwo to distinguish true \tauhadvis from mis-identified \tauhadvis originating from quark- and gluon-initiated jets,
resulting in significant improvement in the rejection of mis-identified \tauhadvis.
The algorithm uses an RNN architecture that is the same as that described in Ref.~\cite{ATL-PHYS-PUB-2019-033}.

Electrons can be mis-identified as \tauhadvis objects, especially in the case of 1-prong \tauhadvis candidates.
A new electron veto algorithm~\cite{ATL-PHYS-PUB-2022-044} for 1-prong and 3-prong \tauhadvis,
also based on an RNN, was developed for \RunThr.
This new algorithm improves electron rejection by approximately a factor of three for a given \tauhadvis efficiency compared with the BDT-based veto used during \RunTwo~\cite{ATL-PHYS-PUB-2015-045}.

A DeepSet neural network (DeepSet NN)~\cite{DBLP:journals/corr/ZaheerKRPSS17} algorithm was developed for \RunThr to both classify the decay modes
and calculate the visible four-momenta of reconstructed $\tau$-lepton candidates~\cite{ATL-PHYS-PUB-2022-044}.
This algorithm can significantly improve the energy calibration of the reconstructed \tauhadvis candidates relative to the BDT-based algorithm used during \RunTwo~\cite{PERF-2014-06}, improving the resolution by almost 50\%.
The algorithm exploits the reconstruction of individual charged and neutral hadrons of the $\tau$-lepton decays using decay kinematics,
track impact parameters and calorimeter energy cluster properties for $\tau$-lepton tracks, nearby conversion tracks, $\pi^{0}$ candidates and \emph{photon shots} (local energy maxima in the first layer of the electromagnetic calorimeter, associated with photons from the candidate decay of a $\pi^{0}$).

\subsubsection{Heavy ions}
\label{sec:hiReco}

Heavy ion collision reconstruction differs from the standard proton--proton reconstruction due to the large number of underlying event (UE) particles produced in the ion collisions.
The standard calorimeter energy clustering in heavy ion reconstruction begins from 0.1$\times \frac{\pi}{32}$ `towers' assembled from geometric combinations of cells, rather than from the cells themselves~\cite{ATLAS-CONF-2015-016}.
The same algorithms as described above are used to reconstruct tracks in heavy ion events, but often different, somewhat more restrictive requirements are applied during physics analyses to reduce fake rates.
For heavy ion data-taking the jet, electron and photon reconstruction sequences are run only after the UE is `removed' from the calorimeter.
During this process the energy deposition from the UE is subtracted from the energy of reconstructed calorimeter energy clusters,
and containers for these subtracted clusters are created.
The UE subtraction process and subsequent heavy ion jet reconstruction procedure is detailed in Refs.~\cite{HION-2011-02,ATLAS-CONF-2015-016}.

The average energy density associated with the UE is calculated as a function of $\eta$ using all calorimeter layers for jets and per-layer for electrons and photons.
It is modulated in $\phi$ to account for anisotropic flow present in heavy ion collisions and corrected for detector non-uniformities.
This calculation excludes \emph{seed jets} from the average, where these seed jets are jets reconstructed using the \antikt algorithm with energy above a given threshold.
From this calculation the UE is subtracted from the seed jets.
The energy calculation, exclusion and subtraction procedure is applied iteratively using different jet definitions.
Each of these iterations proceeds according to the following steps:

\begin{enumerate}
\item Identification of seed jets.
\item Computation of UE parameters~\cite{ATLAS-CONF-2015-016}.
\item Application of the UE subtraction to the towers in each jet, and recalculation of the jet four-momenta.
\item Application of the energy scale calibration procedure.
\end{enumerate}

The resulting jets are UE-subtracted and fully calibrated, permitting their use in the construction of seed jets in the next
iteration. Three iterations are done in total for jet, electron and photon reconstruction.
The criteria for seed jets built with \antikt radius parameter $R$ at each iteration are as follows:

\begin{enumerate}
\item Seed jets are built using $R=0.2$ and must have a maximum tower $\ET>3$~\GeV and a ratio of maximum tower $\ET$ to mean tower $\ET$ greater than 4 (all without UE subtraction applied).
\item Seed jets are built using $R=0.2$ and must have calibrated $\pt>25$~\GeV after subtracting the UE, as determined during the first iteration. Track jets with $\pt>7$~\GeV can also be included.
\item For the final iteration, seed jets are built using $R=0.4$ and must have calibrated $\pt>25$~\GeV after subtracting the UE, as determined during the second iteration.
\end{enumerate}

%
%
%
%
%
%
%
%
%
%
%
%
%
%
%
%
%
%
%
%
%
%
%
%
%
%
%
%
%
%
%
%
%
%
%
%
%
%
%
%
%
%
%
%
%
%
%
%
%
%
%
%
%
%
%
%
%
%
%
%
%
%
%
%
%
%
%
%
%
%
%
%
%
%
%
%
%
%
%
%
%
%
%
%
%
%
%
%
%
%
%
%
%
%
%
%
%
%
%
%
%
%
%
%
%
%
%
%
%
%
%
%
%
%
%
%
%
%
%
%
%
%
%
%
%
%
%
%

%
%
%
%
%
%
%
%
%
%
%
%
%
%
%
%
%
%
%
%
%
%
%
%
%
%
%
%
%
%
%
%
%
%
%
%
%
%
%
%
%
%
%
%
%
%
%
%
%
%
%
%
%
%
%
%
%
%
%
%
%
%
%
%
%
%
%
%
%
%
%
%
%
%
%
%
%
%
%
%
%
%
%
%
%
%
%
%
%
%
%
%
%
%
%
%
%
%
%
%
%
%
%
%
%
%
%
%
%
%
%

%

%


\subsection{Derivations}
\label{sec:derivations}


\subsubsection{Definition and role of derived data}
\label{sec:derivationintro}

A feature common to many high energy physics analyses is the use of intermediate-sized data types at some stage of the analysis procedure. Typically, these files are derived from the output of the reconstruction (AOD in ATLAS) and may have the following features:
\begin{enumerate}
\item Their size is usually around a few percent to a few per mil of the input data;
\item They usually contain all of the information necessary for smearing, scaling, selection, calibration and other operations on reconstructed objects (collectively known in ATLAS as combined performance operations), and to determine the systematic uncertainties related to these operations;
\item They are used privately by physicists or groups of physicists to produce custom data files (usually \textsc{ROOT} ntuples);
\item They are typically modified and reproduced more than once for a given version of the input, and may be read several or many times by the analysis teams as they produce different private ntuples;
\item They may be aimed at one analysis or perhaps a group of related analyses (for example sharing the same final state).
\end{enumerate}

The second point above is particularly important since an optimal understanding of the reconstruction, which feeds into the combined performance recommendations and calibrations, tends only to be achieved after many months of study of the data and MC simulation. Moreover, different domains of the reconstruction update their recommendations at different times, so it is usual practice to store all the information needed to allow the application of these recommendations during user analysis.

ATLAS defines four standard operations for building derived data files (see also Section~\ref{sec:edm}):

\begin{itemize}
\item \emph{Skimming} is the removal of whole events, based on some criteria related to the features of the event;
\item \emph{Thinning} is the removal of individual objects within an event, based on some criteria related to the features of the object;
\item \emph{Slimming} is the removal of variables within a given object type, uniformly across all objects of that type and all events (unlike the other operations, slimming does not depend on any event/object properties because the same variables are removed for every event and object);
\item \emph{Augmentation} involves adding information in the form of new variables (\emph{decorations}) or new objects, which in some way summarise aspects of the reconstructed data, allowing much larger volumes of data to be dropped.
\end{itemize}

At the start of \RunTwo, ATLAS introduced centrally produced intermediate data files written in the xAOD format (see Section~\ref{sec:edm}), formally called Derived AODs (DAODs) but widely known across the collaboration as \emph{derivations}. These were defined and managed by the physics and combined performance groups and were usually tailored to specific analyses. In almost all cases the events were skimmed, with the skim rate varying depending on the physics analysis and whether the input events were real or simulated. At times during \RunTwo there were more than 100 such formats in active use. These overlapped heavily for simulated events, leading to excessive disk space use, and managing such a large profusion of formats became increasingly difficult. Consequently, ATLAS has revised the model for \RunThr, and this is described in the following sections.

\paragraph{Unskimmed derived data}

The main innovation in the \RunThr analysis model is the introduction of two new unskimmed data types written in the xAOD format known as \texttt{DAOD\_PHYS} and \texttt{DAOD\_PHYSLITE}. These are conceptually similar to the MiniAOD and NanoAOD of CMS~\cite{CMSnanoaod}.

\texttt{DAOD\_PHYS} contains all of the object types and variables required by the combined performance tools to apply recommendations and calibrations (the \emph{common slimming} content) and consequently is very similar to the existing DAODs, except that it is unskimmed and therefore can be used by a wide range of analyses. Despite containing all of the essential analysis variables, it is smaller than $50$ $(40)$~kB/event for simulated (data) events, which is less than $10\%$ of the AOD size (see Section~\ref{sec:datasizes}). Most of the size reduction is achieved by removing all tracks with \pt below 10~\GeV that are not associated with leptons or jets. Further reductions are obtained by dropping most of the MC truth record in favour of summary information for the substantive truth particles (detectable leptons, gauge bosons, hypothetical new particles and $b$-hadrons), and also by removing trigger objects, leaving behind only the trigger decision and information indicating which offline objects were responsible for firing each trigger. Essentially any analysis that does not require low-\pt tracking information can use \texttt{DAOD\_PHYS}, which only excludes B-physics and long-lived particle searches. Consequently, it is expected that $80\%$ of the research output of ATLAS will be able to use the format.

\texttt{DAOD\_PHYSLITE} is smaller still, with much of the common slimming content also dropped. Instead, the tools for applying recommendations and calibrations are run as the format is built, and the calibrated objects are written directly into the format in place of the uncalibrated objects produced by the reconstruction. The calibrations are subject to instrumental systematic uncertainties, so the central values are recorded, and the variables required to estimate the uncertainties are retained for use by analysts. \texttt{DAOD\_PHYSLITE} is around $15(10)$~kB/event for simulated (data) events. Assuming that all of the calibrations are available during its production, \texttt{DAOD\_PHYSLITE} has the same functionality as \texttt{DAOD\_PHYS}, the only added value of the larger format being the ability to re-apply recommendations and calibrations at analysis level, which would be relevant in the event of a combined performance update. \texttt{DAOD\_PHYSLITE} can also be produced from \texttt{DAOD\_PHYS}, and this workflow, which is at least six times faster than the usual workflow from AOD due to the smaller size of the input and the absence of the jet and $\tau$-lepton reconstruction (done as \texttt{DAOD\_PHYS} is built), is expected to be highly significant in \RunFour.

\begin{figure}[tbp]
\centering
\subfloat[]{
\includegraphics[width=0.45\textwidth,valign=c]{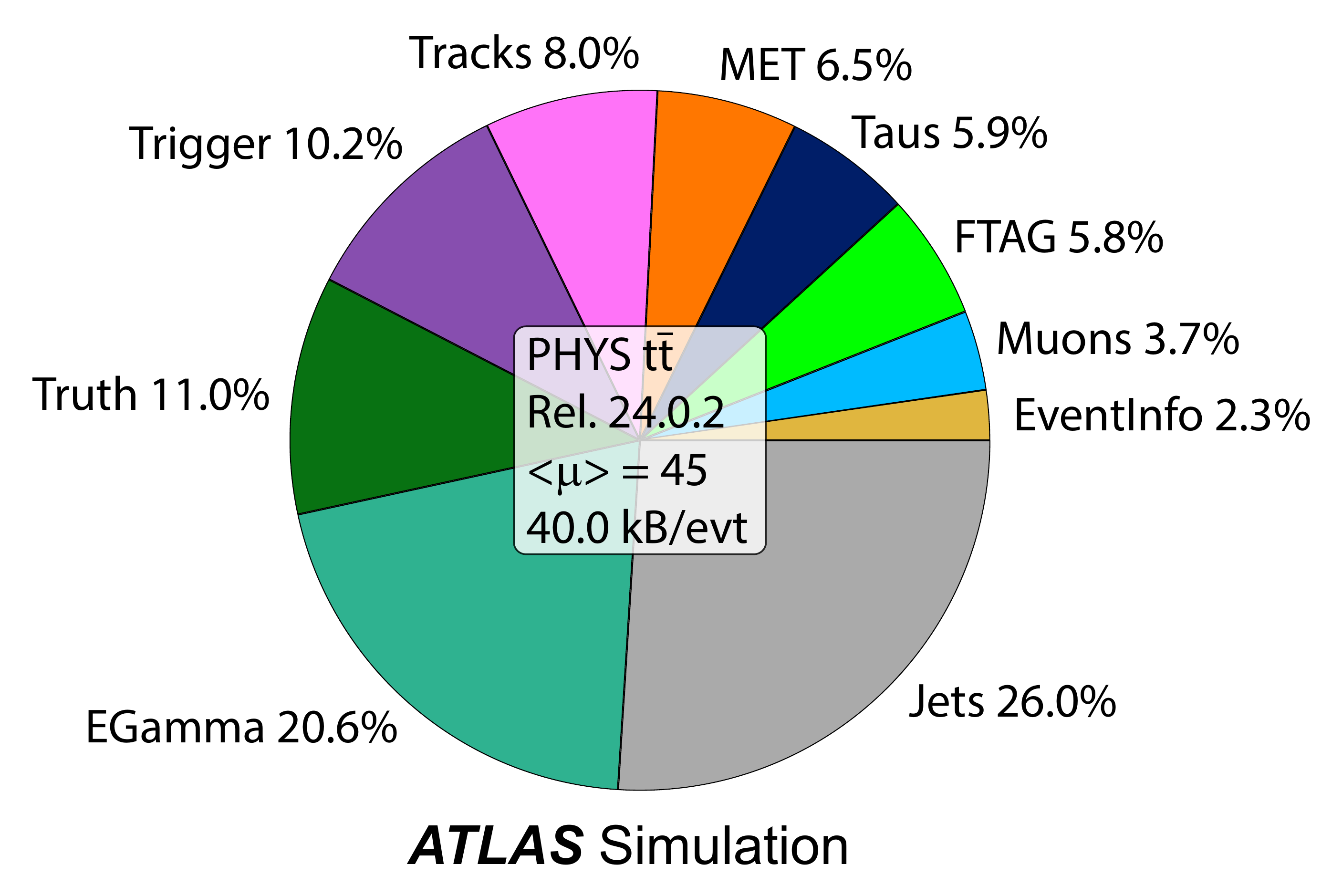}
}
\subfloat[]{
\includegraphics[width=0.53\textwidth,valign=c]{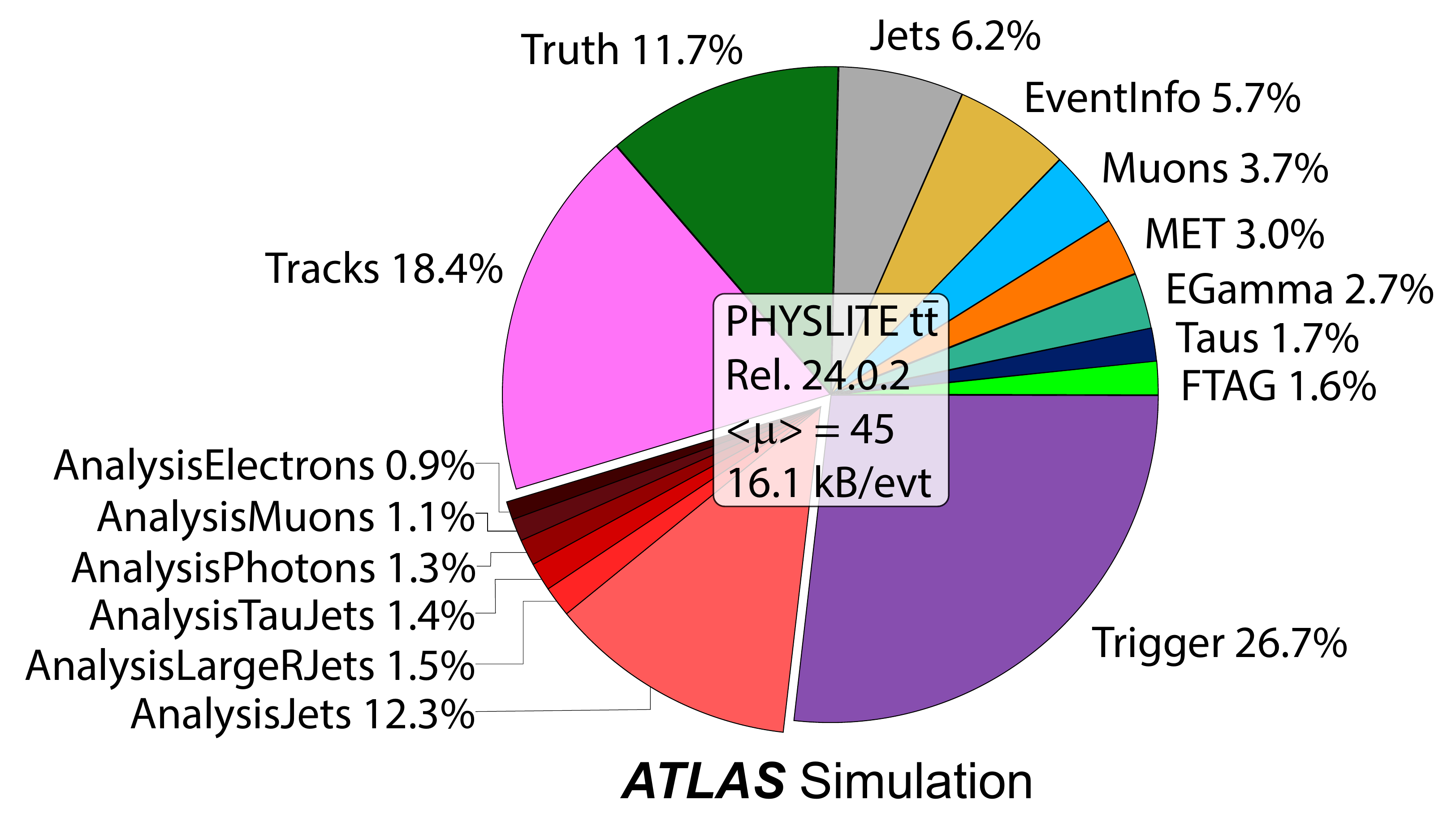}
}
\caption{\label{fig:deriv:pies}The composition of a \RunThr $t\bar{t}$ sample in (a) \texttt{DAOD\_PHYS} and (b) \texttt{DAOD\_PHYSLITE} format. The disk size from \texttt{DAOD\_PHYS} is dominated by the jet collections, while \texttt{DAOD\_PHYSLITE} is dominated by the trigger information. The containers that hold the calibrated analysis objects in \texttt{DAOD\_PHYSLITE} are depicted in red with an offset.}
\end{figure}

Figure~\ref{fig:deriv:pies} shows the composition of \texttt{DAOD\_PHYS} and \texttt{DAOD\_PHYSLITE} for a \RunThr $t\bar{t}$ MC simulation sample, which includes pile-up with, on average, 45 interactions. \texttt{DAOD\_PHYSLITE} is not a fixed format; it will change over the course of \RunThr, with the aim to make it as useful as possible for analysers. There are also ongoing efforts to reduce the size of \texttt{DAOD\_PHYSLITE} further.

\paragraph{Skimmed derived data and event sample augmentation}
\label{sec:deriv:augmentation}

Many ATLAS activities require more information than is available in \texttt{DAOD\_PHYS(LITE)}. These include combined performance analyses, where the recommendations and calibrations referred to in the previous section are derived. Physics analyses requiring low-\pt tracks, such as B-physics and some long-lived particle searches, are also unable to use the two new data types due to the aggressive thinning of inner detector tracks. Such analyses will consequently continue to use the \RunTwo analysis model; that is, they will define DAODs containing all of the variables and objects needed for their tasks, and will compensate for the bigger per-event size by removing events that are not relevant to their analyses. It is unlikely that the combined performance groups will need to process all of the data and MC simulation to determine the calibrations and recommendations. Due to these size reduction measures, and the far fewer DAOD formats, it is expected that production of these residual skimmed formats will be sustainable even during the HL-LHC era.

An inefficiency arises in this model when there is significant overlap between the additional derivations and the common \texttt{DAOD\_PHYS} content, combined with a large skim fraction. To address this issue and reduce data duplication, ATLAS has developed a feature called \emph{Event Sample Augmentation}~\cite{PVGCHEPProceedings}. This feature enables the extension of a base format (e.g.\ \texttt{DAOD\_PHYS}) with supplementary data obtained from a secondary skimmed format, but only for the subset of events that meet the skimming criteria. In the Event Sample Augmentation configuration, the secondary format is incorporated as a \textsc{ROOT} friend tree into the base format and stored within the same file (storage in a separate file is also possible). This design allows seamless access to the supplemental data during downstream data processing, eliminating the need for redundant information to be stored to disk. Moreover, since the augmentations are stored in separate trees, they do not compromise I/O performance for clients who only require the baseline DAOD content.

In one benchmark scenario, a long-lived particle analysis required the inclusion of extra containers on top of the baseline \texttt{DAOD\_PHYS} content, leading to an average event size increase of approximately 40\%. In the \RunTwo model, this situation would cause a significant 56\% rise in disk space usage due to significant redundant data. However, with the implementation of Event Sample Augmentation, incorporating the additional data for just 40\% of the events results in a mere 16\% increase in overall disk usage. This approach proves notably more efficient than either the \RunTwo model or the inclusion of the extra content in \texttt{DAOD\_PHYS} directly. The Event Sample Augmentation thus provides an effective solution that enables analyses to use customized data while mitigating resource impacts.

\subsubsection{Derivation framework software}
\label{sec:deriv_software}

The software used to produce derived data products (the \emph{derivation framework}) is
built on the software used in \RunTwo. It is implemented within the Athena framework,
and so shares many similarities with reconstruction. In fact, to save disk space
in the AOD, some parts of reconstruction are normally run during derivation building.
In particular, jet reconstruction (see Section~\ref{sec:reco:jets}) and the
reconstruction of objects that depend on jets (flavour tagging, \met, and $\tau$-leptons;
see Sections~\ref{sec:reco:met}--\ref{sec:tauReco}) are normally included in derivation-making.
This ensures that any specialisation (e.g.\ for particular grooming techniques to be
applied to large-radius jets) can be provided to an analysis without the need to save all
possible combinations directly in the AOD. The derivation framework consists of an Athena
Algorithm called a \emph{kernel} that drives the event loop. Each derived data product
implements such an Algorithm, and passes it a series of Tools for skimming, thinning and
augmentation, which it calls in turn to produce the new data type. Multiple data
products can be produced in the same job from the same input (\emph{train production}),
even with different skimming selections.

Alongside the other data production activities, the
configuration layer of the derivation framework was recently updated to use the new
Component Accumulator (see Section~\ref{sec:core:configuration}).
The derivation framework can only be run as serial or
multi-process (AthenaMP, see Section~\ref{sec:core}) jobs; multithreading support
(AthenaMT) is in development to be deployed before \RunFour.

To avoid having to merge separate output files from forked workers or processes of a
derivation job, ATLAS developed a |SharedWriter| process that collects data objects
from all workers and writes them to a single output file. In \RunTwo, only the
|SharedWriter| itself would interact with \textsc{ROOT} for file writing, ensuring that the
container entry numbers were synchronized. This means that when the |SharedWriter| received
data from a worker, it needed to modify the data somewhat (e.g.\ to account for the number
of events already written to the output file) before recording them. Similarly, because in
\RunTwo the container entry number
was used as external reference, objects were not relocatable, for example by fast merging
techniques. Since it eliminated the merging step (which had to be run serially),
the |SharedWriter| approach significantly sped up derivation production. But as the single
|SharedWriter| has to compress all data, the scalability with output data volume and the
number of workers could become limiting. Therefore, a new version of the |SharedWriter|
was developed for \RunThr to support unskimmed formats (i.e.\ a larger data volume) and
potentially higher core counts. The new |SharedWriter| relies on a change to the ATLAS
persistence navigational infrastructure: For \RunThr, the persistent references were
changed so that rather than relying on the container entry number they use a unique
identifier that is stored within the container itself. This infrastructure allows fast
merging without invalidating existing references. In this new design, the workers write
their data to a \textsc{ROOT} memory file that is sent to a single |SharedWriter|. Thus, the
worker processes can compress data in parallel, and for a moderate increase in memory
consumption (due to compression buffers) a very significant speed up and much better
scalability is achieved~\cite{SerhanMete:20223X}. The performance improvement is also
shown in Section~\ref{sec:deriv_performance}.


\subsection{Forward detectors}
\label{sec:forward}

The four forward systems are treated specially throughout the software workflow, from simulation through to data analysis.
These generally have stand-alone \Geant{}-based simulation configurations for comparisons to test beam results and
the understanding of detector calibrations.

LUCID--2 is used primarily for luminosity measurements~\cite{DAPR-2021-01}.
Although the detector can be simulated and the signals digitised and reconstructed as a part of the standard
software workflow, this is not done for most MC simulation samples. The analysis for luminosity proceeds from
bespoke data formats, as individual photomultiplier tube signals are analysed. The simulation of LUCID--2 has
been used most for studies of the impact of changes to the detector geometry (e.g.\ changes to the beampipe)
on the ATLAS luminosity measurements.

Far-forward particles are normally not simulated to save CPU time. The far-forward detector systems
(ZDC, AFP, and ALFA) also require simulating material and beam pipe regions not normally included in the ATLAS
detector description, including some regions under the responsibility of the LHC machine group.
For ALFA and AFP, forward-pointing particles are propagated through a simple simulation of the LHC optics
before being injected into specialised simulations of the forward detectors. For analysis purposes, a simple
fast simulation is often used to approximately model the response of these detectors directly from the
forward particle momenta. For events entering these analyses, beam divergence and crossing angles are
included in the simulation.

The Zero-Degree Calorimeter (ZDC) is primarily used for triggering and analysis of heavy ion collisions. A detailed
\Geant{}-based simulation is used to model the response of the detector to single neutrons and neutral pions.
These simulations can be computationally intensive, because the particles incident on the ZDC are often several \TeV.
The simulation software (including digitisation) is being reviewed to reduce its resource needs.
Nevertheless, for some events the same forward-transport mechanisms as with AFP and ALFA will be used, particularly in
the analysis of data from the new Reaction Plane Detector~\cite{RPD_PHD} (a part of the ZDC that is new in \RunThr).


\subsection{Phase-II upgrade support}
\label{sec:upgrade}

The planned Phase-II upgrade of the ATLAS experiment for the HL-LHC requires robust, performant software several years before data-taking to study the impact of changes to the detector design, and to understand and estimate the physics performance of the new detector. The necessary developments, described in this section, were integrated into the existing \RunThr software; some major software developments are still planned before \RunFour, and these are described in Section~\ref{sec:outlook}. Required features include the ability to simulate the Phase-II detector geometry, digitise the simulated HITs, and reconstruct physics objects, all including the new and improved detector systems. Once the concepts are proven and approved, the software must also support new developments and features targeting the HL-LHC to be ready for data-taking in \RunFour. This is also important for the HL-LHC road map (see Section~\ref{sec:outlook}), where support for the Phase-II detector is fundamental to achieve the milestones defined on the set timescales.

There are several major changes foreseen as a part of the Phase-II upgrades~\cite{ATLAS-TDR-29}.
The entire inner tracker will be replaced with a new system, ITk, with significantly expanded coverage in $\lvert\eta\rvert$.
A new timing detector, the High-Granularity Timing Detector (HGTD), will be installed in the forward region.
Significant improvements are planned for readout systems and trigger systems throughout the detector.

Historically, ATLAS chose to provide a dedicated `upgrade release', branched from the main release that was used for \RunTwo data-taking in early 2016. This release was decoupled from the main development branch, mainly to simplify conceptual testing and R\&D projects without worrying about possible interference with on-going data-taking. Inevitably the two releases diverged to the extent that it was no longer possible to compare the current state-of-the-art data reconstruction performance and the future expected detector performance. Estimates had to rely on old algorithms no longer in use by ATLAS analysers at the end of \RunTwo.
The infrastructure updates that took place after the upgrade release was defined also created difficulties, as the upgrade and main releases used different version control systems, build systems, for example.
After the approval of all ATLAS Phase-II TDRs, the urgent need to provide simulated results ceased and significant effort could be put into integrating upgrade software into the main development branch again. The main release now supports both the present and upgrade detectors, providing a solid ground for future developments working towards \RunFour data-taking, while keeping up-to-date with advancements made during \RunThr.

During the conceptual design process of the Phase-II detectors, the layouts rapidly changed, which meant the need to implement new or modified geometries quickly in simulation. The framework for building the ATLAS geometry (GeoModel, see Section~\ref{sec:dd}) is robust but somewhat cumbersome, and not designed for quick iterations. A new tool to easily build the geometries via an XML format, called GeoModelXML, was developed for building the ITk Strip geometry. This tool proved useful and easier to work with, and was adopted by a large fraction of the sub-detectors of ATLAS to prepare for the HL-LHC. %

Another major challenge was to incorporate into the software the new HGTD, which will replace the Minimum Bias Trigger Scintillators (MBTS) during \RunFour. In the simulated geometry, the MBTS is situated and built within the calorimeter envelope/volume. Initially, the HGTD was added in the same place as the MBTS in the simulation framework. This came with few issues that were discovered during development. For instance, the calorimeter volume is treated differently in regards to treating the truth particle record: a significant fraction of information is not stored by default. This created issues for digitising the HGTD HITs and for performance studies. Moreover, there was no support in the EDM for the timing information to be propagated from the hits to the track particles. The time distribution of pile-up proton--proton collisions must also be modelled. Proper solutions to these problems are now in place in the main development release.

Digitisation for the Phase-II tracking detectors is heavily based on the software utilised for the present pixel and SCT detectors, with adaptations to the new granularity and expected operation conditions in terms of low and high voltage, and charge thresholds. This software was developed for the original detectors more than 10 years ago and was not optimised for the high pile-up conditions expected at HL-LHC, which brings with them several complications. The resource usage in terms of CPU, but especially in memory, is far from sustainable in the long-term. For example, the radiation damage modelling used in the pixel detector digitisation (see Section~\ref{digi:sec:pixel}) is satisfactory for current radiation levels, but for the radiation levels expected during the HL-LHC is far too resource-intensive. Efforts are ramping up to revamp the code and several milestones were defined in the HL-LHC roadmap to make pile-up digitisation fully multithreading compatible and to reduce the overall memory usage.

Similarly, as with digitisation, the upgrade reconstruction software is derived from the existing ATLAS algorithms for all physics object domains, although it was adapted to the larger $\eta$-coverage of the ITk. However, track reconstruction was studied in detail during several years and is the most advanced among the domains. A prototype~\cite{ATL-PHYS-PUB-2019-041} already meets the HL-LHC CPU resource requirements.


%
\section{Software integration, evaluation, and validation}
\label{sec:processing}

Before any real or simulated data is used within physics analyses, the quality of the software and configuration used must be thoroughly vetted.
The validation proceeds through several steps, only some of which are shared between data and MC simulation.
An extensive suite of data quality tools is applied to check the real data; these are described in Section~\ref{sec:DQ}.
The same tools are often used both online (i.e.\ in real time as the data are taken) and offline (later, after the data have been recorded) to look for issues with the detector or the data itself, as well as reconstruction and conditions (e.g.\ calibration, noise masking, or alignment) problems that might arise.
A separate procedure, with separate tools, is used to validate configurations of the MC simulation chain; this procedure is described in Section~\ref{sec:validation}.
The computing performance of the software, for example in terms of CPU time, memory consumption, and output file size, is constantly monitored.
The performance is described in Section~\ref{sec:perf}.
When a new MC simulation or data (re)processing campaign is to be undertaken, all of these tools are used to carefully validate the software and configuration to be used.
The steps required for these campaigns to be launched are laid out in Section~\ref{sec:campaigns}.

\subsection{Data quality monitoring of collision data}
\label{sec:DQ}

A detailed description of the ATLAS data quality operation and performance for \RunTwo can be found in Ref.~\cite{DAPR-2018-01}. This section focuses on the software and computing support for data quality.

During data-taking the quality of recorded data may occasionally be compromised due to, for example, hardware failures, noisy detector elements, or configuration problems.
To ensure only high-quality data are selected for use in downstream analysis,
an extensive suite of data quality (DQ) monitoring and assessment tools are applied to each ATLAS physics run.

DQ checks begin in real time, on-site in the ATLAS Control Room (ACR).
Here dedicated \textit{shifters} are on duty around the clock to monitor the status of the detector and the data itself at various points in the data flow.
Once a run has completed and the RAW data have been recorded, a two-stage offline data quality process begins~\cite{DAPR-2018-01}.
The first stage takes place during the calibration loop, and the second stage makes use of the results from the prompt bulk reconstruction (which includes all promptly processed streams),
as indicated in Section~\ref{sec:dataflow}.
The end product of the data quality assessment is a \textit{good runs list} (GRL): a list containing, run-by-run, all luminosity blocks certified for physics analysis (see Section~\ref{sec:eventSelection}).
Depending on their precise needs, different GRLs might be used by different analyses of the same data-taking period.

Time-dependent information about, for example, detector and trigger status or run configuration are stored in the ATLAS conditions database (see Section~\ref{subsec:databases:conditions}).
DQ issues are flagged by the storing of \emph{defects} in the defect database~\cite{defectDB}, a subcomponent of the conditions database.
Defects are set for a given run according to the outcome of the data quality assessment, though they are sometimes uploaded automatically based on feedback from other supporting infrastructure.
If a defect is \emph{present} for a given luminosity block,
and that defect is severe enough to prevent the inclusion of those data in physics analysis, the corresponding luminosity block is absent in the resulting GRL.

During data-taking several applications are used to monitor the status of conditions and to record those in the conditions database.
The Detector Control System (DCS)~\cite{DCS} records the operational status of detector hardware components, including e.g.\ component temperatures.
Another application, GNAM~\cite{Ref-GNAM,GNAM}, monitors the detector status at several stages of the data flow and generates monitoring histograms.
Events passing through the hardware trigger are monitored via the trigger system.
This includes monitoring events not recorded to disk, thereby ensuring interesting events are not unexpectedly discarded.
Histograms are produced at the high-level trigger computing farm to monitor HLT operation.
A subset of collision events are also passed through the full reconstruction chain at this stage,
to produce histograms that can be used to monitor high-level information and the quality of reconstructed physics objects.
The Data Quality Monitoring Framework (DQMF)~\cite{DQMF} provides algorithms to do automated checks on the histogram outputs.
A key requirement of the DQMF is that it is lightweight enough to comply with strict time constraints for data processing.
The DQMF is used in both the online and offline DQ chains, which means it must be compatible with both the online and offline software environments.

The Information Service (IS) retrieves monitoring data from the Trigger and Data Acquisition (TDAQ)~\cite{TDAQ} systems for temporary storage so that they can be shared between various applications.
The IS can also temporarily host histograms generated by GNAM and DQMF results.
Histograms stored in the IS can be retrieved and rendered by dedicated display tools.
The two main display tools used in the ACR are the Data Quality Monitoring Display (DQMD)~\cite{DQMD} and Online Histogram Presenter (OHP)~\cite{OHP},
with DQMD providing a global view of data and detector status in a structured hierarchy, and OHP allowing the display of any published histogram.
Event displays (see Section~\ref{sec:eventdisplay}) are also used for online DQ monitoring in the ACR, as discussed in Section~\ref{sec:eventDisplaysOnline}.

A common framework is used to fill and manage DQ monitoring histograms in both the online software trigger and the offline reconstruction code.
This framework provides an interface to users in which client code only provides the values to plot,
deferring the detailed specification of the histograms themselves to the Athena runtime configuration.
In this way, histograms can be added and modified easily without changing any compiled code.
The user code is completely insulated from the representation of the histograms or filling,
rebinning, and other operations on them.
This layer of abstraction also allows the entire system to be safely multithreaded in a controlled manner as all access and changes proceed through tightly controlled APIs.
This architecture will also permit straightforward migration should the histogramming backend be changed in the future.
Several optimisations are provided by this code~\cite{Bordulev:2021fxm},
such as the ability to `vectorise' filling operations by providing many data points at once in a single library call,
and providing several variables at the same time to trigger the filling of multiple histograms.

%
%
%
%
%
%
%
%
%
%
%
%
%
%
%
%
%
%
%
%
%
%
%
%
%
%
%
%
%
%
%
%
%
%


\subsection{Validation}
\label{sec:validation}

The Physics Validation Group (PVG) is tasked with investigating changes to the core code base, be it at the level of event generation, simulation or reconstruction. Since new MC or data reprocessing campaigns (described in Section~\ref{sec:campaigns}) require significant resources (in terms of both computing and person power), validation is a crucial step. New derivation formats or major updates to the existing derivation formats also require extensive validations, but these are carried out by a different group within ATLAS, the Analysis Model group.

Specifically, PVG is tasked with understanding whether such changes affect the accuracy of the physics modelling in MC simulation and detector data events. Validation of detector data (as opposed to simulated data) is done only occasionally, for example to compare setups for reprocessings after changes to reconstruction algorithms or conditions. If changes to any physics observables are seen, PVG determines whether these changes are expected and acceptable. If the changes are unexpected then the group tracks down the causes of these issues and works with experts to understand and resolve the issue to restore a sensible physics description. The group consists of two conveners, and one or two validators for each physics object being studied: tracking, electrons, photons, muons, $\tau$-leptons, TopoClusters, particle flow, jets, flavour tagging, and \met. Experts from each combined performance group also often take part in these validation efforts, with $\mathcal{O}$(100) validation tasks being carried out each year. Validation itself therefore requires substantial CPU and human assets.

Validation tasks are defined starting with a request to validate a certain feature (e.g.\ a simulation code change). Nightly testing helps safeguard against changes from simple coding bugs or unintended consequences (see Section~\ref{subsec:genInfra:releaseTesting}); validation tasks are normally created for more significant, intentional changes. A \textsc{Jira}~\cite{Jira} ticket (see Section~\ref{sec:infrastructure}) is created with a brief description of the task, from which the PVG conveners work with the relevant experts/conveners to define the production configurations that will evaluate these changes. A reference is defined from a known and stable code base from which the changes can be validated. The test is then defined starting from the reference and changing the specific feature that is under investigation. It is common for multiple references and tests to be defined within a single validation task, to ensure that all aspects of the relevant changes are investigated (e.g.\ reconstruction might be tested both with and without pile-up). While normally the test are done in a stable, numbered release, occasionally physics validations are done using nightly releases (see Section~\ref{subsec:genInfra:buildSystem}) for the test configuration. For very special cases, nightlies can also be used for the reference configuration, for example when a release with a validated feature has not been built yet. This is particularly useful when development is proceeding rapidly, or a significant change must be deployed and then backed-out until it is fully validated.

Once the references and tests are defined, production configurations are created and the PVG conveners launch jobs to produce a set of physics validation samples (\emph{Sample-A}). Sample-A is a list of about a dozen relevant MC simulation samples that effectively populate and test the different objects used to complete the validation task, including both complete physics events and single particles. For example, the jet validation usually uses both a \ttbar\ sample and, separately, a high-\pt\ dijet sample. Similarly, flavour tagging will also use the \ttbar\ sample, and additionally a $Z'\rightarrow q\bar{q}$ sample with a $Z'$ mass of 4~\TeV to test flavour tagging at high-\pt\. The electron (photon) validation includes samples of single electrons (photons). Each sample is produced with 100,000 events to avoid large statistical uncertainties that would make small changes hard to observe consistently.

Once the samples are ready, each group of physics validators runs specialised analysis code over the relevant samples to produce a standard set of $\mathcal{O}$(100) histograms of quantities relevant to their respective physics object. These plots are uploaded to a webpage for easy viewing, where physics validators scrutinise the reference in comparison to the test, together with the respective object experts. Interesting or discrepant results are presented in a weekly PVG meeting, where the conveners together with most validators and experts can collectively scrutinise and discuss the results.

Ultimately, physics validators mark the task as either green (all good), yellow (discrepancies, but may be understood or need further study), or red (clear issues). If all objects report the task to be green, then PVG signs off the task and the ticket is closed. If the task is marked by any object as yellow or red, then further follow-up is done with experts, and either the results are ultimately understood or a new reference and test are defined and the physics validation cycle repeats to look into these differences further.

To get an even broader view, every 6--12 months a round of Physics Analysis Validation (PAV) is done, in which analysis teams from the Higgs, Exotics, HDBS, Standard Model, SUSY and Top physics groups run their analysis code over a standard set of MC simulation and detector data samples, comparing a well-known reference to a recent test version which has the integrated changes from numerous validation tasks. This effort is vital as it brings even more scrutiny, crucially in numerous corners of kinematic and physics usage phase-space, using standard analysis tools. While standard physics validation covers common ATLAS data formats, PAV also ensures that validation is done for the entire workflow down to final analysis outputs.

Leading up to \RunThr, there was a large and in-depth validation effort, starting from the \RunTwo\ code base and sequentially checking both optimisations and changes due to the Phase-I upgrade. This validation effort started in October 2021 with the validation of the first \RunThr\ geometry tag and continued right up to July 2022 when the MC21a production campaign was launched on a large scale (see Section~\ref{sec:campaigns}). The major milestones along the way were the validation of the various conditions updates that involved most of the sub-detectors, \Geant{} updates in terms of version and speed optimisation, MC Overlay against the standard pile-up, and the variable beam spot (see Section~\ref{sec:sim}). Some of these tasks required multiple validation rounds as their outcomes were not initially in-line with expectations.

Some of the main issues and causes of delay that were experienced during this validation effort were unexpected changes between software releases and incorrect settings that led to crashes and unphysical behaviours (e.g.\ unreasonably large changes in performance) or other kinds of inaccuracies. In the case of the former a careful validation of the single releases was sometimes needed to find the specific changes that altered the validation chain, to restore the expected agreement.

Given the experience from the \RunThr\ validation effort, some improvements are explored for future large-scale validation efforts. Those include robust checks of any individual changes before validating them combinedly in a validation campaign, which can help catching problems at a low level and will reduce the number of validation iterations. Extensive low-level automated checks should be done to better track release-by-release issues that arise in production. It should also be ensured that production campaigns precisely match the configurations used for a detailed physics validation. Sufficient input from experts, who can disentangle smaller issues within large overall expected changes, will ensure nothing subtle is missed.


\subsection{Software performance}
\label{sec:perf}

\subsubsection{Introduction and configuration}

This section describes the computing performance of the \RunThr software. In all cases, tests are run on a bare-metal machine with two AMD $\mathrm{EPYC}^{\mathrm{TM}}$ 7302 16-core processors configured to be in the \emph{Performance} mode with Simultaneous Multi-Threading (SMT) disabled. To disentangle its effects, clock frequency boosting is disabled and the clock frequency is capped at the base frequency of 3.0~GHz. The machine has 252~GB of available memory. In all results, memory is measured by the proportional set size, and time by the wall clock processing time unless stated otherwise.

Insofar as it is possible, the job configurations mirror what is used in the production system by ATLAS. The number of events to be processed concurrently is set to the number of available threads. The computing performance results for each step of the MC simulation processing workflow are provided in their respective subsections below. The MC simulation workloads use \ttbar-production events. In all cases, a constant number of events per thread or worker is processed (i.e.\ these are tests of weak scaling).

\subsubsection{Monte Carlo processing chain}

\paragraph{MC simulation}

As described in Section~\ref{sec:sim}, during LS2 the \Geant{} Optimisation Task Force was launched with the aim of reducing the \Geant{} simulation CPU time in \RunThr by more than 30\% without making any compromise on the physics accuracy. In this section, the simulation is benchmarked for the three latest ATLAS MC simulation campaigns: MC20 (the last ATLAS MC campaign for \RunTwo), MC21 (the first ATLAS MC Campaign for \RunThr) and MC23 (the latest ATLAS MC Campaign for \RunThr).

\begin{figure}[tbp]
\centering
\subfloat[]{
\label{fig:perf:simdigi_memory}
\includegraphics[width=0.48\textwidth,valign=c]{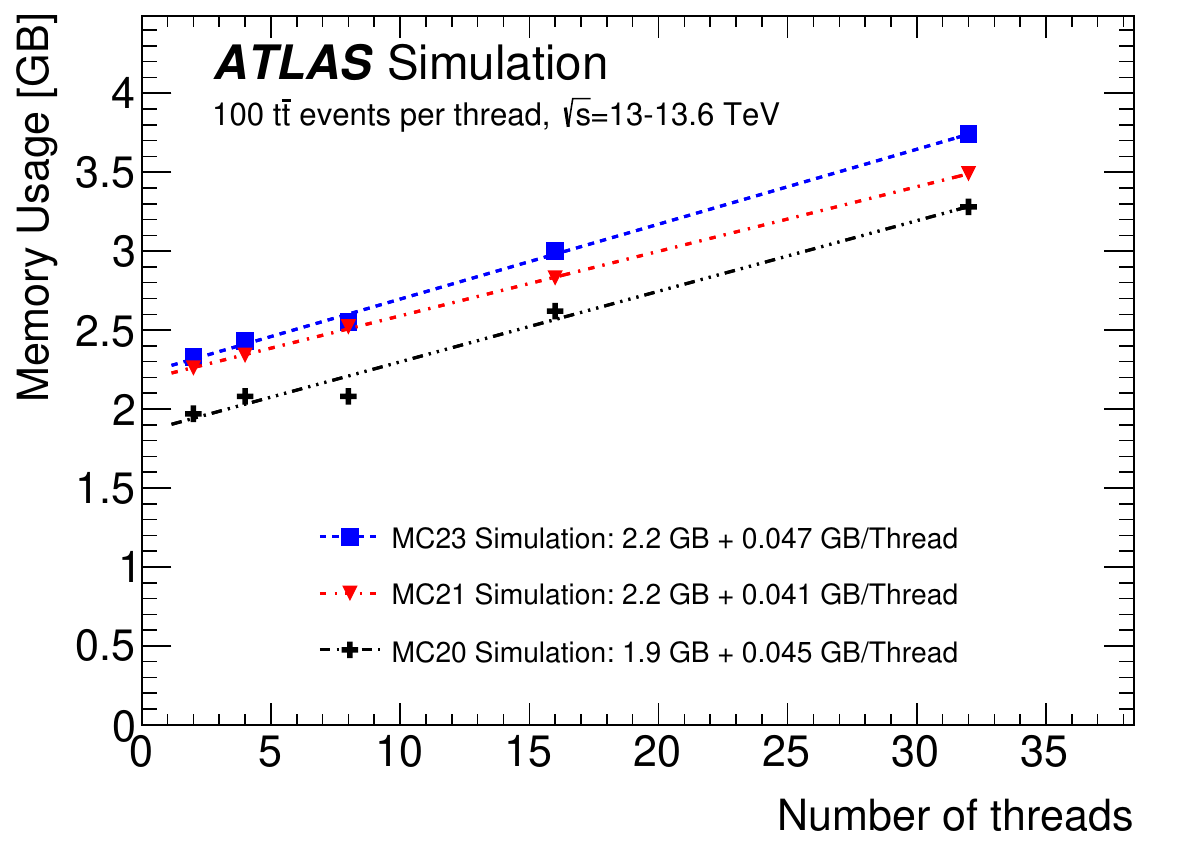}
}
\subfloat[]{
\label{fig:perf:simdigi_throughput}
\includegraphics[width=0.48\textwidth,valign=c]{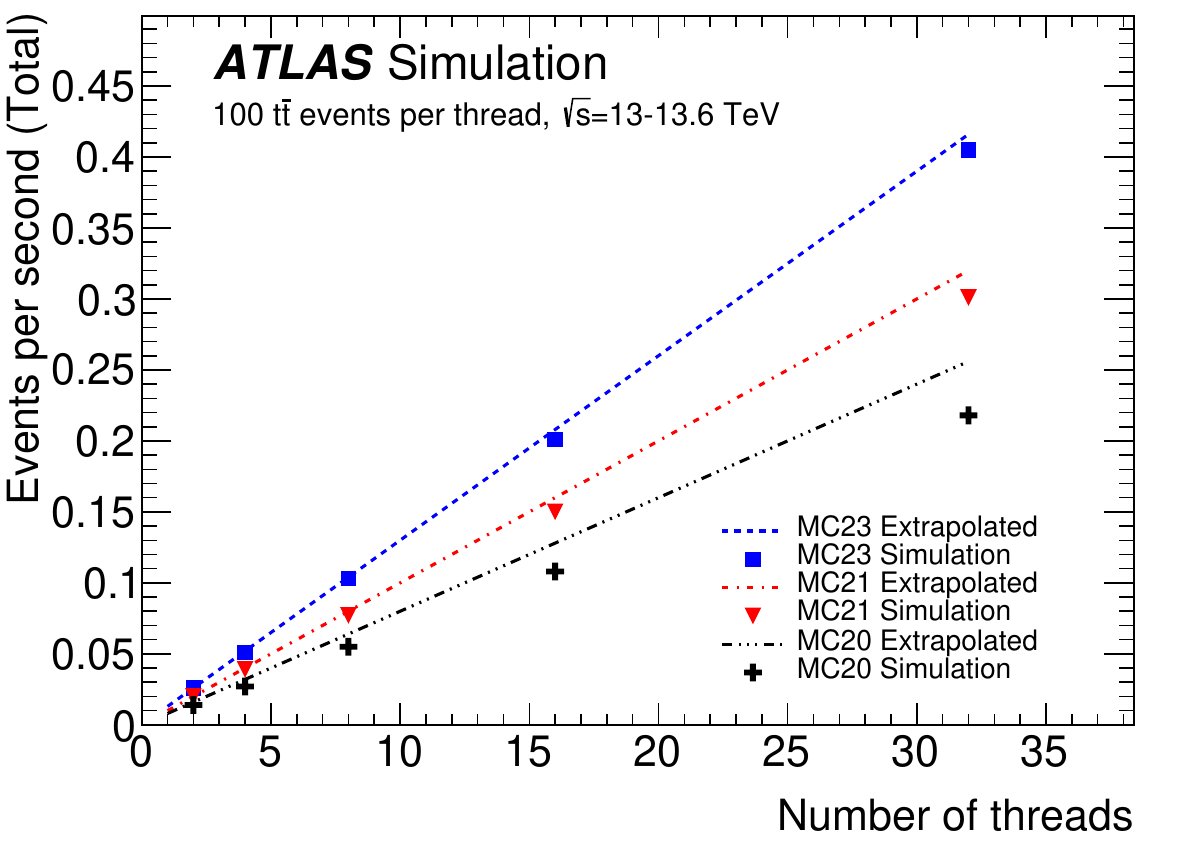}
}
\caption{\label{fig:perf:simdigi}(a) Memory usage as a function of number of threads and (b) event throughput as a function of number of threads for three recent MC simulation campaigns. In the memory figure linear fits are superimposed, and in the throughput figure extrapolations of the single-thread results are shown to guide the eye.}
\end{figure}

Figure~\ref{fig:perf:simdigi_memory} shows the memory usage, which scales well with the number of threads and similarly across the three campaigns. In these jobs, Athena release 22.0.92 is used, with 100 events per thread. The increase in the memory footprint is expected. It is explained partly by the different centre-of-mass energy of the input events in \RunTwo (13~\TeV) and \RunThr (13.6~\TeV), some changes in the geometry and beam conditions and, most significantly, by the introduction of a technical optimisation that concerns the way Athena code is linked with \Geant{}~\cite{Marcon:2813807} (see Section~\ref{sec:sim:staticlinking}, this optimisation reduces CPU in turn).

Figure~\ref{fig:perf:simdigi_throughput} shows the event throughput scaling of these simulation configurations in the same benchmark jobs. The scaling is almost ideal; an extrapolation of the single-thread performance is provided for reference. In fact, ideal performance is not expected to be linear on modern processors because, for example, the core frequency is automatically changed to optimise instruction throughput against power (see, for example, Ref.~\cite{ATL-SOFT-PUB-2021-002}). The CPU hyper-threading is also expected to result in somewhat lower throughput. Overall, the three workflows scale in a similar way. The throughput increased from MC20 to MC21 by 27\%--32\%, and from MC21 to MC23 by 45\%--47\%. These improvements are the results from the CPU optimisations in MC21 and MC23 with the \Geant{} optimisations that are detailed in Section~\ref{sec:sim}; a CPU speed-up of 33\%--38\% is observed in MC21 relative to the MC20 setup and the speed-up increases to 50\% in MC23.

\paragraph{Overlay, trigger simulation and reconstruction for MC}

\begin{figure}[tbp]
\centering
\subfloat[]{
\label{fig:perf:full_chain_memory}
\includegraphics[width=0.48\textwidth,valign=c]{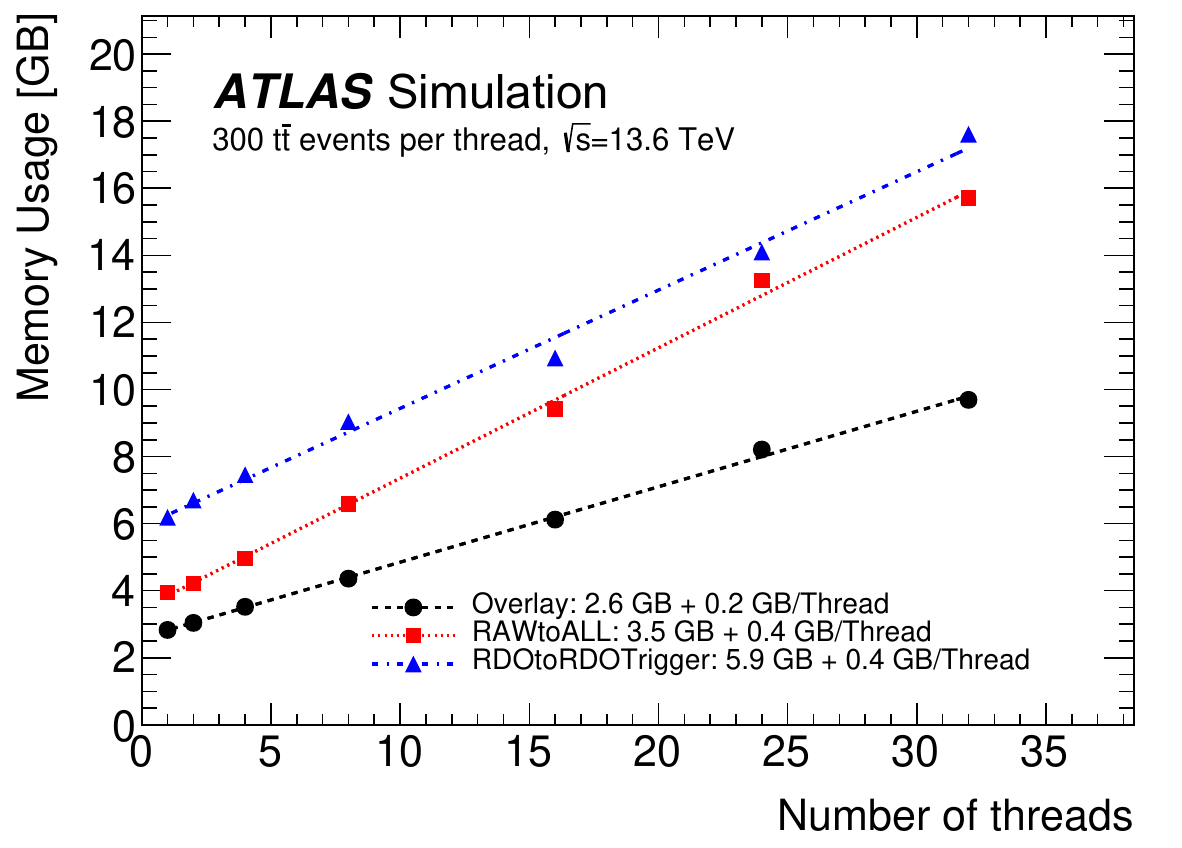}
}
\subfloat[]{
\label{fig:perf:full_chain_throughput}
\includegraphics[width=0.48\textwidth,valign=c]{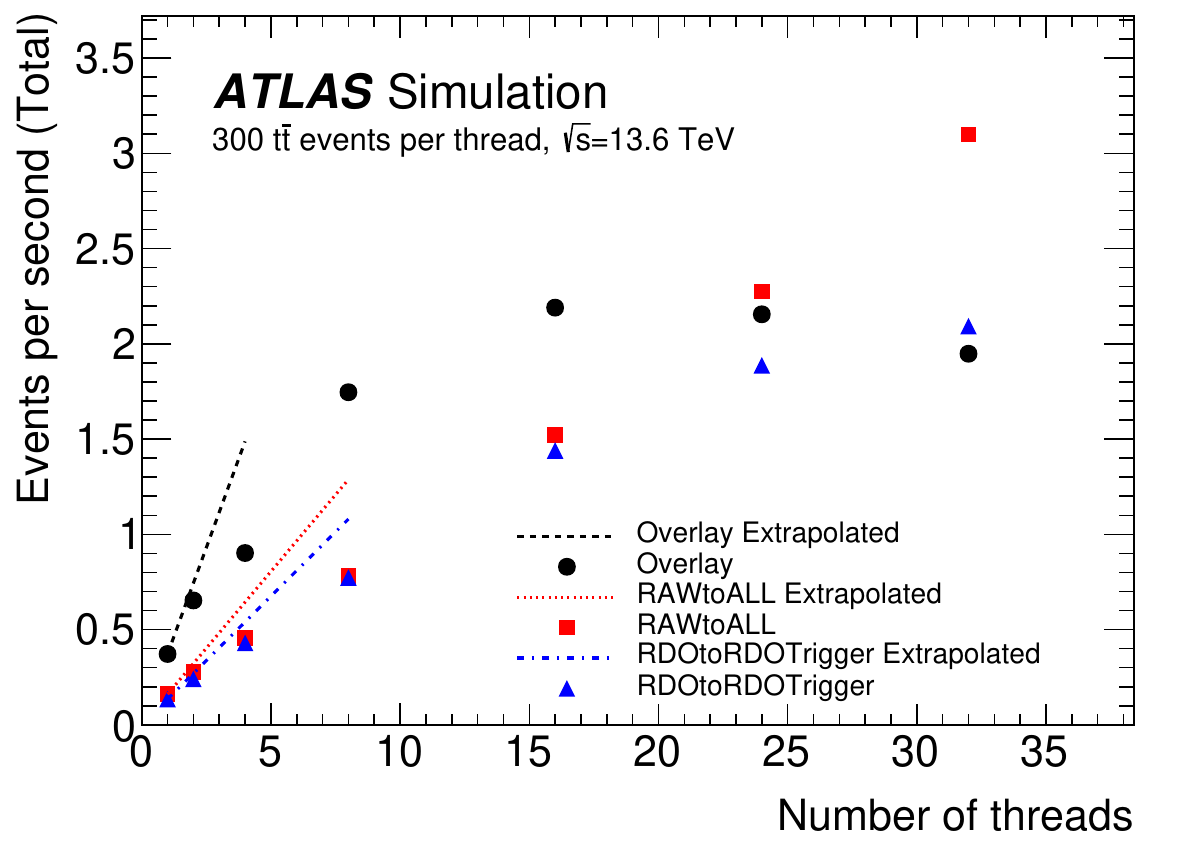}
}
\caption{\label{fig:perf:full_chain}(a) Memory usage as a function of number of threads and (b) event throughput as a function of number of threads for several MC simulation workloads in MC23. In the memory figure linear fits are superimposed, and in the throughput figure extrapolations of the single-thread results are shown to guide the eye.}
\end{figure}

Figure~\ref{fig:perf:full_chain_memory} shows the memory usage of several workloads normally run together in MC simulation production. The MC23 production configuration is used with Athena release 23.0.53. The specific MC campaign of this performance study is MC23c, which corresponds to an average pile-up of 56 interactions. The first step is MC Overlay, where the effects of pile-up are simulated by overlaying hard-scattering events with pre-mixed RDOs to model the desired $\langle\mu\rangle$ profile and other effects (see Section~\ref{digi:sec:overlay}). The result of this step is fed into the second step, offline trigger simulation, labelled RDOtoRDOTrigger. The resulting file is then passed through the reconstruction, labelled RAWtoALL, where physics objects are created (see Section~\ref{sec:reco}) and stored in an AOD. The memory usage increases by 200--400~MB per additional thread, and each 8-thread configuration stays well below the standard 2~GB per core available on the Grid.

Figure~\ref{fig:perf:full_chain_throughput} shows the event throughput scaling of these MC simulation production workloads in the same benchmark job. Standard production configurations use eight threads, and thus optimisation efforts were focused on this regime. Indeed, up to eight threads the scaling is relatively linear in all of the three steps. At this point, MC Overlay reaches its Amdahl's law~\cite{Amdahl} plateau and the improvement due to additional threads is significantly reduced. This is a subject for future improvement. The same happens for offline trigger simulation above 24 threads, whereas reconstruction shows almost ideal improvement throughout (within the caveats discussed above). Nevertheless, the production setup is well-optimised for the most-used hardware configurations on standard Grid production queues that offer eight cores per job.

\subsubsection{Reconstruction of collision data}

\begin{figure}[tbp]
\centering
\subfloat[]{
\label{fig:perf:coll_reco_memory}
\includegraphics[width=0.48\textwidth,valign=c]{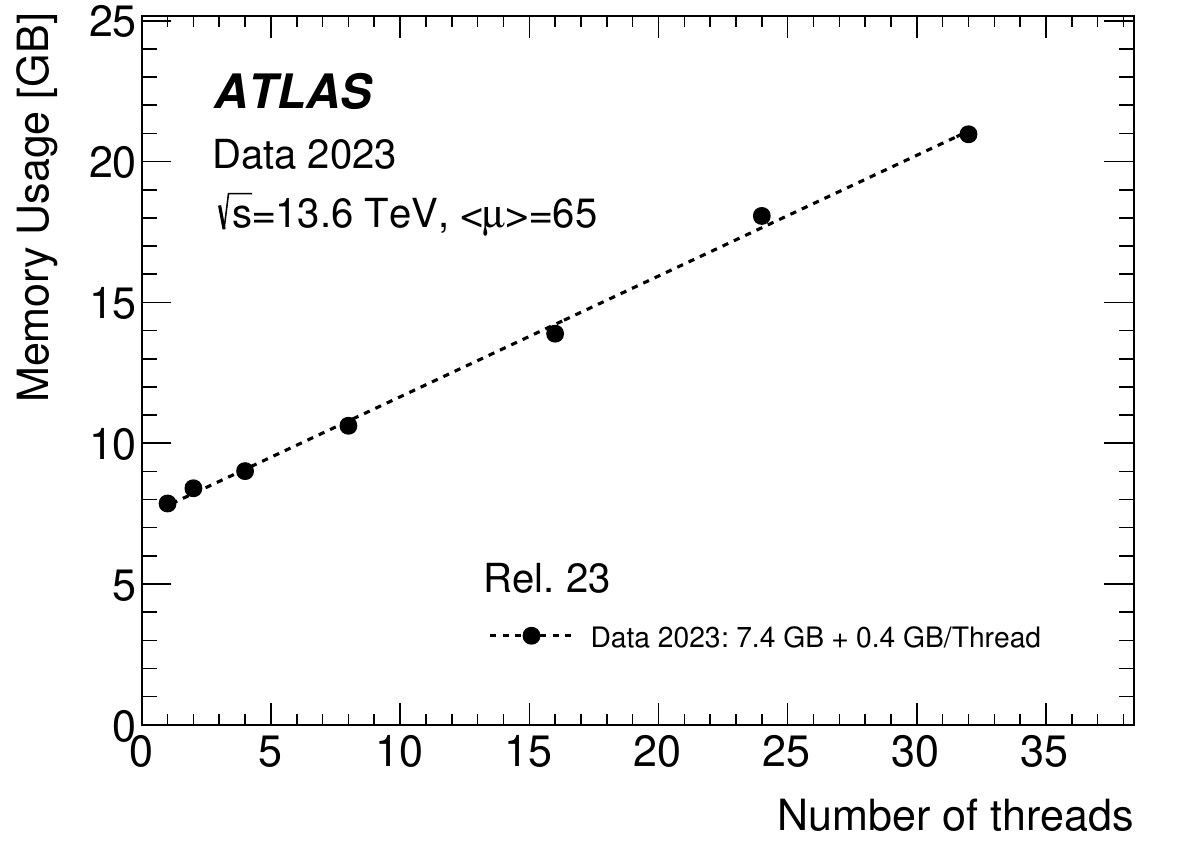}
}
\subfloat[]{
\label{fig:perf:coll_reco_throughput}
\includegraphics[width=0.48\textwidth,valign=c]{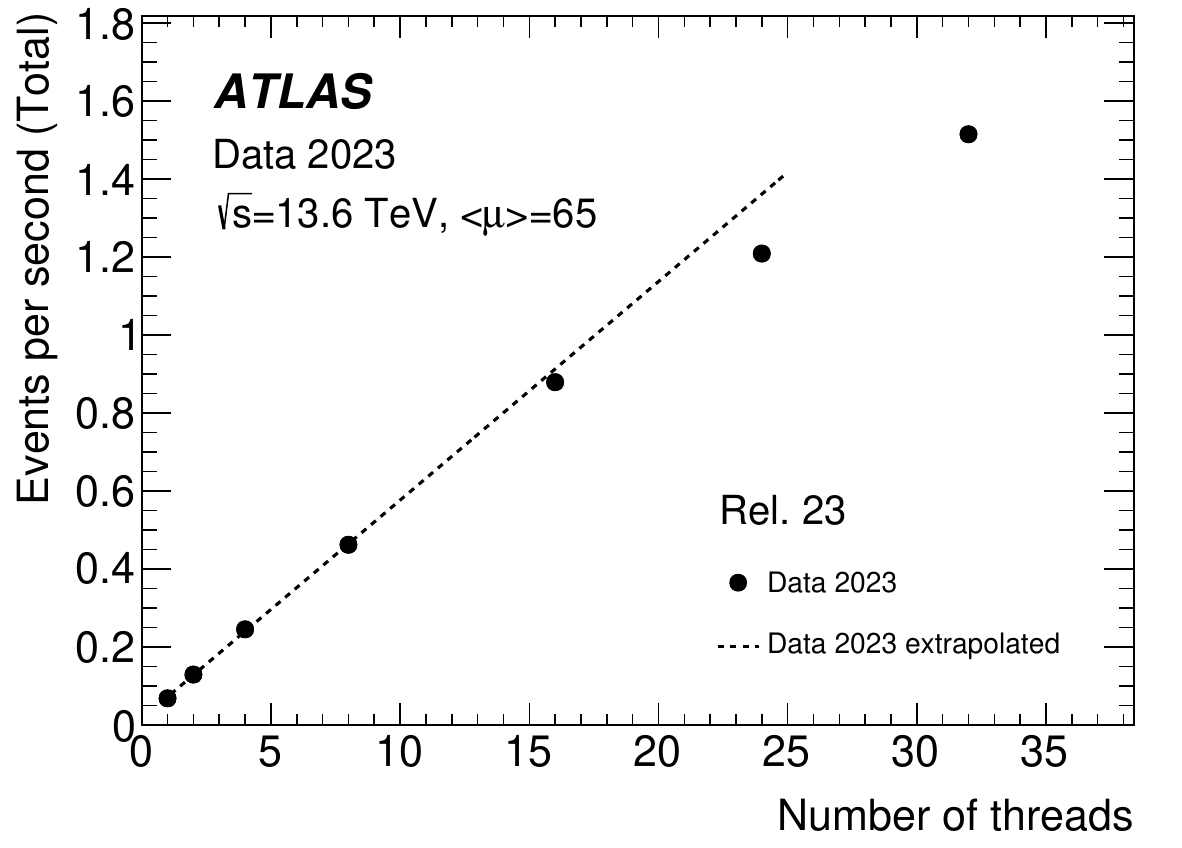}
}
\caption{\label{fig:perf:reco}(a) Memory usage as a function of number of threads and (b) event throughput as a function of number of threads for 2023 detector data reconstruction. The average pile-up $\langle\mu\rangle$ was 65. In the memory figure a linear fit is superimposed.}
\end{figure}

Figure~\ref{fig:perf:coll_reco_memory} shows the memory usage as a function of the number of threads, using input proton--proton collision data collected in 2023. Athena release 23.0.53 is used for these tests, with 150 events per thread. The corresponding average numbers of interactions per bunch crossing, $\langle\mu\rangle=65$. As expected, the memory usage grows linearly with the number of worker threads. However, it is always well below the standard 2~GB per core available on the Grid for 8-core configurations, meaning that data reconstruction jobs can successfully be executed on any Grid node. The memory is significantly larger than that for the reconstruction of MC (see Figure~\ref{fig:perf:full_chain_memory}) because data reconstruction includes additional data quality monitoring tools that involves the creation of many histograms.

Figure~\ref{fig:perf:coll_reco_throughput} shows the event throughput as a function of the number of threads in the same jobs. For the production configuration with eight threads, the data throughput is measured to be 0.5 events per second. The scaling is close to ideal, with small deviations in the region with many threads, but the throughput continuously improves until all available cores on the node are utilised. Although there is room for improvement, the reconstruction job is no longer memory-bound in production configurations, a notorious problem in previous versions of the Athena software.

\subsubsection{Derivation production}
\label{sec:deriv_performance}

\begin{figure}[tbp]
\centering
\subfloat[]{
\label{fig:perf:deriv_memory}
\includegraphics[width=0.48\textwidth,valign=c]{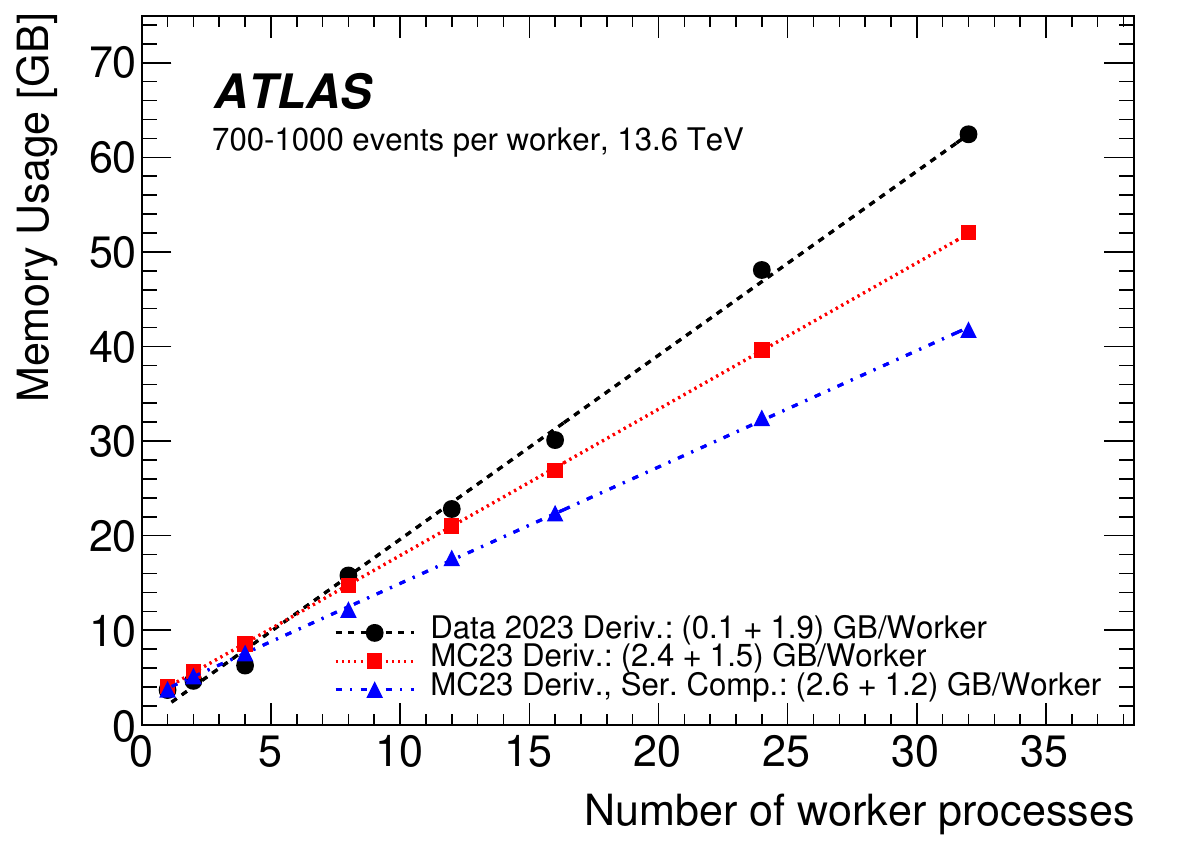}
}
\subfloat[]{
\label{fig:perf:deriv_throughput}
\includegraphics[width=0.48\textwidth,valign=c]{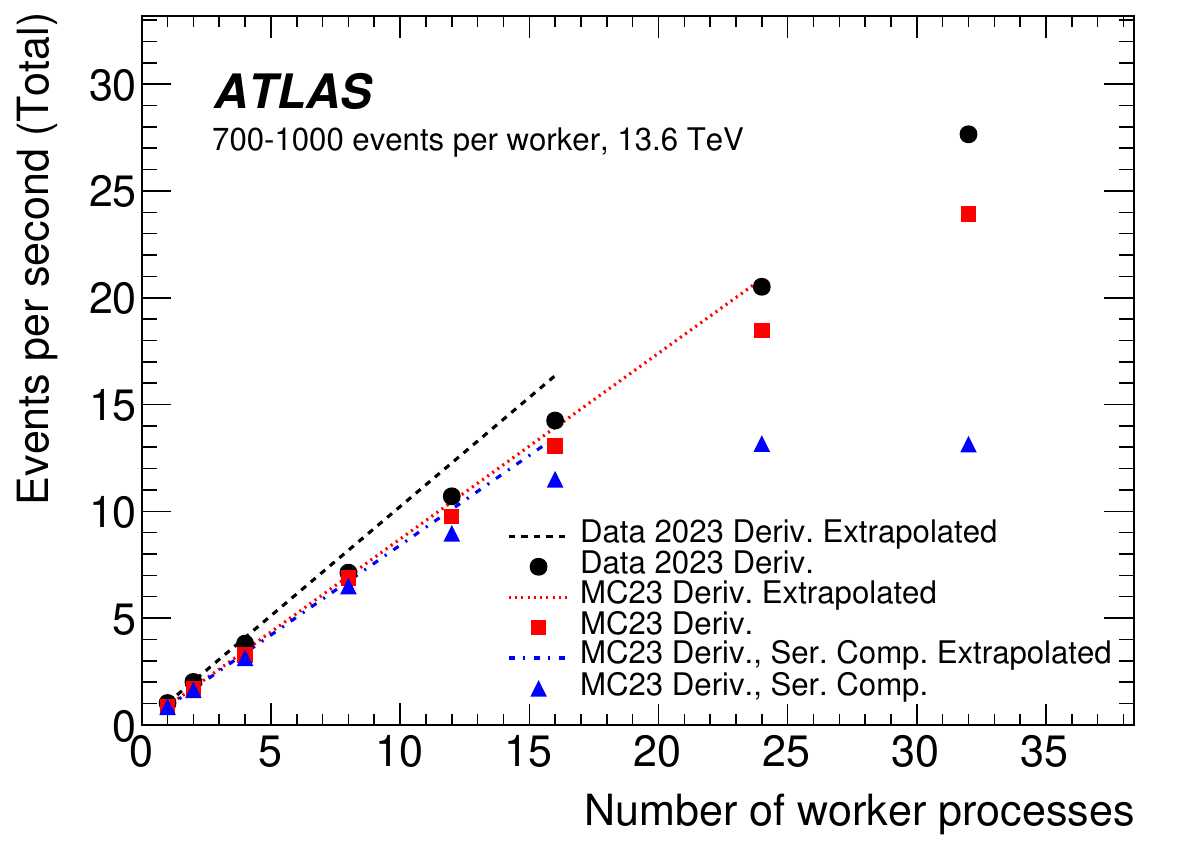}
}
\caption{\label{fig:perf:deriv}(a) Memory usage as a function of number of workers and (b) event throughput as a function of number of workers for derivation production using 2023 data and MC simulation. The MC simulation workloads use \ttbar-production events. Also shown is derivation production for MC simulation with serial compression (`Ser. Comp.').  In the memory figure linear fits are superimposed, and in the throughput figure extrapolations of the single-worker results are shown to guide the eye.}
\end{figure}

Figure~\ref{fig:perf:deriv_memory} shows the memory usage as a function of the number of worker processes in three different multi-process derivation production configurations. These performance tests are all done with Athena release 24.0.12, mirroring production configurations as closely as possible. The data run was mostly in the range of $62<\langle\mu\rangle<67$. Derivation production based on simulated \ttbar-production events is shown with serial compression (see Section~\ref{sec:deriv_software}) and the \RunThr-standard parallel compression. As expected from the additional compression buffers, the memory requirement for parallel compression is higher. The memory usage for derivation production using 2023 detector data as input is also shown. The memory usage is slightly lower for low numbers of workers because fewer algorithms run on data (e.g.\ those operating on generator records run only on MC simulation); the scaling with the number of workers is slightly worse owing to subtle differences in the configuration of the forking (i.e. which memory pages are fully shared between processes).

Figure~\ref{fig:perf:deriv_throughput} shows the event throughput as a function of the number of worker processes in the same jobs. 1000 events per worker are included for MC simulation, and 700 events per worker are included for detector data. Here the advantage of parallel compression is clearly visible: there is dramatically better scaling for high numbers of workers when parallel compression is enabled. The derivation production for detector data runs slightly faster than that of MC simulation owing again to the lack of algorithms and tools running on generator records. With parallel compression in MC simulation, the scaling is very close to ideal beyond 16 workers.

\subsubsection{Disk sizes of formats and containers}
\label{sec:datasizes}

\begin{figure}[t]
\centering
\includegraphics[width=0.48\textwidth]{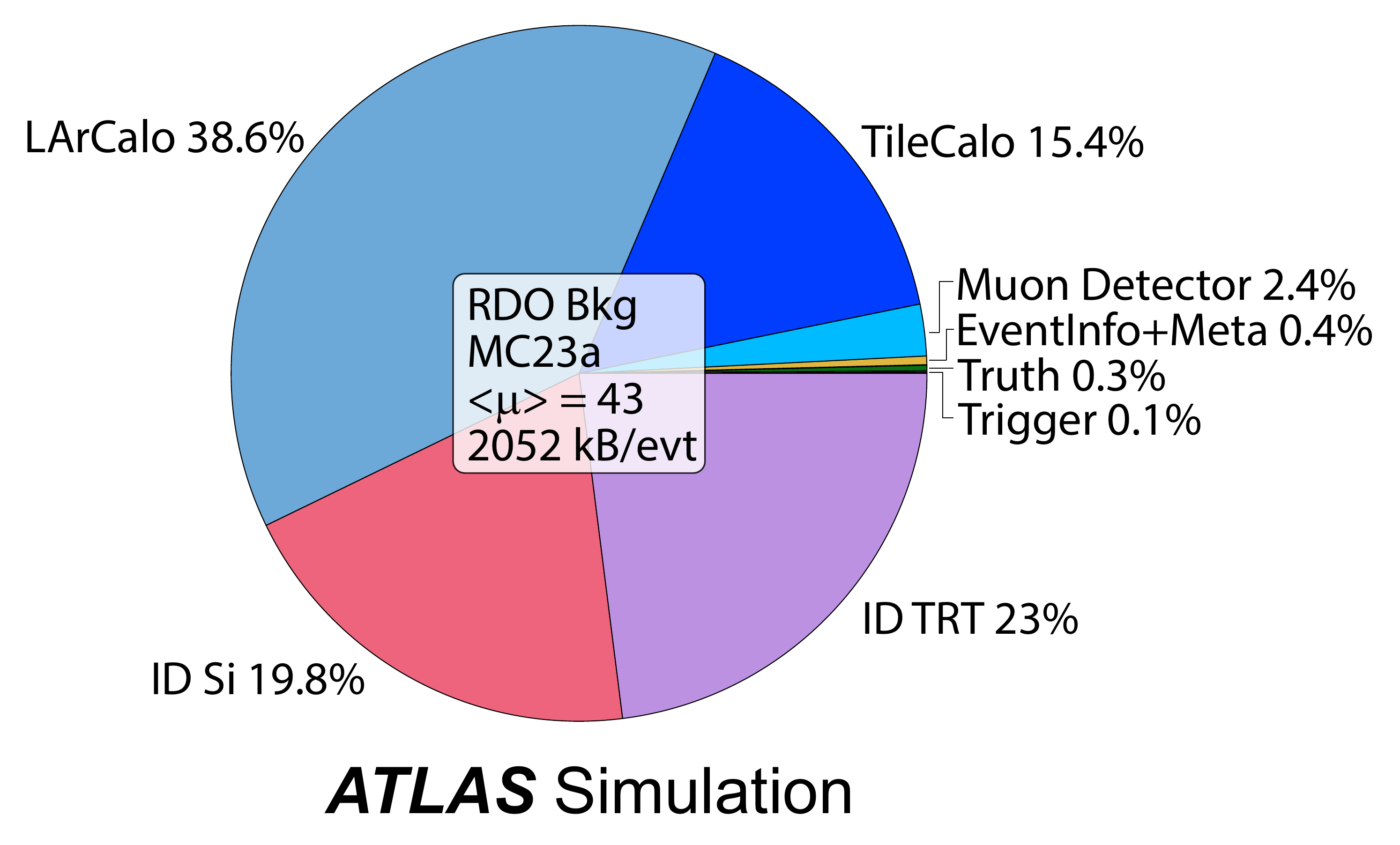}
\caption{The composition of a simulated \RunThr background RDO file that stores pile-up information, which is later overlayed on hard-scatter events. The pile-up corresponds to $\langle\mu\rangle=43$, which approximates the run conditions of data taking in 2022. Here `ID Si' includes both the silicon pixels and strips.}
\label{fig:perf:pie_RDO}
\end{figure}

\begin{figure}[b]
\centering
\subfloat[]{
\includegraphics[width=0.48\textwidth,valign=c]{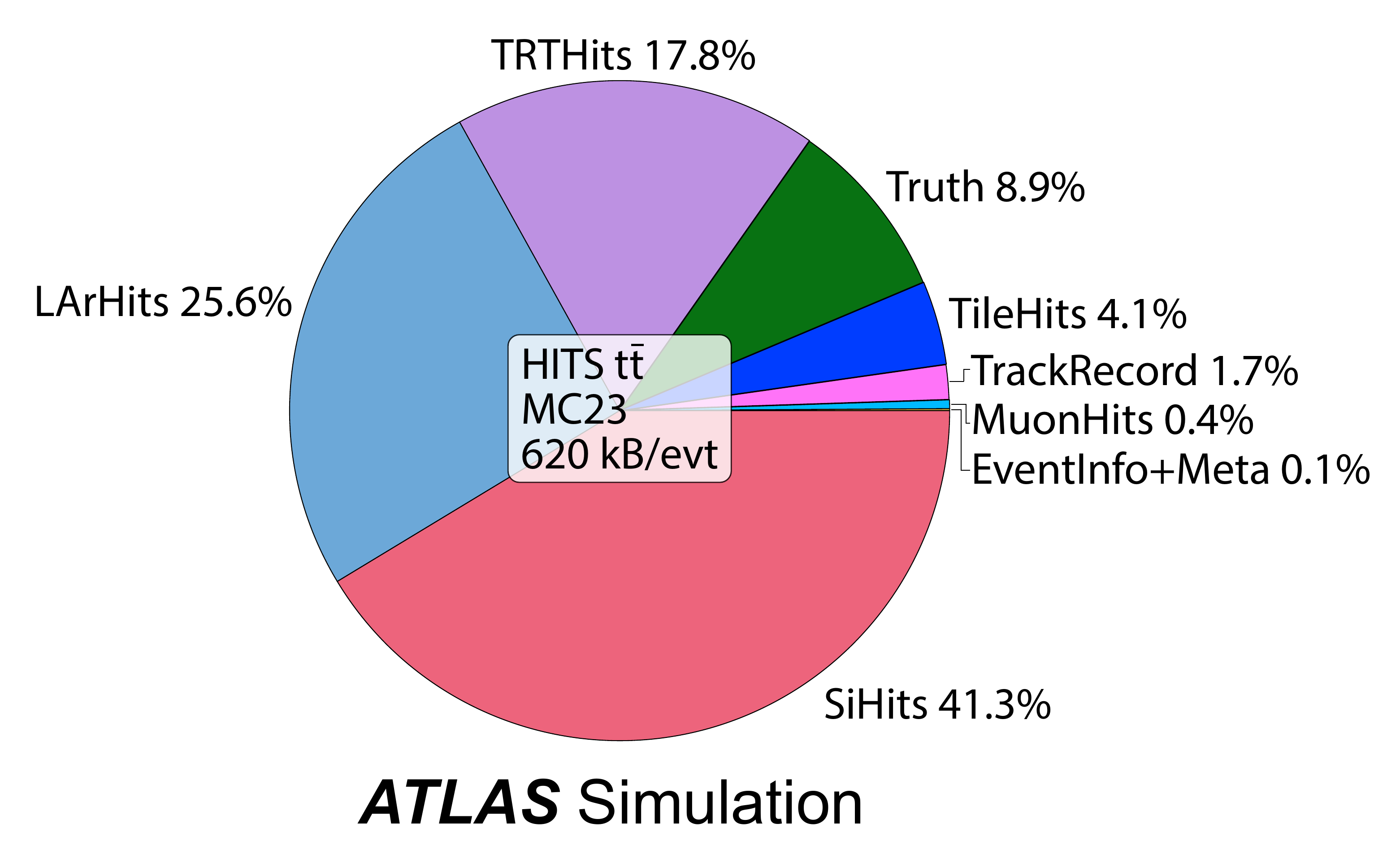}
}
\subfloat[]{
\includegraphics[width=0.48\textwidth,valign=c]{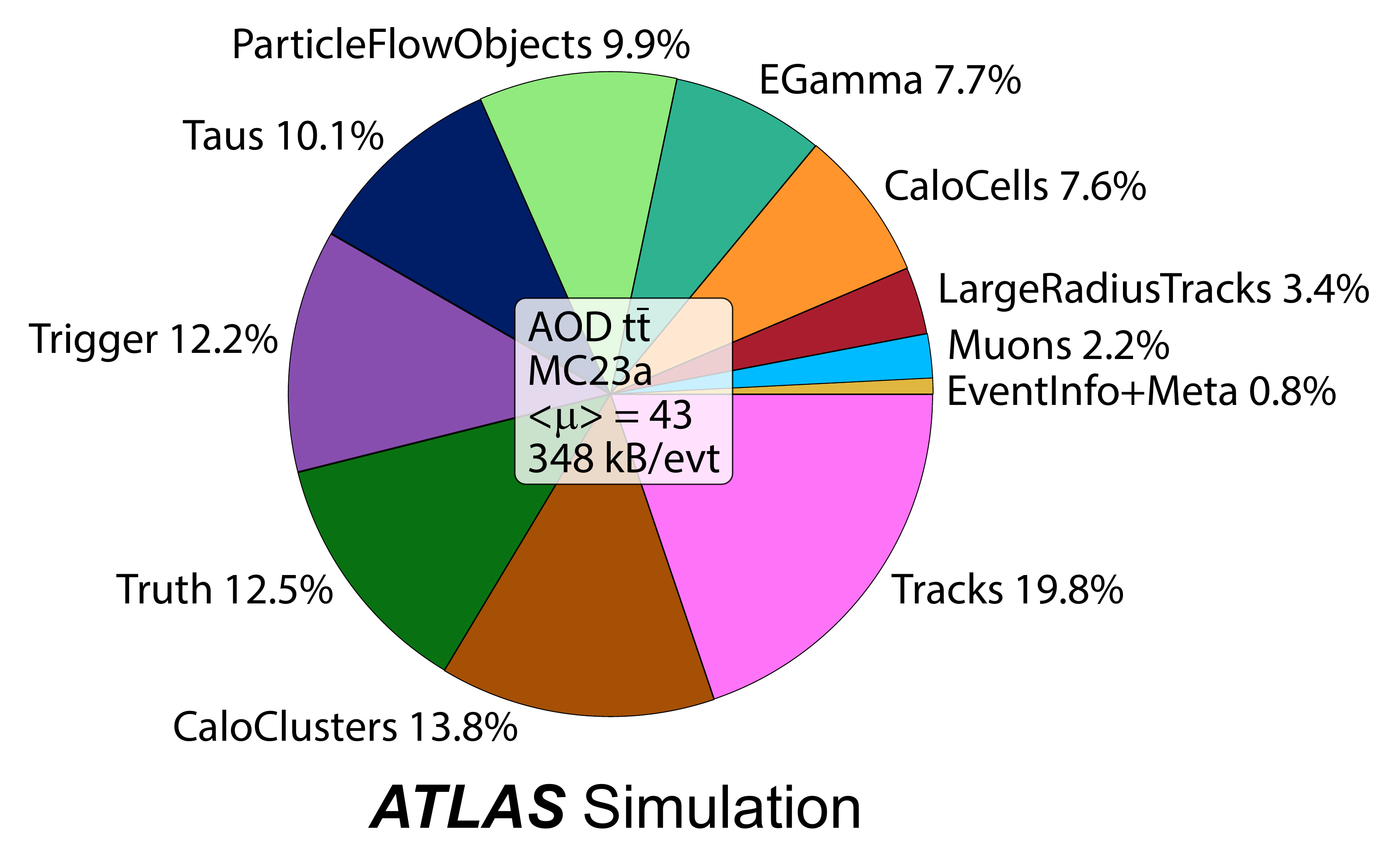}
}
\caption{\label{fig:perf:HIT_AOD}The composition of a \RunThr (a) HITS file and (b) AOD file. The process simulated in both cases is $\ttbar$ with a single lepton from a $W$-boson decay. The AOD includes hard-scatter information and also simulated pile-up. The pile-up corresponds to $\langle\mu\rangle=43$, which approximates the run conditions of data taking in 2022.}
\end{figure}

The average disk sizes of the standard formats for MC simulation datasets under several representative sets of pile-up conditions are listed in Table~\ref{fig:perf:eventsizes}. From MC20 to MC21, the `standard' $\ttbar$-production sample was changed; however, this has only a minor impact on the file size (at the 5\%-level). The dominant part of the change in HIT file size between MC20 and MC21 is due to the full simulation optimisations described in Section~\ref{sec:sim}. Changes like the range cut application to all EM processes and the neutron and photon Russian Roulette reduce the number of steps taken in the simulation, and particularly the number of steps taken at large times. This has a significant impact on the calorimeter output, reducing it by almost a factor of two. From MC21 to MC23 the compression algorithm was changed from ZLIB to LZMA, resulting in a 10\%--15\% reduction in file size after compression. Both these changes affected both the Full and Fast Simulation outputs. The RDO and AOD sizes increased over time due to the larger amount of pile-up. The composition of a background RDO (which stores pile-up information only, to be later overlayed on hard-scatter events) is displayed in Figure~\ref{fig:perf:pie_RDO}, and the compositions of a $\ttbar$-production FullSim HITS file and AOD file are shown in Figure~\ref{fig:perf:HIT_AOD}. The RDO file is dominated by calorimeter and inner detector information. The largest fraction in the HIT file is that of Silicon HITs (recorded in the pixel and SCT detectors). The AOD is later reduced to a DAOD (derived AOD, see Section~\ref{sec:derivations}), which is the format used for most physics analyses. The AOD does not contain any reconstructed jet collections, because jets are built only at the derivation step using the low-level inputs stored in the AOD. This approach helps to reduce the file size of the AOD.

\begin{table}
\begin{center}
\caption{\label{fig:perf:eventsizes}Average size of an MC simulation $\ttbar$ event, for various MC production campaigns. The data taking year represented by each campaign and the average number of proton--proton interactions, $\langle\mu\rangle$, are reported. All the MC20 sub-campaigns use the same simulation configuration at $\sqrt{s}=13$~\TeV; the MC23 sub-campaign simulation configurations differ only by the beamspot, which has a negligible effect on the output file size. Both MC21 and MC23a represent the 2022 data conditions at $\sqrt{s}=13.6$~\TeV, but the former refers to production performed before and during data taking, while the latter represents production done in 2023 with an updated software release and conditions matching those of the data taking (rather than a prediction). The HITS files are the output of the \Geant{}-based or \AF{} simulation of the hard-scatter process; the RDO files are the output of digitisation and MC Overlay, and store the contribution of pile-up; and the AOD files are the output of reconstruction of objects needed for physics analysis. `N.A.' indicates combinations that were not validated for physics analysis use.
}
\begin{tabular}{|l|c|c|c|c|c|c|}
\hline
Campaign                   & MC20a    & MC20d & MC20e & MC21 & MC23a & MC23c \\
\hline
Representing data taken in & 2015--16 & 2017  & 2018  & 2022 & 2022  & 2023 \\
$\langle\mu\rangle$        & 23.9     & 37.8  & 36.1  & 44.3 & 42.8  &  55.6  \\
Full Simulation HITS (kB/event)  & \multicolumn{3}{c|}{906} & 711  &    \multicolumn{2}{c|}{620}  \\
Fast Simulation HITS (kB/event)  & \multicolumn{3}{c|}{801} & N.A. &    \multicolumn{2}{c|}{591}  \\
RDO (kB/event)   & 1648    & 1897  & 1878  & 2073 & 2052  & 2240 \\
AOD (kB/event)   &  271    &  364  &  354  &  375 &  348  &  430 \\
\hline
\end{tabular}
\end{center}
\end{table}

\FloatBarrier


\subsection{Preparation for new campaigns}
\label{sec:campaigns}

Major campaigns for improvements to reconstruction (of either data or MC simulation) or detector simulation require significant resources and are only undertaken after careful preparation and validation. The data and MC simulation must be kept consistent to minimise analysis corrections, and therefore often the reprocessing of one mandates the reprocessing of the other. Typically, two major data (re)processing and MC simulation campaigns accompany each LHC Run: one that is undertaken continuously as the data taking proceeds, and one at the end of the Run, gathering the best-known geometry and conditions and final optimisation of the reconstruction. The time required for these campaigns is dominated by preparation time; the time for the processing itself is typically dominated by fixing rare crashes and similar issues that affect a few events, to ensure that every data event is reprocessed and provided to analysis.

\subsubsection{Monte Carlo simulation campaign preparations}

The preparation for a new MC simulation campaign involves many inter-dependent steps that must be performed and checked before large-scale production is undertaken. The full sequency of changes described here is required for updates of the detector simulation and its reconstruction; some new MC simulation campaigns involve only changes to the reconstruction of the simulated events, and the first of these steps can be skipped.

First among these is the preparation of the simulation software, including the choice of \Geant{} version. Because changes to the \Geant{} version often result in changes to physics observables (even with the same configuration), the version must be fixed for the duration of an MC simulation campaign. Normally, some large (e.g.\ 10~million event) sample of \ttbar events is simulated to detect any remaining, rare potential bugs, and several other samples are run with modest numbers of events to check for computing performance issues (speed, memory consumption, and so on). Small patches are normally allowed after this stage, built as part of an ATLAS-specific \Geant{} release, but the official \Geant{} patch version is fixed.

In parallel with the fixing of the \Geant{} version, the initial geometry and detector conditions must be prepared. Only coarse (large-scale) detector alignments, the beam spot position and distribution, and the gas mixtures used in detectors (e.g.\ in the TRT the gas mixture may change during the run and is therefore recorded as a part of the conditions rather than as a part of the geometry) enter the simulation; other conditions like detailed maps of disabled channels only enter in the reconstruction and therefore can be finalized later on. Small improvements to the geometry might arrive close to the time of production, but most physics validation (see Section~\ref{sec:validation}) is done with a close-to-final geometry.

With the geometry and \Geant{} versions in place, the calorimeter sampling fractions can be checked, which calibrate the electromagnetic-scale response of the calorimeters. These are normally calculated using bespoke single-particle samples, where the particles are simulated starting from within the uniform sampling region of the calorimeter. The sampling fractions are dependent on the \Geant{} version, owing primarily to the details of the treatment of range cuts in various processes (see Section~\ref{sec:sim}). Although these are in principle purely geometric quantities, they have also been observed to have dependence on particle incident angle in some detector geometries.

The next step towards production is the preparation of OFCs for the calorimeter (see Section~\ref{sec:digi}). The calorimeter OFCs are based on noise pedestals that are pile-up dependent, and therefore a $\langle\mu\rangle$ value must be selected for their calculation. This $\langle\mu\rangle$ value need not match the anticipated pile-up conditions as discrepancies can be re-calibrated and high values result in the suppression of some physics signal, but generally a close, round value is used (e.g. $\langle\mu\rangle=40$). The OFCs (and sampling fractions) are calculated assuming perfect calorimeter detector conditions (e.g.\ without disabled channels and with nominal high-voltage conditions).

With these steps complete, frozen showers (see Section~\ref{sec:sim:frozenshowers}) can be prepared for the MC simulation campaign. Calibrations also begin at this stage with the preparation of some TopoCluster calibrations, calculated based on single particle simulation. These each depend only weakly on final detector conditions and non-calorimeter geometry.

At this stage, the geometry and those detector conditions important to the simulation can be finalized, and physics validation of the simulation itself can begin. Often the first validations are done using the reconstruction from the previous MC simulation campaign, since it is well-understood. This procedure allows the isolation of simulation changes, although changes that require new calibrations (e.g.\ changes in the hadronic energy scale in the calorimeter) must be inspected with care by experts.

With the simulation validated, the digitisation is the next workload to finalize. This includes updates to digitisation algorithms (e.g.\ for improved noise models or response treatments) and conditions that must be applied during digitisation (e.g.\ abnormal high-voltage in the calorimeter). These changes are normally validated with and without pile-up, and both the direct digitisation of hard-scatter and pile-up as well as the MC Overlay workflow must be validated. Normally these validations proceed with a preliminary version of the reconstruction to be used in the MC simulation campaign, because often in production MC Overlay and reconstruction are run in a single job on the Grid.

Finally, the geometry, conditions and configuration to be used for reconstruction must be finalised and validated. As a part of the Sample-A production run during physics validation (see Section~\ref{sec:validation}), computing performance issues can also be checked on the Grid. The final validation and production normally begins once the last updates to the detector and accelerator conditions are included; this can include, for example, a final update to the $\langle\mu\rangle$ distribution used in the sample, which then requires an updated RDO production for MC Overlay before the reconstruction configuration is finalised. Often, only once the full chain is validated and the sample production begins can samples be produced for the training of the various types of fast simulation, which then requires its own update and validation process.

The production itself often begins from samples required for the derivation of calibrations and systematic uncertainties. These must be available before data analysis can begin. Sometimes, \emph{transfer} uncertainties are used, wherein the uncertainty prescription from a previous MC simulation campaign is applied, along with extra systematic uncertainties related to the known or expected changes between the two campaigns. These temporary uncertainties allow faster data analysis uptake in new software releases.

Depending on the number and scale of changes expected, this entire procedure from start to end may take more than a year. This was the case in the lead-up to \RunThr, where the thread-safety of the software was being regularly checked on increasingly large-scale samples, new detector elements from the Phase-I upgrades had to be introduced and the corresponding reconstruction software developed and commissioned, and significant uncertainty around the accelerator conditions remained until significant data were recorded. For this reason, often the initial MC simulation campaign in a Run is used as a prototype for preparatory work, validation, early data tests, and very early analysis. A subsequent campaign includes a more realistic picture of the data taking and is used for most data analyses. This was the case, for example, for MC15 and MC16 (the latter has run for seven years), and more recently for MC21 and MC23. For the same reason, a small sub-campaign is often undertaken based on estimates of the coming year's data-taking conditions, and a longer campaign follows the data-taking with conditions that are known to be representative of the year. This was the case, for example, for MC16c (before 2017 data-taking and meant to be representative of the 2017 data) and MC16d (after 2017).

\subsubsection{Data reprocessing preparations}
\label{sec:campaigns-data}

To begin a campaign to reprocess data, it is necessary first to identify what improvements will be made and then to validate each of the improvements via the process discussed in Sections~\ref{sec:DQ} and~\ref{sec:validation}. The improvements broadly fall into three categories: updates to the detector conditions data, updates to the detector geometry, and updates to reconstruction algorithms.

Conditions data updates~\cite{Bohler:2015emb} incorporate improvements in the knowledge of the time-dependent status of the detector during periods of operation. Typical examples include improved knowledge of detector alignment, detector noise and instantaneous luminosity delivered by the LHC. This information is stored in the ATLAS conditions database~\cite{Verducci:2008zzb}, an \textsc{Oracle}\textregistered{} database hosting a COOL technology schema~\cite{USLHC:2012ewx} (see Section~\ref{subsec:databases:conditions}). Each detector system may provide updated conditions in advance of a reprocessing. Only coarse (large-scale) alignment is applied in the detector simulation; small alignment corrections, common in the inner detector and muon spectrometer, can be updated without modifying the simulated geometry. Similarly, updates to disabled or noisy channel maps do not require an update of the MC simulation.

Detector geometry updates incorporate improvements in knowledge of the detector geometry over a long period of stability. Typical examples are updates to maps of \emph{dead material}, the inactive parts of the detector that are not read-out, such as supporting material of subdetectors or the cryostat walls. The updates may also include corrections to the positioning or internal geometry of sub-detectors assumed in event reconstruction. Significant updates to the geometry may require a new simulation campaign to maintain good agreement with the data.

Algorithm updates incorporate improvements to the algorithms used to reconstruct the events. New tracking and vertexing algorithms, as well as updates to lepton identification and algorithms are typical examples. The updates may also include fixes to deal with crashes or floating point exceptions observed in events that were previously problematic to reconstruct. Many of these updates require corresponding changes to the reconstruction of the MC simulation; some, however, might involve improvements to the rejection of backgrounds that only occur in real detector data, in which case only the detector data must be updated.

The validation of these changes uses the Data Quality tools described in Section~\ref{sec:DQ}. Most often, a small sub-set of data are first processed to test for technical issues. Later, a larger sub-set of the data are processed and passed through a Data Quality validation process to check for issues with various updates and ensure that the changes are as expected. Particularly because conditions are period-specific, this larger sub-set includes runs from several different conditions periods. Only after all of these checks are completed is the reprocessing of the remainder of the data undertaken. Afterwards there may be updates to the Good Runs List to include additional data that could be declared `good' thanks to improvements in the conditions or reconstruction.


%
\section{Infrastructure and databases}
\label{sec:geninfrastructure}

This section describes the significant central infrastructure required to support the services, systems, and developers of the collaboration.
Section~\ref{sec:infrastructure} introduces the infrastructure maintained to support the software development cycle.
The database systems used throughout the experiment are described in Section~\ref{sec:databases}.
Finally, the development and use of metadata, much of which relies on these systems, is described in Section~\ref{sec:metadata}.

\subsection{Software engineering process and infrastructure}
\label{sec:infrastructure}

The software developed for event simulation, reconstruction and analysis, as well as for detector calibration and alignment, has to run on the computing facilities that are available to the collaboration members; these facilities are not necessarily uniform in their hardware architecture, nor their operating systems and environments.
To ensure that ATLAS software is robust and portable, it is necessary to provide a common process and infrastructure to integrate all new code with the existing code base, and then build and test changes.
A central ticket and workflow system is provided using \textsc{Jira}~\cite{Jira}, which provides functionality for bug reports, improvements, and tasks; for the preparation of MC simulation requests before they enter the production system; for tracking physics validation (see Section~\ref{sec:validation}); and for the tracking of long-term development. GitLab integration is used to ensure code updates and tickets are cross-linked.

\subsubsection{Software structure and compilation system}
\label{subsec:genInfra:cmake}

\begin{figure}[tbp]
\begin{center}
\includegraphics[width=0.7\textwidth]{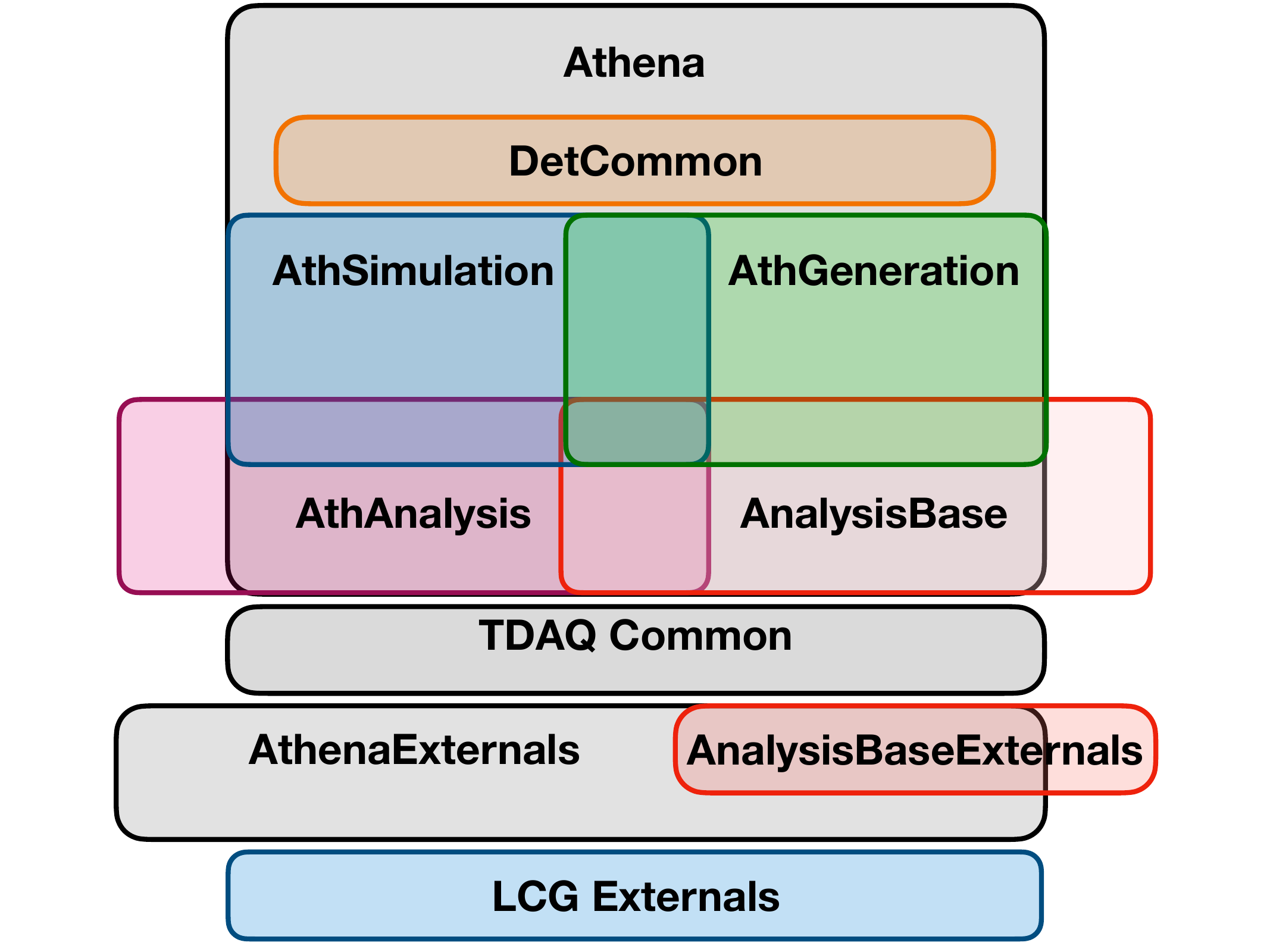}
\end{center}
\caption{Schematic overview of the ATLAS software projects and external dependencies.
The overlaps between project areas indicate overlaps in code included in the projects.
The widths of AnalysisBaseExternals and AthenaExternals are wider to indicate their support for the entirety of the AnalysisBase and AthAnalysis projects, respectively.}
\label{fig:atlasSoftware}
\centering
\end{figure}

The ATLAS software, shown in Figure~\ref{fig:atlasSoftware}, is organized into a project for general use (Athena) and several projects for specific purposes like event generation (AthGeneration), detector simulation (AthSimulation), and data analysis (the AthAnalysis and AnalysisBase projects).
These rely on several projects like `TDAQ Common' (common trigger and data acquisition software) and several external packages (e.g.\ event generators, \textsc{ROOT}, or \Geant{}).
Each project is further sub-divided into packages with much more specific purposes (e.g.\ pixel detector digitisation, or muon momentum calibration); there are about 2000 packages in the ATLAS software.

ATLAS uses the \textsc{CMake} build system~\cite{cmake} for building all of its own code, in conjunction with most
of the HEP software projects that the experiment uses. \textsc{CMake} is a very flexible system supporting Linux,
MacOS and Windows hosts, x86 and ARM CPU (aarch64) architectures, \textsc{GNU Make}, and \textsc{Ninja} and additional low-level build
systems, with a rich high-level language allowing support for the complex software building steps required
for the ATLAS reconstruction, simulation, trigger, and analysis code.

Thanks to the flexibility of \textsc{CMake} and to harmonise how different parts of the code are built, ATLAS-specific
functions and macros are used throughout the build configuration of the offline and trigger software for
setting up the building of libraries and executables, and for dealing with the installation of scripts and
data files. For the offline software, that basic \textsc{CMake} infrastructure is maintained in the |AtlasCMake|
package~\cite{AtlasCMake}.

One fundamental requirement for the build system is to allow the users to build small parts of the software
against a pre-built full software release (a \emph{base project}), as described in Section~\ref{subsec:genInfra:runEnvironment}. This is
achieved with imported targets in \textsc{CMake}. Using |AtlasCMake|'s facilities the user is able to
build packages and libraries as part of a project that are already part of the base project. This is
achieved by only importing the targets of the base project into the build of the current project that are
not present in the current project already.

The \textsc{HEP\_OSlibs} meta package~\cite{heposlibs} provides the software package dependencies for the Athena
build and runtime environment for the two currently-used operating systems (CentOS 7 and Alma Linux 9; see
Section~\ref{subsec:genInfra:buildSystem}). The LCG software stack~\cite{LCG},
maintained by the CERN EP-SFT group in collaboration with several large LHC experiments like ATLAS,
provides several hundred external packages that include recent versions of commonly used
packages as well as HEP-specific tools and MC event generators (see Section~\ref{sec:evgen}). There are
usually two or three major LCG software releases per year and development builds are available every night.
The major LCG versions are normally released with a new major stable release of \textsc{ROOT}~\cite{ROOT}. Minor LCG version
updates are done every few weeks and usually contain updates of MC event generators or bug-fixes of other
external packages. The LCG software stack is built for x86\_64 and aarch64 architectures and several recent
versions of the GCC and Clang compilers and C++ standards. As of autumn 2023, LCG version 104b with \textsc{ROOT}
6.28/08 is used in Athena, compiled using GCC 13 with the C++20 language standard.

\emph{External software}, software used by the ATLAS software that is maintained separately from the
experimental software and is not provided by an LCG package, is built using \textsc{CMake} as part of an ATLAS external
project. As shown in Figure~\ref{fig:atlasSoftware}, every ATLAS offline software project (Athena,
AthSimulation, etc.) has a corresponding external project (AthenaExternals, AthSimulationExternals, etc.),
upon which it is built. This includes packages like \textsc{Gaudi}~\cite{GAUDI}, and allows a more rapid update of
external packages that receive rather frequent version updates or are not sufficiently common to be included
in the LCG stack directly. This separation is not present in the TDAQ projects: TDAQ Common builds a combination
of external and ATLAS packages as part of the same project. Some parts of this TDAQ software are also
required in the ATLAS offline data processing, for example to read the ATLAS RAW data.

Any software build that uses \textsc{CMake} is separated into configuration, build, installation and packaging steps.
Configuring the build of small projects like analysis projects, or the WorkDir project used for developing
code in the Athena repository, takes just a few seconds, while configuring the build of the largest project,
Athena, takes a few minutes.

Each package is typically built into one or more small dynamic libraries that are loaded on-demand at runtime.
This helps ensure a minimal memory footprint from library loading despite the wide variety of jobs that can be
run within Athena. It also allows the user to override specific libraries provided in the release or by the system,
for example by pre-loading a different memory allocation library or math library. There is a small CPU overhead
associated with dynamic library loading, discussed further in Section~\ref{sec:sim:staticlinking}.

The packaging and distribution of the ATLAS software is achieved using RPM packages built using the \textsc{CPack}
tool within \textsc{CMake}~\cite{cpack}. The RPMs are configured to depend on each other such that the installation of a full Athena
release can be done by requesting the installation of just the top-most Athena RPM package. This top-most
package in turn ensures that all dependent RPMs, including those provided by LCG and packages providing
various data files (e.g.\ for \Geant{}) are installed in the correct place automatically.

\subsubsection{Build system, CI, nightly and stable releases}
\label{subsec:genInfra:buildSystem}

The ATLAS Nightly and Continuous Integration (CI) Systems provide a modern software development workflow for developers, feature fast development cycles and assure confidence in new software deployments.
Both the systems are Jenkins-based~\cite{jenkins} and have a long evolution history~\cite{Elmsheuser_2017}.
Interconnected with the ATLAS GitLab code repository~\cite{athenaGitlab}, the CI system performs up to 100 multi-project software builds daily, probing code changes proposed in GitLab merge requests.
A comprehensive test suite of unit and short integration tests runs for each CI job.
The Nightly System (separate from CI) performs daily builds of all ATLAS software projects from the code repository (called `nightlies' because the largest projects are built overnight), often on several platforms.
It maintains a multi-stream, parallel development environment with many simultaneous branches.
The Nightly System probes how the changes from accepted merge requests work together.
In addition, it helps to support migrations to new platforms and compilers and verify patches to external tools.

The CI and Nightly jobs run on the 1400-core build farm, including both real hardware and virtual nodes.
Various \emph{operational intelligence} techniques such as incremental compilations, selective testing, operations parallelization, and dimensionality reduction result in efficient resource use and faster delivery of results.
The CI and Nightly monitoring system provides dynamic information about build and test results and installation status.
It is based on the Nightlies Database residing in the ATLAS database production cluster dedicated to offline analysis (ATLR, see Section~\ref{subsec:databases:infrastruture})~\cite{Dimitrov_2012}.
The ATLAS Nightlies Database is the source of dynamic content for the nightlies dashboards hosted on the BigPanDA web application~\cite{Alekseev_2018} (see Section~\ref{subsec:monitoring:dashboards}).

The systems build software releases from the GitLab main branch and from dedicated branches for the online high-level trigger, reconstruction and simulation of \RunThr data, and legacy branches for \RunTwo reconstruction, simulation and analysis.
Regular `sweeps' are used to copy changes made in various branches to the main branch.
Special development nightlies are available that use development versions of \textsc{ROOT}, LCG builds, or other externals.
These ensure that when preparing for significant changes (e.g.\ a new operating system, GCC version, or C++ language standard), the changes can be thoroughly vetted without putting at risk the main development stream.

When the compilation and test steps of the nightly build are successfully completed, \textsc{CPack}~\cite{cpack} is used to produce RPMs, which are then uploaded to EOS~\cite{eos}.
This acts as a staging area for the subsequent installation step that fetches them and, using \textsc{ayum}~\cite{ayum}, an ATLAS-specific wrapper around \textsc{yum}, installs the nightly releases in the atlas-nightlies repository on the CernVM FileSystem (CVMFS)~\cite{cvmfs}, and thereby makes them globally accessible to developers.
CVMFS is the service by which ATLAS distributes all of its development and production software, and auxiliary datasets.

The installation software~\cite{installation} is executed locally on a publishing node (officially, a Release Manager) that communicates with a backend Gateway node.
Having multiple publisher nodes allows concurrent transactions, increasing the overall publication rate of the repository and improving the speed at which nightly builds become available.

Currently the total size of the atlas-nightlies repository is 10~TB, with approximately 4~TB in use.
In 2020, the repository data was migrated to \textsc{Ceph}-based S3 object storage, resulting in a performance enhancement over the previous \textsc{aufs} system (advanced multi-layered unification system).
All nightly releases (currently about 35) built in ATLAS and using different architectures (binaries, operating systems and compilers) are deployed on CVMFS in the form of RPMs.
The installation time ranges from 3 to 60 minutes, depending on the number of external packages needed and the format of the build (e.g.\ whether debugging symbols are included) and type of release (e.g.\ DetCommon/AnalysisBase compared with full Athena).
The installations are kept for 30 days by default, with the possibility of extensions for developers to test newer code against older nightlies or if a physics validation is performed with a nightly release (see Section~\ref{sec:validation}).

The nightly releases are rigorously tested in the ATLAS Release Tester (ART) Grid-based framework~\cite{art} described in Section~\ref{subsec:genInfra:releaseTesting}.
When the set development goals are achieved, a successful nightly release is transformed into a stable release by the team of ATLAS offline release shifters.
Stable releases have unique numeric identifiers and indefinite lifetime.
The numbering of stable releases follows the pattern A.B.C(.D), where A is the major release version that changes at most once a year, B generally indicates the purpose of the release (e.g. 2 for analysis, or 6 for generation; these numbers are often historical), C indicates the minor release version and is typically increased about once a week for a new stable release, and D is an optional patch version in case a stable release must be patched. Patching is most often used so that a stable release, for example one to be used for trigger simulation, which must therefore match the release used during data taking, must be patched to fix a rare bug, add a feature for MC simulation, or read a new file version.

All production releases until fall 2023 were built on the CentOS 7 Linux operating system~\cite{centos7} on the x86\_64 and aarch64 architectures using the \textsc{CMake} build system described above. A migration to Alma Linux 9~\cite{alma9} of all infrastructure was completed by the end of 2023, though several legacy releases continue to be based on CentOS 7. %

The default compiler for all \RunThr releases is at present GCC version 11.2.0~\cite{gcc11.2} with the \textsc{binutils} 2.37~\cite{binUtils} and the C++17 standard.
In parallel there is a nightly release build from the GitLab main branch using the Clang 16.0.3 compiler~\cite{clang} on the x86\_64 architecture.
Maintaining builds with at least two compilers has helped identify a variety of minor issues in the software and led to increased robustness.
For the releases built on Alma Linux 9, the recent stable version of GCC, version 13.1, with the C++20 language standard and \textsc{binutils} 2.40 are used.
Nightly builds on the aarch64 architecture exist both on the CentOS 7 and Alma Linux 9 operating systems with the GCC 11.2 and 13.1 compiler versions respectively.
Since summer 2023, the main branch releases are built for the microarchitecture x86-64-v2 and newer~\cite{microarchitectureLevels}, which brings support among other things for vector instructions up to Streaming SIMD Extensions 4.2 (SSE4.2).
Newer microarchitecture levels are not yet supported by a sufficiently large fraction of machines available on the WLCG Grid to make adoption beneficial.

\subsubsection{Release testing}
\label{subsec:genInfra:releaseTesting}

The ART system~\cite{art} is designed to run test jobs on the Grid after an ATLAS nightly release is built.
The choice was made to exploit the Grid as a back-end as it offers a huge resource pool, suitable for a deep set of integration tests, and running the tests could be delegated to the highly scalable ATLAS production system (\PanDA, see Section~\ref{sec:prodsys}).
The challenge of enabling the Grid as a test environment is met with the CVMFS file system for the software and input data files. Test jobs are submitted to the Grid by the GitLab-CI system, which itself is triggered at the end of a release build.
Jobs can be adorned with special headers that direct how to run the specific test, allowing many options to be customised.
These options might include which releases are to be tested, input- and output-file specifications, and the number of cores to be used for multi-process or multithreaded running, for example.
The GitLab-CI waits for the exit status, and output files are copied back from the Grid to an EOS area accessible by the users.
All GitLab-CI jobs run in ART virtual machines using \textsc{Docker} images~\cite{docker} for their ATLAS setup. ART jobs can be tracked by using the \PanDA monitoring system.

ART can also be used to run short test jobs locally.
They use the same ART command-line interface, where the back-end is replaced to access a local machine for job submission rather than the Grid.
This feature allows developers to ensure their tests work correctly before adding them to the system.
In both the Grid and local machine options, running and result copying are completely parallelised.

\subsubsection{Software quality}
\label{sec:genInfra:quality}

To assist in the development of high-quality code, ATLAS has maintained a Coding Standard~\cite{ATL-SOFT-2002-001}.
The goal of the standard is to ensure some consistency in code style (e.g.\ naming conventions for variables and files) so that
developers can quickly understand a piece of ATLAS code with which they are not familiar, while
acknowledging that with thousands of developers working over tens of years, not all style issues must be rigorously
enforced (e.g. spacing and bracket placement is not enforced). The standard is divided into requirements, which are
checked using a static code checker implemented as a GCC plugin~\cite{chep18athenamt}, and recommendations that are
taught but not enforced.

A wide variety of static code-checking tools are used, including \textsc{Coverity}~\cite{Coverity},
\textsc{CppCheck}~\cite{CppCheck}, and \textsc{lizard}~\cite{lizard} for C++, and \textsc{flake8}~\cite{flake8} for
\textsc{Python}. Both \textsc{flake8} and \textsc{CppCheck} (when available) are integrated into the build
system so that compilation warnings are raised if code problems are identified.

The primary mechanism used to enforce code quality and consistency is via multi-level reviews of merge requests.
Because of the size of the code base, `tidying' the entirety of ATLAS code would be impractical. Instead, when
touching a particular piece of code, developers are encouraged to gently improve the code around their changes,
help identify and improve missing documentation or code weakness, and only introduce new code that is of high quality.
For most merge requests, a first-level reviewer looks over the changes to ensure that they are sensible and well
explained in the merge request, and that all of the automated tests have succeeded. If the merge request is
particularly complex, a second-level (expert) reviewer might be asked
to examine the code further. Some domain-specific review is often performed, as the CI system automatically alerts
some experts to changes to code within their domain. Finally, a release coordinator is responsible for a last check
before the changes are merged into the central repository.

\subsubsection{Development and run environment}
\label{subsec:genInfra:runEnvironment}

As described in Section~\ref{subsec:genInfra:buildSystem}, the current default runtime environment is the CentOS 7 Linux operating system.
The AFS~\cite{openafs} and EOS~\cite{eos} file systems are used during the software nightly builds to access larger calibration files that are then integrated into the respective software releases.

Although primarily designed for clusters of Linux processors with open external connectivity that allows the software to be imported through CVMFS and the conditions data accessed through \textsc{FroNTier} (see Section~\ref{subsec:databases:conditions}), ATLAS software must be able to run on different hardware architectures and different operating system versions.
A notable example is High-Performance Computers (HPCs), which may have large numbers of processors but may lack external connectivity.
In that case the software and conditions data need to be packaged into a container that is then uploaded through a gateway to the HPC; input and output data files also go through the same gateway.

The \textsc{ATLASLocalRootBase} (ALRB)~\cite{ALRB} suite of shell scripts, whether accessed from CVMFS or locally installed on a laptop or HPC site, provides a uniform look and feel to configure many tools with their dependencies in the correct order by encapsulating the details.
It is also a wrapper to start containers with the necessary environments and mount points for runtime \textsc{Docker}~\cite{docker}, \textsc{Singularity}~\cite{singularity}, \textsc{Apptainer}~\cite{apptainer} and \textsc{Shifter}~\cite{shifter}.
It provides mechanisms to test tools and validate containers before deployment.
Diagnostics are also available in ALRB for user support and for tutorials.
It works for \textsc{bash}/\textsc{zsh} shells, x86\_64 and aarch64 architectures, and RHEL-derived operating systems with versions 5--9; it is \textsc{Python} 3 ready and supports containers on other Linux flavours and MacOS.

The \textsc{AtlasSetup} (or simply \textsc{asetup}) script~\cite{ALRB} provides users a convenient and fast way (a few seconds) to locate the required release and configure its environment for stable and nightly releases.
For the latest nightly releases, \textsc{asetup} checks the release completeness because a nightly release is built in a few steps with a time gap of about an hour.
It only requires users to specify a few short tags, like the project (e.g. Athena or AthGeneration) and stable version number or day of the month for a nightly release.
There are a few layers of configuration for \textsc{asetup}, and \textsc{asetup} can print out the origin of a given configuration parameter.
The original shell environment is saved before the first \textsc{asetup} run, so that \textsc{asetup} can configure another release environment in the same session, even in subshells.
The details of the required release are saved into the working directory, such that the same release under that working directory can be restored easily and quickly the next time.

The \textsc{asetup} script works for \textsc{bash}/\textsc{zsh} shells, with \textsc{Python} 2 (including old \textsc{Python} 2 in SLC5) and \textsc{Python} 3, and in both \textsc{Singularity}/\textsc{Apptainer} and \textsc{Docker} containers.
It supports releases in architectures of i686, x86\_64 and aarch64 for RHEL-derived operating systems from SLC 5 to Alma Linux 9.
In the case of a conflict between the setup release environment and system commands, \textsc{asetup} helps provide a solution via wrappers and aliases.
A suite of regression tests comes with \textsc{asetup} to validate new versions, as \textsc{asetup} can run in different RHEL-derived Linux operating systems through \textsc{Singularity}/\textsc{Apptainer} and \textsc{Docker} containers from CVMFS.
The regression tests include more than 100 tests for the CentOS 7 platform.

Most contributors to the ATLAS software only need to work on a subset of packages (usually one or two packages) and do not require checking out the whole Athena GitLab repository.
A wrapper command, \textsc{git-atlas}~\cite{git-atlas}, handles the consistent checking out of the packages the developer needs using the sparse checkout functionality, as well as an interface to the developer's fork, which is necessary for submitting merge requests.
A user needs only to specify the package name (e.g.\ |PixelDigitization|), and the code from the relevant package area (in this case, |InnerDetector/InDetDigitization/PixelDigitization|) is added to the existing sparse checkout.
This configuration enables local partial builds against the nightlies, significantly speeding up development cycles.
Since it aimed to minimise the disruption to (non-core) feature developers, the \textsc{git-atlas} command was very useful to many developers during the transition from SVN to GIT, as it did not expose the complex commands required for sparse checkouts to the end users.

\subsubsection{Central services}
\label{subsec:genInfra:centralServices}

The Central Services Operations (CSops) team is central to all ATLAS activities.
Its role is to support a wide-ranging set of projects given the great variety of services needed by ATLAS computing.
The team is responsible for the configuration, deployment, and operation of different services on the CERN IT infrastructure.
They also act as the interface between CERN IT and the service managers of different projects and ensure that security and good computing practices are maintained.

CSops is responsible for the management of over 630 machines with more than 4,700 CPU cores, spread across 60 projects including \PanDA~\cite{panda} (see Section~\ref{sec:prodsys:panda}), \Rucio~\cite{rucio} (see Section~\ref{subsec:ddm:rucio}), the services described Section~\ref{subsec:genInfra:buildSystem}, and others.
These machines are hosted in the CERN-IT \textsc{Openstack}~\cite{openstack} infrastructure and the configurations are managed by \textsc{Puppet}~\cite{puppet}, which allows a significant reduction of complicated central operation tasks, while ensuring the high availability and scalability of all ATLAS central services.

These machines can be split up into three main groups: those for ADC, those for ATLAS nightly software building, and the miscellaneous nodes.
The last group contains machines from many different detector and operations groups in ATLAS and include critical web services, logbooks, monitoring and other services that are used by both the offline and online teams in ATLAS.

More than half of the machines deployed in \textsc{Openstack} are maintained with \textsc{Puppet} manifests.
This helps to ensure that security and configurations are reproducible on a large scale.
All machines that are controlled with \textsc{Puppet} can generally be destroyed and recreated within 45 minutes.
There are 139 different configuration manifests in place that are used to define the major configurations.
Specific changes to each configuration can be made with YAML~\cite{yaml} files that are used by the \textsc{Puppet} Database to fine tune individual hosts.
All of these manifests and configurations are stored in CERN GitLab repositories to aid with rapid deployment and keep track of changes to configurations over time.

Together with ADC and the ATLAS TDAQ system administrators, CSops also helps to maintain the Simulation at Point~1 (Sim@P1, see Section~\ref{sec:prodsys:simP1}) project, which is an opportunistic cloud that makes use of the ATLAS TDAQ HLT computing farm for offline grid workflows.
The Sim@P1 project is able to spawn over 2,000 virtual machines using in excess of 111,000 CPU cores that can be deployed in under 15 minutes; however, this is usually done over an hour for maintainability and to avoid unnecessary pressure on critical systems.


\subsection{Databases}
\label{sec:databases}

\subsubsection{Database infrastructure}
\label{subsec:databases:infrastruture}

ATLAS relies on several services for the management of non--event data; for example data derived from the run configuration, detector calibration and alignment, slow control system of the experiment, metadata associated with events, raw data files or data and workload management systems.
Most of these services profit from a database infrastructure that is maintained by the CERN IT department and is based on \textsc{Oracle}\textregistered{} technology.
Other applications include analytics clusters utilising the \textsc{Hadoop}~\cite{hadoop} file system and \textsc{Elasticsearch}~\cite{elasticSearch}.
The same infrastructure supports other database applications like those for ATLAS publications and membership information~\cite{GlanceFence}.

Some conditions data, including information about the detector state (e.g.~temperature, enabled and disabled modules, and the trigger configuration) are recorded in real time and automatically inserted into the conditions database. In some cases these data may undergo pre-processing to reduce their volume and simplify subsequent use. Significant additional conditions data are inserted into the database by experts and calibration tools that process the detector data later on.

\begin{figure}[tbp]
\includegraphics[width=\textwidth]{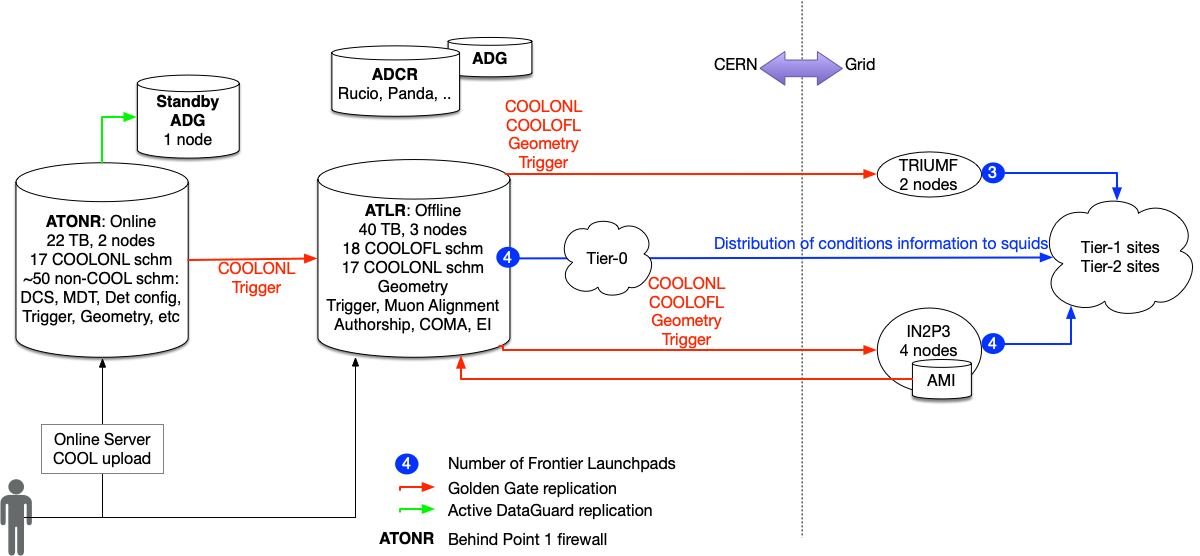}
\caption{The database infrastructure employed by ATLAS before \RunThr. The person icon represents the injection of conditions information into the database by either an expert or an application.}
\label{fig:databasesBeforeRun3}
\centering
\end{figure}

The original (\RunTwo) relational database infrastructure, as shown in Figure~\ref{fig:databasesBeforeRun3}, consisted of three production \textsc{Oracle}\textregistered{} clusters deployed and maintained by CERN IT: ATONR for Point~1 (P1) usage, ATLR as a general purpose cluster and ADCR for data and workload management systems like \Rucio and \PanDA (see Section~\ref{sec:distcomp}).
In this configuration, for data-processing purposes a subset of the ATLR data (which is limited to conditions, geometry and some trigger information) is regularly replicated to two Tier-1 sites, TRIUMF and CC-IN2P3.
ATONR was and still is within the firewall at P1, protecting it from accidental external use.

\begin{figure}[tbp]
\includegraphics[width=\textwidth]{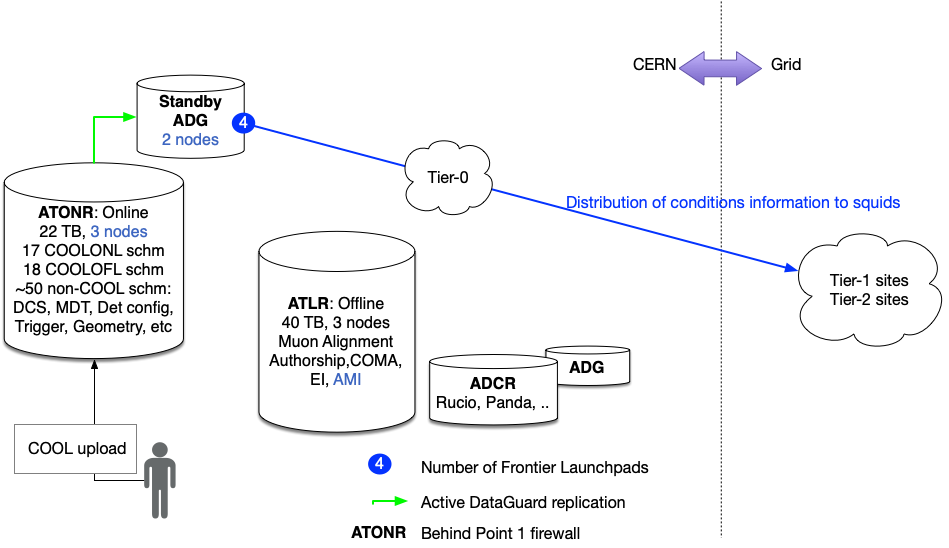}
\caption{The database infrastructure employed by ATLAS for \RunThr. The person icon represents the injection of conditions information into the database by either an expert or an application.}
\label{fig:databasesRun3}
\centering
\end{figure}

During the long shutdown between \RunTwo and \RunThr, the relational database infrastructure was consolidated in view of changes to the \textsc{Oracle}\textregistered{} licensing model at CERN.
ATLAS therefore progressively migrated to a new infrastructure, now in place and illustrated in Figure~\ref{fig:databasesRun3}, completing the migration in early 2024.
The cost and the complexity of the system was reduced by eliminating \textsc{Oracle}\textregistered{} instances at the Tier-1 sites, and to consolidate the CERN resources of the P1 cluster ATONR and its standby \textsc{Active Data Guard} (ADG) instance that are accessible from outside of the P1 network in a read-only mode.
An important step towards consolidation was the migration of the ATLAS Metadata Interface (AMI) database~\cite{ami} (see Section~\ref{sec:metadata}), which was historically based at CC-IN2P3, to CERN by the end of 2021.
Although the independent infrastructure had offered some advantages in terms of resilience, the advantages of the tighter integration, central support, and centralised authentication mechanisms available at CERN made the migration beneficial.
The migration of the conditions data in the offline schemas into ATONR is now complete as well; to preserve isolation of Point~1 during data taking a special proxy server was developed to permit updates of conditions in the offline COOL schemas.
Additional nodes were added into the ATONR cluster to accommodate additional workflows originating from the General Purpose Network, which were previously utilizing ATLR.
This expansion ensures the preservation of the existing resources for data-taking workflows at Point~1.
On the ADCR cluster, new more powerful hardware was deployed to support the heavy workflows coming from the data management system.
The most delicate parts of this migration had to be performed during pauses in data taking (either short technical stops or end of year shutdowns), but the migration could be done during Run 3.

\subsubsection{Conditions data distribution}
\label{subsec:databases:conditions}

The conditions data are distributed using several technologies.
The main service in place for access to conditions is \textsc{FroNTier}~\cite{Frontier}, a web application that redirects SQL requests to \textsc{Oracle}\textregistered{} databases and provides a caching layer implemented using Squid proxies.
The system is in use since \RunOne, and the main difference in \RunThr is its deployment architecture as mentioned in the previous section; the \textsc{FroNTier} launchpads are deployed only at CERN and provide access via the ATONR\_ADG cluster.
The network access to conditions via FroNTier may not be possible, especially for certain HPCs lacking outbound network connectivity (see Section~\ref{sec:prodsys:hpc}).
In such cases, an alternative approach is available; data can be extracted from \textsc{Oracle}\textregistered{} and distributed through simple SQLite files.
In this case there is the possibility to extract data from \textsc{Oracle}\textregistered{} and distribute them via simple SQLite files.
The tools for the easy generation of an SQLite file with a selection of a subset of conditions data are gathered in a GitLab repository~\cite{dbrelease}.
The SQLite files can then be distributed in a file system or via a \textsc{Docker} container provided their volume are reasonable.


\subsection{Metadata handling}
\label{sec:metadata}

The ATLAS experiment has developed three systems that build upon information from other ATLAS systems to provide unique metadata services to the experiment at the three fundamental levels of granularity of ATLAS data: services at the event, run, and dataset levels are provided by the EventIndex, COMA, and AMI, respectively.
Each of these systems has evolved considerably since \RunOne in mutual cooperation with many systems and services throughout ATLAS (having connections to aid data collection and service optimisation).
Furthermore, they each have evolved into related areas to provide additional services as described in this section.

The EventIndex~\cite{Barberis:2022tsf} is the metadata catalogue of all ATLAS events.
For each real or simulated event in the primary processing formats (e.g. RAW and AOD), it stores the main event identification parameters (run and event number, trigger stream, luminosity block), the trigger record and the provenance information, i.e.\ the Globally Unique Identifiers (GUIDs) of the files that contain it.
This information is used to check the completeness and correctness of the processing campaigns, as the number of output events must match the number of input events and each event must be found once and only once in the output streams, and to monitor the overlaps between trigger chains or between offline selection streams.
The provenance information is used to extract one or several events that are selected during any analysis and produce event displays (see Section~\ref{sec:eventdisplay}), to study details of the calibration, alignment and reconstruction algorithms, or for special analysis workflows.
In this last instance, physics analyses that need specific reconstruction algorithms to be applied on specific event samples (up to one million events) can use EventIndex provenance information and the Event Picking Service~\cite{EventPickingService} to extract those events from the bulk of the RAW data and process them separately.
The EventIndex metadata are collected by Grid jobs that scan all newly produced data files.
These metadata are then sent to a central storage system at CERN, implemented using \textsc{Hadoop}~\cite{hadoop}, \textsc{HBase}~\cite{hBase}, and \textsc{Phoenix}~\cite{phoenix}, an interface layer on top of \textsc{HBase} that allows SQL access.
For \RunThr, the system was extensively revised and re-implemented~\cite{ATL-SOFT-PROC-2023-027}.
The EventIndex stores several hundred billion real and simulated event records and grows linearly with data production rates.

The COMA system (Conditions/Configuration Metadata for ATLAS)~\cite{Gallas_2012} collects run-level conditions and configuration metadata
mainly from the Conditions database, as well as essential run-related information from other systems: the HLT farm, Tier-0 site, Trigger, AMI, and EventIndex data repositories and file systems such as good runs list (GRL) XML files (see Section~\ref{sec:DQ}).
Early in \RunOne, COMA became the main repository to store the ATLAS Data Periods, which are official sets of run numbers during physics data-taking with common detector or machine conditions.
These ATLAS Periods form the basis to group sets of runs for many purposes for ATLAS data processing and reprocessing as well as data analysis.
COMA provides both web-based and command line interfaces featuring collected data and unique related derived quantities,
as well as aggregated information across many runs (e.g.\ runs in specific projects, periods and GRLs) such as event counts (by stream and trigger), data volumes, luminosity and LHC beam-related quantities.
To ease the management of the ATLAS Conditions database, the COMA repository was expanded~\cite{Gallas_2014} to collect metadata about conditions data structure, payloads and related metrics, providing reports to help experts understand  conditions database organisation, content, and usage.
COMA systems have undergone continuous refinements in data collection and client utilities into \RunThr.

The ATLAS Metadata Interface (AMI)~\cite{ami} is a software ecosystem dedicated to scientific metadata that provides two critical services; management of metadata for all datasets produced by the experiment and bookkeeping of all the parameters defining the data processing workflows.
The AMI task server aggregates metadata information from several sources (e.g.\ \Rucio and \PanDA) via specific tasks and stores them in the AMI database.
Metadata sets are recorded for each processing step a dataset undergoes and the `filiation links' (provenance information) between processing steps of the same datasets are saved.

In parallel with ongoing operations, the AMI environment is constantly evolving: an internal rewrite of both the core server and Web interfaces were done for \RunThr to provide improved stability, scalability and flexibility when addressing the metadata needs of the users.
A new administration interface for the AMI task server was written to easily isolate, monitor and manage tasks.
In addition, the interface to make super containers (a set of datasets for a given data taking period or year, or any other logically-grouped set of datasets) was rewritten.
A prototype was developed and tested for the revision of the AMI-tags that define all the parameters of a processing workflow; it will be finalised in coordination with the Tier-0 and database teams.

Users most often access AMI via Web applications to manage and display metadata.
These are also available via command-line interfaces for easy use in scripts or data analyses.
AMI also recently developed a whiteboard system that allows users to set arbitrary tags onto a dataset to be able to retrieve them in a more efficient manner.
These tags are used to identify samples that are standard samples to be used in data analysis; they can also be used to identify all datasets in use by a single analysis, for example.

In addition to these external metadata systems, ATLAS employs \emph{in-file} metadata to transfer information
from one processing step to the next. These data include information about the configuration of the job
that produced the data (e.g.\ the parton distribution function used during event generation, or the geometry
used during simulation). They can be used for automatic configuration of subsequent workflow steps. For
example, the reconstruction automatically initializes the correct geometry based on information in the input
file. This automatic configuration can be overridden, or the information can be explicitly provided if it is
not present. The in-file metadata can also be used for consistency checks, for example by checking that
files to be combined in MC Overlay (see Section~\ref{digi:sec:overlay}) were created with a consistent
geometry. For real data, the in-file metadata includes information about which luminosity blocks were
processed, to help ensure that the entire set of events from each luminosity block is correctly processed
in an analysis. During derivation production (see Section~\ref{sec:derivationintro}), some of the in-file
metadata are converted into a simple \textsc{ROOT}-readable format. These can then be used in analyses, for example for
the configuration of calibration and uncertainty tools. Including this in-file metadata helps reduce the
necessity of database connections in analysis jobs and enhances the ability to run analysis with only simple
\textsc{ROOT} or similar software.


%
\section{Distributed computing}
\label{sec:distcomp}

ATLAS Distributed Computing (ADC) comprises the hardware, software and operations needed to support distributed processing, simulation and analysis of ATLAS data and to support the evolving needs of the experiment.
This can be broken down into two closely related areas, which operate side by side: Distributed Data Management (DDM), covering all aspects of storage, transfer, and access to the (collision and simulated) data, and the Workflow Management System (WFMS), which handles request, task and job definition, and manages the various workloads performed on the ATLAS data, for both large scale production tasks and user analysis.
The first site through which all data move, the Tier-0 site, uses special configurations and software due to the stringent operational requirements of the site.
This site is described in Section~\ref{subsec:prodsys:tier0}.
The remaining Grid sites use common software for both the data and workflow management.
The DDM model employed by ATLAS is presented in Section~\ref{sec:ddm}, followed by a description of the many aspects of the WFMS in Section~\ref{sec:prodsys}.
The data monitoring and analytics systems described in Section~\ref{sec:analytics} naturally cover both of these areas.


\subsection{The Tier~0 site}
\label{subsec:prodsys:tier0}

%

The ATLAS Tier-0 site comprises the hardware, software and operations needed to support the \emph{prompt} processing of the data produced by the ATLAS detector.
This prompt processing mainly consists of the first-pass reconstruction workflow and many calibration workflows (see Section~\ref{sec:dataflow}).
The Tier-0 site also provides support for ad-hoc reprocessing, and is frequently used for various commissioning tasks.
Reprocessing and commissioning tasks may include the recall of data from tape storage if necessary.

The Tier-0 site uses the same conceptual models of \emph{datasets} (i.e.\ collections of files) being processed by \emph{tasks} (i.e.\ collections of jobs executed on some external batch service) as DDM (see Section~\ref{sec:ddm}) and WFMS (see Section~\ref{sec:prodsys}).
However, it uses different workflow management software from most Grid sites for the prompt processing of data.

\subsubsection{Tier-0 resources}

The Tier-0 site consists of about 40,000 cores in the CERN \textsc{HTCondor}-managed batch system.
The cluster is shared with Grid processing, which takes over the resources if they are not needed for Tier-0 operations (see also Section~\ref{sec:prodsys:opportunistic}).
The Tier-0 site has about 2~PB of EOS disk space available to temporarily store intermediate datasets, and about 100~GB of shared AFS disk space to store all other (non-dataset) files needed for operation.

In addition, a single-core virtual machine is used to run the software that drives Tier-0 site operations.
Another single-core virtual machine is used to host the web service allowing both the monitoring and operation of the system.

\subsubsection{Tier-0 software}

The Tier-0 software is designed to depend on as few external services as possible.
This not only minimises the chance of service interruptions caused by them, but also the effort required to follow their evolution over time.
The services the Tier-0 software does depend on tend to be stable and highly reliable.
For the software, there is a similar strategy to depend on as little third-party software as possible, for the same reasons.

The Tier-0 software system is a modular, self-contained system comprising a workload management system (comparable to ProdSys2--\PanDA, as described in Section~\ref{sec:prodsys}), a data management system (comparable to \textsc{Rucio}, as described in Section~\ref{sec:ddm}), a monitoring system (probes, data collectors, alarms, etc.), and a comprehensive web interface for monitoring and operations.

The Tier-0 workload management system consists of two entities: one operating on the dataset/task level (the Tier-0 Manager, or TOM), and one operating on the file/job level (the Supervisor).
Both follow a component-oriented software paradigm; each entity is composed of approximately ten components that collaborate to provide the desired functionality.
This approach not only allows different implementations of the same component (e.g.\ interfacing to different batch system flavours, or to different database backend flavours), but also allows the sharing of components between the two entities (e.g.\ the logging component).
In addition, some components provide a plug-in mechanism that allows loading of alternative implementations for particular key algorithms.

This combined component-oriented and plug-in approach has not only facilitated a smooth evolution of the system over the last 15 years, but also allowed another CERN experiment (NA62) to use the ATLAS Tier-0 software system with minimal effort~\cite{tier0na62}.

\paragraph{Tier-0 manager}

Each Tier-0 Manager (TOM) instance/process runs a set of configured plug-in tasks, called \emph{tomprocesses}, at regular intervals in an infinite loop.
Each TOM is single-threaded, ensuring strict serialisation.
The main task of the TOMs is to implement the Tier-0 reconstruction and calibration workflows.
The paradigm followed here is that of a dataset blackboard: files created on EOS and published in a handshake catalogue by the online system are organised by a tomprocess into RAW datasets and put on a virtual blackboard.
Other tomprocesses regularly scan the blackboard for new datasets fulfilling certain criteria and, if triggered, automatically define tasks that process these datasets into one or more new output datasets.
For \RunThr, about 100 such dataset-processing tomprocesses are defined, each configured with about 20 parameters.

Taking advantage of the generic architecture of the TOM system, most other Tier-0 activities are also run as tomprocesses.
Examples include the registration of permanent output datasets and their files with \Rucio, the management of the temporary EOS disk space containing the transient datasets, and the management of the temporary AFS disk space.

\paragraph{Supervisor}

As the name suggests, the task of a Supervisor process is to supervise the execution of the jobs defined by the T0 Manager(s).
Each supervisor process is single-threaded, and multiple processes can be configured and deployed simultaneously.
The standard configuration for \RunThr, however, only needs to deploy a single instance, which manages the about 10,000 jobs running simultaneously on the \textsc{HTCondor} batch system.

\paragraph{Web control interface}

The Tier-0 web interface, called \emph{conTZole}, is a classical \textsc{Javascript}/\textsc{Django}/\textsc{Apache} single-page web application.
It allows comprehensive monitoring of the state of the Tier-0 system, with both live and historical information.
Subject to authentication and authorisation, it also allows the Tier-0 operators to dynamically change the Tier-0 configuration, and third-party users to request ad-hoc processing of selected datasets.


\subsection{Distributed data management}
\label{sec:ddm}

\begin{figure}[tbp]
\includegraphics[width=\textwidth]{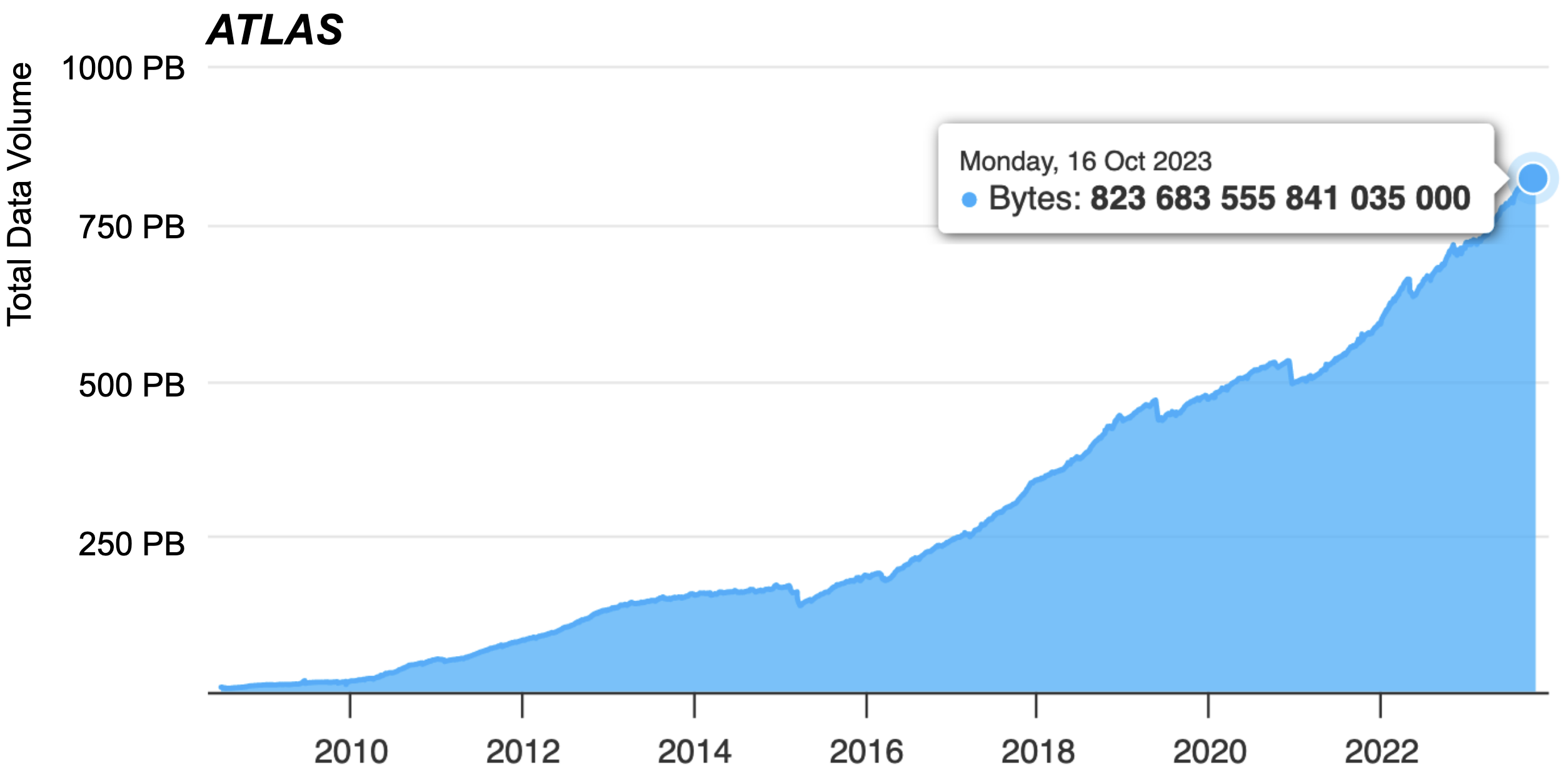}
\caption{The evolution of the ATLAS total data volume over the last 15 years, since before \RunOne.}
\label{fig:rucioData}
\centering
\end{figure}

ATLAS data are distributed over a worldwide network of data centres, also called sites, under the umbrella of the WLCG~\cite{wlcg}.
These sites are categorised into Tiers in a semi-rigid hierarchy with various capacities, duties, and responsibilities.
CERN is the origin of detector data and thus the single Tier-0 centre.
There are 11 ATLAS Tier-1 sites that are connected via dedicated national research and education networks (NRENs) with typically 10 to 100~GBit optical private networks~\cite{LHCOPN}.
These Tier-1 sites provide disk and tape storage and are charged with the perpetual archival of detector data. Around 70 ATLAS Tier-2 sites, typically hosted by national universities and laboratories, provide disk storage that is used for data processing and user analysis.
Several additional computing centres complement the storage and computing available to ATLAS, including opportunistic sites (see Section~\ref{sec:prodsys:opportunistic}) and sites dedicated to data analysis (see Section~\ref{sec:analysis:infrastructure}).
As shown in Figure~\ref{fig:rucioData}, as of 2023 ATLAS has more than 800~PB of resident data.
The total volume is split roughly equally between tape and disk.
All these data are organised, managed, transferred, accessed, and accounted for via the \Rucio distributed data management system (see Section~\ref{subsec:ddm:rucio}).

\subsubsection{\Rucio}
\label{subsec:ddm:rucio}

\Rucio~\cite{rucio} is a flexible and modular software framework to build data management federations.
It allows seamless integration of scientific and commercial storage, and their network systems.
The data are stored in a single global namespace and can contain any payload.
Facilities hosting the storage can be distributed at multiple locations, and belong to different administrative domains.
\Rucio was designed with more than a decade of operational experience in very large-scale data management, building from the work of its predecessor DQ2~\cite{DQ2}.

As a technology, \Rucio is location-aware and manages data in heterogeneous storage solutions and environments.
It allows the creation, location, transfer, deletion, annotation, and access of and to the data.
The major, standout feature is the orchestration of dataflows.
These include both high-level dataflows, such as institutional or experiment policies, and low-level dataflows, such as specific user requests.
Its main users are the \PanDA and ProdSys2 workflow management systems, including the \PanDA Pilot job processing framework, as described in Section~\ref{sec:prodsys}.
Individual users have command-line interface and web-based access through dedicated clients, both for access to data and for the manipulation of rules for data movement and storage.

At ATLAS mid-2022 scale, the interaction rate of operations with the \Rucio system is typically beyond 200~Hz and often reaches 500~Hz, from distributed clients across the world interacting with the single \Rucio instance at CERN.
This includes diverse operations such as registering new files, searching for data, downloading files for processing jobs, scheduling files for deletion,\footnote{See Section~\ref{subsec:ddm:datapolicies} for the reason behind scheduling files for deletion, rather than deleting them immediately.} or modifying metadata.
\Rucio handles on average 1.5 million file transfers per day, with a peak rate of three million files per day.
On average around 8~PB of data is moved per day from job and user upload/download, peaking at 12~PB per day.
Transfers between data centres accounts for 2~PB per day on average, peaking at 4~PB per day.
The high-availability deployment of \Rucio is done via \textsc{Kubernetes}~\cite{kubernetes} in the CERN data centre and is built atop free, open-source technologies.
Additionally, extensive monitoring solutions were built and are the cornerstone of daily operations.

\Rucio was principally developed by and for the needs of the ATLAS experiment, with a view to eventually making it community open source software.
It has since matured into the de-facto general solution for scientific data management.
Now, it is developed, used, and supported by a multitude of different communities from diverse scientific fields, from neutrino experiments and dark matter searches, to astronomical observatories.
As software, it is both free and open source, provided under the \textsc{Apache} 2.0 License~\cite{apacheLicence}.

\subsubsection{Dataset nomenclature}
\label{subsec:nomenclature}

ATLAS datasets follow strict nomenclature rules to ease findability and ensure that dataset contents can be quickly identified. The names can be generated algorithmically, are unique, and are case-insensitive, although the original case is preserved throughout the system.

The dataset name is a series of fields separated with a `.'. For MC simulation, the fields are:
\begin{enumerate}
\item The project, a short indicator of the MC simulation campaign and, when relevant, centre-of-mass energy ($<15$ characters);
\item A numerical dataset identifier ($<8$ characters);
\item A short description of the physics described by the dataset ($<50$ characters);
\item The production step that generated the dataset (e.g.\ `simul' for simulation or `evgen' for event generation; $<15$ characters);
\item The data type (e.g.\ `EVNT', `HITS', or `AOD'; $<15$ characters);
\item A series of processing tags, called AMI tags (see Sections~\ref{sec:metadata} and \ref{sec:core:configuration}), indicating the configuration of the software that was used for each production step in the creation of the dataset, separated by underscores. Each step is represented by a letter, followed by the numerical index of the configuration, stored in the AMI database (e.g.\ `e1234' for an event generation configuration; $<32$ characters in total); and
\item Optionally, a production task index and sub-task index.
\end{enumerate}
An example of an MC simulation dataset name is\\
`mc15\_13TeV.300402.Pythia8B\_A14\_CTEQ6L1\_Bs\_mu3p5mu3p5.recon.AOD.e4397\_s2608\_r6869'. In this case, `mc15\_13TeV' is the project; 300402 is the dataset identifier; the production step is `recon' (reconstruction); the data type is `AOD'; and the configuration tags are `e4397', `s2608', and `r6869' for event generation (e), simulation (s), and reconstruction (r). The description means that the sample was generated with the \textsc{Pythia8B} event generator (see Section~\ref{sec:evgen}) using the A14 tune of underlying event and hadronisation parameters in \PYTHIA and the CTEQ6L1 parton distribution function set. The sample contains $\Bs\rightarrow\mu\mu$ decays, where both the muons are required to have $\pt>3.5$~\GeV. Although not trivial to decode, with some experience these labels are enough to understand the contents of the sample.

For detector data, the fields are:
\begin{enumerate}
\item The project, indicating the year and, when relevant, the centre-of-mass energy for collision data;
\item The unique run number for the dataset, established by the data acquisition software;
\item The name of the data stream (e.g.\ `physics\_Main' for the most commonly analysed data stream~\cite{ATLAS-TRIG-2022-03});
\item The production step that generated the data, similar to MC simulation;
\item The data type, similar to MC simulation;
\item A series of processing tags, similar to MC simulation; and
\item Optionally, a production task index and sub-task index, similar to MC simulation.
\end{enumerate}
One example is `data15\_13TeV.00284484.physics\_Main.merge.AOD.f644\_m1518'. In this case, `data15\_13TeV' is the project; 284484 is the run number; the data are from the physics main stream; the data are merged (following the reconstruction, in this case); the data format is `AOD'; and the production tags are `f644' and `m1518' for prompt reconstruction at the Tier-0 site (f) and merging (m).

User- or group-defined datasets are required to follow a less-strict set of rules; they must begin with `user.' followed by the user name, or `group.' followed by the group name.

\Rucio, used to organize the data, requires each dataset to be defined within a scope. For MC simulation and detector data, the scope is identical to the project in the dataset name. For group and user data, the scope is `group.' followed by the group name, or `user.' followed by the user name. Datasets can also be grouped into containers (groups of like datasets) within \Rucio; containers follow the same naming conventions where possible.\footnote{For example, a container of multiple data runs follows the same general scheme, but does not specify a unique run number if it contains many runs. It may instead specify a Period (see Section~\ref{sec:metadata}) based on which the runs are grouped.}

These conventions were established before the start of data collection and were only carefully updated since. This ensures that any ATLAS dataset in the entire storage system can be quickly and easily identified by any users familiar with the naming conventions.

\subsubsection{Data policies and life cycle}
\label{subsec:ddm:datapolicies}

One key feature of ATLAS data is immutability.
Datasets are locked once production is complete, and no additional files can be added, nor removed except in rare cases of data loss.
Changes are not allowed; if a fix must be applied, an additional processing step is undertaken and the pre- and post-fix data can be retained in the system independently if necessary.
This ensures that all data produced centrally (whether MC simulation datasets or real detector data, at any stage of processing) can be reproduced at any time, and the full chain of provenance is known.

\begin{figure}[tbp]
\includegraphics[width=\textwidth]{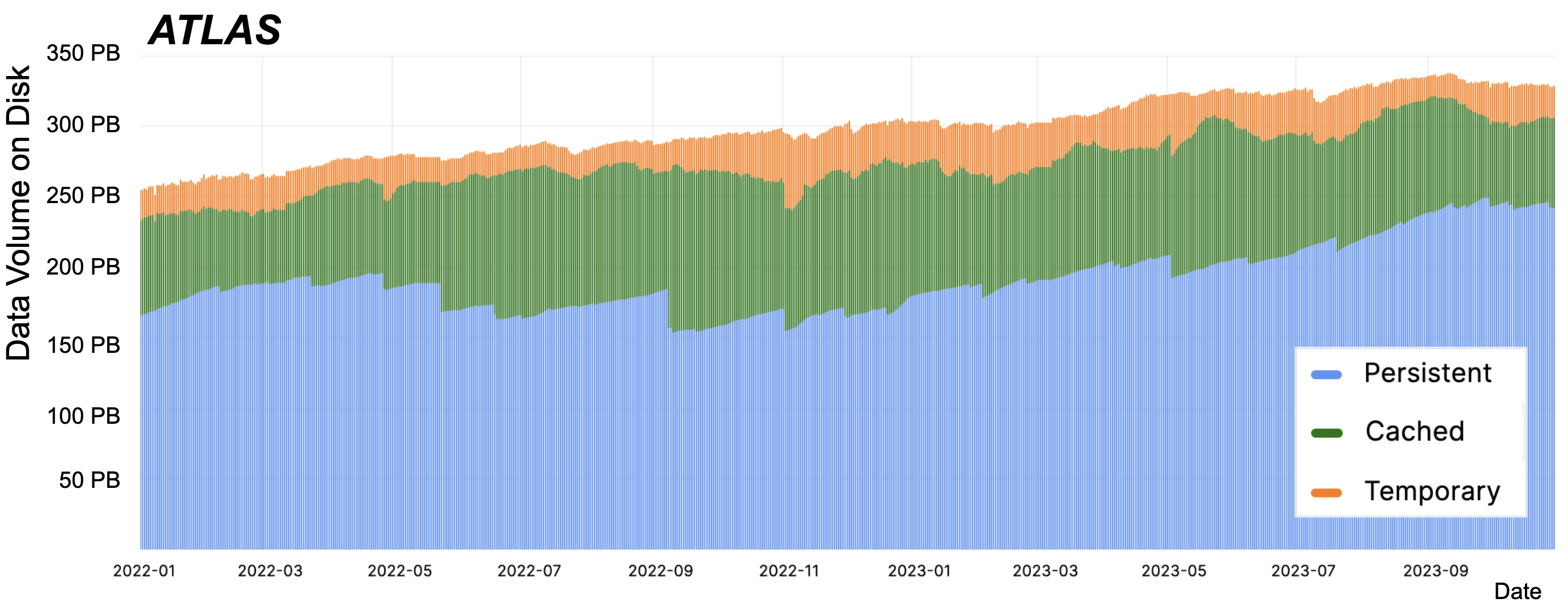}
\caption{The distribution among persistent, temporary, and cached of ATLAS data stored on disk in 2022--2023.}
\label{fig:diskUsage}
\centering
\end{figure}

Data on disk and tape are managed in slightly different ways.
Disk-resident data are divided into three categories:
\begin{itemize}
\item \emph{Persistent} data are pinned on disk for an indefinite period of time,
\item \emph{Temporary} data are pinned for a limited period of time, and
\item \emph{Cached} data are not pinned.
\end{itemize}
When non-user-analysis data are produced by Grid jobs they are copied to their final location and pinned there as persistent data.
These data cannot be deleted except through one of the mechanisms described later in this section.
Data that are moved around by the production system to be used as input for jobs (including data that are replicated from tape) are pinned on disk for a limited time and are thus labelled temporary.
Once this time runs out, the data are considered cached and are eligible for deletion.
The total volume of data stored on disk by ATLAS and the distribution within these different categories is shown in Figure~\ref{fig:diskUsage}.

Each site hosting ATLAS data defines a quota of space available for use by ATLAS, and publishes frequently the capacity, used and free space.
\Rucio sets a watermark just below the defined capacity and aims to keep the used space at this watermark.
If the site reports used space over the watermark, \Rucio deletes cache data in order of least recent access until the watermark is reached.
In this way, ATLAS uses as much disk space as is possible, unpopular data are automatically expunged from disk and popular data are likely to stay and be reused.

A site can be assigned as a \emph{nucleus} when it can reliably aggregate output by \emph{satellite} sites.
The nucleus designation is made manually by operators in the Computer Resource Information Catalogue, CRIC~\cite{cric},
and can change over time. Job output data are transferred from the satellite sites and aggregated at the nucleus
for the full task. The output is kept at the nucleus for archival or further processing. Brokering ensures that
jobs are spread across satellites, while taking the available free storage space of the nuclei into account.

Data on tape are managed in a much less dynamic way, as the medium is not designed for rapid turnover.
Obsolete data on tape are marked as cached, but they are only deleted in organised deletion campaigns coordinated with the sites, which take place at most once per year.
This is because space freed from deleted data is usually only reclaimed when tape cassettes are \emph{repacked} (i.e.\ the remaining data on each cassette are packed into a new one).

The policies for data replication and lifetime are designed to optimise the availability and redundancy of data for processing and analysis, whilst ensuring that the limited storage capacity is effectively used.
The data replication factor and lifetime of the data are heavily dependent on how actively the data are used and their reproducibility in the case of loss.
The policies defining whether to store data on disk or tape also depend on the access patterns and expected lifetime of the data.
Raw data from the detector, for example, are always copied both to tape storage at CERN and one Tier-1 centre, because they are read infrequently and to ensure a negligible risk of data loss (only one raw data file was lost since the start of ATLAS operations).
On the other hand, DAOD datasets used for analyses have at least two copies on disk so those data are always available, but are not replicated to tape because new versions are frequently produced and hence each version has a limited lifetime.

The replication of new data to match the policies is handled by \Rucio subscriptions, which automatically create copies of data in the required places.
Archiving to tape of data produced on the Grid (HITS and AOD, and reprocessed data AOD) is done through a separate asynchronous process that delays the archival until one month after the data are produced, to avoid bad data being written to tape that are then immediately deleted.

An important shift in the ATLAS data model in recent years is from viewing tape as a pure archive or backup to be used in limited cases to a more active use of tape as the main data store, and disk as a cache to which data are staged from tape temporarily and then deleted.
This change required detailed investigation into how sites' tape storage was organised and developments in ATLAS services for optimally reading data back from tape.
This work was carried out under the auspices of the Data Carousel project~\cite{dataCarousel} (see Section~\ref{sec:prodsys:dataCarousel}), which was successfully implemented during LS2.
The amount of data staged from tape increased from 20~PB in 2018 to 130~PB in 2021, without disrupting writing to tape (which increased from 47~PB in 2018 to 75~PB in 2021).

Another important handle in controlling what data are cached on disk is the automatic deletion of unused data from disk when there is an archival copy on tape.
This mechanism can be tuned to be more or less aggressive according to how much cache space is available globally, which processing campaigns are planned and so on.

The life cycle of data is controlled by the \emph{lifetime model}, through which data are deleted from all centrally-managed storage a certain period of time after they are produced.
The lifetime depends on the type of data and their expected use; following the examples above, raw data have an infinite lifetime and DAOD data have a lifetime of six months.
The lifetime may be extended in two ways. When data are accessed (i.e.\ used as input to a \PanDA job or manually downloaded using \Rucio), their lifetime is extended (usually by six or twelve months).
Alternatively, an analysis team may ask for an extension at the point when data become eligible for deletion, for example if the team is midway through a publication procedure and the data may be needed to recreate results.
The lifetime model has worked well to control the disk and tape space used by ATLAS, allowing obsolete data to be easily removed to free up space for more copies of actively used data.
However, with the length of individual ATLAS analysis efforts often stretching to many years, the amount of data that is not active but was requested to be kept `just in case' is steadily growing and is now roughly 10~PB.
Recent studies have shown that only a fraction of these data are accessed after a lifetime extension request.
Therefore, work is ongoing and will continue throughout \RunThr to identify alternatives to keeping such data pinned on disk, such as moving to `cold' storage or streamlining procedures to allow recreation of data on demand.

This lifecycle is applied to data on general ATLAS resources; some resources exist outside of these rules. Analysis
groups (e.g.\ the Standard Model or Higgs group) have dedicated \emph{group disk} pools where data supporting
ongoing analyses can be stored without regard to the lifetime model. Similarly, institutes have local group disk
areas where data supporting the local data analysis can be stored. Users or group space managers are responsible for
creating rules to transfer data to these disk areas and for deleting data from them.

\subsubsection{WLCG data challenges}
\label{subsec:ddm:datachallenges}

In anticipation of HL-LHC data rates, the WLCG has decided to run increasingly challenging stress tests of the data infrastructure, including disk, network, and the tape systems.
The plan is to run these tests every two years until 100\% of the expected HL-LHC data rate is reached by 2027.
Two target rates are defined: the \emph{minimal} target, which is the rate expected with a rigid hierarchical computing model, and the \emph{flexible} target, which is the rate expected with flexible computing models allowing transfers across and between all Tiers dynamically.
The latter model is similar to the current computing model.

\begin{figure}[tbp]
\includegraphics[width=\textwidth]{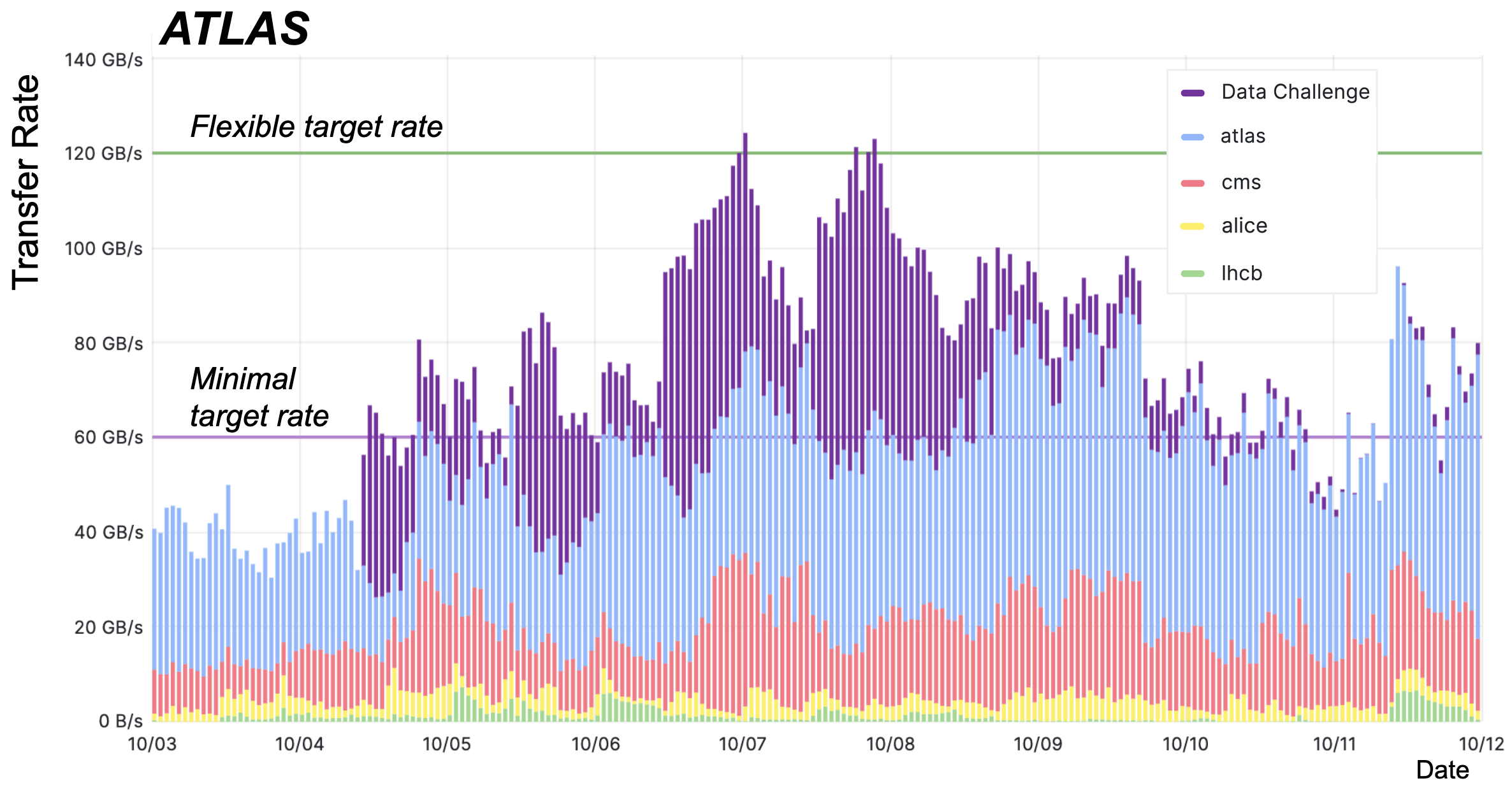}
\caption{In purple, additional `Data Challenge' traffic injected to reach target rates during the October 2021 Data Challenge, along with the standard operational traffic from the four LHC experiments.}
\label{fig:dataChallenge}
\centering
\end{figure}

In October 2021, the first disk and network challenge took place with the aim of achieving a sustained overall transfer throughput of 10\% of the expected HL-LHC data rate, with the minimal target corresponding to the expected \RunThr rate~\cite{campana_simone_2021_5532452}.
Consequently, this data challenge also served as a commissioning test for \RunThr.
The results are shown in Figure~\ref{fig:dataChallenge}, showing that the minimal target was easily met throughout the duration of the challenge and at peak rate the `flexible' target was achieved.

Dedicated tape challenges focused on the capability of the Tier-0 and Tier-1 tape systems to handle the expected rate of \RunThr RAW data export, which is up to 8~GB/s from all physics streams combined.
These tape challenges were successful, and these rates were met and even exceeded without any operational incidents during \RunThr data taking.
Different streams are distributed across different sites, but all data within each stream are sent to a single site for a single run.
The largest stream is produced at a typical rate of 3.5~GB/s, and the final tape challenge in March 2022 showed that all Tier-1 sites were capable of handling this data rate.

In mid-2022, the planning for the follow-up data challenge has started, with a projected date in early 2024.
The major features to be tested are the new token authentication mechanisms (see Section~\ref{sec:ddm:tokensAndHepscore}), software-defined networks, and much-improved monitoring.
Potential new dataflows introduced by HPC centres (see Section~\ref{sec:prodsys:hpc}) and dedicated analysis facilities (see Section~\ref{sec:analysis:infrastructure}) will be folded into the plan once their usage is more clearly articulated within the community.

\subsubsection{Further WLCG infrastructure changes during \RunThr}
\label{sec:ddm:tokensAndHepscore}

A virtual organisation-based (VO-based) security architecture, using X.509 certificates, has been a reliable solution for authentication and authorisation in ATLAS, but has also showed usability issues and required ad-hoc services and libraries in the Grid middleware.
The need to move beyond the VO-based scheme was recognised as an important objective in WLCG Authentication and Authorisation Infrastructure (AAI)~\cite{aai}, to overcome the usability issues of the current AAI and embrace recent advancements in web technologies.
A token-based AAI was implemented in ATLAS using the \textsc{Indigo} Identity and Access Management service~\cite{iiam}, fully compliant with OIDC/\textsc{OAuth2.0} and capable of identity federations among scientific and academic identity providers.
The deployment of this new AAI infrastructure across services is ongoing.

Until 2023, all WLCG experiments utilised the \textsc{HEPSpec06}~\cite{hepspec06} benchmark for accounting and pledges.
However, this 32-bit benchmark is no longer a reliable indicator of HEP workloads, whilst the underlying SPEC-CPU 2006 benchmark~\cite{spec2006} is no longer
supported since 2018.
Moreover, there is a coming need for a benchmark that caters to non-x86 architectures such as ARM processors and GPUs.
To address these issues, a task force was established by the WLCG in November 2020, to identify a new benchmark based on the current workloads of the LHC experiments and a transition plan.
A new benchmark, \textsc{HEPScore}, which includes both x86 and ARM processors, was developed~\cite{hepscore} and deployed in April 2023.
The experiments now use this new benchmark for resource requests, and sites are expected to score new hardware purchases using this benchmark instead of \textsc{HEPSpec06}.
The benchmark is made up of a variety of workloads~\cite{hepscore23} from all four LHC experiments and Belle II; the initial version is \textsc{HEPScore23}.
The transition to \textsc{HEPScore23} will enable more accurate accounting and pledges, providing a more realistic representation of HEP workloads.
Furthermore, \textsc{HEPScore23} was normalised to the old \textsc{HEPSpec06} benchmark, so that sites do not have to re-benchmark existing systems, and only use it to measure newly purchased hardware.

\subsubsection{DDM R\&D projects}
\label{subsec:ddm:rad}

In addition to the ongoing development work and data challenges, which come with their own research and development communities, several distributed data management-specific R\&D projects were started: most importantly, integration with commercial clouds, dynamic data handling, and distributed caching.

The first of these R\&D projects, underway in \RunThr, is the integration of \Rucio with commercial clouds, specifically Google Cloud~\cite{gcs}, Amazon Web Services~\cite{aws}, and Seal Storage Technologies~\cite{seal}.
The integration of such cloud services is reaching a stage where these types of storage can be included in the distributed data management system beyond simple demonstrators.
The main benefit is that funding agencies and institutes can achieve more flexible storage installations (including, potentially, the purchase of cloud-based resources), while \Rucio takes care of abstracting the peculiarities from the scientists.

A second R\&D project is revamping the policies that drive ATLAS dataflows, mostly given by experiment data agreements, processing requirements, and MoUs, as well as operational and infrastructural needs.
To address this, a working group was formed to address dynamic data handling in a coherent and consistent way.
The working group will study and tune ATLAS dataflows, most importantly placement of new data, rebalancing between data centres, deletion of obsolete and unused data, and data replication both for production and analysis.
The eventual goal is to reduce workload execution time, reduce data access time and make better use of available storage through improved placement, movement, addition, and deletion of replicas of ATLAS data, under the hard constraint of limited storage space.

Finally, a third R\&D project deals with distributed caching to reduce wide-area-network (WAN, site-to-site) traffic, reduce latency to the analysis software, and eventually to increase CPU efficiency.
This is an integration of \Rucio with the \textsc{Xcache} system~\cite{xcache} to assign processing jobs not only with the common `job to data' paradigm, but to allow a more flexible approach that includes significant staging of data to sites with extra computing capacity.
During LS2 a prototype deployment was put in place.
First observations have already provided insights, including increased cache use and reduced WAN traffic.
During \RunThr this mechanism will be exercised with real analysis use cases and, if proven successful, will be integrated into \Rucio.


\subsection{Workflow management}
\label{sec:prodsys}

\begin{figure}[tbp]
\includegraphics[width=\textwidth]{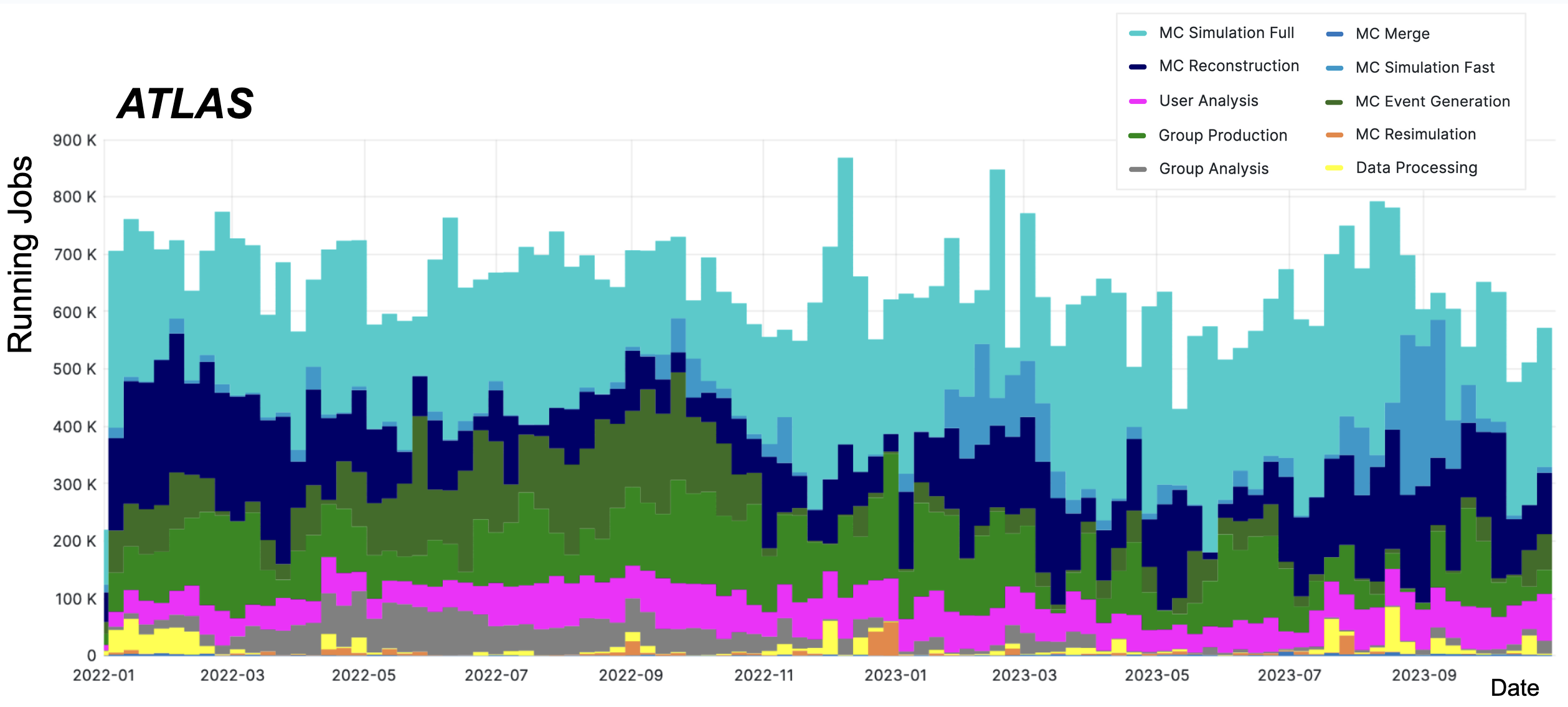}
\caption{Running ATLAS jobs since the beginning of 2022, grouped by activity.}
\label{fig:runningJobsByActivity}
\centering
\end{figure}

A large variety of workflows are required by the ATLAS experiment, including data processing (and reprocessing), MC event generation, simulation and reconstruction, and derivation data production for physics groups (see Section \ref{sec:transforms} for details).
This is in addition to data analysis conducted by physics groups and individual users.
Figure~\ref{fig:runningJobsByActivity} shows the number of running ATLAS jobs since the beginning of 2022, for each of the different activities.

\begin{figure}[tbp]
\includegraphics[width=\textwidth]{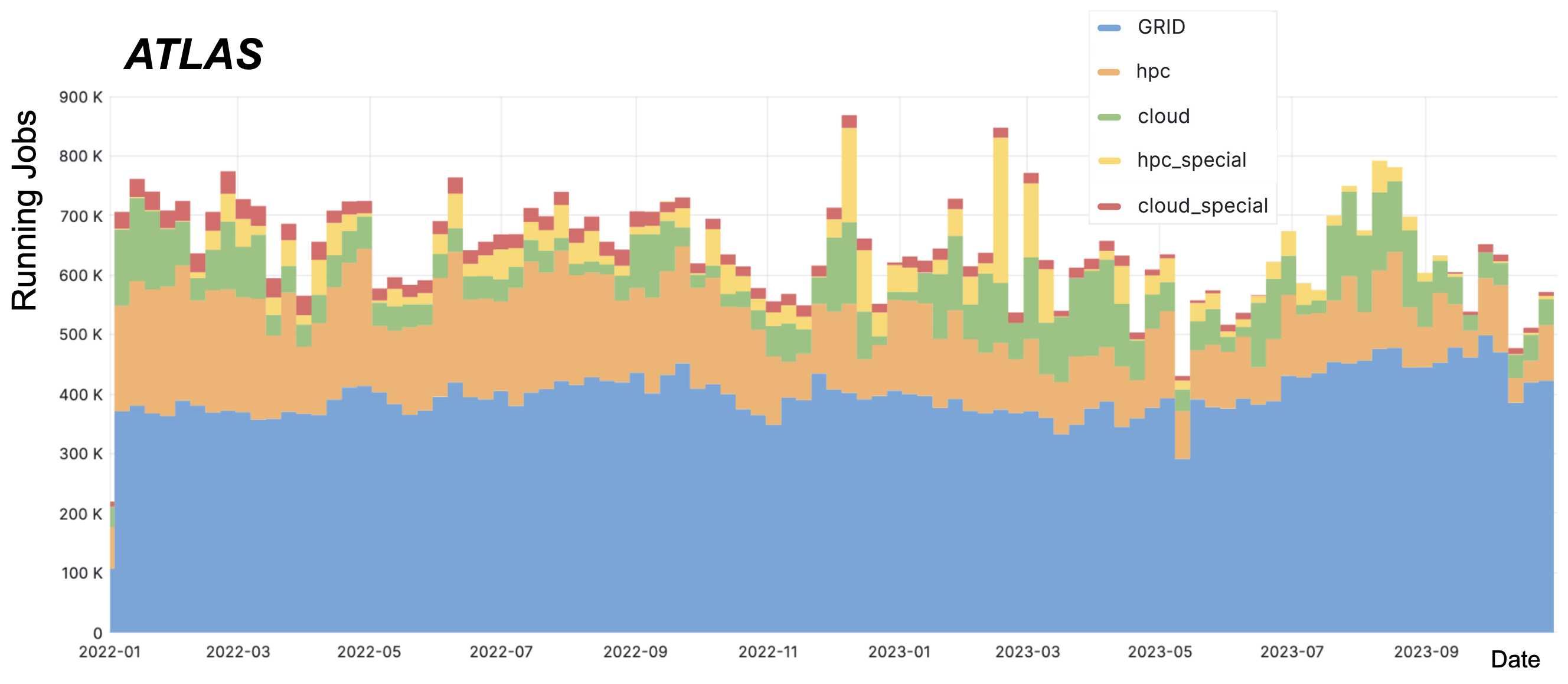}
\caption{Running ATLAS jobs since the beginning of 2022 grouped by resource type.}
\label{fig:runningJobsByResourceType}
\centering
\end{figure}

To fulfil this workload demand, the ATLAS experiment uses more than 250 computer centres worldwide, primarily made up of the WLCG Grid sites described in the previous section, but also including HPC centres and national, academic, and commercial cloud computing resources.
Figure~\ref{fig:runningJobsByResourceType} again shows the number of running ATLAS jobs since the beginning of 2022, but this time grouped by resource type.
It can be seen that although the contribution from the Grid dominates, as expected, there are many running jobs on opportunistic resources such as HPCs and cloud, the latter contribution mainly coming from Sim@P1\footnote{Because originally Sim@P1 was managed using OpenStack, it was accounted for as part of the cloud resources. Although this is no longer the case, for historical consistency it remains in this accounting category.} (see Section~\ref{sec:prodsys:simP1}).
Further contributions are present from assigning dedicated production tasks to both HPCs (`hpc\_special') and clouds (`cloud\_special'), where the latter is mainly made up of BOINC~\cite{boinc1,boinc2} job submissions and jobs running on the Tier-0 site.
These additional resources are described in Section~\ref{sec:prodsys:opportunistic}.

\subsubsection{A universal workflow management system for ATLAS}
\label{sec:prodsys:prodsysPanda}

The processing, simulation and analysis of data from modern HEP experiments requires the orchestration of multiple diverse computing facilities.
This section describes the methods developed and the approaches taken to arrive at a universal workflow management system for ATLAS, ProdSys2--\PanDA~\cite{prodsys2,panda}, which has been used since the beginning of 2014.
New elements such as iDDS~\cite{idds1} and Harvester~\cite{harvester} have since been developed and brought into production before the beginning of LHC \RunThr to address the heterogeneity of resources and the introduction of new data processing scenarios such as the Data Carousel~\cite{dataCarousel}~(Section \ref{sec:prodsys:dataCarousel}).

\paragraph{Design concept}
\label{sec:prodsys:design}

When designing a suitable workflow management system for a distributed and heterogeneous computing infrastructure, there are many aspects to consider.
The primary role of such a system is the management of the various production and analysis workloads, as well as defining how the resources are to be used and the application of appropriate fair--share policies.
The system should also work regardless of the type of infrastructure and level of heterogeneity and be dynamic enough for possible changes in the available resources.
The various resources used and the associated access protocols, are described using CRIC~\cite{cric}.
User identification and security considerations are handled via the use of X.509 certificates, which are to be replaced by tokens during \RunThr as described in Section~\ref{sec:ddm:tokensAndHepscore}.

After analysing the various classes of workflows, which are described in Sections~\ref{sec:transforms} and \ref{sec:processing}, the essential components of the production system were identified and a logical data model built.
The following entities are defined:
\begin{itemize}

\item A \emph{request} is the upper level of abstraction, and is made of tasks of one type. A typical example would be the (re)processing of all data for a certain period or a collection of MC simulations with common parameters, which are to be used together to compare to a given period of data. A request may comprise multiple steps, where event generation is one step, for example. Each step on each input dataset is represented as a separate task.

\item A \emph{task} is an entity for passing parameters to the payload submission system, and is composed of jobs. The result of the task is a set (or several sets) of files, organised in datasets, typically with one dataset per task per output file type.

\item A \emph{job} is a single executable workload, where each task consists of one or many jobs (up to ten thousand). A job is executed on a single Grid computing element, opportunistic CPU or worker node. A job may have input data and writes the result of the work to an output file or files.

\end{itemize}

\begin{figure}[tbp]
\includegraphics[width=\textwidth]{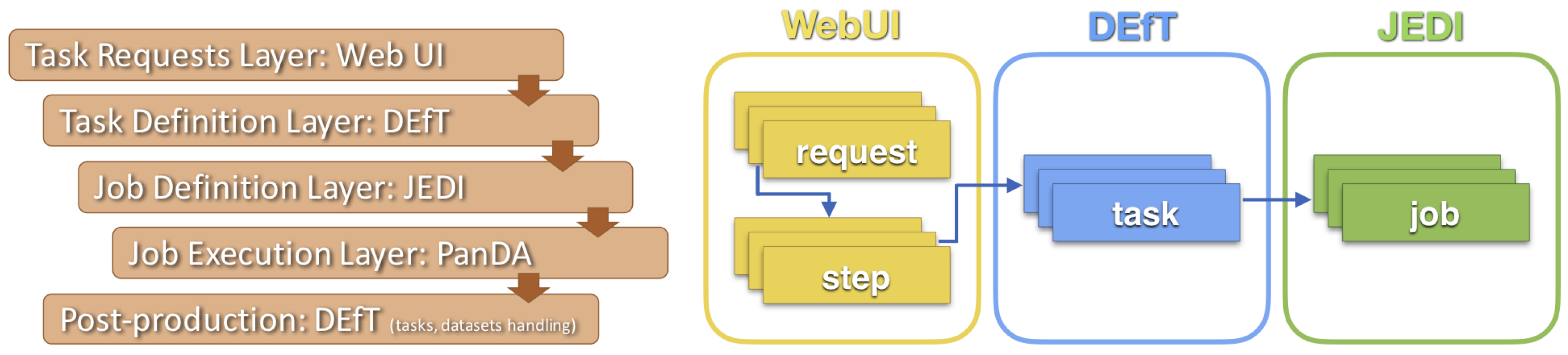}
\caption{The several levels of the ATLAS workflow management system. Figure from Ref.~\cite{AdvancedAnalytics}.}
\label{fig:wfmsLayers}
\centering
\end{figure}

The system architecture is designed to ensure continuous and optimal access of the scientific community to computing resources, which is achieved with an extensible layered architecture.
Figure~\ref{fig:wfmsLayers} shows the different levels of the workflow management system, where the relationship between the relevant entities is schematically described on the right~\cite{AdvancedAnalytics}.
The implementation of these levels as part of the actual system, shown in the left of Figure~\ref{fig:wfmsLayers}, is described in the following.

\paragraph{ProdSys2 web user interface}
\label{sec:prodsys:prodsys}

The ProdSys2~\cite{prodsys2} Web User Interface (Web-UI) is used by the ATLAS production managers to interact with the task management system.
A production manager represents a group activity in the ATLAS experiment and defines workflows for their group: Data (re)processing, MC production (event generation, simulation and reconstruction), and Derivation (DAOD) production.
Each activity has its own workflow and specific task requirements.

The heart of a work process with the Web-UI is the request. The production manager creates a request using the Web-UI, and defines its parameters and input datasets.
For each input dataset, a sequential set of steps to be executed is defined, which is translated by the system into a sequence of tasks to be executed.
A single step might correspond to event generation alone, with detector simulation as a second step following it.
The data model features the logical division of the request into horizontal \emph{slices} defining all the steps beginning from a single input dataset.
Each slice may comprise several sequential steps in one production workflow, each executed as a task, which itself is divided into jobs.

The Web-UI is also used to monitor the progress of running tasks and adjust them as required.
Active bookkeeping is provided, metadata is stored, including user-defined tags, and the aggregation of task statistics is done, which may be used for the fine-tuning of running and historical task analysis.
Advanced error handling and reporting is included to help quickly understand the root of any problem and to
fix it by redefining and resubmitting the task.
The chaining together of slices in separate requests is also possible and successfully used, for example when connecting MC event generation, simulation and reconstruction together with derivation production.

\paragraph{DEfT}
\label{sec:prodsys:deft}

The Database Engine for Tasks (DEfT) is a top-level sub-system and is the engine beneath the Web-UI.
DEfT accepts task requests via a dedicated user interface or from prepared lists (e.g.\ a text file, or Google or Excel document).
DEfT processes requests and is responsible for the formation of processing steps, tasks, and input data and parameters.

ProdSys2 sets default parameters via DEfT, for example memory and CPU limits, which are often not defined by the user.
ATLAS has many default parameters for complex workflows, which may be viewed and changed via a special ProdSys2 interface.
After parameters are set, the system checks for their consistency and compatibility.
For example, it is verified that the input dataset has enough events or that the parameters defined in the task are valid for the ATLAS software release to be used.
A further check is performed to see if similar tasks were defined and successfully executed, to avoid event duplication.
During task definition, if ProdSys2 detects that some of the events in an input dataset were already used for a given configuration, DEfT sets an offset (in files when input datasets are used, or in event numbers and random number seeds for steps without input data like event generation) and the new task uses only unique events.

\paragraph{JEDI}
\label{sec:prodsys:JEDI}

The Job Execution and Definition Interface, or JEDI, is a middle-level sub-system that receives formalised job descriptions from DEfT.
JEDI dynamically determines the number of jobs for each task and is responsible for launching and executing individual jobs.
JEDI verifies information about the data via \Rucio and about the job queues via CRIC.
A common database is used by JEDI and \PanDA (see Section~\ref{sec:prodsys:panda}) to store information about the status of jobs and tasks.

\paragraph{\PanDA}
\label{sec:prodsys:panda}

\PanDA~\cite{panda} is the engine of the underlying system and the most complex layer. It determines which resources are used when for each of the jobs, receives information from Pilot jobs (see below) and the CRIC information system, and manages the progress of jobs.
The various \PanDA components are described below.

A database of jobs and tasks is used system-wide, storing comprehensive static and dynamic information and meta information about all jobs and tasks defined, running, and completed in the system, including the history of their execution and errors that have occurred in the process.

\emph{Pilot} jobs are used to collect information about the state of computing resources and requirements of a task.
Jobs are submitted to successfully activated and verified Pilots by a \PanDA server based on resource selection criteria.
Late binding of jobs to the execution location prevents delays and failures and maximises the flexibility of resource allocation for a job, based on the dynamic state of processing resources and job priorities.
A Pilot is the main `isolating layer' for WFMS, encapsulating complex heterogeneous environments and the Grid interfaces and tools with which the WFMS interacts.
Pilot jobs also serve to identify the resource requirements of a task and adjust subsequent job definitions, for example to request additional memory.
In the case of failures, pilots ensure that only a small fraction of the task runs, and the task is marked as failed before wasting significant resources.

The \emph{Intelligent Data Delivery Service} (iDDS)~\cite{idds1} is an additional layer developed to orchestrate workflow management and distributed data management systems to optimise resource usage in various workflows.
The input data to a task are dynamically transformed, so that the data pre-processing, delivery, and main processing in each workflow are decoupled, which allows them to run asynchronously.
iDDS was introduced in 2020 to address inefficiencies in the Data Carousel (see Section~\ref{sec:prodsys:dataCarousel}), which previously required many input events due to constraints in the workflow, creating delays in bulk reprocessing campaigns.
iDDS propagates the detailed information about the input data status from \Rucio to JEDI, allowing the incremental release of tasks so that processing can begin even if input data are only partially staged-in.
iDDS is now also used in new workflows such as some AI/ML workflows~\cite{idds2,idds3}, and supports several workflow definition languages.

\emph{Harvester}~\cite{harvester} mediates the control and information flow between \PanDA and the resources to enable more intelligent workload management and dynamic resource provisioning based on detailed knowledge of resource capabilities and their real-time state.
Harvester was designed around a modular structure to separate core functions and resource-specific plugins, simplifying the operation with heterogeneous resources and providing a uniform monitoring view.

\PanDA also includes functionality to automatically retry or re-broker jobs. Sometimes, site failures can be
identified and jobs can be re-directed to an alternative site that has a copy of the input data. Additionally,
a catalogue of known errors is kept, defining rules for the number of retry attempts that should be made when a
specific error is encountered. With this functionality, jobs that are suffering from system or site failures can
be retried and moved around the Grid, while jobs with clear failures that will not converge can be stopped before
significant resources are consumed.

The concept of \emph{global shares} is used to set the amount of resources available instantaneously for a certain activity, for example MC simulation, as a fraction of the total amount of resources available to ATLAS.
Global shares have recently been updated to be measured in the new benchmark, \textsc{HEPScore23} (see Section~\ref{sec:ddm:tokensAndHepscore}); previously, they were measured in \textsc{HEPSpec06}, the previous standard unit for core-power benchmarking used in HEP community.
A nested structure of global shares is used, where siblings have the preference to occupy unused shares, before the unused share is taken by higher levels.
For example, 75\% of the resources might be assigned to Production (Level~1 share); of those, 25\% might be assigned to Simulation (Level~2 share); and of those, 20\% might be assigned to the MC16 simulation campaign (Level~3 share). If MC16 does not fully occupy its fraction of the resources, idle resources would be preferentially reallocated first to run other simulation tasks, then to run other production tasks, and then to run any available tasks.
Within a share, jobs and tasks are assigned priorities, allowing urgent tasks to run quickly, ahead of large, low-priority tasks.
Tasks and jobs are assigned a global share at creation time, and jobs are sorted by priority and creation time within a particular global share.

\begin{figure}[tbp]
\includegraphics[width=\textwidth]{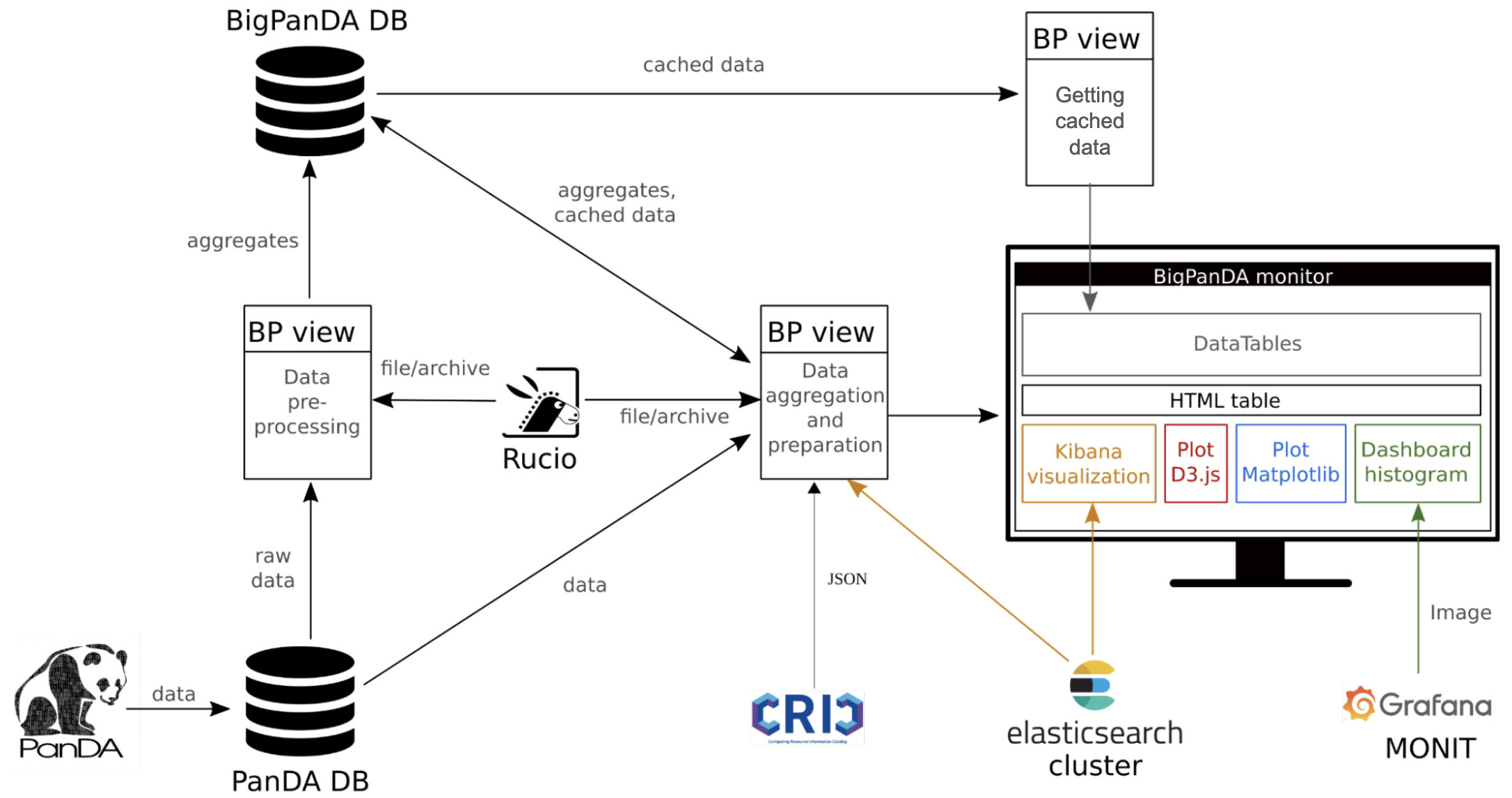}
\caption{Data flows in the BigPanDA monitoring system.}
\label{fig:bigPandaDataflows}
\centering
\end{figure}

The \emph{BigPanDA} monitoring system~\cite{Alekseev_2018} is a \textsc{Django}-based~\cite{django} web application consisting of a set of standard monitoring pages and separate modules that can be plugged in or removed on demand.
Data are collected from a variety of different sources including the \PanDA database, the CRIC information system, payload logs of jobs from \Rucio, \PanDA/JEDI logs from the \textsc{ElasticSearch} cluster, and the \textsc{Monit Grafana} instance where accounting data are available (see Section~\ref{subsec:monitoring:adcmonitoring}).
The data flow of these sources is illustrated in Figure~\ref{fig:bigPandaDataflows}.
More information about monitoring is provided in Section~\ref{sec:analytics}.
\subsubsection{Opportunistic resources}
\label{sec:prodsys:opportunistic}

One of the challenges of the ATLAS production system from the beginning has been the integration of a wide variety of computing resources whilst at the same time hiding the heterogeneity from the users.
The initial computing model was based on the Grid~\cite{CERN-LHCC-2005-024}, developed for the LHC experiments
and realised via homogeneous resources situated in the computing centres distributed worldwide.
This homogenisation was based on a list of software and hardware requirements to be met by the centres, where any remaining differences were absorbed by the associated Grid middleware.

More recently, new and often more exotic types of resources have become available to ATLAS, requiring adaptation and evolution of the production system to integrate them.
Among these resources are the Tier-0 site when not in use for operations (see Section~\ref{subsec:prodsys:tier0}), HPCs, various cloud resource providers and other \emph{opportunistic resources},\footnote{The name `opportunistic' is meant to distinguish these resources from those upon which the experiment relies to deliver its primary physics programme. The WLCG Grid sites pledge resources annually to the experiment to satisfy that physics programme. Recently, because of the scale of these opportunistic resources and their consistent availability, some amount of these resources were accounted for in the core resource needs, to more fairly request the necessary resources from the Grid sites. Another term for these resources that is in use, therefore, is \emph{unpledged}, to be distinguished from \emph{pledged} resources at WLCG Grid sites and \emph{beyond-pledge} resources that Grid sites deliver in addition to their pledged resources.} such as those running volunteer computing initiatives like `ATLAS@Home'~\cite{atlasAtHome1,atlasAtHome2}.

Such resources may not be dedicated specifically to ATLAS and therefore miss critical component infrastructure, are often constrained in their use in that they are optimal only for some workflows, and may not be permanently available, depending on the allocated hours or budget.
A measure of the success of the integration of these special resources is the production system's ability to effectively broker jobs to them, dispatching the appropriate job type, and to detect changes in resource availability in real time.
These various additional resources are described in the following.

\paragraph{Use of high performance computing resources}
\label{sec:prodsys:hpc}

ATLAS has a long history of exploiting the potential of High Performance Computing (HPC) centres to provide additional computing cores for ATLAS workloads.
Employing HPCs presents significant challenges, not least that access to such systems is usually strictly limited, so that connections to the nodes themselves are heavily restricted and installation of local software is tightly controlled.
Furthermore, the CPU architecture may not be suitable for ATLAS software, the expected job structure and number may be atypical of an ATLAS workflow, and network issues may arise due to geographic location.

Nevertheless, ATLAS has successfully used a series of HPCs for more than a decade, many of which appeared or continue to appear near the top of the Top500 list of supercomputers~\cite{top500}.
These included Cori~\cite{cori} at the National Energy Research Scientific Computing Center (NERSC), Titan~\cite{titan} at Oak Ridge National Laboratory, both in the USA, SuperMUC-NG~\cite{supermuc} at the Leibniz Supercomputing Centre in Germany, and Toubkal~\cite{toubkal} at the African Supercomputing Centre in Morocco.
Typically, these machines ran dedicated ATLAS Full Simulation tasks, delivered to the site with all required inputs and database information, allowing them to run in an isolated way via the use of edge services at each machine.
In the case of Titan, this was mainly used in back-filling mode, where ATLAS exploited spare cores not used by existing tasks to run ATLAS Fast Simulation jobs.

More recently, Perlmutter~\cite{perlmutter} at NERSC, Karolina~\cite{karolina} at the IT4Innovations National Supercomputing Center in Czechia and in particular Vega~\cite{vega} at the Institute of Information Science in Slovenia have provided significant additional computing resources to ATLAS.
Karolina and Vega are part of the EuroHPC project~\cite{eurohpc}.
The more widespread use and acceptance of native CVMFS and the adoption of containerised workflows by ATLAS has enabled a more generic use model of HPCs for ATLAS.
All production workflows now run on most HPCs, allowing such machines to be fully integrated as additional resources, essentially no different to a standard Grid site.
This has meant reduced operational load, more flexibility and periods where the number of available CPU cores available to ATLAS has more doubled relative to the WLCG pledged resources.
Some HPC centres are sufficiently well integrated that they can run as part of standard WLCG sites, like MareNostrum~\cite{marenostrum} at the Barcelona Supercomputing Center in Spain.
The large-scale use of these resources is expected to continue throughout \RunThr and beyond.

\paragraph{Opportunistic use of the HLT farm: Sim@P1}
\label{sec:prodsys:simP1}

The ATLAS Trigger and Data Acquisition (TDAQ) high-level trigger (HLT) computing farm~\cite{tdaqTDR} is a critical part of online data taking for the ATLAS experiment.
It facilitates the final selection of events to be stored for further physics analysis based on full event reconstruction~\cite{ATLAS-TRIG-2022-03}.
It consists of 145,000 computing cores spread amongst three different flavours of hypervisors with RAM between 0.9~and 1.1~GB/core.
The nodes are located within the private network of the ATLAS experiment, which is separated from the internet and are accessible to ATLAS Distributed Computing via a VLAN on a data link layer through two 100 Gbps connections.
Networking and RAM per core are the two limiting factors that define the type of workflows that can be run efficiently on this resource.

Since 2013, ATLAS has been running MC full simulation in longer periods of inactivity of the LHC within the \emph{Simulation at Point One} (Sim@P1) project on these resources~\cite{simAtP1}.
Recently it was shown that they can also successfully be used for other workflows, such as MC reconstruction, whenever necessary, albeit with lower efficiency.

Current studies~\cite{glushkovAtP1} show that thanks to improvements in the new ATLAS fast simulation, \AF{} (see Section~\ref{sim:fastsim}), and optimisation of the job submission configuration, ATLAS is able to process many simulated events within one hour.
This speed allows Sim@P1 to process simulation events between LHC \RunThr fills when the inter-fill break lasts longer than one hour.
Furthermore, with the move of the project to \textsc{Kubernetes}, it may be possible to run Sim@P1 in parallel to the trigger whenever part of the compute resources are not needed by the ATLAS online system.

\paragraph{Volunteer computing: ATLAS@Home}
\label{sec:prodsys:boinc}

ATLAS@Home~\cite{atlasAtHome1,atlasAtHome2} is a volunteer computing project based on BOINC~\cite{boinc1,boinc2} for utilising the free CPU cycles of volunteer computers around the world for MC simulation of the ATLAS experiment.
ATLAS@Home was one of the first volunteer computing initiatives in HEP.

This resource is fully integrated into the Grid computing infrastructure of the ATLAS experiment through \PanDA.
Detector simulation tasks assigned to this resource are assigned to a single \PanDA queue that covers all the volunteer resources.
The jobs from the tasks are pulled by an ARC Control Tower~\cite{arcControlTower}, and then the control tower submits the jobs to an ARC compute element, which forwards the jobs to the BOINC server.
The ARC Control Tower and Compute Element handle all interactions with \PanDA and Grid services and hence no credentials are required on the volunteer hosts.
BOINC clients request jobs from the BOINC server and process them whenever they have idle CPU cycles.

In addition to members of the public, ATLAS@Home was used in a back-filling mode at several Grid sites to make full use of CPU not normally used due to job or scheduling inefficiencies~\cite{backfillingAtlasAtHome}.
BOINC has become a very reliable computing resource for the ATLAS experiment; major simulation tasks are run, and it contributes about 1\% of the CPU available to ATLAS computing daily.

\subsubsection{The data carousel}
\label{sec:prodsys:dataCarousel}

The evolution of the computing facilities and the way storage is organised and consolidated will play a key role in how the LHC experiments will address the possible shortage of resources in the HL-LHC era.
In particular, to reduce storage costs to the experiments at a time when the data volume is expected to significantly increase, it is anticipated that the use of tape may be expanded.
To address this data handling challenge, the Data Carousel project~\cite{dataCarousel} was established to study the feasibility of directly receiving input data from tape for various ATLAS workflows.

\begin{figure}[tbp]
\includegraphics[width=\textwidth]{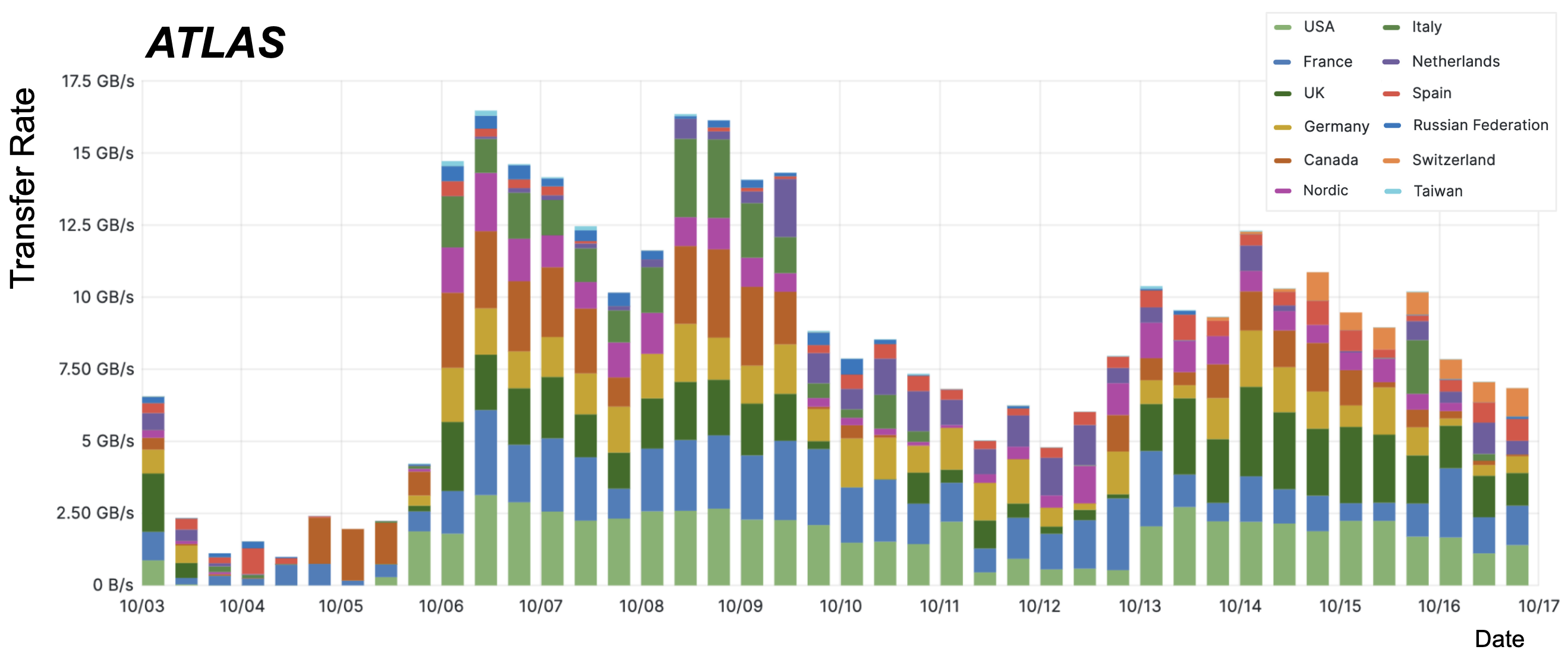}
\caption{The data-staging pattern over a two-week period corresponding to the Data Challenge (see Section~\ref{subsec:ddm:datachallenges}) across many Tier-1 sites by the Data Carousel.}
\label{fig:dataCarousel}
\centering
\end{figure}

The Data Carousel is the result of a successful orchestration between the workflow management system ProdSys2--\PanDA, the distributed data management system \Rucio, and the tape services at the Tier-1 sites.
It enables a bulk production campaign, with input data resident on tape, to be executed by staging and promptly processing a sliding window of a fraction of the input onto buffer disk such that only a percentage of the data is pinned on disk at any one time.
The production system follows site-specific preferred staging profiles, provided by the Tier-0 and each Tier-1 site, which define the upper and lower limits of concurrent staging requests that can be handled.
The typical staging pattern over a two-week period can be seen in Figure~\ref{fig:dataCarousel}, where the peak data staging performance reaches over 15~GB/s and the colours represent different Tier-1 sites.

Staging inputs from tape rather than having them resident on (more expensive) disk allows the dedication of significant disk space for more popular data such as DAODs.
Without Data Carousel, a large fraction of HITS or AODs would need to be kept permanently on disk to run regular processing campaigns.
In the case of data reprocessing campaigns, this would require complete (and large) RAW datasets to be pre-staged onto disk before they could begin.
The Data Carousel model therefore brings significant disk space savings for ATLAS, where, for example, less than half of the AODs are on disk at any one time.

To promptly process the staged data and to improve turnaround time, iDDS (see Section~\ref{sec:prodsys:panda}), was developed and integrated with the existing system.
The collaboration between the Data Carousel and iDDS R\&D projects is an excellent example of early HL-LHC R\&D delivery and commissioning for LHC \RunThr.

\subsubsection{WFMS R\&D projects}
\label{subsec:prodsys:rad}

In addition to the distributed data management research and development projects described in Section~\ref{subsec:ddm:rad}, several areas related to ATLAS workflow management are being investigated in further R\&D initiatives.

The integration of additional HPC resources is a priority for ATLAS, so that the experiment can benefit from new machines in the US, in Europe via the EuroHPC project, or anywhere they become available.
Whilst an easier integration than a decade ago is possible thanks to the evolution described in Section~\ref{sec:prodsys:hpc}, each new machine still presents a different challenge.
Discussions are continuously underway to identify possible new HPCs of interest, whether in production or in planning, to understand if they might provide significant resources to the experiment.

An evaluation of commercial cloud resources complements the HPC effort, and includes the commissioning of an ATLAS site at Google~\cite{Megino:2024htf}, together with a Total Cost of Ownership analysis~\cite{ATLAS:2024gjt}.
Both HPC and commercial cloud resources provide an opportunity to investigate and integrate non-x86 resources before such technologies are provided by WLCG pledged resources.
This includes both GPUs and ARM architectures, which are expected to be more prevalent in the future (see Sections~\ref{sec:ddm:tokensAndHepscore} and~\ref{sec:outlook}).

Related to the dynamic data handling R\&D project described in Section~\ref{subsec:ddm:rad}, the future creation, storage and lifetime of DAODs is under investigation.
The benefits and consequences of the adoption of a model where DAODs are recreated on demand will be evaluated.


\subsection{Data monitoring and analytics}
\label{sec:analytics}

The distributed computing systems produce a wealth of data that can be used to monitor the health of the many servers and services, and at the same time investigate the ways collaboration members use these services and find ways to optimise the overall resource usage.

\subsubsection{ADC monitoring}
\label{subsec:monitoring:adcmonitoring}

The monitoring infrastructure collects information from the workflow management system \PanDA~\cite{panda} and the data management system \Rucio~\cite{rucio}, complements it with static information about site configurations, aggregates it into time bins and stores it in appropriate data storage systems.

\begin{figure}[tbp]
\includegraphics[width=\textwidth]{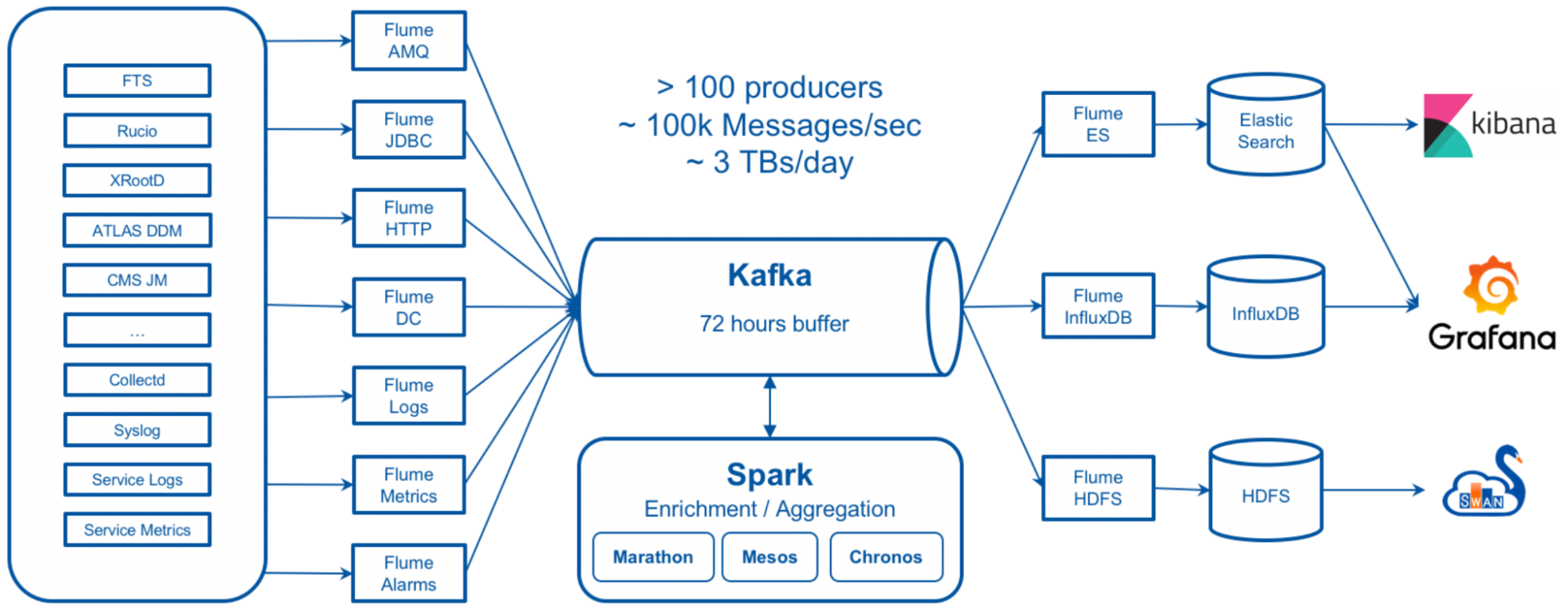}
\caption{The information flow through the monitoring infrastructure provided by the CERN IT \textsc{Monit} team. Figure from Ref.~\cite{MonitWebDoc}.}
\label{fig:monitoring1}
\centering
\end{figure}

Figure~\ref{fig:monitoring1} shows a schema of the data flow through the main monitoring system based on \textsc{Kafka}~\cite{kafka}, using the infrastructure provided by the CERN IT \textsc{Monit} team~\cite{Aimar_2017,MonitWebDoc}.
Several data sources are queried or send periodically information to the central monitoring system; here the information is processed and aggregated using \textsc{Apache Spark}~\cite{spark} jobs, and finally the records are stored in \textsc{ElasticSearch}~\cite{elasticSearch}, \textsc{InfluxDB}~\cite{influxdb} or \textsc{HDFS}~\cite{hdfs} systems (depending on their type).
These data stores can be used as sources of data for visualisation in \textsc{Grafana}~\cite{grafana} or \textsc{Kibana}~\cite{kibana} dashboards, or treated interactively using the CERN \textsc{Swan} suite~\cite{swan}.

\subsubsection{Dashboards}
\label{subsec:monitoring:dashboards}

\begin{figure}[tbp]
\includegraphics[width=\textwidth]{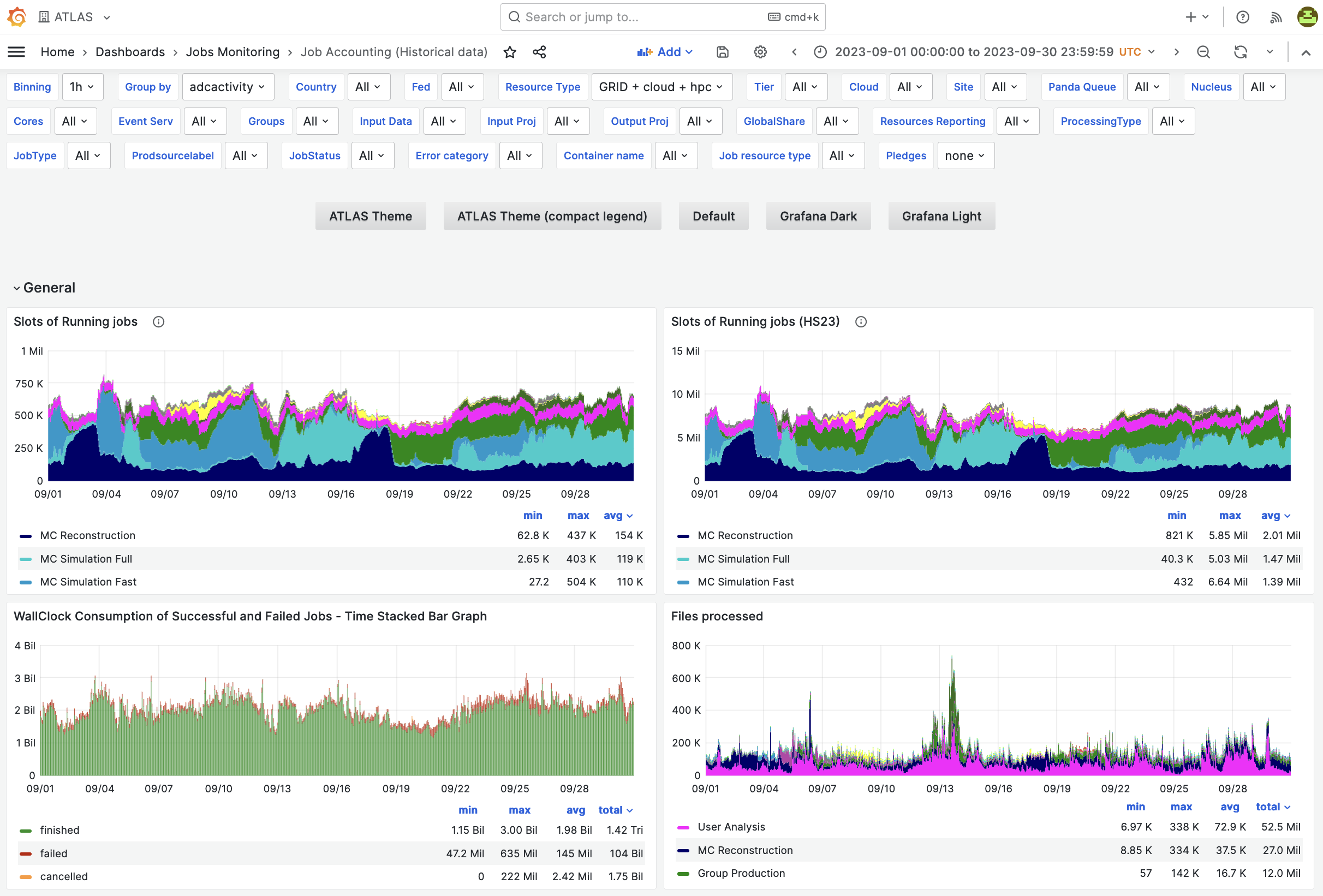}
\caption{An example of a customisable dashboard showing the number of jobs and CPU resources used during a set time period.}
\label{fig:monitoring2}
\centering
\end{figure}

The same data may be used to feed dashboards that have several purposes.
For example, information about successful and failed jobs grouped in time bins of 10 minutes and displayed over a few hours can be useful to detect faults in software or site operations, or grouped in bins of one week and displayed over several years can be used to prepare accounting reports.
Several dashboards~\cite{atlasDashboards} were developed to cover the main needs, from short-term operation monitoring to long-term accounting, for the main ATLAS systems, \PanDA and \Rucio, as well as for the operation of WLCG sites that support ATLAS and data transfers between them.
Auxiliary dashboards cover specific needs and other smaller systems.
All dashboards can be customised in real time, changing the time range, the time binning, the quantities to be displayed and the data grouping.
Figure~\ref{fig:monitoring2} shows, as an example, a screenshot of the top part of the job accounting dashboard for a period of one month, in 1-hour bins, grouped by ATLAS activity; the pull-down menus at the top of the page allow filtering the data and customising the displays, and many other charts are displayed in the lower parts of the page.

The BigPanDA monitoring suite~\cite{Alekseev_2018} is a web application that provides various processing capabilities and representations of the \PanDA system state.
Analysing hundreds of millions of computation entities, such as tasks or jobs, BigPanDA monitoring builds reports with several scales and levels of abstraction in real time.
The reports allow users to understand specific failures or observe the broad picture by tracking the computation performance or the progress of a whole production campaign.
BigPanDA is a core component of the \PanDA system, commissioned in the middle of 2014, and is now the primary source of information for ATLAS users about the state of their computations and a key source of information for shifters, site operators and production managers.

\subsubsection{ADC analytics}
\label{subsec:monitoring:adcanalytics}

An ADC Analytics coordination activity was started to better link the diverse efforts in ADC, raise awareness of the existing projects, and share experience with tools and available data.
In addition, the effort aims to improve the link between users (e.g.\ physicists doing data analysis) and developers of analytics solutions.
Current activities cover a wide range of topics, including optimising the use of \textsc{FroNTier}, anomaly detection in networks, data popularity, studies of access patterns at the file and tree level, prediction of the time to completion of tasks, development toward alerts for managers of operations, anomaly detection in data management, and the composition of dashboards that allow effective analysis of operational incidents and issues.

\subsubsection{\textsc{ElasticSearch} clusters for analytics}
\label{subsec:monitoring:es}

The \textsc{ElasticSearch} infrastructures at CERN and at the University of Chicago (UC) form the backbone of many monitoring and analytics activities in ADC.
At CERN, the cluster is a 20-core two-node \textsc{Openstack} \textsc{Kubernetes} cluster that collects WFMS data and ships it to the \textsc{ElasticSearch} infrastructure at UC.
The 29-node cluster at UC serves as the infrastructure for the `ATLAS Alarm and Alert System' and provides resources to the new ADC analytics platform~\cite{atlasML}.
This platform will offer a common front-end to different ML solutions in analytics and monitoring and serve as a registry for analytics projects.

New data sources were added to \textsc{ElasticSearch} over the years.
For example, xAOD data-access monitoring data enables the study of access patterns in great detail, with the potential to speed up data access by tuning the configuration of |TTreeCache| in real time, based on the profile of similar jobs.
To keep Squid monitoring data (see Section~\ref{subsec:databases:conditions}) with high precision for a longer time period and allow analytics tasks to combine the data with other datasets, the monitoring data is now sent via \textsc{Logstash} to \textsc{ElasticSearch}.
To discover unused files in disk storage, access information was exported from \textsc{dCache}~\cite{dCache} metadata databases to the \textsc{ElasticSearch} infrastructure. Two sites are currently part of the first implementation of this project.

%
%
%
%
%
%
%
%
%
%
%
%


%
\section{Analysis tools, event displays, and tutorials}
\label{sec:downstream}

Performing physics analysis is the final step of the data and MC processing chain, with the aim to publish groundbreaking results using stringent scientific methods.
A variety of tools are available that assist physicists with their analyses.
Among these, event displays are special tools for visualizing recorded or simulated events and the detector for analysis purposes, and they can also be useful for visual inspection and troubleshooting.
This section describes these tools and also the efforts to train physicists to use them.

\subsection{Analysis tools}
\label{sec:analysis}

Analysis tools enable the data processing steps from centrally provided derivation formats for analysis to final results for publication and public dissemination. Many analyses are carried out in ATLAS, with several hundred ongoing at any one time, which cover a broad range of topics and focus on a wide range of signatures. As a result, analysis tools and workflows are necessarily diverse, flexible, and focused on analysis-specific needs. This section covers commonly used tools and methods and highlights examples of how some analysis-specific needs are accommodated.

Analysis workflows in ATLAS often entail the following steps, described in more detail below:
\begin{enumerate}
\item Processing from central derivations, which may include applying corrections to analysis objects (calibrations), steps towards data reduction, and systematic variations.
\item Analysis design and development, which may include the creation of additional intermediate data formats and processing steps.
\item Statistical analysis to quantify the correspondence between experimental observations and theoretical predictions.
\item Creation of data products (e.g.\ figures, tables, likelihood functions) for publications and other publicly available material, analysis preservation, and reinterpretation.
\end{enumerate}

Carrying out an analysis typically relies on a combination of centrally provided code, community supported tools, and analysis-specific software. In addition, several computing and storage resources are available for processing of analysis data, ranging from the Grid to local machines and personal computers.

\subsubsection{Analysis data formats and workflows}

The starting point for an analysis workflow is the central derivation formats for analysis, described in Section~\ref{sec:derivations}. The model for \RunThr is characterised by the introduction of a new common central derivation format for analysis, \verb|DAOD_PHYS|, whose goal is to support most analyses in \RunThr~\cite{Elmsheuser:2708664}. This format contains information for all physics objects (muons, electrons, photons, hadronically-decaying $\tau$-leptons, jets including identification and flavour tagging, and missing transverse momentum), which makes it possible to do calibrations and study systematic variations. This information allows significant flexibility in object definitions, such that analysts can optimise the choice of objects for their specific analysis needs. The format also has content related to tracking and vertexing, the trigger, and information from the event generator record for simulated samples.

As was already done during \RunTwo, additional derivation formats are produced for a few specific needs:

\textit{Combined Performance (CP) studies:} Specialized formats are used to complete detailed studies of object performance and to design new reconstruction and identification algorithms. These formats are developed by the CP groups whose task is to deliver the recommendations for the different analysis objects. These specialized derivation formats typically contain detailed information relevant to the object(s) in question and are produced for a few key samples. The recommendations are then propagated to the central recommendations and included in \verb|DAOD_PHYS| or other formats for analysis.

\textit{Analyses using non-standard objects or methods:} Analyses with specialized processing needs, such as searches for long-lived particles, often use dedicated derivation formats that include the relevant information required for the analysis. These derivation formats typically are subjected to heavy skimming, reducing the fraction of events to the level of a few percent to reduce the storage needs. As these methods advance, the goal is to integrate these skimmed events into the central format to minimize both the number of formats and the number of analyses relying on special formats (see Section~\ref{sec:deriv:augmentation}). Having this two-stage setup allows for flexibility in pursuing new developments for analysis.

\textit{MC event generator information:} Derivation formats containing more detailed information about the MC event generation process, called \verb|TRUTH| formats, are used for specialized tasks including the validation of event generator configurations and MC simulation samples, detailed classification based on object origin, and the calculation of the acceptance for a selection. Standard analysis formats like \verb|DAOD_PHYS| include significant truth information as well, including links between reconstructed objects and the truth particles to which they most closely correspond, as well as classification of the truth particles in terms of their origin (e.g.\ lepton from a hadron decay or from a $Z$ boson; truth particle jet containing a $B$-hadron or not).

All derivation formats are typically further skimmed and slimmed by analysers using a set of analysis tools that comprise an analysis framework. In the case of formats for analysis workflows, objects are generally calibrated, after which common object selections are applied using a set of tools provided by the CP groups, referred to as `CP Tools'. Derivation formats also retain additional in-file metadata (see also Section~\ref{sec:metadata}), describing, for example, the number of events before skimming is applied, to allow the correct normalisation of an MC simulation sample and to check for dataset completeness.

Another new development for \RunThr is the introduction of a smaller derivation format, \verb|DAOD_PHYSLITE| (see Section~\ref{sec:derivationintro}), which contains calibrated physics objects, obtained after applying the CP Tools. It is intended to support most analyses in the future (i.e.\ for \RunFour and beyond). This format is already being deployed for development and early adoption in \RunThr, and was used for published data analyses~\cite{STDM-2022-17}. The \verb|DAOD_PHYSLITE| format is intended to be produced from \verb|DAOD_PHYS| and can be centrally produced with frequent updates, typically every few weeks or months as needed.

\subsubsection{Object calibrations and systematic uncertainties}

ATLAS provides software releases that are dedicated for analysis and include all the relevant tools, including the tools from CP groups. These are the AnalysisBase releases and AthAnalysis. Conceptually, AthAnalysis was designed to provide an algorithm scheduling framework as close to that of Athena as possible, while reducing the amount of code that must be distributed. AnalysisBase was designed to be a light project external to Athena, containing only the bare bones necessary to run an analysis (e.g. the CP Tools and Algorithms themselves). Significant effort was invested to ensure that Tools are \emph{dual-use}, meaning that they can be run either within Athena and AthAnalysis or within AnalysisBase releases; this avoids risks of code duplication between the two projects. The AnalysisBase software stack is somewhat simpler (e.g. it does not rely on LCG releases and instead explicitly tracks all its external dependencies), but several components must therefore be re-implemented within it (e.g. Algorithms, Tools, and messaging base-classes are all re-written in AnalysisBase compared with those in AthAnalysis). One of the key advantages of all these software releases is the wide availability of the software, which is available on all Grid computing sites and user machines via \textsc{CVMFS}~\cite{DeSalvo:1448195}. The inputs required for CP Tools, such as conditions information for object calibration, are also stored on \textsc{CVMFS}. Software images of AthAnalysis and AnalysisBase releases are additionally available to download to local machines on a GitLab registry~\cite{athenaGitlab}.

Analysis frameworks are used to run on derivations and can read xAOD objects and apply CP tools that are used to calibrate, correct simulation to match the expected performance in data, and select the physics objects used for most physics analyses. These tools also provide estimates of the systematic uncertainties in the performance of the objects (such as the efficiency and resolution). There are a few centrally available frameworks, including some that are ATLAS-specific. Most often, analyses running using AnalysisBase rely on the |EventLoop| framework, a framework widely used in ATLAS that processes single events at a time, helps with parallel processing of events by providing a job configuration that is sent out to worker nodes either in a batch or Grid system, and handles the merging of outputs after processing. Analyses running in AthAnalysis most often directly use the Athena framework and corresponding base-classes. Several physics groups have developed specific frameworks, often based on |EventLoop|. Analysis frameworks are typically based on \textsc{ROOT}~\cite{Antcheva:2009zz}, which is used for I/O and provides many other capabilities.

To improve the sharing of code and harmonize user analyses, common `CP Algorithms' are increasingly being used in analysis frameworks in ATLAS. These CP Algorithms provide a wrapper around the CP Tools that configures and schedules them in such a way that analysts do not need to write additional code to use them, but rather can incorporate them into the rest of the analysis code. The common CP Algorithms are used to produce the \verb|DAOD_PHYSLITE| format and are also a common layer in several analysis frameworks.

In addition, several analysis frameworks are designed to account for the computational infrastructure available and aid parallel processing, for example relying on |TSelector| in the \textsc{PROOF} environment~\cite{Iwaszkiewicz:1116551}. There is also an increasing use of \textsc{Python} for analysis, particularly for the user interface such as with \textsc{PyROOT}~\cite{Galli:2752850} and \textsc{Dask}~\cite{rocklin2015dask} for parallel processing. There have also been several recent developments in functional and distributed approaches to data analysis in \textsc{ROOT}~\cite{Amadio:2664841} and an increasing interest in using tools and methods from data science (see for example Refs.~\cite{CMS:2020kpn,Adamec_2021}).

\subsubsection{Event selection and data reduction}
\label{sec:eventSelection}
Dedicated, common analysis tools are used to select the data for analysis, account for trigger selections and modelling, model the pile-up and so on.

\textit{Data quality checks:} Soon after data-taking, each luminosity block (see Section~\ref{sec:atlasOper}) of data is certified for physics analysis. The quality of the data in each luminosity block is encoded with data quality flags that are set in the data quality monitoring and stored in a database. Depending on the specific needs of each analysis, the data in luminosity blocks with undesired or bad quality need to be filtered out. A good runs list (GRL) is an XML file that lists the ranges of good luminosity blocks in each run. A tool is provided for analyses to check if events are contained in a luminosity block within a desired range. Data quality processes at ATLAS are detailed in Ref.~\cite{DAPR-2018-01} and in Section~\ref{sec:DQ}.

\textit{Event cleaning:} A few events are also not recommended for analysis due to specific detector, readout, or software issues, as described in Ref.~\cite{DAPR-2018-01}. These problematic events, which may be corrupted or incomplete, are flagged in the analysis data formats for removal. In addition, common tools were implemented for the removal of non-collision backgrounds (e.g. beam-halo passing through the detector)~\cite{DAPR-2012-01}.

\textit{Luminosity:} Another ingredient needed for an analysis is the integrated luminosity. A tool is available to calculate the luminosity corresponding to a specific analysis depending on two inputs: the luminosity blocks processed from the GRL, and the trigger(s) to be applied in the analysis. The trigger is then used to calculate the fraction of the luminosity that ATLAS recorded, which is called the trigger \emph{live-time}, and the \emph{prescale} of the trigger used to select events. Prescaled triggers are those that do not accept all events (e.g.\ a trigger with a prescale of 50 only accepts one in 50 events) and the value of the prescale often changes with changing luminosity conditions. Prescales can only change along luminosity block boundaries, and the prescale of each trigger in each luminosity block is available in a database. Most analyses rely on triggers without a prescale applied (unprescaled triggers), but the luminosity tool checks and calculates the luminosity by adding all the selected luminosity blocks and scaling the luminosity each according to the prescale of the trigger of interest~\cite{DAPR-2021-01}.

\textit{Pile-up modelling:} Simulated samples are produced with an estimate for the pile-up in the corresponding data. It is difficult to predict the exact conditions of the LHC, including how much luminosity is delivered for each pile-up value; some trigger selections induce biases in the pile-up distribution (i.e. there is not a single pile-up distribution for all analyses); and systematic uncertainties in the luminosity calibration and other sources lead to uncertain modelling of pile-up that require the evaluation of systematic uncertainties in the distribution. Therefore, a reweighting procedure is used to match the distribution of pile-up in MC simulation to that of data. A tool is used to derive event weights that correct differences between MC simulation and data for the distribution of instantaneous luminosity and trigger prescale conditions. The modelling of pile-up affects several key variables such as the reconstruction efficiency and isolation of objects.

\textit{Trigger modelling:} The data for analysis is collected with a suite of triggers. The trigger information in a sample is packed to save space; therefore, tools are provided to identify events that satisfied trigger selections based on the human-readable trigger names, and to describe the conditions under which the event satisfied the trigger selection. To match the MC simulation and the data, each analysis requires that both satisfy the requirements of the trigger logic for one or multiple triggers. While the MC simulation tries to describe the expected behaviour of the triggers, there might be differences between the logic implemented in the MC simulation and the one used for the actual data-taking. Sometimes, a trigger used online might not be available in the MC simulation, and a proxy must be used. Corrections to the efficiency for the trigger selection are derived as event weights that may depend on several inputs such as the kinematics of the reconstructed objects (e.g. where in the detector the object falls; see for example Ref.~\cite{TRIG-2018-01}). A tool is provided to determine the appropriate scaling factors that an analyst should apply to the MC simulation to reproduce the efficiency in data. The choice of trigger is typically associated with several event selection requirements for an analysis. Analysers might wish to associate a reconstructed object to one that resulted in the trigger decision (often called \emph{trigger matching}, see for example Ref.~\cite{TRIG-2018-01}), and this information is also available.

\textit{Overlap removal:} Several algorithms are used to identify objects of a particular type, and a single object in the detector can be identified as multiple types of objects (e.g.,\ hadronic taus or electrons may also be identified as jets). To carry out an analysis, a choice must be made between the types of objects to interpret the content of the event. This choice is not universal: some analyses may wish to favour leptons over jets; others may wish to favour jets over leptons, for example. The removal of the duplication of an object is called \emph{overlap removal}. Several schemes for overlap removal are supported centrally in ATLAS (see for example Ref.~\cite{EGAM-2019-01}) and the implementation for analysis is available in a tool.

\subsubsection{Simulation modelling and uncertainties}
\label{sec:uncertaintyTools}

Most analyses rely on MC simulation, which is often used to verify the understanding of the detector performance, to model the process of interest (signal) and several background processes, to validate data-driven methods, and to quantify systematic uncertainties by changing the settings used in the simulations or using alternative methods. As described previously, calibration and uncertainty recommendations are implemented in common tools in AnalysisBase and AthAnalysis software releases that are used by almost all analyses to improve the modelling of objects and event properties (see for example Ref.~\cite{METSoftware}). The use of MC simulation for analysis also heavily relies on metadata (see also Section~\ref{sec:metadata}), which includes information such as a unique integer MC \emph{dataset identifier} that can be used to identify a sample (i.e. a specific event generator configuration) and the number of events processed in a file, which is needed to normalise the samples to the expected cross section and luminosity. Theoretical uncertainties from the knowledge of the parton distribution functions, scale variations, alternative generators, parton showers, tunes, etc., are available either through alternative samples or stored as event weights in the samples. Insofar as is possible, recommendations are made for specific sample combinations, most precise available cross-section calculations for normalisation, and weights to apply for the best possible estimate of a Standard Model process and its uncertainties. However, each event generator suffers from some (often unique) set of modelling inaccuracies, and some analyses are uniquely sensitive to the modelling of particular observables; therefore, sometimes alternative prescriptions must be identified.

In addition, many analyses rely on multivariate classifiers, such as boosted decision trees or neural networks, to complete tasks such as discriminating between signal and background. These classifiers or methods typically rely on MC simulation for the training and performance validation (see also Section~\ref{sec:ml}).

\subsubsection{End-stage analysis and statistical interpretation}

Several tools are used for end-stage analysis to do additional data reduction steps, derive selections for analysis regions; compute derived variables, including multivariate discriminants; produce histograms; compare data and expected backgrounds; and prepare inputs for statistical tools. In addition to the many capabilities available in \textsc{ROOT}~\cite{ROOT},
there are several projects within \textsc{scikit-hep}~\cite{Rodrigues:2020syo} that provide similar functionality: file-handling tools like \textsc{uproot}(+\textsc{coffea})~\cite{Pivarski_Uproot_2017}, histogramming tools like \textsc{boost-histogram}~\cite{boostHistogram} and \textsc{hist}~\cite{Schreiner_hist}, data manipulation tools like \textsc{awkward array}~\cite{Pivarski_Awkward_Array_2018}, and tools for statistical analysis like \textsc{iminuit}~\cite{iminuit}.

Several tools are available to implement the most common statistical tests used at the LHC experiments. Experimental results are formulated in a statistical language, so a measurement is a parameter estimate, a discovery is a hypothesis test, and for a physics model parameterised by theoretical parameters (such as hypothetical masses and couplings for new particles), excluded and allowed regions are defined as confidence intervals. For most analyses, once the goal and statistical model are defined, the statistical procedures are encoded in the \textsc{RooStats} project~\cite{moneta2011roostats}, which is based on the \textsc{RooFit} modelling language~\cite{RooFit}. Several of the statistical analysis frameworks that are widely used in ATLAS are based on \textsc{HistFactory}, a tool to build parameterised probability density functions in the \textsc{RooFit}/\textsc{RooStats} framework from \textsc{ROOT} histograms~\cite{Cranmer:1456844}, such as \textsc{HistFitter}~\cite{Baak:2014wma} and \textsc{TRexFitter}~\cite{TRexFitter}. These frameworks provide a user-friendly interface and help with common tasks, such as profiling of nuisance parameters used to encode uncertainties, carry out validation checks, and provide an interpretation of the results. A more modern implementation of \textsc{HistFactory} in \textsc{Python} using tensors and automatic differentiation is available in \textsc{pyHF}~\cite{Feickert:2022Rb}.

\subsubsection{Preparation of results, preservation and reinterpretations}

The final key steps in an analysis are to prepare the results for publication and sharing with the broader scientific community. In addition to providing the figures and tables in public pages, many ATLAS results are available in HEPData~\cite{Maguire:2017ypu}, which is an open-source, publicly available repository for high-energy physics results. In addition to providing the results in figures and tables in digitised format, some ATLAS analyses have recently also provided the likelihood function in HEPdata~\cite{ATL-PHYS-PUB-2019-029}. Several other tools have recently become available to help with the reuse and preservation of analyses. Examples that have widespread use in ATLAS, particularly for searches, include \textsc{Simple Analysis}~\cite{ATL-PHYS-PUB-2022-017}, \textsc{RECAST}~\cite{Cranmer:2010hk}, and \textsc{Reana}~\cite{Simko:2018zzz}. Another example, extensively used for measurements, is the \textsc{Rivet} framework~\cite{Bierlich:2019rhm}, which is used by theorists and experimentalists for a range of studies including understanding and improving the modelling of event generators.

\subsubsection{Analysis infrastructure}
\label{sec:analysis:infrastructure}

The specific resources used for analysis depend on the size of the derivation used, the specific workflow, and sometimes the analysis needs for example for a final event selection. Some analyses that are very selective can run on local computing clusters or even laptops. However, the most common analysis workflows involve running on derivations on the Grid or sometimes on a local computing cluster. The final selections, analysis optimisations, and the statistical interpretation are often done on a local cluster or sometimes on a personal computer. The final stages of analyses, such as preparing final figures, are often done on local resources such as a personal computer or interactive linux system like lxplus available at CERN. It is relatively straightforward to run the ATLAS software using \textsc{CVMFS}. There are several options for data access from several sites (see Section~\ref{subsec:ddm:datapolicies}), typically with standard \textsc{ROOT} I/O~\cite{ROOT}, with or without \textsc{XCache}~\cite{xcache}, or the \Rucio redirector~\cite{rucio}, both over the LAN and the WAN.

More recently, there is a renewed interest in developing dedicated infrastructure for analysis (often referred to as \emph{analysis facilities}), particularly when considering future needs at the HL-LHC. There is also an interest in exploring the potential of technologies such as the cloud and partnerships with industry, for example with Google and Amazon, that enable rapid and large scaling-up and specialised resources for analysis. Already, existing national analysis facilities are being expanded with additional resources and capabilities (e.g.\ \textsc{Jupyter}~\cite{Jupyter} support and GPU resources) to support modern data analysis techniques and tools. Local computing clusters at individual institutes already represent significant computing resources; according to a recent estimate, together they are comparable to or larger than the largest Grid sites.\footnote{These resources are not included among those discussed in Section~\ref{sec:distcomp}.} The availability and harmonisation of these resources is important not just for the optimisation of computing resources towards the HL-LHC, but also for the equity of the collaboration and its institutes.


\subsection{Event displays}
\label{sec:eventdisplay}

Interactive data visualisation is a key component in HEP experiments, where it is used at each step of the data pipeline (simulation, reconstruction, analysis, and so forth) to inspect and explore different types of data interactively: event data (such as tracks, hits, or energy deposits), detector geometry (such as passive and sensitive volumes), magnetic field (e.g.\ direction and magnitude) or conditions data. These visualization tools are often used offline (for example in physics analysis), and can also be used online for the detector operation. They can be utilised to prepare high--quality images to show the experimental signature of particular physics processes (used for example for outreach purposes) or to inspect the detector's response under given experimental conditions, such as beam splashes~\cite{vp1-beamsplash}.
These images are collectively called \textit{Event Displays}, and all public ATLAS event displays can be found at the ATLAS Event Display Public Results webpage~\cite{publicED}.

The ATLAS Collaboration has developed several tools to visualise events~\cite{Bianchi_EventDisplays_2017}, each of them addressing different needs and targeting different use cases and end users.
For the \RunThr data-taking period, three visualisation tools were updated and are actively developed and maintained: \textsc{VP1} (Virtual Point 1)~\cite{Kittelmann_2010}, \textsc{Atlantis}~\cite{atlantis,Konstantinidis:865603} and \textsc{PhoenixATLAS}~\cite{phoenixATLAShome, phoenixATLASgit}, as well as the \textsc{JiveXML} data exporter tool and the Online Event Displays machinery, all of which are briefly described below.
A further tool was developed for \RunThr, the GeoModel Explorer or \textsc{gmex}, which is derived from \textsc{VP1} and specialises in the visualisation of the detector geometry, as described in Section~\ref{sec:dd}.

\subsubsection{Software design principles}
\label{sec:eventdisplaysSoftware}
When developing visualisation tools, there are two main paradigms that applications can follow: full integration of the visualisation tools into the main experiment's software framework or standalone operation outside of the framework.
The tools that follow the former paradigm can access all experimental data in a native way, directly from the experiment's software framework, but are limited by the technology and platform boundaries imposed by the framework itself.
The applications following the standalone paradigm are free to choose any technology or tool that suit best their need, but have to design intermediate data exchange formats and develop and maintain tools to export data from the experiment's software framework.

\subsubsection{\textsc{VP1}}
\label{sec:vp1}
\textsc{VP1}~\cite{Kittelmann_2010} is the visualisation tool integrated into the ATLAS experiment's framework, Athena.
As such, it can directly access experiment data, without the need for intermediate data formats, and can re--use the same tools as are used in reconstruction and simulation workflows.
\textsc{VP1} provides C++ based interactive 3D graphics, and can be extended with custom plugins to visualise ATLAS data.
As it also reads the detector description information from Athena directly, it shows precisely the same detector geometry that is used in the simulation and reconstruction, and is able to show any geometry configuration of the detector.
Furthermore, as \textsc{VP1} is a module that can be run together with all other Athena applications, it can make use of all other |AlgTools| in Athena (see Section~\ref{sec:transforms}), and can also be used to visualise transient data, such as inner detector tracks or jets, while they are reconstructed by the dedicated algorithms.

\begin{figure}[h]
\centering
\includegraphics[width=\textwidth]{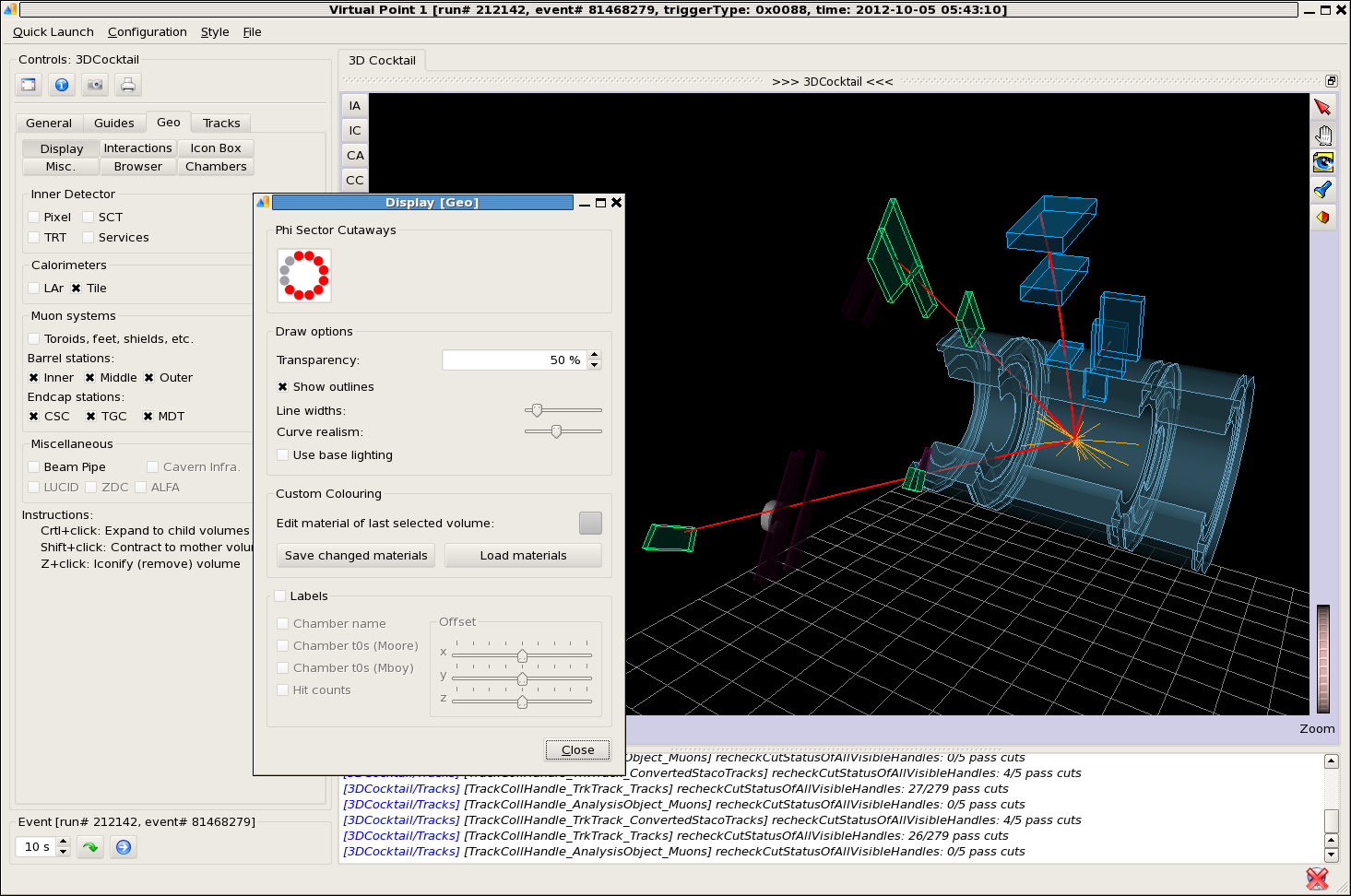}
\caption{The \textsc{VP1} event display. The image shows the main window and one of the many widgets that let users change settings, set cuts, and customize the visualisation. In this example, the main window shows some of the physics objects from a candidate Higgs boson production event in real data: the tracks reconstructed in the inner detector (orange lines), as well as the four muons (red) and their associated muon chambers (blue and green boxes). The widget shows the settings to customize the visualisation of the detector geometry.}
\label{fig:vp1}
\end{figure}

\textsc{VP1} is developed on top of the \textsc{Coin}~\cite{coin} 3D graphics C++ framework and uses the \textsc{Qt}~\cite{qt5} framework as the graphical user interface (GUI) layer, with the \textsc{SoQt}~\cite{soqt} library as the glue package between the 3D and GUI layers.
One of the main tasks in preparation for \RunThr was the integration of \textsc{VP1} in the ATLAS Control Room and the \textit{Online Event Displays} machinery, as described in Section~\ref{sec:eventDisplaysOnline}.
Figure~\ref{fig:vp1} shows the \textsc{VP1} user interface.

\subsubsection{\textsc{Atlantis}}
\label{sec:atlantis}

\begin{figure}[tb]
\centering
\includegraphics[width=\textwidth]{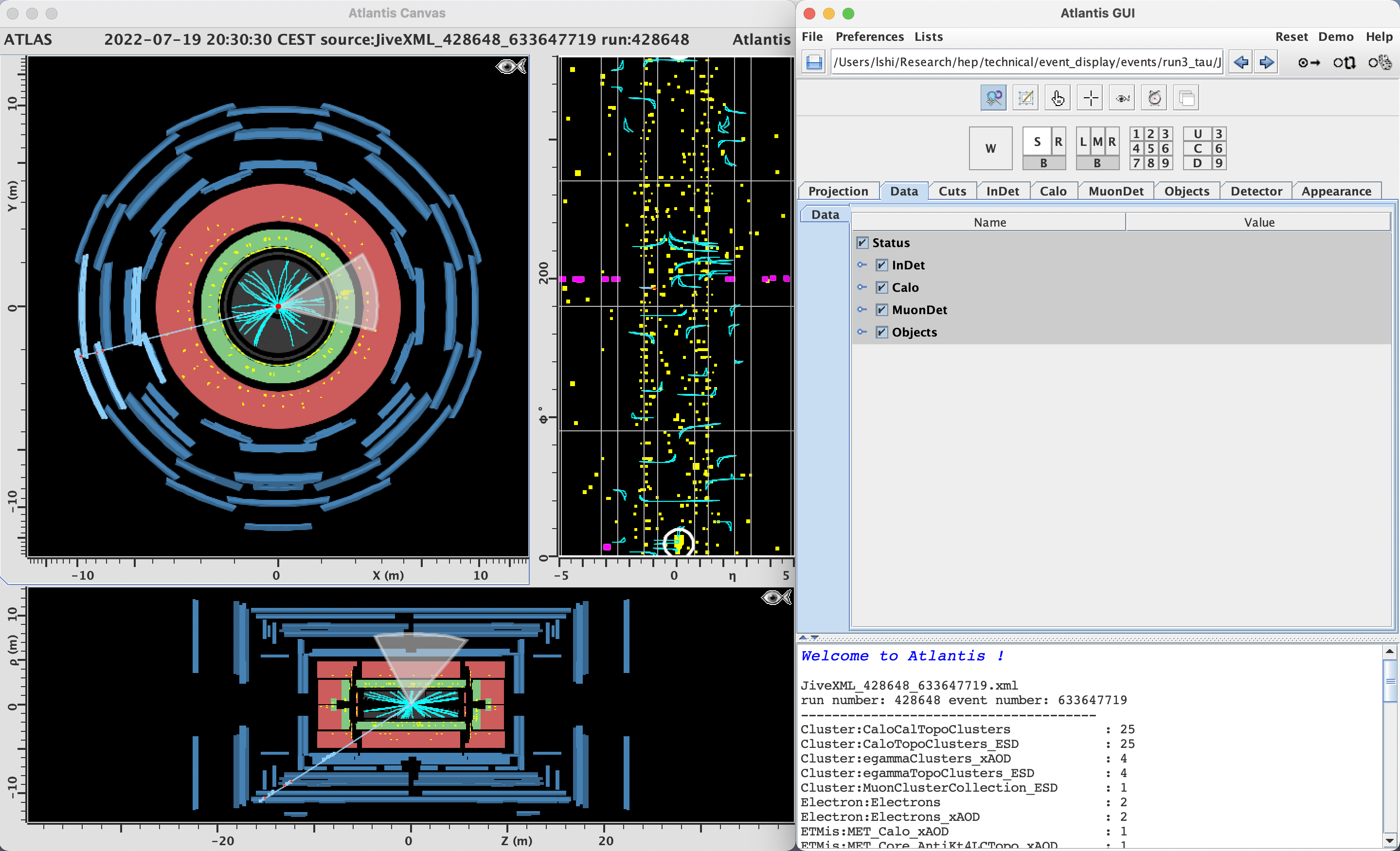}
\caption{The \textsc{Atlantis} event display. The main \textsc{Atlantis} canvas on the left allows multiple projections to be displayed simultaneously. In this example, a $Z\to\tau\tau$ candidate event in real data is displayed in the $y$--$x$, $\phi$--$\eta$ and $\rho$--$z$ projections, with one $\tau$-lepton decaying leptonically into a muon (the light blue track) and the other $\tau$-lepton decaying hadronically, shown as a jet (the white cone). The \textsc{Atlantis} GUI on the right provides a tabbed panel of menus for display manipulation.}
\label{fig:atlantis}
\end{figure}

\textsc{Atlantis}~\cite{atlantis,Konstantinidis:865603} is a stand-alone JAVA event display for the ATLAS experiment.
It provides a collection of specialised, data--oriented projections in two or three dimensions to visualise physics processes and monitor the performance of all ATLAS sub-detectors.
These projections and modifications to the projections (e.g.\ fish-eye or regional zoom) allow users to explore the data in a way that a strictly physical representation of the detector might not allow.
For example, a fish-eye view allows hits in both the tracker and the calorimeter to be simultaneously visible.
Figure~\ref{fig:atlantis} shows the user interface of the \textsc{Atlantis} event display.
\textsc{Atlantis} runs independent of Athena, making it easier to install on different platforms.
It uses a simplified geometry of the ATLAS detector in XML~\cite{xml} format, and the event data are also read from XML files.
Both the geometry XML and event data XML are produced by a dedicated algorithm, \textsc{JiveXML}, which is detailed in Section~\ref{sec:jivexml}.
Customized versions of \textsc{Atlantis} are used in the MINERVA~\cite{minerva} and HYPATIA~\cite{hypathia} projects as educational tools in master-classes for high school students.

\subsubsection{\textsc{PhoenixATLAS}}
\label{sec:phoenixAtlas}
\begin{figure}[tb]
\centering
\includegraphics[width=\textwidth]{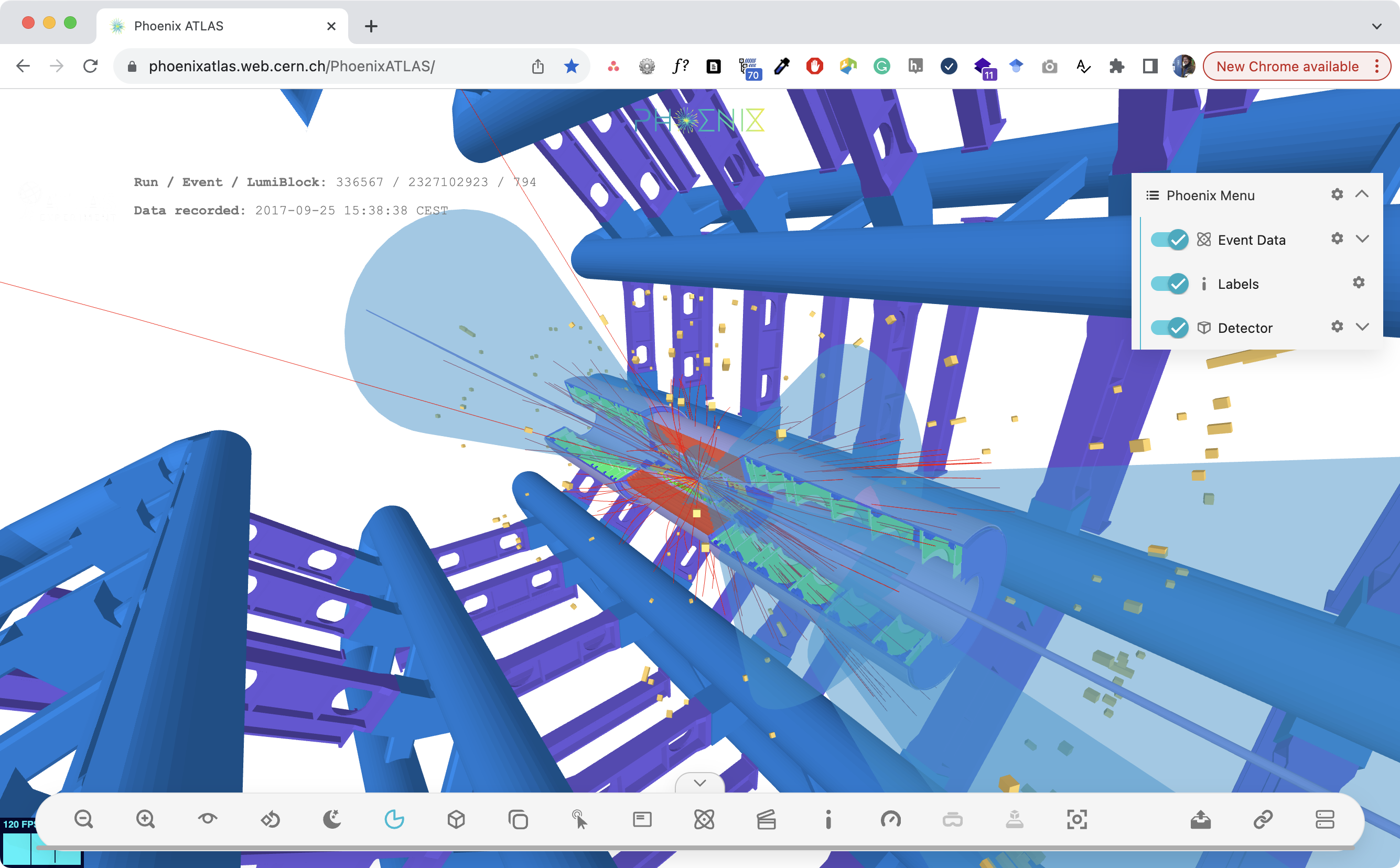}
\caption{The \textsc{PhoenixATLAS} event display. The image shows the main user interface, with the buttons and the drop--down menus that let users change the settings and customize the visualisation. In the image, several physics objects (such as jets, tracks, and hits) are shown on top of a view of the barrel toroid magnet (blue/purple bars).}
\label{fig:phoenix}
\end{figure}

\textsc{PhoenixATLAS} is the ATLAS web-based event display intended to easily visualise ATLAS events using a web browser.
It is a \textsc{TypeScript}~\cite{typescript-lang} application and uses the \textsc{Phoenix}~\cite{phoenixHSFgit} event display library to read and process the events, with a dedicated \textsc{JiveXML} converter providing access to the ATLAS XML data format.
An Athena Algorithm to dump ATLAS event data to \textsc{Phoenix}'s native JSON data format is also available.
\textsc{PhoenixATLAS} can show all the main reconstructed analysis objects and provides an intuitive interface to add cuts, slice away geometry and change the colours and various properties of what is displayed.
Additionally, there is support for virtual reality and augmented reality on appropriate display technology.
Figure~\ref{fig:phoenix} shows a screenshot of \textsc{PhoenixATLAS}.

Preparation for \RunThr has involved several tasks: adding the ability to show different geometries for different runs, validating \textsc{PhoenixATLAS} output by ensuring it matches that of \textsc{VP1} and \textsc{Atlantis}, improved cut (requirements like minimum $\pt$ or $\eta$ restrictions) functionality and support for more analysis object types.
Another recent and significant project was preparing \textsc{PhoenixATLAS} to show live events and adding links to these on the ATLAS live page~\cite{atlaslive}.

\subsubsection{\textsc{JiveXML}}
\label{sec:jivexml}
\textsc{JiveXML} is a C++ event converter interface between the ATLAS event displays and the ATLAS reconstruction data.
It consists of a series of retriever algorithms, running in Athena and converting fully reconstructed events to XML format.
The event data XML files can be viewed in both \textsc{Atlantis} and \textsc{PhoenixATLAS}.
\textsc{JiveXML} can extract the detector geometry to produce a geometry XML file that is then used by \textsc{Atlantis} to display a simplified ATLAS detector.
\textsc{JiveXML} can also act as an XML-RPC server to send files directly to \textsc{Atlantis}.

\subsubsection{Online event displays}
\label{sec:eventDisplaysOnline}
\begin{figure}[tb]
\centering
\includegraphics[width=\textwidth]{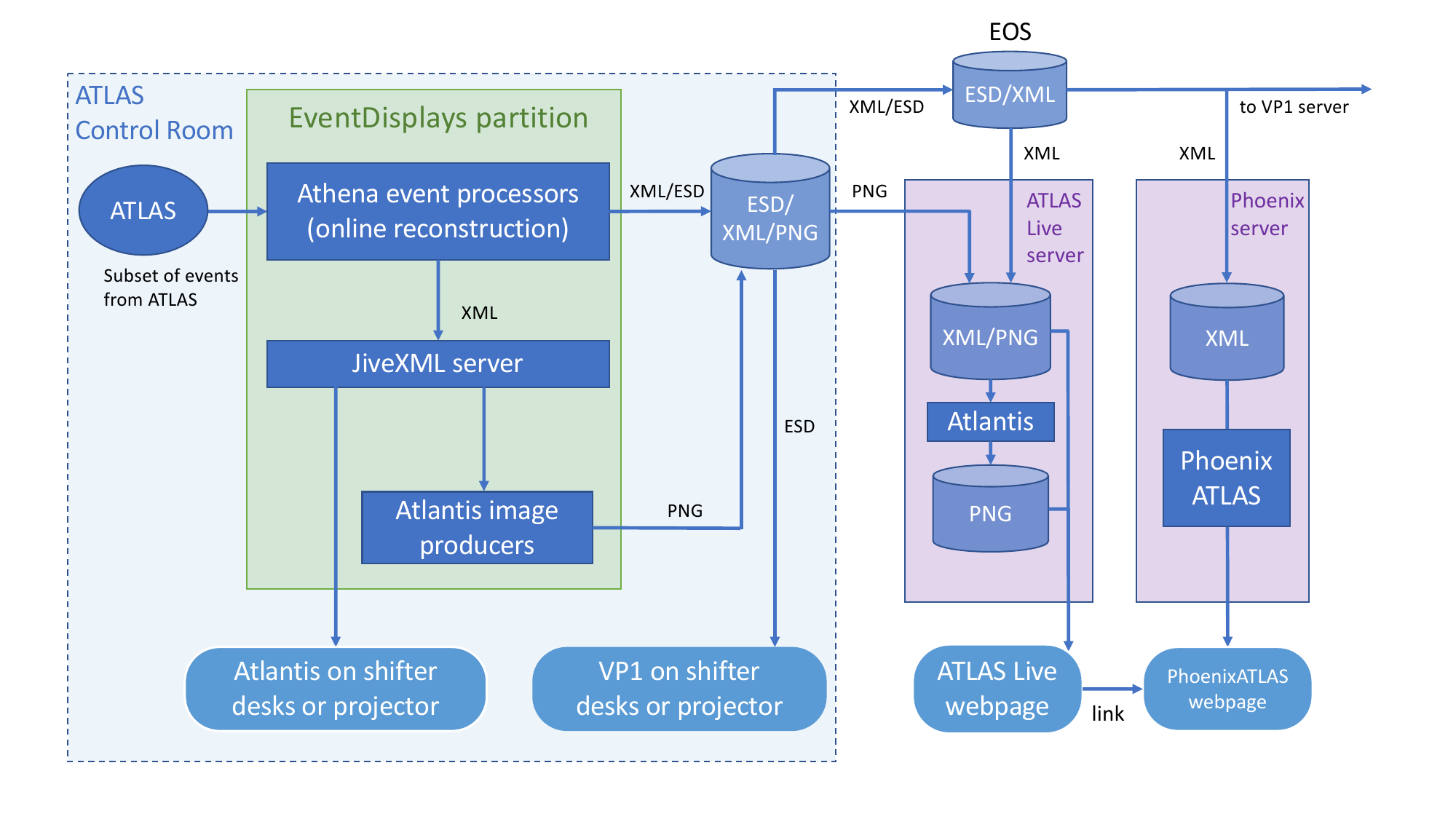}
\caption{The online event display workflow.}
\label{fig:onlineED}
\end{figure}

Event displays are used during ATLAS data taking to provide visual feedback to the operations team on the detector's performance.
Figure~\ref{fig:onlineED} shows a chart of the online event display workflow.
The core of the online event displays is hosted in an \emph{Event Displays partition}~\cite{ATLAS-TDR-16}, which controls a set of applications as a group that can function independently during data taking.
Several Athena event-processor applications run the online reconstruction on a subset of events recorded by ATLAS in real time, producing XML and ESD files and writing them to disk.
A \textsc{JiveXML} server application runs in the partition to serve the reconstructed events in XML format to several instances of \textsc{Atlantis}, including two instances of \textsc{Atlantis} running as image producer applications in the partition that render a subset of events to share with the CERN Control Centre and the general public, and several instances of \textsc{Atlantis} running outside the partition in the control room: the one running on the Data Quality shifter's desk allowing the shift crew to assess the data online, and one projected on the control room's wall for quick feedback. These event displays have proven useful for quickly identifying issues like dysfunctional regions of the detector, even when low-level monitoring does not indicate a serious problem.
The ESD files on disk can be read by instances of \textsc{VP1} to produce interactive 3D event displays on the shifters' desks and the control room's wall.

The XML and ESD files on disk are also transferred to EOS, where different event displays can access the event data and render them on offline servers.
The ATLAS Live web page~\cite{atlaslive} displays the images rendered by \textsc{Atlantis} with the corresponding XML files for downloading, as well as providing links to the same events on the \textsc{PhoenixATLAS} web page, to allow a wider range of interested physicists to study these events.

\subsection{Tutorials and education}
\label{sec:training}

\subsubsection{Historical software tutorials}

The ATLAS Collaboration has offered specialized training for various aspects of the available software and workflows since 2004. Beginning in 2008, ATLAS began holding regular centralized tutorial events that comprehensively cover the necessary tools for offline physics analysis. The target audience for the tutorials is early PhD students and other, more senior individuals who are new to the collaboration or who would benefit from learning the details of software analysis. Tutorial events are held three to four times per year, on average. Initially, the event consisted of 2.5 days of material, and it has since expanded to have four to five days of material. Since its inception, the curriculum evolved to be increasingly pedagogical with a focus on practical applications for physics analysis. These general tutorials, along with several subject-specific tutorials, are publicly available online~\cite{AtlasSoftwareDocs}.

The educational material was designed to showcase the latest available software using examples of usage in data analysis. Until 2020, tutorial events were held in-person at CERN without a remote connection, consisting of lectures and hands-on sessions providing examples of the tools and methods introduced in the lectures. Starting in 2017, the software tutorial was coupled to Induction Day, a day-long event that introduces new members of ATLAS to critical aspects of the collaboration, its organization, and work in experimental high-energy physics. The Induction Day is held in person at CERN with a video connection for remote participants. Following the Induction Day sessions, experts are made available to assist participants in setting up their CERN and ATLAS computing accounts. The process of setting up the necessary accounts is complex and often requires the intervention of expert advice, beyond what is available from the tutorial organizers and tutors. This centralized help session makes the account setup process relatively simple for the participants. In a typical event, the Induction Day and account setup session are held on a Monday and the software training session is held during the remainder of the week. Holding the events in a single week makes it easier for participants who travel to CERN.

\subsubsection{Tutorials during COVID-19}

Due to the COVID-19 global pandemic and subsequent safety measures, the ATLAS software training transitioned to an asynchronous, remote format. The in-person lectures were replaced by pre-recorded lectures that participants could watch at their convenience. The recorded lectures were optimised for a remote audience with the lecturer speaking to the camera with slides. Closed captioning was professionally provided for the videos to improve accessibility. An approximately hour-long live question-and-answer session was scheduled each day during which participants could interact with experts for a discussion of any questions that arose while watching the recorded lectures. The question-and-answer session was scheduled to maximize accessibility for registered participants around the world. Throughout each day, tutors were available on a Discord server to provide in-time assistance for the hands-on exercises. It was found that Discord provides an optimal experience allowing group discussions as well as text- and voice-based breakout rooms for one-on-one help. This format was highly successful in making the training resources (material, exercises, and access to experts) available to participants around the world.

In 2022 as COVID-19 safety measures were loosened, in-person events at CERN were once again allowed. In June 2022, Induction Day and software tutorials were held in a hybrid format to provide the benefits of in-person attendance at CERN and remote asynchronous accessibility for participants not at CERN. For the software training, in-person participants were provided with live lectures (typically in-person, with some being remote) and access to in-person tutors. Remote participants could either remotely listen to the live lectures or watch the recorded lectures from the online version of the tutorial and could interact with tutors on the Discord server. It was found that those attending in-person had a high level of engagement, while those attending remotely, either synchronously or asynchronously, had a significantly lower level of engagement. This is possibly due to subconscious bias of the organizers, lecturers and tutors towards focusing primarily on in-person participants. It is also noteworthy that many individuals participated remotely. This experience with a hybrid format was crucial for making future tutorials as widely accessible as possible. In 2023, the collaboration transitioned to offering separate in-person and remote options, offset by a few weeks, to try to improve the tutorial experience for both the groups.

\subsubsection{\RunThr training format}

Since Fall 2023, the software training was revamped to pedagogically demonstrate the use of ATLAS software in the major steps of a physics analysis workflow. The format closely follows the full \RunTwo same-flavour dilepton scalar leptoquark (LQ) analysis~\cite{EXOT-2019-13}, from MC simulation production to statistical analysis and setting limits. To allow participants to experience the numerous steps in the analysis in a condensed period of four days, the procedure is significantly simplified and time-consuming computing steps are bypassed such that complete output files are provided to students after each step. Additionally, analysis optimisation procedures, background estimation methods, and systematic uncertainty evaluation steps are minimized to provide examples of each without being as rigorous and time consuming as they would be for a published analysis. The tutorial is designed assuming familiarity of participants with C++ and \textsc{Python}.

The tutorial begins with MC simulation (see also Section~\ref{sec:evgen}). Participants generate LQ MC simulation samples using leading-order \MADGRAPH~\cite{Alwall:2014hca} with \PYTHIA~\cite{Sjostrand:2014zea} showering with inclusive decays. Generator filters are implemented to focus on the same-flavour dilepton final state and analysis code is provided to validate the simulated kinematics and decay modes. To minimize computation time, participants produce only a few hundred events. Next, participants are given information about the procedure to request centrally produced MC simulation samples and how to search for and access available samples.

The next step in the tutorial focuses on producing simple \textsc{ROOT} ntuples from derivations (see Section~\ref{sec:derivationintro}).
Signal samples with $\mathcal{O}$(10,000) events are provided in a DAOD format. First, the infrastructure to read DAODs and write information to ntuples is presented, with the option of using either |EventLoop| or Athena/AthAnalysis. Next, CP Algorithms are introduced with the example of using the good runs list (see Section~\ref{sec:DQ}) and pile-up reweighting. CP algorithms are then used to access, calibrate, and store physics objects (e.g.\ electrons, jets, and missing transverse energy) for event reconstruction and analysis. Finally, CP algorithms are used to access trigger decision information and to select events satisfying a set of single lepton triggers. Participants are then introduced to using batch systems and the Grid to process the ntuple production jobs.

At this point, participants are provided with a complete set of ntuples for analysis. The ntuples contain significantly more variables than those produced in the previous part of the tutorial and consist of numerous signal MC simulation samples, a complete set of background MC simulation samples, and detector data. The first exercise using the ntuples is to plot dilepton masses in data to see the $Z$-boson and $\mathrm{J/}\Psi$ peaks, which is used to constrain the normalisation of the estimated background from $Z$-boson production with jets. This is done using \textsc{NumPy}, \textsc{uproot}, \textsc{awkward array}, \textsc{coffea}, and \textsc{MatPlotLib}~\cite{Hunter:2007,MatPlotLib} through a \textsc{Jupyter}~\cite{Jupyter} notebook interface. Next, the same interface is used to compare kinematic distributions between signal and background samples. A signal significance figure of merit is used to find an optimal threshold for a single kinematic variable. Following this cut-based optimisation, an alternative approach is also presented, using Boosted Decision Trees (BDTs) to discriminate signal from background. The BDT material is also presented through a Jupyter notebook interface. While the work done is not sufficient for a published analysis, it is designed to give participants some experience with common analysis steps and available software. Although the tutorial focuses heavily on \textsc{Python}-based tools (targetting mostly, but not exclusively, younger colleagues), ROOT is still widely used in practice for analysis in the collaboration.

Following the analysis design and optimisation section, systematic uncertainties are introduced. Participants return to the ntuple production step and make use of the CP Algorithms to evaluate and save systematic variations of the physics objects. From a technical perspective, this section would be better suited before the analysis optimisation step. However, in a typical analysis development, systematic uncertainties are neglected or greatly simplified until the analysis strategy is mature and the optimisations are well-advanced. Therefore, this section ordering is chosen to reflect actual analysis workflows.

Finally, participants complete a statistical analysis and set limits on the LQ signal. \textsc{TRexFitter}~\cite{TRexFitter} is used for the fit, first using Asimov data and then with detector data. The fit is done with the reconstructed LQ mass and the BDT discriminant. Some background normalisations are allowed to float in the fit and a few systematic uncertainties are included. Finally, expected and observed upper limits on the cross section times branching ratio are set for a range of signal mass points.

A feedback survey is provided to the participants before the final session in the tutorial event.  The participants are requested to submit their responses before the end of the session. This results in a much higher response rate than circulating a survey to the participants after the event has ended. The results of the surveys are taken into account when modifying the material and format for future events.

\subsubsection{Retention rates and offsite events}

Most of the tutorial events are held at CERN for reasons of cost and simple logistics. However, due to the nature of the daily routine at CERN, these sessions often have a low retention rate. Participants generally have various meetings and other obligations while on site, and therefore do not attend all sessions. The new format in which one step follows from the next to form a complete analysis was found to strongly encourage participants to attend all sessions, resulting in an improved retention rate. Throughout the years, several training events were held at other sites, often with financial assistance for participants. This isolation from the daily life at CERN results in nearly perfect retention.

\subsubsection{Other available instructional resources}

In addition to the analysis software tutorial events, ATLAS offers a variety of other training materials and events. Self-guided tutorials are provided for \textsc{GitLab}, the Grid, Athena development, specific tools such as \textsc{Visual Studio Code} and \textsc{Docker}, and various specific aspects of ATLAS software such as tracking and flavour-tagging tools. This material is provided via internal \textsc{Twiki} pages, AtlasSoftwareDocs~\cite{AtlasSoftwareDocs}, and dedicated websites such as Ref.~\cite{VP1Website}. The AtlasSoftwareDocs pages also host numerous other resources, including training guides and instruction manuals for code review and building releases. Furthermore, numerous mailing lists are available that allow individuals to quickly contact experts for technical support. In addition to these resources for asynchronous training, ATLAS also organizes a variety of other tutorial events, in-person at CERN, at offsite locations, or remotely. These cover topics ranging from Athena development to machine learning and introductions to newly adopted collaboration-wide tools such as \textsc{GitLab}.

Several other training initiatives not specific to ATLAS but useful for high-energy physicists were developed worldwide~\cite{TrainingAndOnboarding}.
An attempt is made to avoid duplication when good learning modules are already available for specific tools (e.g. C++ and \textsc{Python}).


%
\section{Outlook highlights}
\label{sec:outlook}

The ATLAS experiment is preparing for a major upgrade during the next long shutdown, LS3, when the accelerator complex at CERN will also be upgraded. The outcome of these upgrades together are known as the high-luminosity LHC (HL-LHC). The HL-LHC will deliver 3--4 times more proton--proton collisions to ATLAS per second, reaching $\langle\mu\rangle=140$ in \RunFour and $\langle\mu\rangle=200$ in Run 5. The detector will have several parts completely replaced and some new components added to deal with this higher rate, and the readout system will be upgraded to record about 10,000 events per second, with an event size about three times larger than that of today, resulting in overall data processing requirements that are about 10 times greater than today~\cite{ATLAS-TDR-29}.

To prepare for the HL-LHC, ATLAS Software and Computing have developed a Conceptual Design Report~\cite{HLLHCCDR} and a Roadmap~\cite{Roadmap} detailing the challenges towards the HL-LHC and the milestones and deliverables required to meet those challenges. Major development is expected throughout the software and computing systems, improving concurrency, adopting modern approaches to certain simulation and reconstruction challenges, integrating new tools for event generators, machine learning, and data analysis, and re-working some parts of the processing and data handling systems to cope with an order of magnitude more data (see also Section~\ref{sec:upgrade}).

These upgrade research and development projects include the incorporation of accelerators (GPUs, FPGAs, and other types) into several different parts of the software. The online trigger system is investigating the use of hardware accelerators for a variety of purposes, and offline developments are significantly advanced in event generation, simulation, reconstruction, and data analysis. The Athena infrastructure itself already supports the use of accelerators, and limited applications including fast calorimeter simulation~\cite{FastCaloSimGPU} and topological clustering of energy in the calorimeter~\cite{ClusteringGPU} were already validated, with significant development around charged particle tracking ongoing~\cite{ACTSGPU,TrackGNN}. The outstanding question is whether the savings in time and electrical power will be sufficient to merit the investment to deploy hardware accelerators on worldwide Grid sites. A firm decision about the use of accelerators both online and offline is expected around 2025--26.

One key question is what will happen to the existing data from Runs 1, 2, and 3 during the HL-LHC era. The ATLAS Collaboration has committed to saving all the RAW data from the experiment and to releasing some of the data\footnote{The \RunOne data will not be released in this way without significant additional effort and attention. These data have not been reprocessed in almost 10 years, despite a significant effort at the beginning of \RunTwo. The challenge of ensuring that all of the \RunTwo software was able to support the different geometry, conditions, and configuration of the \RunOne detector proved too onerous a task.} to the public after a period of embargo: 25\% of the data will be released five years after each Run, 50\% will be released 10 years after each run, and all the data will be released by the end of the lifetime of the collaboration. These public releases of data will be in the \texttt{DAOD\_PHYSLITE} format described in Section~\ref{sec:derivations}, and will be accompanied by a set of MC simulation datasets sufficient to complete real data analysis. Other, smaller datasets will be regularly released for specific purposes. Several were already released on the CERN Open Data Portal~\cite{CODP} for use in educational settings and for exploration of machine learning techniques.

An effort is now underway to understand how the collaboration should treat the Runs 1, 2, and 3 data during the HL-LHC internally, from relying on the open datasets and published data artifacts as the only connection to earlier Runs, to retaining the ability to fully reconstruct the older data in modern releases. The latter path presents significant challenges: not only must the software be able to reproduce older geometries and configurations, but conditions data must be brought forward into new database infrastructure, calibrations must be provided for old runs, the issue of how to run a static piece of trigger software (that which was run during the corresponding data-taking period) on top of new MC simulation must be addressed, and so on. The development is a major undertaking, almost equivalent to supporting two experiments simultaneously.


%
\FloatBarrier
\section{Summary}
\label{sec:summary}

The ATLAS experiment is supported by a complex software and computing system that provides extensive functionality, flexibility, and performance in support of about 100 published data analyses every year. Many of these systems were re-developed in recent years, following 25 years of experience in the experiment and a variety of advances in modern computing, including multithreading, database systems, and open-source software solutions. This paper describes the modern production software at the beginning of \RunThr of the LHC.

The computing infrastructure of the experiment is broad, flexible, and highly distributed, with a hub at CERN where the real detector data are first processed and examined.
The simulated and real data proceed through a series of well-defined transformation steps, where the simulation closely mirrors the behaviour of the detector, and all processing is built on the same core infrastructure.
The core software of the experiment has evolved considerably over the years, now supporting multithreading and sporting an entirely new, more standard and more maintainable configuration system.
Robust support for machine learning techniques was built into the system, and the data model allows extreme flexibility in outputs.

The simulated and detector data processing steps integrate gold-standard tools for event generation and detailed detector simulation.
At the same time, many bespoke steps like fast simulation, digitization, and reconstruction have developed complexity to deal with new detectors and conditions, to provide greater veracity with limited computing resources, and to make the best possible use of new software technologies like machine learning.
The first common analysis step, derivation making, in which analysis teams have direct influence over the processing of the data is rapidly evolving in preparation for the HL-LHC.
All these processing steps are integrated into a unified workflow system, with the flexibility to optionally include forward detector systems, to change the detector layout to account for upgrades, or to reconfigure the reconstruction for different operational conditions.

Significant validation is done in preparation of new productions of MC simulation or reprocessings of detector data, with monitoring and staged productions ensuring minimal waste in the case an issue is identified.
Extensive infrastructure has also been built-up to support developers and users, from flexible builds, to nightly builds and testing, to scripts supporting consistent environments that ensure work is relocatable.
Databases and metadata systems ensure that configuration, condition, and result histories are preserved and can be used both for monitoring and to automate some parts of subsequent software configurations and productions.

The simulated and real detector data, as well as the jobs processing the data, are spread over many distributed computing sites in a system that is sufficiently flexible to integrate many complex new resources.
Automation and industry-standard tools were integrated to assist in and reduce operation efforts.
Downstream, a suite of software was developed for user analysis that ensures robustness, useability, and flexibility to maximize the physics output of the experiment.
New users are continuously introduced to the system and trained on the use of modern data analysis tools in processing the complex detector data.

Development within software and computing is continuing in preparation for the HL-LHC, and a 10-year plan was laid out in some detail. These plans provide for further modernization of some systems, explore the incorporation of new tools like more advanced machine learning techniques and hardware accelerators, and reinforce the infrastructure that will be ready to deal with an order-of-magnitude increase in data volume. Even as a mature experiment, there is constant, healthy pressure for improvements and optimisations, pushing ATLAS to meet the coming challenges.

The current schedule foresees the ATLAS experiment and the HL-LHC running through at least 2041. By that time, it is likely that the software and computing landscape will have changed significantly. With sustainable, robust, performant, and adaptable infrastructure, the experiment will be able to continue to deliver high-quality physics analyses for the decades to come.


\section*{Acknowledgements}

%
%

%
%

We thank CERN for the very successful operation of the LHC and its injectors, as well as the support staff at
CERN and at our institutions worldwide without whom ATLAS could not be operated efficiently.

The crucial computing support from all WLCG partners is acknowledged gratefully, in particular from CERN, the ATLAS Tier-1 facilities at TRIUMF/SFU (Canada), NDGF (Denmark, Norway, Sweden), CC-IN2P3 (France), KIT/GridKA (Germany), INFN-CNAF (Italy), NL-T1 (Netherlands), PIC (Spain), RAL (UK) and BNL (USA), the Tier-2 facilities worldwide and large non-WLCG resource providers. Major contributors of computing resources are listed in Ref.~\cite{ATL-SOFT-PUB-2023-001}.

We gratefully acknowledge the support of ANPCyT, Argentina; YerPhI, Armenia; ARC, Australia; BMWFW and FWF, Austria; ANAS, Azerbaijan; CNPq and FAPESP, Brazil; NSERC, NRC and CFI, Canada; CERN; ANID, Chile; CAS, MOST and NSFC, China; Minciencias, Colombia; MEYS CR, Czech Republic; DNRF and DNSRC, Denmark; IN2P3-CNRS and CEA-DRF/IRFU, France; SRNSFG, Georgia; BMBF, HGF and MPG, Germany; GSRI, Greece; RGC and Hong Kong SAR, China; ISF and Benoziyo Center, Israel; INFN, Italy; MEXT and JSPS, Japan; CNRST, Morocco; NWO, Netherlands; RCN, Norway; MNiSW, Poland; FCT, Portugal; MNE/IFA, Romania; MESTD, Serbia; MSSR, Slovakia; ARIS and MVZI, Slovenia; DSI/NRF, South Africa; MICIU/AEI, Spain; SRC and Wallenberg Foundation, Sweden; SERI, SNSF and Cantons of Bern and Geneva, Switzerland; NSTC, Taipei; TENMAK, T\"urkiye; STFC/UKRI, United Kingdom; DOE and NSF, United States of America.

Individual groups and members have received support from BCKDF, CANARIE, CRC and DRAC, Canada; PRIMUS 21/SCI/017, CERN-CZ and FORTE, Czech Republic; COST, ERC, ERDF, Horizon 2020, ICSC-NextGenerationEU and Marie Sk{\l}odowska-Curie Actions, European Union; Investissements d'Avenir Labex, Investissements d'Avenir Idex and ANR, France; DFG and AvH Foundation, Germany; Herakleitos, Thales and Aristeia programmes co-financed by EU-ESF and the Greek NSRF, Greece; BSF-NSF and MINERVA, Israel; Norwegian Financial Mechanism 2014-2021, Norway; NCN and NAWA, Poland; La Caixa Banking Foundation, CERCA Programme Generalitat de Catalunya and PROMETEO and GenT Programmes Generalitat Valenciana, Spain; G\"{o}ran Gustafssons Stiftelse, Sweden; The Royal Society and Leverhulme Trust, United Kingdom.

In addition, individual members wish to acknowledge support from CERN: European Organization for Nuclear Research (CERN PJAS); Chile: Agencia Nacional de Investigaci\'on y Desarrollo (FONDECYT 1190886, FONDECYT 1210400, FONDECYT 1230812, FONDECYT 1230987); China: Chinese Ministry of Science and Technology (MOST-2023YFA1605700, MOST-2023YFA1609300), National Natural Science Foundation of China (NSFC - 12175119, NSFC 12275265, NSFC-12075060); Czech Republic: Czech Science Foundation (GACR - 24-11373S), Ministry of Education Youth and Sports (FORTE CZ.02.01.01/00/22\_008/0004632), PRIMUS Research Programme (PRIMUS/21/SCI/017); EU: H2020 European Research Council (ERC - 101002463); European Union: European Research Council (ERC - 948254, ERC 101089007), Horizon 2020 Framework Programme (MUCCA - CHIST-ERA-19-XAI-00), European Union, Future Artificial Intelligence Research (FAIR-NextGenerationEU PE00000013), Italian Center for High Performance Computing, Big Data and Quantum Computing (ICSC, NextGenerationEU); France: Agence Nationale de la Recherche (ANR-20-CE31-0013, ANR-21-CE31-0013, ANR-21-CE31-0022, ANR-22-EDIR-0002), Investissements d'Avenir Labex (ANR-11-LABX-0012); Germany: Baden-Württemberg Stiftung (BW Stiftung-Postdoc Eliteprogramme), Deutsche Forschungsgemeinschaft (DFG - 469666862, DFG - CR 312/5-2); Italy: Istituto Nazionale di Fisica Nucleare (ICSC, NextGenerationEU), Ministero dell'Università e della Ricerca (PRIN - 20223N7F8K - PNRR M4.C2.1.1); Japan: Japan Society for the Promotion of Science (JSPS KAKENHI JP21H05085, JSPS KAKENHI JP22H01227, JSPS KAKENHI JP22H04944, JSPS KAKENHI JP22KK0227); Netherlands: Netherlands Organisation for Scientific Research (NWO Veni 2020 - VI.Veni.202.179); Norway: Research Council of Norway (RCN-314472); Poland: Polish National Agency for Academic Exchange (PPN/PPO/2020/1/00002/U/00001), Polish National Science Centre (NCN 2021/42/E/ST2/00350, NCN OPUS nr 2022/47/B/ST2/03059, NCN UMO-2019/34/E/ST2/00393, NCN \& H2020 MSCA 945339, UMO-2020/37/B/ST2/01043, UMO-2021/40/C/ST2/00187, UMO-2022/47/O/ST2/00148); Slovenia: Slovenian Research Agency (ARIS grant J1-3010); Spain: Generalitat Valenciana (Artemisa, FEDER, IDIFEDER/2018/048), Ministry of Science and Innovation (MCIN \& NextGenEU PCI2022-135018-2, MICIN \& FEDER PID2021-125273NB, RYC2019-028510-I, RYC2020-030254-I, RYC2021-031273-I, RYC2022-038164-I), PROMETEO and GenT Programmes Generalitat Valenciana (CIDEGENT/2019/023, CIDEGENT/2019/027); Sweden: Swedish Research Council (Swedish Research Council 2023-04654, VR 2018-00482, VR 2022-03845, VR 2022-04683, VR 2023-03403, VR grant 2021-03651), Knut and Alice Wallenberg Foundation (KAW 2018.0157, KAW 2018.0458, KAW 2019.0447, KAW 2022.0358); Switzerland: Swiss National Science Foundation (SNSF - PCEFP2\_194658); United Kingdom: Leverhulme Trust (Leverhulme Trust RPG-2020-004), Royal Society (NIF-R1-231091); United States of America: U.S. Department of Energy (ECA DE-AC02-76SF00515), Neubauer Family Foundation.

%
%


%
%
%
\printbibliography
\clearpage

\begin{flushleft} 
\hypersetup{urlcolor=black} 
{\Large The ATLAS Collaboration}

\bigskip

\AtlasOrcid[0000-0002-6665-4934]{G.~Aad}$^\textrm{\scriptsize 104}$,
\AtlasOrcid[0000-0001-7616-1554]{E.~Aakvaag}$^\textrm{\scriptsize 16}$,
\AtlasOrcid[0000-0002-5888-2734]{B.~Abbott}$^\textrm{\scriptsize 122}$,
\AtlasOrcid[0000-0002-1002-1652]{K.~Abeling}$^\textrm{\scriptsize 56}$,
\AtlasOrcid[0000-0001-5763-2760]{N.J.~Abicht}$^\textrm{\scriptsize 49}$,
\AtlasOrcid[0000-0002-8496-9294]{S.H.~Abidi}$^\textrm{\scriptsize 29}$,
\AtlasOrcid[0009-0003-6578-220X]{M.~Aboelela}$^\textrm{\scriptsize 44}$,
\AtlasOrcid[0000-0002-9987-2292]{A.~Aboulhorma}$^\textrm{\scriptsize 35e}$,
\AtlasOrcid[0000-0001-5329-6640]{H.~Abramowicz}$^\textrm{\scriptsize 153}$,
\AtlasOrcid[0000-0002-1599-2896]{H.~Abreu}$^\textrm{\scriptsize 152}$
% ...
\bigskip
\\

% ...
$^{16}$Department for Physics and Technology, University of Bergen, Bergen; Norway.\\
% ...
$^{29}$Physics Department, Brookhaven National Laboratory, Upton NY; United States of America.\\
% ...
$^{35}$$^{(a)}$Facult\'e des Sciences Ain Chock, R\'eseau Universitaire de Physique des Hautes Energies - Universit\'e Hassan II, Casablanca;$^{(b)}$Facult\'{e} des Sciences, Universit\'{e} Ibn-Tofail, K\'{e}nitra;$^{(c)}$Facult\'e des Sciences Semlalia, Universit\'e Cadi Ayyad, LPHEA-Marrakech;$^{(d)}$LPMR, Facult\'e des Sciences, Universit\'e Mohamed Premier, Oujda;$^{(e)}$Facult\'e des sciences, Universit\'e Mohammed V, Rabat;$^{(f)}$Institute of Applied Physics, Mohammed VI Polytechnic University, Ben Guerir; Morocco.\\
% ...
$^{44}$Physics Department, Southern Methodist University, Dallas TX; United States of America.\\
% ...
$^{49}$Fakult\"{a}t Physik , Technische Universit{\"a}t Dortmund, Dortmund; Germany.\\
% ...
$^{56}$II. Physikalisches Institut, Georg-August-Universit\"{a}t G\"ottingen, G\"ottingen; Germany.\\
% ...
$^{104}$CPPM, Aix-Marseille Universit\'e, CNRS/IN2P3, Marseille; France.\\
% ...
$^{122}$Homer L. Dodge Department of Physics and Astronomy, University of Oklahoma, Norman OK; United States of America.\\
% ...
$^{152}$Department of Physics, Technion, Israel Institute of Technology, Haifa; Israel.\\
$^{153}$Raymond and Beverly Sackler School of Physics and Astronomy, Tel Aviv University, Tel Aviv; Israel.\\
% ...
$^{a}$ Also Affiliated with an institute covered by a cooperation agreement with CERN.\\
$^{b}$ Also at An-Najah National University, Nablus; Palestine.\\
$^{c}$ Also at Borough of Manhattan Community College, City University of New York, New York NY; United States of America.\\
$^{d}$ Also at Center for Interdisciplinary Research and Innovation (CIRI-AUTH), Thessaloniki; Greece.\\
$^{e}$ Also at Centro Studi e Ricerche Enrico Fermi; Italy.\\
$^{f}$ Also at CERN, Geneva; Switzerland.\\
% ...

\end{flushleft}


%

%
%
\clearpage
\appendix

\end{document}